\documentclass[english,aps,prd,showpacs,superscriptaddress,nofootinbib]{revtex4}
\usepackage{amsmath,graphicx,epsfig,amssymb,dsfont,mathtools,}
\usepackage[usenames]{color}
\usepackage{ulem} 
\usepackage{bigstrut}
\usepackage{slashed}
\usepackage{multirow}
\usepackage{makecell}
\usepackage{verbatim}
\newcommand {\bseq}{\begin{subequations}}
\newcommand {\eseq}{\end{subequations}}
\usepackage{amsmath,graphicx,color,epsfig}
  \allowdisplaybreaks[1]

\allowdisplaybreaks

\begin{document}
\title{Weak Decays of   Stable Doubly Heavy Tetraquark States}

\author{
Ye Xing~\footnote{Email:xingye\_guang@sjtu.edu.cn}
}
\affiliation{
 INPAC,  Shanghai Key Laboratory for Particle Physics and Cosmology, MOE Key Laboratory for Particle Physics, School of Physics and Astronomy, \\ Shanghai Jiao Tong University, Shanghai  200240, China}
 \author{
Ruilin Zhu~\footnote{Email:rlzhu@njnu.edu.cn}
}
\affiliation{
Department of Physics and Institute of Theoretical Physics,
Nanjing Normal University, Nanjing, Jiangsu 210023, China
 }
\begin{abstract}
In the light of the recent discovery of the $\Xi^{++}_{cc}$ by the  LHCb collaboration, we study the stable doubly heavy tetraquarks.
These states  are compact exotic hadrons which can be approximated as   the diquark-anti-diquark correlations. In the flavor SU(3) symmetry, they form a SU(3) triplet or anti-sextet. The spectra of the stable doubly heavy tetraquark states are predicted by Sakharov-Zeldovich formula. We find that the $T^{-}_{bb\bar{u}\bar{d}}({\bf  3})$ is about 73MeV below the $BB^*$ threshold at SET I. We then study the semileptonic and nonleptonic weak decays of the stable doubly heavy tetraquark states.  The   doubly heavy tetraquark  decay amplitudes are parametrized in terms of flavor SU(3)-irreducible parts.  Ratios between decay widths of different channels are  also derived.   At the end, we collect  the Cabibbo allowed two-body and three-body decay channels, which are most promising  to search for the stable doubly heavy tetraquark states at LHCb and Belle II experiments.
\end{abstract}
\maketitle

\section{Introduction}
Up to now most hadrons found by experiments can be well established as quark-antiquark pair and triquarks configurations~\cite{Patrignani:2016xqp}.
Based on the principle of color confining, the multiquark color singlet states such as $qq\bar{q}\bar{q}$ (tetraquarks) and $qqqq\bar{q}$ (pentaquarks) can also exist.
On the  experimental aspect, many multiquark candidates have been observed even though their physical figures are still not established.
The most aged of these exotic resonances is the neutral $X(3872)$ discovered in $B^\pm\to K^\pm X (X\to J/\psi\pi^+\pi^-)$ by Belle in 2003~\cite{Choi:2003ue}. Four years later, the Belle
Collaboration observed  a charged hidden charm  tetraquark candidate, i.e. $Z^+(4430)$~\cite{Choi:2007wga}.
In 2013, the BESIII Collaboration discovered $Z_c^+(3900)$ through the channel $Y(4260) \to \pi^-\pi^+J/\psi$ ~\cite{Ablikim:2013mio}, which directly hadronic decays into $J/\psi\pi^+$, and then implies that it shall be  a  meson with quark contents $c\bar{c}u\bar{d}$.
In 2015, the LHCb Collaboration discovered that two exotic baryons $P_c(4380)$ and $P_c(4450)$ hadronic decaying into $J/\psi p$, which
are candidates for pentaquark states and shall be  a  baryon with quark contents $c\bar{c}uu\bar{d}$~\cite{Aaij:2015tga}.

The LHCb Collaboration have recently observed the doubly charmed baryon $\Xi_{cc}^{++}$ in the $\Lambda_c^+K^-\pi^+\pi^+$ invariant mass spectrum, whose mass is measured to be $3621.40\pm0.72(stat.)\pm 0.27(syst.)\pm 0.14(\Lambda_c^+)$ MeV~\cite{Aaij:2017ueg}. This discovery has   attracted wide attentions from
both the theoretical and experimental sides in high energy physics. From the diquark-based model, the doubly heavy  quarks can provide a static color source as the attractive diquark in the color ${\bf \bar{3}}$ representation.  The attractive heavy diquark and the light quark in the color ${\bf 3}$ representation then form a color singlet hadron.  Thus it is natural to conceive the  doubly heavy tetraquark states with attractive heavy diquark and attractive light diquark. From the basic principles of QCD,
the long-distance interactions among light quarks and gluons has a characteristic scale of the order of 300MeV.
 When the two heavy quarks attract each other and their  separation is smaller enough than the separation to the light quark,
 the two heavy quarks interact with a perturbative one-gluon-exchange Coulomb-like potential. When the two heavy quarks have a
 large separation, the four quarks will form two weakly interacting mesons. It is an important issue to be discussed about whether
 or not the stable   doubly heavy tetraquark states exist. When they steadily exist, it is another important issue on how to detect them.

On the theoretical aspect, the mass spectra of the doubly heavy tetraquark states have been studied in many literatures~\cite{Ader:1981db,Ebert:2007rn,Karliner:2017qjm,Eichten:2017ffp,Francis:2016hui,Bicudo:2016ooe,Mehen:2017nrh,
Du:2012wp,Esposito:2013fma,Chen:2013aba,Guerrieri:2014nxa,Cheung:2017tnt,Wang:2017dtg,Yao:2018zze,Yan:2018gik,Ali:2018ifm}.
Most of them supported the existence of the doubly heavy tetraquark states, however, they predicted differently the  spectra of the doubly heavy tetraquark states in these works. The structures of the doubly heavy tetraquark states were also different in their descriptions.
Unlike the $Qq\bar{Q}\bar{q}$ system which can be classified into four kinds of four quark structures~\cite{Peters:2017hon}, the structures of the $QQ\bar{q}\bar{q}$ system are  relatively simple.
Take the $bb\bar{q}\bar{q}$ for example, the four quark structures may be classified into two groups: one is treated as a bound  state made of a loosely bound $BB$ meson pair or two far separated and essentially  weak interacting $B$ mesons;
the other one is treated as a bound state made of a heavy diquark with color anti-triplet and  a light anti-diquark with color triplet.

Theoretical description of doubly heavy tetraquark states decays is few in current studies. Whether or not the QCD factorization is valid in the doubly heavy tetraquark states decays is an open question. An alternative and model-independent approach is to employ the flavor SU(3) symmetry, which has been successfully applied into the $B$ meson  and the heavy baryon decays~\cite{Savage:1989ub,Gronau:1995hm,He:1998rq,He:2000ys,Chiang:2004nm,Li:2007bh,Wang:2009azc,Cheng:2011qh,
Hsiao:2015iiu,Lu:2016ogy,He:2016xvd,Wang:2017mqp,Wang:2017azm,Hu:2017dzi,Shi:2017dto,Wang:2018utj,He:2018php,Zhu:2018epc,Agaev:2018vag}. In this paper, we will investigate the amplitudes and decay widths of doubly heavy tetraquark states under the  flavor SU(3) symmetry.

The paper will be presented as follows. In Sec.~\ref{sec:particle_multiplet}, we classify the doubly heavy tetraquark states  into  an SU(3) triplet or anti-sextet according to the decomposition of $\bar{3}\otimes \bar{3}=3\oplus \bar{6}$. Other related baryons and mesons are also listed in SU(3) flavor
symmetry.
 In Sec.~\ref{sec:spectra},  we give the spectra of the doubly heavy tetraquark states. Their stability properties are essential for the discovery and will be discussed. In Sec.~\ref{sec:semileptonic} and ~\ref{sec:bbq_nonleptonic}, we mainly study
 the semileptonic and nonleptonic  weak decays of the stable doubly heavy tetraquarks. The decay amplitudes are explored with the SU(3) flavor
symmetry. The ratios between the decay widthes of  different decay channels  are predicted.  We summarize and conclude in the end.

\section{Particle Multiplets}
\label{sec:particle_multiplet}
Following the flavor SU(3) group,  the doubly heavy tetraquark states and their decay products can be grouped into the particle multiplets.

In principle, doubly heavy tetraquark states with the $QQ\bar{q}\bar{q}$ are similar to the $\bar{Q}\bar{Q}qq$. We are focusing on the $QQ\bar{q}\bar{q}$ states for simplication. The doubly heavy tetraquark ($QQ\bar{q}\bar{q}$) can form a SU(3) triplet or anti-sextet by the decomposition of $\bar{3}\otimes \bar{3}=3\oplus \bar{6}$.
The triplet has the expression
\begin{small}
\begin{eqnarray}
T_{cc3}= \left(\begin{array}{ccc} 0 & T_{cc\bar{u}\bar{d}}^{+}  &  T_{cc\bar{u}\bar{s}}^{+}  \\ -T_{cc\bar{u}\bar{d}}^+  &  0 &  T_{cc\bar{d}\bar{s}}^{++} \\ - T_{cc\bar{u}\bar{s}}^{+}   & -T_{cc\bar{d}\bar{s}}^{++}  & 0
  \end{array} \right),
T_{bc3}= \left(\begin{array}{ccc} 0 & T_{bc\bar{u}\bar{d}}^{+}  &  T_{bc\bar{u}\bar{s}}^{+}  \\ -T_{bc\bar{u}\bar{d}}^+  &  0 &  T_{bc\bar{d}\bar{s}}^{++} \\ - T_{bc\bar{u}\bar{s}}^{+}   & -T_{bc\bar{d}\bar{s}}^{++}  & 0
  \end{array} \right),
T_{bb3}= \left(\begin{array}{ccc} 0 & T_{bb\bar{u}\bar{d}}^{+}  &  T_{bb\bar{u}\bar{s}}^{+}  \\ -T_{bb\bar{u}\bar{d}}^+  &  0 &  T_{bb\bar{d}\bar{s}}^{++} \\ - T_{bb\bar{u}\bar{s}}^{+}   & -T_{bb\bar{d}\bar{s}}^{++}  & 0
  \end{array} \right).
\end{eqnarray}
\end{small}
While the doubly heavy tetraquark in anti-sextet can usually strong decay into the triplets and  are not stable, then we will not consider them here.

When we study the  weak decays of the doubly heavy tetraquarks under the flavor SU(3) symmetry, we should classify the products. The charmed  bottom baryons can form a SU(3) triplet with $F_{bc} = ( \Xi^+_{bc}(bcu) , \Xi^0_{bc}(bcd), \Omega^0_{bc}(bcs))$.
The charmed anti-baryons are classified into a triplet and an anti-sextet
\begin{eqnarray}
 F_{\bf{\bar{c} 3}}= \left(\begin{array}{ccc} 0 & \Lambda_{\bar{c}}^-  &  \Xi_{\bar{c}}^-  \\ -\Lambda_{\bar{c}}^- & 0 & \overline \Xi_{\bar{c}}^0 \\ -\Xi_{\bar{c}}^-   &  -\overline \Xi_{\bar{c}}^0  & 0
  \end{array} \right), \;\;
 F_{\bf{\bar{c}\bar{6}}} = \left(\begin{array}{ccc} \Sigma_{\bar{c}}^{--} &  \frac{1}{\sqrt{2}}\Sigma_{\bar{c}}^-   & \frac{1}{\sqrt{2}} \Xi_{\bar{c}}^{\prime-}\\
  \frac{1}{\sqrt{2}}\Sigma_{\bar{c}}^-& \overline \Sigma_{\bar{c}}^{0} & \frac{1}{\sqrt{2}} \overline \Xi_{\bar{c}}^{\prime0} \\
  \frac{1}{\sqrt{2}} \Xi_{\bar{c}}^{\prime-}   &  \frac{1}{\sqrt{2}} \overline \Xi_{\bar{c}}^{\prime0}  & \overline \Omega_{\bar{c}}^0
  \end{array} \right)\,.
\end{eqnarray}
Similarly, the singly charmed baryons $F_{\bf{c\bar 3}}$ and $ F_{\bf{c6}}$ can be described in the same way, whose explicit expressions can be
found in Refs.~\cite{Wang:2017azm,Wang:2018utj}.

\begin{figure}
\begin{center}
\includegraphics[scale=0.6]{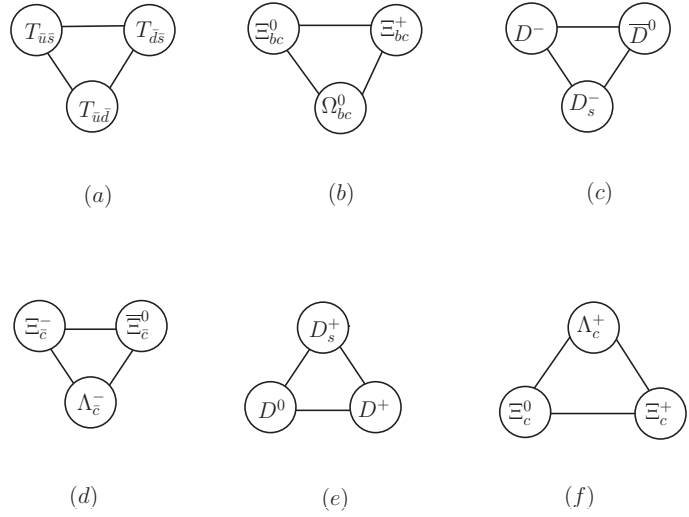}
\end{center}
\caption{Flavor SU(3) weight diagrams for the triplet and anti-triplet hadrons. }\label{fig:particle multiplet-1}
\end{figure}
\begin{figure}
\begin{center}
\includegraphics[scale=0.6]{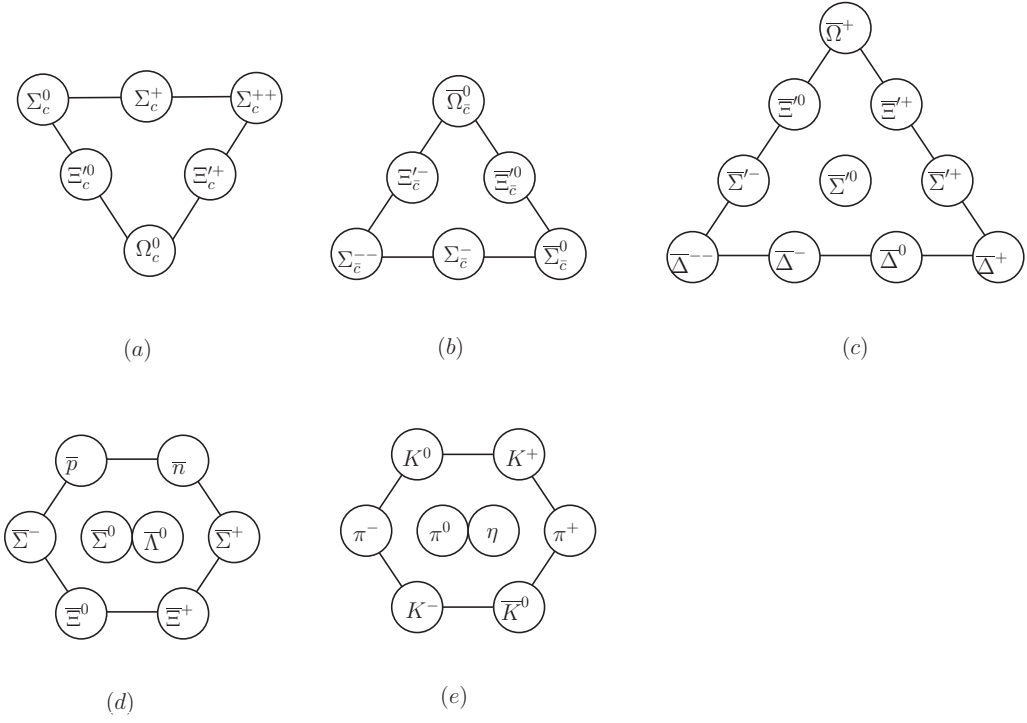}
\end{center}
\caption{Flavor SU(3) weight diagrams for the sextet, anti-sextet, anti-decuplet, anti-octet and octet hadrons.  }
\label{fig:particle multiplet-2}
\end{figure}
The anti-baryons with light quarks can be classified into an octet and an anti-decuplet. We write the octet as
\begin{eqnarray}
F_8= \left(\begin{array}{ccc} \frac{1}{\sqrt{2}}\overline \Sigma^0+\frac{1}{\sqrt{6}}\overline \Lambda^0 & \overline \Sigma^+  &  \overline \Xi^+   \\ \overline \Sigma^-  &  -\frac{1}{\sqrt{2}}\overline \Sigma^0+\frac{1}{\sqrt{6}}\overline \Lambda^0 & \overline \Xi^0 \\ \overline p   & \overline n  & -\sqrt{\frac{2}{3}}\overline \Lambda^0
  \end{array} \right),
\end{eqnarray}
while the anti-decuplet can be written as~\cite{Wang:2017mqp,Wang:2017azm}
\begin{eqnarray}
(F_{\overline {10}})^{111} &=&  \overline \Delta^{--},\;\;\; (F_{\overline {10}})^{112}= (F_{\overline {10}})^{121}=(F_{\overline {10}})^{211}= \frac{1}{\sqrt3}  \overline \Delta^{-},\nonumber\\
(F_{\overline {10}})^{222} &=&  \overline \Delta^{+},\;\;\; (F_{\overline {10}})^{122}= (F_{\overline {10}})^{212}=(F_{\overline {10}})^{221}= \frac{1}{\sqrt3} \overline \Delta^0, \nonumber\\
(F_{\overline {10}})^{113} &=& (F_{\overline {10}})^{131}=(F_{\overline {10}})^{311}= \frac{1}{\sqrt3} \overline \Sigma^{\prime-},\;\;(F_{\overline {10}})^{223} = (F_{\overline {10}})^{232}=(F_{\overline {10}})^{322}= \frac{1}{\sqrt3} \overline \Sigma^{\prime+},\nonumber\\
(F_{\overline {10}})^{123} &=& (F_{\overline {10}})^{132}=(F_{\overline {10}})^{213}=(F_{\overline {10}})^{231}=(F_{\overline {10}})^{312}=(F_{\overline {10}})^{321}= \frac{1}{\sqrt6} \overline \Sigma^{\prime0},\nonumber\\
(F_{\overline {10}})^{133} &=& (F_{\overline {10}})^{313}=(F_{\overline {10}})^{331}= \frac{1}{\sqrt3} \overline \Xi^{\prime0},\;\;(F_{\overline {10}})^{233} = (F_{\overline {10}})^{323}=(F_{\overline {10}})^{332}= \frac{1}{\sqrt3}  \overline \Xi^{\prime+}, \nonumber\\
(F_{\overline {10}})^{333}&=& \overline \Omega^+.
\end{eqnarray}

The light pseudo-scalar mesons belong to an octet and a singlet. The explicit expression of the octet can be found in Refs.~\cite{Wang:2017azm,Wang:2018utj}.
The flavor singlet $\eta_1$ will not be considered here for simplicity.
For the heavy flavor mesons, we can write them as
$B_i=\left( B^-,  \overline B^0, \overline B^0_s  \right)$ and $B^i=\left(  B^+,  B^0, B^0_s  \right)$.
Similarly, the charmed mesons can be written as
$D_i=\left( D^0,  D^+,  D^+_s   \right)$ and $D^i=\left(\overline D^0, D^-,  D^-_s  \right)$. To a summary of the particle multiplets,
we plot the flavor SU(3) weight diagrams for hadrons with different representations in Fig.~\ref{fig:particle multiplet-1} and Fig.~\ref{fig:particle multiplet-2}.

\section{Spectra of the doubly heavy tetraquarks}
\label{sec:spectra}

The wave function of a tetraquark  consists of four parts: space-coordinate,  flavor, color, and spin subspaces,
 \begin{eqnarray}
\Psi(Q,Q',\bar{q},\bar{q}')&=&R(\bold{x}_1,\bold{x}_2,\bold{x}_3,\bold{x}_4)\otimes \chi_f(f_1,f_2,f_3,f_4)
\nonumber\\&&\otimes \chi_\lambda(\lambda_1,\lambda_2,\lambda_3,\lambda_4)\otimes \chi_s(s_1,s_2,s_3,s_4)\,,
\end{eqnarray}
where $R(\bold{x}_i)$, $\chi_f(f_i)$,  $\chi_\lambda(\lambda_i)$, and $\chi_s(s_i)$ denote the radial, flavor,  color, and spin wave functions, respectively. The sub-labels 1, 2, 3, 4 in the above equation denote $Q$, $Q'$, $\bar{q}$, $\bar{q'}$, respectively.

For the two quark system, there are eight distinct diquark multiplets in flavor $\otimes$ color $\otimes$ spin space. According to the Pauli exclusion principle, the diquark-antidiquark configuration $ [QQ'][\bar{q}\bar{q'}]$ of doubly heavy tetraquark state only has four possible topologies~\footnote{ One should note that the two quarks and two antiquarks can be encoded by three different ways of color basis. One of them is
the color basis of $\{|\bar {\bf {3}}_{12},{\bf 3}_{34}\rangle,~|{\bf 6}_{12},\bar {\bf {6}}_{34}\rangle\}$, and the other two are
 $\{| {\bf {1}}_{13},{\bf 1}_{24}\rangle,~|{\bf 8}_{13}, {\bf {8}}_{24}\rangle\}$ and  $\{| {\bf {1}}_{14},{\bf 1}_{23}\rangle,~|{\bf 8}_{14}, {\bf {8}}_{23}\rangle\}$ respectively. Through decomposition, the singlet-singlet and octet-octet basis can be described by the triplet-antitriplet and antisextet-sextet base.}
\bseq
 \begin{eqnarray}
|{\bf 1}_f(S), {\bf 3}_f(A) \rangle \otimes |{\bf \bar 3}_c(A) ,{\bf 3}_c(A) \rangle  \otimes |{1}_s(S), {0}_s (A)\rangle\,,&&\label{dq3a}\\
|{\bf 1}_f(S), {\bf 3}_f (A)\rangle \otimes |{\bf 6}_c(S) ,{\bf \bar 6}_c(S) \rangle  \otimes |{0}_s(A), {1}_s (S)\rangle\,,&&\label{dq6}\\
|{\bf 1}_f(S), {\bf \bar 6}_f (S)\rangle \otimes |{\bf \bar 3}_c(A) ,{\bf 3}_c(A) \rangle  \otimes |{1}_s(S), {1}_s (S)\rangle\,,&&\label{dq3b}\\
|{\bf 1}_f(S), {\bf \bar 6}_f (S)\rangle \otimes |{\bf 6}_c(S) ,{\bf \bar 6}_c(S)  \rangle  \otimes |{0}_s(A), {0}_s(A) \rangle\,,&&\label{dq6b}
\end{eqnarray}
\eseq
where the sub-label $f,~c,~s$ denote the flavor, color, spin spaces, respectively. $S$ and $A$ denote the symmetric and antisymmetric  properties. Each half bracket denotes the diquark configuration. For example, $|{\bf 1}_f(S), {\bf 3}_f(A) \rangle$ denotes the diquark $[QQ']$ is the singlet
in flavor space and thus is symmetric (S), while the diquark $[\bar{q}\bar{q'}]$ is the triplet in flavor space and thus is antisymmetric (A).
Consider that the color sextet diquarks have larger color electrostatic energy thus is not a well-favored configuration, and the odd parity diquark operators will vanish in the single mode configuration.   The diquarks $ |{\bf 3}_c(A)\rangle  \otimes |{0}_s (A)\rangle$ and  $|{\bf \bar 3}_c(A)  \rangle  \otimes |{1}_s(S)\rangle$ in Eqs. (\ref{dq3a}) and (\ref{dq3b}) are the ``scalar" and ``axial-vector" diquarks  respectively,  as the ``good" and ``bad" diquarks named by Jaffe~\cite{Jaffe:2004ph}.  According to Jaffe's diquark, the ``good" and ``bad" diquarks have  mass splitting.  Approximately, for the up and down diquark, the  mass difference between the ``good" and ``bad" diquarks is around $210$MeV, and for the up and strange diquark, the  mass difference becomes around $150$MeV~\cite{Jaffe:2004ph}. While for the charm and light diquark, the mass difference is around $60$MeV~\cite{Ali:2011ug}.  One can assume the diquark mass difference for the doubly heavy quark  is around $50$MeV.
Other configurations of the diquark are ``worse" diquarks.
For simplication, we will not consider these ``worse" diquarks in the prediction of the spectra.

The constituent quark models have a robust power to predict hadron spectra, especially for the $S$-wave states. In these quark models, the hadrons
are  bound states composed of the constituent quarks.  In Sakharov-Zeldovich formula, the interaction Hamiltonian through the color-spin interaction is given by~\cite{Jaffe:2004ph,DeRujula:1975qlm}
 \begin{eqnarray}
{\cal H}_{\mathrm{color~spin}}&=&\sum_{i<j} (-\frac{3}{8})\frac{C^{ij}}{m_i m_j}\vec{\lambda}_i\cdot\vec{\lambda}_j \vec{s}_i\cdot\vec{s}_j\,,\label{Mh}
\end{eqnarray}
where the overall strength can be given as $C^{ij}=v^{ij}\langle\delta(r_{ij})\rangle$ with the coupling $v^{ij}$ and the strength of the radial wave function at zero separation $\langle\delta(r_{ij})\rangle$ which is dependent on the hadron constituent quark flavors. The $\vec{\lambda}_i$ is the Gell-Mann matrix for color SU(3) group, and $\vec{s}_i=\vec{\sigma}_i/2$ is the quark spin operator  with the Pauli matrix $\vec{\sigma}_i$.  The effects of color-Coulomb potential and the color  confining potential are included in the fitted quark mass.

\begin{table}
\caption{Matrix elements $\vec{\lambda}_i\cdot\vec{\lambda}_j ~\vec{s}_i\cdot\vec{s}_j$ for two quarks in color $\bar{\bf 3}, {\bf 6}, {\bf 1},  {\bf 8}$ configurations. }\label{tab:colorm}
\begin{tabular}{|c|cccc|}
\hline
  & $\bar{\bf 3}$  & ${\bf 6}$  & ${\bf 1}$ & ${\bf 8}$
\\\hline
$\langle\vec{\lambda}_i\cdot\vec{\lambda}_j\rangle$ & $-\frac{8}{3}$  & $\frac{4}{3}$  & $-\frac{16}{3}$ & $\frac{2}{3}$
\\\hline
$\langle\vec{\lambda}_i\cdot\vec{\lambda}_j ~\vec{s}_i\cdot\vec{s}_j\rangle ~~(s=0)$ & $2$  & $-1$  & $4$ & $-\frac{1}{2}$
\\\hline
$\langle\vec{\lambda}_i\cdot\vec{\lambda}_j ~\vec{s}_i\cdot\vec{s}_j\rangle~~(s=1)$ & $-\frac{2}{3}$  & $\frac{1}{3}$  & $-\frac{4}{3}$ & $\frac{1}{6}$\\
\hline
\end{tabular}
\end{table}

The parameters in Eq.~(\ref{Mh}) can be fitted by the hadron spectra~\cite{Lee:2007tn}, which have been given in Tab.~\ref{tab:extracts}. The overall factor $C^{ij}/(m_im_j)$  can be extracted from the hadron mass differences. From Tab.~\ref{tab:extracts}, one gets the below results: $C^{qq}/m_u^2=193$MeV, $C^{sq}/(m_u m_s)=118$MeV,  $C^{cq}/(m_u m_c)=23$MeV, $C^{bq}/(m_u m_b)=2.3$MeV for the diquark configuration;
$C^{q\bar{q}}/m_u^2=318$MeV, $C^{s\bar{q}}/(m_u m_s)=199$MeV, $C^{c\bar{q}}/(m_u m_c)=69$MeV, $C^{b\bar{q}}/(m_u m_b)=23$MeV, $C^{s\bar{s}}/m_s^2=118$MeV, $C^{c\bar{s}}/(m_s m_c)=72$MeV, $C^{b\bar{s}}/(m_s m_b)=24$MeV, $C^{c\bar{c}}/m_c^2=57$MeV, $C^{b\bar{b}}/m_b^2=31$MeV for the quark-antiquark configuration. These fitting are consistent with the previous literatures~\cite{Maiani:2004vq,Ali:2011ug,Wang:2016tsi,Wang:2017vnc}.
 The effective quark masses are fitted as:  $m_{u,d}=305$MeV, $m_{s}=490$MeV, $m_{c}=1670$MeV, and $m_{b}=5008$MeV for SET I (fitted from the mesons~\cite{Maiani:2004vq,Ali:2011ug});   $m_{u,d}=330$MeV, $m_{s}=500$MeV, $m_{c}=1550$MeV, and $m_{b}=4880$MeV for SET III (widely used in quark potential model~\cite{Ebert:2005nc}); $m_{u,d}=362$MeV, $m_{s}=546$MeV, $m_{c}=1721$MeV, and $m_{b}=5050$MeV for SET II (fitted from the baryons~\cite{Maiani:2004vq,Ali:2011ug}). For the tetraquark, it is approximate to use the quark mass parameter values extracted from the mesons.

\begin{table}
\caption{Fitting the  overall factors $C^{ij}/(m_i m_j)$ from  the baryon and meson spectra. The fitted results for these factors become: $C^{qq}/m_u^2=193$MeV, $C^{sq}/(m_u m_s)=118$MeV,  $C^{cq}/(m_u m_c)=23$MeV, $C^{bq}/(m_u m_b)=2.3$MeV for the diquark configuration;
$C^{q\bar{q}}/m_u^2=318$MeV, $C^{s\bar{q}}/(m_u m_s)=199$MeV, $C^{c\bar{q}}/(m_u m_c)=69$MeV, $C^{b\bar{q}}/(m_u m_b)=23$MeV, $C^{s\bar{s}}/m_s^2=118$MeV, $C^{c\bar{s}}/(m_s m_c)=72$MeV, $C^{b\bar{s}}/(m_s m_b)=24$MeV, $C^{c\bar{c}}/m_c^2=57$MeV, $C^{b\bar{b}}/m_b^2=31$MeV for the quark-antiquark configuration.  These fitted results are consistent the previous literatures~\cite{Maiani:2004vq,Ali:2011ug,Wang:2016tsi,Wang:2017vnc}.}\label{tab:extracts}
\begin{tabular}{|c|cccc|}
\hline
 Mass & $M_\Delta-M_n$  & $M_\Sigma-M_\Lambda$  & $M_{\Sigma_c}-M_{\Lambda_c}$ & $M_{\Sigma_b}-M_{\Lambda_b}$
\\\hline
Form. & $3C^{qq}/(2m_u^2)$  & ~$C^{qq}/m_u^2-C^{sq}/(m_s m_u)$ ~ & $C^{qq}/m_u^2- C^{cq}/(m_cm_u)$~
& $C^{qq}/m_u^2-C^{bq}/(m_bm_u)$
\\\hline
Exp.~\cite{Patrignani:2016xqp} & 290MeV  & 75MeV  & 170MeV &191MeV
\\\hline\hline
Mass& $M_\rho-M_\pi$  & $M_{K^*}-M_K$  & $M_{D^*}-M_{D}$ & $M_{B^*}-M_{B}$
\\\hline
Form. & $2C^{q\bar{q}}/m_u^2$  & $2C^{s\bar{q}}/( m_s m_u)$   & $2C^{c\bar{q}}/(m_c m_u)$   & $2C^{b\bar{q}}/(m_b m_u)$
\\\hline
Exp.~\cite{Patrignani:2016xqp} & 635MeV  & 397MeV  & 137MeV & 46MeV
\\\hline\hline
Mass& $M_\omega-M_\eta$  & $M_{D_s^*}-M_{D_s}$  & $M_{B_s^*}-M_{B_s}$ &
\\\hline
Form. & $2C^{s\bar{s}}/m_s^2$  & $2C^{c\bar{s}}/(m_c m_s)$  & $2C^{b\bar{s}}/(m_b m_s)$   &
\\\hline
Exp.~\cite{Patrignani:2016xqp} & 235MeV  & 144MeV  & 48MeV &
\\\hline\hline
Mass& $M_{J/\psi}-M_{\eta_c}$ & $M_{\Upsilon}-M_{\eta_b}$  &  &
\\\hline
Form. & $2C^{c\bar{c}}/m_c^2$  & $2C^{b\bar{b}}/m_b^2$  &   &
\\\hline
Exp.~\cite{Patrignani:2016xqp}& 113MeV  & 61MeV  &  &
\\\hline
\end{tabular}
\end{table}
The spectra of the triplet doubly charm tetraquark for SET I are determined as
\begin{align}\label{mass1}
 m(T^{+}_{cc\bar{u}\bar{d}}({\bf  3}))&= 3.86 {\rm GeV} , & J^P=1^+ ,
 \\
  m(T^{+}_{cc\bar{u}\bar{s}}({\bf  3}))=  m(T^{++}_{cc\bar{d}\bar{s}}({\bf \bar 3}))&= 4.10 {\rm GeV} , & J^P=1^+ .
\end{align}
In above, $T^{+}_{cc\bar{u}\bar{d}}({\bf  3})$ lies the spectrum about 16MeV below the $DD^*$ threshold and about 120MeV above the
$DD$ threshold. However, the $T^{+}_{cc\bar{u}\bar{d}}({\bf  3})$ is an axial-vector meson and can not directly hadronic decay to $DD$. Thus $T^{+}_{cc\bar{u}\bar{d}}({\bf  3})$ with spin-parity $1^+$ may be a stable tetraquark, but the situation is not optimistic if we consider the large uncertainty from the good and bad diquark's mass splitting.  The binding energy 16MeV may be polished if the good and bad diquark's mass splitting increases 16MeV.  $T^{+}_{cc\bar{u}\bar{s}}({\bf  3})$ and $T^{++}_{cc\bar{d}\bar{s}}({\bf  3})$ lie the spectrum about 124MeV above the $D_sD^*$ threshold and about 118MeV above the
$D_s^*D$ threshold. These two states can  hadronic decay thus are not stable.

The spectra of the  triplet charm-beauty tetraquark for SET I are determined as
\begin{align}\label{mass2}
 m(T^{0}_{bc\bar{u}\bar{d}}({\bf  3}))&= 7.20 {\rm GeV} , & J^P=1^+ ,
 \\
  m(T^{0}_{bc\bar{u}\bar{s}}({\bf  3}))=  m(T^{+}_{bc\bar{d}\bar{s}}({\bf \bar 3}))&= 7.43 {\rm GeV} , & J^P=1^+ .
\end{align}
In this kind, $T^{0}_{bc\bar{u}\bar{d}}({\bf  3})$ lies the spectrum about 86MeV below the $BD^*$ threshold  but about 5.8MeV above the
$B^*D$ threshold.  Thus $T^{0}_{bc\bar{u}\bar{d}}({\bf  3})$ can hadronic decay to $B^*D$ and has a large decay width. $T^{0}_{bc\bar{u}\bar{s}}({\bf  3})$ and $T^{+}_{bc\bar{d}\bar{s}}({\bf  3})$ lie the spectrum about 56MeV above the $B_sD^*$ threshold and about 137MeV above the
$B^*D_s$ threshold. These two states are also not stable.

The spectra of the  triplet doubly bottom tetraquark for SET I are determined as
\begin{align}\label{mass3}
 m(T^{-}_{bb\bar{u}\bar{d}}({\bf  3}))&= 10.53{\rm GeV} , & J^P=1^+ ,
 \\
  m(T^{-}_{bb\bar{u}\bar{s}}({\bf 3}))=  m(T^{0}_{bb\bar{d}\bar{s}}({\bf \bar 3}))&= 10.77 {\rm GeV} , & J^P=1^+ .
\end{align}
For bottom sector, $T^{-}_{bb\bar{u}\bar{d}}({\bf  3})$ lies the spectrum about 73MeV below the $BB^*$ threshold.  Thus $T^{-}_{bb\bar{u}\bar{d}}({\bf  3})$ with spin-parity $1^+$ is a stable tetraquark, which shall be tested in experiment. $T^{-}_{bb\bar{u}\bar{s}}({\bf  3})$ and $T^{0}_{bb\bar{d}\bar{s}}({\bf  3})$ lie the spectrum about 78MeV above the $B_sB^*$ threshold and about 75MeV above the
$B_s^*B$ threshold, which are not stable.

 Considering the uncertainties from the quark masses and the  heavy diquark mass splitting, we adopt another input for these parameters. When we adopt the quark masses with SET II and increase the heavy diquark mass splitting by $100$MeV, we can easily get the tetraquark mass differences and then predict the doubly heavy quark tetraquark spectra. The mass of $T^{+}_{cc\bar{u}\bar{d}}({\bf  3})$ will been reduced by 90MeV, while $T^{+}_{cc\bar{u}\bar{s}}({\bf  3})$ and $T^{++}_{cc\bar{d}\bar{s}}({\bf  3})$ will be reduced by 105MeV. The conclusions of the stability discussions of them are unchanged.
For the  triplet charm-beauty tetraquark with SET II, the mass of $T^{0}_{bc\bar{u}\bar{d}}({\bf  3})$ will be reduced by 98MeV, while $T^{0}_{bc\bar{u}\bar{s}}({\bf  3})$ and $T^{+}_{bc\bar{d}\bar{s}}({\bf  3})$ will be reduced by 113MeV. Then $T^{0}_{bc\bar{u}\bar{d}}({\bf  3})$ is about 179MeV below the $B D^*$ threshold and about
87MeV below the $B^*D$ threshold, which indicates that $T^{0}_{bc\bar{u}\bar{d}}({\bf  3})$ may be a stable state.
For the  triplet doubly bottom tetraquark with SET II, the mass of $T^{-}_{bb\bar{u}\bar{d}}({\bf  3})$ will be reduced by 86MeV, while $T^{-}_{bb\bar{u}\bar{s}}({\bf  3})$ and $T^{0}_{bb\bar{d}\bar{s}}({\bf  3})$ will be reduced by 101MeV. Thus $T^{-}_{bb\bar{u}\bar{d}}({\bf  3})$  becomes more stable. $T^{-}_{bb\bar{u}\bar{s}}({\bf  3})$ and $T^{0}_{bb\bar{d}\bar{s}}({\bf  3})$ is about 43MeV below the $B_sB^*$ threshold and about 46MeV below the
$B_s^*B$ threshold, which  become stable. If one uses the quark mass values of SET III and considers the uncertainty of the heavy bad diquark mass,  only one state $T^{-}_{bb\bar{u}\bar{d}}({\bf  3})$ near its hadron threshold. Other states stay away from their hadron threshold.

In Ref.~\cite{Karliner:2017qjm}, Karliner and Rosner predicted a stable doubly bottom tetraquark $T(bb\bar{u}\bar{d})$ with $J^P=1^+$
 at $10.389\pm0.012$GeV.  In Ref.~\cite{Eichten:2017ffp}, Eichten and Quigg predicted two stable doubly bottom tetraquarks $T(bb\bar{u}\bar{d})$ with $J^P=1^+$
 at $10.482$GeV and $T(bb\bar{q}\bar{s})$ with $J^P=1^+$
 at $10.643$GeV. Our work also supports the possibility of a stable doubly bottom tetraquark with  $J^P=1^+$
 at $(10.45-10.53)$GeV.  On the other hand,  Karliner and Rosner predicted a doubly charm tetraquark $T(cc\bar{u}\bar{d})$ with $J^P=1^+$
 at $3.882\pm0.012$GeV, which is 7MeV above the  $DD^*$ threshold in Ref.~\cite{Karliner:2017qjm}. Our work gives the $T^{+}_{cc\bar{u}\bar{d}}({\bf  3})$ state also near the
 $DD^*$ threshold but above that by 16MeV. But one should note that  the theoretical uncertainties for the doubly charm tetraquark become larger than that of the doubly bottom tetraquark.

To hunt for these possible doubly heavy tetraquarks  in flavor triplet, we will study their weak decays properties.  Their  semi-leptonic and nonleptonic  decay amplitudes  can be parametrized in terms of SU(3)-irreducible amplitudes. For completeness, we will investigate the weak two-body, three-body and four-body decays of the  doubly heavy tetraquarks.

\section{Semi-Leptonic decays}
\label{sec:semileptonic}

\subsection{Semileptonic $T_{bb\bar{q}\bar{q}}$  decays}

\subsubsection{Decays into mesons and $\ell \overline \nu_{\ell}$}

\begin{figure}
\begin{center}
\includegraphics[scale=0.6]{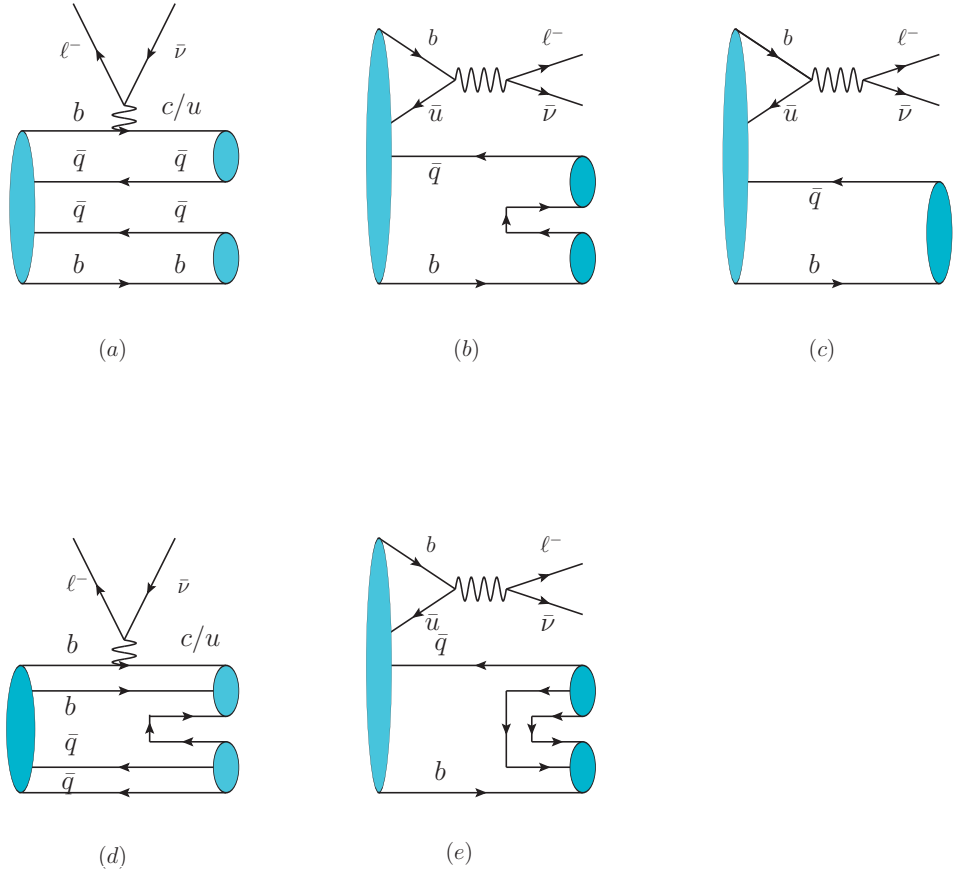}
\end{center}
\caption{Feynman diagrams for semileptonic decays of  doubly bottom tetraquark. Panels (a,b) correspond to the decays into a pair of mesons. In panel (c), there is only one meson in the final states. Panels (d,e) denote the decays into baryonic states. In panels (b,c,e), the two $b\bar u$ quarks in the initial state can annihilate, but such contributions are usually  power suppressed. }\label{fig:Feynman_semileptonic_tetraquark}
\end{figure}

First we study the decays into mesons and $\ell \overline \nu_{\ell}$, where the electro-weak   Hamiltonian is
\begin{eqnarray}
 {\cal H}_{eff}^{e.w.} &=& \frac{G_F}{\sqrt2} \left[V_{q'b} \bar q' \gamma^\mu(1-\gamma_5)b \bar  \ell\gamma_\mu(1-\gamma_5) \nu_{\ell}\right] +h.c.,
\end{eqnarray}
with $q'=u,c$.  The corresponding Feynman diagrams are given in Fig.~\ref{fig:Feynman_semileptonic_tetraquark}.  The transition of $b\to c\ell^-\bar\nu_{\ell}$  belongs to a SU(3) singlet, while the $b\to u\ell^-\bar\nu_{\ell}$ transition belongs to a SU(3) triplet and them can be described as  $H_{3}$ which has the matrix elements $(H_3)^1=V_{ub}$ and $(H_3)^{2(3)}=0$ with  the CKM matrix element $V_{ub}$. Except the   two three-body decay channels shown in Fig.~(\ref{fig:Feynman_semileptonic_tetraquark}c), i.e.
\begin{eqnarray*}
T_{bb\bar{u}\bar{d}}^{-}\to \overline B^0 \ell^-\bar\nu_{\ell}, \;\;\; T_{bb\bar{u}\bar{s}}^{-}\to \overline B_s^0 \ell^-\bar\nu_{\ell},
\end{eqnarray*}
most modes involve four particles in the final states. The semileptonic $T_{bb3}$  decays  can be  described through the hadron-level  Hamiltonian which is written as
\begin{eqnarray}
{\cal H}&=&a_3 (T_{bb3})_{[ij]}  \overline B^{i}\overline D^{j}~\bar \ell\nu_{\ell}+a_4 (T_{bb3})_{[ij]} (H_{  3})^i  \overline B^{k}M^j_k~\bar \ell\nu_{\ell} + a_5 (T_{bb3})_{[ij]} (H_{  3})^k  \overline B^{i}M^j_k~\bar \ell\nu_{\ell}. \label{eq:bbqq_semi}
\end{eqnarray}
Here the $a_i$s are the nonperturbative model-independent parameters. The $a_3$ and $a_5$ will be present  in  Fig.~(\ref{fig:Feynman_semileptonic_tetraquark}a), and $a_4$ is related to the annihilation diagrams in Fig.~(\ref{fig:Feynman_semileptonic_tetraquark}b). Through the Hamiltonian, we can obtain the decay amplitudes for different
decay modes, which are given in Tab.~\ref{tab:Tbb_semileptonic_meson}.

\begin{table}
\caption{Decay amplitudes for doubly bottom tetraquark $T_{bb\bar{q}\bar{q}}$ semileptonic decays into mesons. }\label{tab:Tbb_semileptonic_meson}
\begin{tabular}{|cc|cc|}
\hline
channel  & amplitude  & channel  & amplitude\tabularnewline
\hline
\hline
$T_{bb\bar{u}\bar{s}}^{-}\to B^-   D^+_s \ell^-\bar\nu_{\ell} $ & $ a_3 V_{{cb}}$&
$T_{bb\bar{u}\bar{s}}^{-}\to \overline B^0_s   D^0 \ell^-\bar\nu_{\ell}$ & $ -a_3 V_{{cb}}$\\\hline
$T_{bb\bar{d}\bar{s}}^{0}\to \overline B^0   D^+_s \ell^-\bar\nu_{\ell}$ & $ a_3 V_{{cb}}$&
$T_{bb\bar{d}\bar{s}}^{0}\to \overline B^0_s   D^+ \ell^-\bar\nu_{\ell}$ & $ -a_3 V_{{cb}}$\\\hline
$T_{bb\bar{u}\bar{d}}^{-}\to B^-   D^+ \ell^-\bar\nu_{\ell}$ & $ a_3 V_{{cb}}$&
$T_{bb\bar{u}\bar{d}}^{-}\to \overline B^0   D^0 \ell^-\bar\nu_{\ell}$ & $ -a_3 V_{{cb}}$\\\hline
\hline
channel  & amplitude  & channel  & amplitude\tabularnewline
\hline
\hline
$T_{bb\bar{u}\bar{s}}^{-}\to B^-  K^+  \ell^-\bar\nu_{\ell}$ & $ \left(a_4+a_5\right) V_{{ub}}$&
$T_{bb\bar{u}\bar{s}}^{-}\to \overline B^0  K^0  \ell^-\bar\nu_{\ell}$ & $ a_4 V_{{ub}}$\\\hline
$T_{bb\bar{u}\bar{s}}^{-}\to \overline B^0_s  \pi^0  \ell^-\bar\nu_{\ell}$ & $ -\frac{a_5 }{\sqrt{2}}V_{{ub}}$&
$T_{bb\bar{u}\bar{s}}^{-}\to \overline B^0_s  \eta  \ell^-\bar\nu_{\ell}$ & $ -\frac{\left(2 a_4+a_5\right)}{\sqrt{6}} V_{{ub}}$\\\hline
$T_{bb\bar{d}\bar{s}}^{0}\to \overline B^0  K^+  \ell^-\bar\nu_{\ell}$ & $ a_5 V_{{ub}}$&
$T_{bb\bar{d}\bar{s}}^{0}\to \overline B^0_s  \pi^+  \ell^-\bar\nu_{\ell}$ & $ -a_5 V_{{ub}}$\\\hline
$T_{bb\bar{u}\bar{d}}^{-}\to B^-  \pi^+  \ell^-\bar\nu_{\ell}$ & $ \left(a_4+a_5\right) V_{{ub}}$&
$T_{bb\bar{u}\bar{d}}^{-}\to \overline B^0  \pi^0  \ell^-\bar\nu_{\ell}$ & $ -\frac{\left(a_4+a_5\right) }{\sqrt{2}}V_{{ub}}$\\\hline
$T_{bb\bar{u}\bar{d}}^{-}\to \overline B^0  \eta  \ell^-\bar\nu_{\ell}$ & $ \frac{\left(a_4-a_5\right)}{\sqrt{6}} V_{{ub}}$&
$T_{bb\bar{u}\bar{d}}^{-}\to \overline B^0_s  \overline K^0  \ell^-\bar\nu_{\ell}$ & $ a_4 V_{{ub}}$\\\hline
\hline
\end{tabular}
\end{table}

 To satisfy the SU(3) flavor symmetry, one can ignore the effects of phase space when analyzing their decay widths. From Tab.~\ref{tab:Tbb_semileptonic_meson}, all six decay channels into a $B$ and a $D$ meson have the same decay widths.
Besides, we can get the relations for the decays into a $B$ meson and a light meson:
\begin{eqnarray*}
     \Gamma(T_{bb\bar{u}\bar{s}}^{-}\to \overline B^0_s \pi^0 l^-\bar\nu)= \frac{1}{2}\Gamma(T_{bb\bar{d}\bar{s}}^{0}\to \overline B^0 K^+ l^-\bar\nu)=\frac{1}{2}\Gamma(T_{bb\bar{d}\bar{s}}^{0}\to \overline B^0_s \pi^+ l^-\bar\nu),\\
      \Gamma(T_{bb\bar{u}\bar{d}}^{-}\to B^- \pi^+ l^-\bar\nu)= { }\Gamma(T_{bb\bar{u}\bar{s}}^{-}\to B^- K^+ l^-\bar\nu)=2\Gamma(T_{bb\bar{u}\bar{d}}^{-}\to \overline B^0 \pi^0 l^-\bar\nu),\\
      \Gamma(T_{bb\bar{u}\bar{s}}^{-}\to \overline B^0 K^0 l^-\bar\nu)= { }\Gamma(T_{bb\bar{u}\bar{d}}^{-}\to \overline B^0_s \overline K^0 l^-\bar\nu).
\end{eqnarray*}

\subsubsection{Decays into a bottom baryon, a light anti-baryon and $\ell \overline \nu_{\ell}$}

As shown in the last two panels in Fig.~\ref{fig:Feynman_semileptonic_tetraquark}, the  $T_{bb3}$ can  transit   into a bottom baryon and a light anti-baryon.  Since the decuplet is anti-symmetric for light quarks in flavor space, the two spectator light quarks will not go into the decuplet. We write the  Hamiltonian  as:
  \begin{eqnarray}
  \mathcal{H}&=&b_1 (T_{bb3})_{[ij]} \epsilon^{xjk} (F_8)^l_x (H_3)^i (\overline F_{b\overline{3}})_{[kl]} \bar{\ell} \nu_{\ell}+b_2 (T_{bb3})_{[ij]} \epsilon^{xkl} (F_8)^j_x (H_3)^i (\overline F_{b\overline{3}})_{[kl]} \bar{\ell} \nu_{\ell}\nonumber\\
  &&+b_3 (T_{bb3})_{[ij]} \epsilon^{xij} (F_8)^l_x (H_3)^k (\overline F_{b\overline{3}})_{[kl]} \bar{\ell} \nu_{\ell}
  +b_4 (T_{bb3})_{[ij]} \epsilon^{xil} (F_8)^j_x (H_3)^k (\overline F_{b\overline{3}})_{[kl]} \bar{\ell} \nu_{\ell}\nonumber\\
  &&+b_5 (T_{bb3})_{[ij]} \epsilon^{xjk} (F_8)^l_x (H_3)^i (\overline F_{b6})_{\{kl\}} \bar{\ell} \nu_{\ell}+b_6 (T_{bb3})_{[ij]} \epsilon^{xij} (F_8)^l_x (H_3)^k (\overline F_{b6})_{\{kl\}} \bar{\ell} \nu_{\ell}\nonumber\\
  &&+b_7 (T_{bb3})_{[ij]} \epsilon^{xil} (F_8)^j_x (H_3)^k (\overline F_{b6})_{\{kl\}} \bar{\ell} \nu_{\ell}+b_8 (T_{bb3})_{[ij]}  (F_{\overline{10}})^{\{jkl\}} (H_3)^i (\overline F_{b6})_{\{kl\}} \bar{\ell} \nu_{\ell}.
  \end{eqnarray}
For convenience, we label the decay channels with different final states: class \uppercase\expandafter{\romannumeral1} for an octet baryon plus a heavy triplet   baryon;  class \uppercase\expandafter{\romannumeral2} for an octet baryon plus a heavy sextet baryon;  class \uppercase\expandafter{\romannumeral3} for a decuplet light baryon plus  a heavy sextet baryon. The last type of decays can occur only through the annihilation of $b\bar u$ shown in Fig.~(\ref{fig:Feynman_semileptonic_tetraquark}c). The explicit amplitudes can be found  in Tab.~\ref{tab:bb_Fbar8_Fb}.

  \begin{table}
\caption{Amplitudes for doubly bottom tetraquark $T_{bb\bar{q}\bar{q}}$ decays into a light anti-baryon octet plus a bottom baryon anti-triplet(class \uppercase\expandafter{\romannumeral1}) or sextet(class \uppercase\expandafter{\romannumeral2}),  and a light anti-baryon decuplet plus a bottom baryon sextet (class \uppercase\expandafter{\romannumeral3}) }\label{tab:bb_Fbar8_Fb}\begin{tabular}{|cc|cc|}\hline\hline
class \uppercase\expandafter{\romannumeral1} & amplitude &class \uppercase\expandafter{\romannumeral3} & amplitude \tabularnewline\hline
$T_{bb\bar{u}\bar{s}}^{-}\to \overline \Lambda^0  \Lambda_b^0 \ell^-\bar\nu_{\ell}$ & $ \frac{\left(2 b_1-4 b_2-2 b_3-b_4\right) V_{{ub}}}{\sqrt{6}}$&$T_{bb\bar{u}\bar{s}}^{-}\to \overline \Sigma^{\prime-}  \Sigma_{b}^{+} \ell^-\bar\nu_{\ell}$ & $ \frac{b_8 V_{{ub}}}{\sqrt{3}}$\\\hline
$T_{bb\bar{u}\bar{s}}^{-}\to \overline \Sigma^0  \Lambda_b^0 \ell^-\bar\nu_{\ell}$ & $ \frac{\left(2 b_3+b_4\right) V_{{ub}}}{\sqrt{2}}$&$T_{bb\bar{u}\bar{s}}^{-}\to \overline \Sigma^{\prime0}  \Sigma_{b}^{0} \ell^-\bar\nu_{\ell}$ & $ \frac{b_8 V_{{ub}}}{\sqrt{3}}$\\\hline
$T_{bb\bar{u}\bar{s}}^{-}\to \overline \Xi^+  \Xi_b^- \ell^-\bar\nu_{\ell}$ & $ -\left(b_1-2 b_2\right) V_{{ub}}$&$T_{bb\bar{u}\bar{s}}^{-}\to \overline \Sigma^{\prime+}  \Sigma_{b}^{-} \ell^-\bar\nu_{\ell}$ & $ \frac{b_8 V_{{ub}}}{\sqrt{3}}$\\\hline
$T_{bb\bar{u}\bar{s}}^{-}\to \overline \Xi^0  \Xi_b^0 \ell^-\bar\nu_{\ell}$ & $ \left(b_1-2 b_2-2 b_3-b_4\right) V_{{ub}}$&$T_{bb\bar{u}\bar{s}}^{-}\to \overline \Xi^{\prime0}  \Xi_{b}^{\prime0} \ell^-\bar\nu_{\ell}$ & $ \sqrt{\frac{2}{3}} b_8 V_{{ub}}$\\\hline
$T_{bb\bar{d}\bar{s}}^{0}\to \overline \Sigma^+  \Lambda_b^0 \ell^-\bar\nu_{\ell}$ & $ \left(2 b_3+b_4\right) V_{{ub}}$&$T_{bb\bar{u}\bar{s}}^{-}\to \overline \Xi^{\prime+}  \Xi_{b}^{\prime-} \ell^-\bar\nu_{\ell}$ & $ \sqrt{\frac{2}{3}} b_8 V_{{ub}}$\\\hline
$T_{bb\bar{d}\bar{s}}^{0}\to \overline \Xi^+  \Xi_b^0 \ell^-\bar\nu_{\ell}$ & $ \left(2 b_3+b_4\right) V_{{ub}}$&$T_{bb\bar{u}\bar{s}}^{-}\to \overline \Omega^+  \Omega_{b}^{-} \ell^-\bar\nu_{\ell}$ & $ b_8 V_{{ub}}$\\\hline
$T_{bb\bar{u}\bar{d}}^{-}\to \overline \Lambda^0  \Xi_b^0 \ell^-\bar\nu_{\ell}$ & $ \frac{\left(b_1-2 \left(b_2+2 b_3+b_4\right)\right) V_{{ub}}}{\sqrt{6}}$&$T_{bb\bar{u}\bar{d}}^{-}\to \overline \Delta^{-}  \Sigma_{b}^{+} \ell^-\bar\nu_{\ell}$ & $ \frac{b_8 V_{{ub}}}{\sqrt{3}}$\\\hline
$T_{bb\bar{u}\bar{d}}^{-}\to \overline \Sigma^0  \Xi_b^0 \ell^-\bar\nu_{\ell}$ & $ -\frac{\left(b_1-2 b_2\right) V_{{ub}}}{\sqrt{2}}$&$T_{bb\bar{u}\bar{d}}^{-}\to \overline \Delta^{0}  \Sigma_{b}^{0} \ell^-\bar\nu_{\ell}$ & $ \sqrt{\frac{2}{3}} b_8 V_{{ub}}$\\\hline
$T_{bb\bar{u}\bar{d}}^{-}\to \overline \Sigma^+  \Xi_b^- \ell^-\bar\nu_{\ell}$ & $ -\left(b_1-2 b_2\right) V_{{ub}}$&$T_{bb\bar{u}\bar{d}}^{-}\to \overline \Delta^{+}  \Sigma_{b}^{-} \ell^-\bar\nu_{\ell}$ & $ b_8 V_{{ub}}$\\\hline
$T_{bb\bar{u}\bar{d}}^{-}\to \overline n  \Lambda_b^0 \ell^-\bar\nu_{\ell}$ & $ -\left(b_1-2 b_2-2 b_3-b_4\right) V_{{ub}}$&$T_{bb\bar{u}\bar{d}}^{-}\to \overline \Sigma^{\prime0}  \Xi_{b}^{\prime0} \ell^-\bar\nu_{\ell}$ & $ \frac{b_8 V_{{ub}}}{\sqrt{3}}$\\\hline
&&$T_{bb\bar{u}\bar{d}}^{-}\to \overline \Sigma^{\prime+}  \Xi_{b}^{\prime-} \ell^-\bar\nu_{\ell}$ & $ \sqrt{\frac{2}{3}} b_8 V_{{ub}}$\\\hline
&&$T_{bb\bar{u}\bar{d}}^{-}\to \overline \Xi^{\prime+}  \Omega_{b}^{-} \ell^-\bar\nu_{\ell}$ & $ \frac{b_8 V_{{ub}}}{\sqrt{3}}$\\\hline
\hline
class \uppercase\expandafter{\romannumeral2} & amplitude &class \uppercase\expandafter{\romannumeral2} & amplitude \\\hline
$T_{bb\bar{u}\bar{s}}^{-}\to \overline \Lambda^0  \Sigma_{b}^{0} \ell^-\bar\nu_{\ell}$ & $ -\frac{\left(2 b_6+b_7\right) V_{{ub}}}{2 \sqrt{3}}$&$T_{bb\bar{d}\bar{s}}^{0}\to \overline \Sigma^+  \Sigma_{b}^{0} \ell^-\bar\nu_{\ell}$ & $ \frac{\left(2 b_6+b_7\right) V_{{ub}}}{\sqrt{2}}$\\\hline
$T_{bb\bar{u}\bar{s}}^{-}\to \overline \Sigma^- \Sigma_{b}^{+} \ell^-\bar\nu_{\ell}$ & $ \left(b_5-2 b_6-b_7\right) V_{{ub}}$&$T_{bb\bar{d}\bar{s}}^{0}\to \overline \Xi^+  \Xi_{b}^{\prime0} \ell^-\bar\nu_{\ell}$ & $ \frac{\left(2 b_6+b_7\right) V_{{ub}}}{\sqrt{2}}$\\\hline
$T_{bb\bar{u}\bar{s}}^{-}\to \overline \Sigma^0  \Sigma_{b}^{0} \ell^-\bar\nu_{\ell}$ & $ \frac{1}{6} \left(-6 b_5+6 b_6+3 b_7\right) V_{{ub}}$&$T_{bb\bar{u}\bar{d}}^{-}\to \overline \Lambda^0  \Xi_{b}^{\prime0} \ell^-\bar\nu_{\ell}$ & $ \frac{\left(3 b_5-2 \left(2 b_6+b_7\right)\right) V_{{ub}}}{2 \sqrt{3}}$\\\hline
$T_{bb\bar{u}\bar{s}}^{-}\to \overline \Sigma^+  \Sigma_{b}^{-} \ell^-\bar\nu_{\ell}$ & $ -b_5 V_{{ub}}$&$T_{bb\bar{u}\bar{d}}^{-}\to \overline \Sigma^0  \Xi_{b}^{\prime0} \ell^-\bar\nu_{\ell}$ & $ \frac{b_5 V_{{ub}}}{2}$\\\hline
$T_{bb\bar{u}\bar{s}}^{-}\to \overline \Xi^+  \Xi_{b}^{\prime-} \ell^-\bar\nu_{\ell}$ & $ -\frac{b_5 V_{{ub}}}{\sqrt{2}}$&$T_{bb\bar{u}\bar{d}}^{-}\to \overline \Sigma^+  \Xi_{b}^{\prime-} \ell^-\bar\nu_{\ell}$ & $ \frac{b_5 V_{{ub}}}{\sqrt{2}}$\\\hline
$T_{bb\bar{u}\bar{s}}^{-}\to \overline \Xi^0  \Xi_{b}^{\prime0} \ell^-\bar\nu_{\ell}$ & $ \frac{\left(b_5-2 b_6-b_7\right) V_{{ub}}}{\sqrt{2}}$&$T_{bb\bar{u}\bar{d}}^{-}\to \overline p  \Sigma_{b}^{+} \ell^-\bar\nu_{\ell}$ & $ -\left(b_5-2 b_6-b_7\right) V_{{ub}}$\\\hline
$T_{bb\bar{d}\bar{s}}^{0}\to \overline \Lambda^0  \Sigma_{b}^{+} \ell^-\bar\nu_{\ell}$ & $ \frac{\left(2 b_6+b_7\right) V_{{ub}}}{\sqrt{6}}$&$T_{bb\bar{u}\bar{d}}^{-}\to \overline n  \Sigma_{b}^{0} \ell^-\bar\nu_{\ell}$ & $ -\frac{\left(b_5-2 b_6-b_7\right) V_{{ub}}}{\sqrt{2}}$\\\hline
$T_{bb\bar{d}\bar{s}}^{0}\to \overline \Sigma^0  \Sigma_{b}^{+} \ell^-\bar\nu_{\ell}$ & $ \frac{\left(2 b_6+b_7\right) V_{{ub}}}{\sqrt{2}}$&$T_{bb\bar{u}\bar{d}}^{-}\to \overline \Xi^+  \Omega_{b}^{-} \ell^-\bar\nu_{\ell}$ & $ b_5 V_{{ub}}$\\\hline
\hline
\end{tabular}
\end{table}

From them,  the  relations for class \uppercase\expandafter{\romannumeral1} decays can be found:
\begin{eqnarray*}
    \Gamma(T_{bb\bar{u}\bar{s}}^{-}\to \overline \Sigma^0 \Lambda_b^0l^-\bar\nu)= \frac{1}{2}\Gamma(T_{bb\bar{d}\bar{s}}^{0}\to \overline \Xi^+ \Xi_b^0l^-\bar\nu)=\frac{1}{2}\Gamma(T_{bb\bar{d}\bar{s}}^{0}\to \overline \Sigma^+ \Lambda_b^0l^-\bar\nu),\\
    \Gamma(T_{bb\bar{u}\bar{d}}^{-}\to \overline \Sigma^0 \Xi_b^0l^-\bar\nu)=\frac{1}{2}\Gamma(T_{bb\bar{u}\bar{s}}^{-}\to \overline \Xi^+ \Xi_b^-l^-\bar\nu)=\frac{1}{2}\Gamma(T_{bb\bar{u}\bar{d}}^{-}\to \overline \Sigma^+ \Xi_b^-l^-\bar\nu),\\
    \Gamma(T_{bb\bar{u}\bar{d}}^{-}\to \overline n \Lambda_b^0l^-\bar\nu)= { }\Gamma(T_{bb\bar{u}\bar{s}}^{-}\to \overline \Xi^0 \Xi_b^0l^-\bar\nu).
\end{eqnarray*}
The relations for class \uppercase\expandafter{\romannumeral2} decays become:
\begin{eqnarray*}
    \Gamma(T_{bb\bar{u}\bar{s}}^{-}\to \overline \Lambda^0 \Sigma_{b}^{0}l^-\bar\nu)= \frac{1}{6}\Gamma(T_{bb\bar{d}\bar{s}}^{0}\to \overline \Xi^+ \Xi_{b}^{\prime0}l^-\bar\nu)=\frac{1}{2}\Gamma(T_{bb\bar{d}\bar{s}}^{0}\to \overline \Lambda^0 \Sigma_{b}^{+}l^-\bar\nu)=\frac{1}{6}\Gamma(T_{bb\bar{d}\bar{s}}^{0}\to \overline \Sigma^+ \Sigma_{b}^{0}l^-\bar\nu)\\=\frac{1}{6}\Gamma(T_{bb\bar{d}\bar{s}}^{0}\to \overline \Sigma^0 \Sigma_{b}^{+}l^-\bar\nu),\\
    \Gamma(T_{bb\bar{u}\bar{s}}^{-}\to \overline \Sigma^-\Sigma_{b}^{+}l^-\bar\nu)= 2\Gamma(T_{bb\bar{u}\bar{s}}^{-}\to \overline \Xi^0 \Xi_{b}^{\prime0}l^-\bar\nu)=2\Gamma(T_{bb\bar{u}\bar{d}}^{-}\to \overline n \Sigma_{b}^{0}l^-\bar\nu)=\Gamma(T_{bb\bar{u}\bar{d}}^{-}\to \overline p \Sigma_{b}^{+}l^-\bar\nu),\\
    \Gamma(T_{bb\bar{u}\bar{s}}^{-}\to \overline \Sigma^+ \Sigma_{b}^{-}l^-\bar\nu)= 2\Gamma(T_{bb\bar{u}\bar{s}}^{-}\to \overline \Xi^+ \Xi_{b}^{\prime-}l^-\bar\nu)=4\Gamma(T_{bb\bar{u}\bar{d}}^{-}\to \overline \Sigma^0 \Xi_{b}^{\prime0}l^-\bar\nu)=
2\Gamma(T_{bb\bar{u}\bar{d}}^{-}\to \overline \Sigma^+ \Xi_{b}^{\prime-}l^-\bar\nu)\\=
\Gamma(T_{bb\bar{u}\bar{d}}^{-}\to \overline \Xi^+ \Omega_{b}^{-}l^-\bar\nu).
\end{eqnarray*}
The relations for class \uppercase\expandafter{\romannumeral3} decay widths are
\begin{eqnarray*}
    \Gamma(T_{bb\bar{u}\bar{s}}^{-}\to \overline \Sigma^{\prime-} \Sigma_{b}^{+}l^-\bar\nu)=
    \Gamma(T_{bb\bar{u}\bar{s}}^{-}\to \overline \Sigma^{\prime0} \Sigma_{b}^{0}l^-\bar\nu)=
\Gamma(T_{bb\bar{u}\bar{s}}^{-}\to \overline \Sigma^{\prime+} \Sigma_{b}^{-}l^-\bar\nu)=
\frac{1}{2}\Gamma(T_{bb\bar{u}\bar{s}}^{-}\to \overline \Xi^{\prime0} \Xi_{b}^{\prime0}l^-\bar\nu)\\=
\frac{1}{2}\Gamma(T_{bb\bar{u}\bar{s}}^{-}\to \overline \Xi^{\prime+} \Xi_{b}^{\prime-}l^-\bar\nu)=
\frac{1}{3}\Gamma(T_{bb\bar{u}\bar{s}}^{-}\to \overline \Omega^+ \Omega_{b}^{-}l^-\bar\nu)=
\frac{1}{2}\Gamma(T_{bb\bar{u}\bar{d}}^{-}\to \overline \Delta^{0} \Sigma_{b}^{0}l^-\bar\nu)=
\frac{1}{3}\Gamma(T_{bb\bar{u}\bar{d}}^{-}\to \overline \Delta^{+} \Sigma_{b}^{-}l^-\bar\nu)\\=
\Gamma(T_{bb\bar{u}\bar{d}}^{-}\to \overline \Sigma^{\prime0} \Xi_{b}^{\prime0}l^-\bar\nu)=
\frac{1}{2}\Gamma(T_{bb\bar{u}\bar{d}}^{-}\to \overline \Sigma^{\prime+} \Xi_{b}^{\prime-}l^-\bar\nu)=
\Gamma(T_{bb\bar{u}\bar{d}}^{-}\to \overline \Xi^{\prime+} \Omega_{b}^{-}l^-\bar\nu)=
\Gamma(T_{bb\bar{u}\bar{d}}^{-}\to \overline \Delta^{-} \Sigma_{b}^{+}l^-\bar\nu).
\end{eqnarray*}

\subsubsection{Decays into a charmed and bottomed baryon plus a light anti-baryon and $\ell \overline \nu_{\ell}$}

 $T_{bb3}$ can also decay  into a light octet or anti-decuplet anti-baryon  and a charmed and bottomed baryon for  the $b\to c$ transition. After removing the forbidden constructions, the Hamiltonian becomes:
  \begin{eqnarray}
  \mathcal{H}_{eff}&=&b_9 (T_{bb3})_{[ij]} \epsilon^{xij} (F_8)^k_x (\overline F_{bc})_{k} \bar{\ell} \nu_{\ell}+b_{10} (T_{bb3})_{[ij]} \epsilon^{xik} (F_8)^j_x (\overline F_{bc})_{k} \bar{\ell} \nu_{\ell}.
  \end{eqnarray}
The amplitudes are derived and  given in Tab.~\ref{tab:bb_Fbar_Fbc}.
From them, the relations of the related decay widths are:
\begin{eqnarray*}
    \Gamma(T_{bb\bar{u}\bar{s}}^{-}\to \overline \Lambda^0 \Xi_{bc}^{0}l^-\bar\nu)= \frac{1}{6}\Gamma(T_{bb\bar{u}\bar{s}}^{-}\to \overline \Xi^0 \Omega_{bc}^{0}l^-\bar\nu)=
\frac{1}{6}\Gamma(T_{bb\bar{d}\bar{s}}^{0}\to \overline \Xi^+ \Omega_{bc}^{0}l^-\bar\nu)=
\frac{1}{4}\Gamma(T_{bb\bar{u}\bar{d}}^{-}\to \overline \Lambda^0 \Omega_{bc}^{0}l^-\bar\nu)\\=
\frac{1}{6}\Gamma(T_{bb\bar{u}\bar{s}}^{-}\to \overline \Sigma^-\Xi_{bc}^{+}l^-\bar\nu)=
\frac{1}{3}\Gamma(T_{bb\bar{u}\bar{s}}^{-}\to \overline \Sigma^0 \Xi_{bc}^{0}l^-\bar\nu)=
\frac{1}{6}\Gamma(T_{bb\bar{d}\bar{s}}^{0}\to \overline \Sigma^+ \Xi_{bc}^{0}l^-\bar\nu)=
\frac{1}{6}\Gamma(T_{bb\bar{u}\bar{d}}^{-}\to \overline n \Xi_{bc}^{0}l^-\bar\nu)\\=
\Gamma(T_{bb\bar{d}\bar{s}}^{0}\to \overline \Lambda^0 \Xi_{bc}^{+}l^-\bar\nu)=
\frac{1}{3}\Gamma(T_{bb\bar{d}\bar{s}}^{0}\to \overline \Sigma^0 \Xi_{bc}^{+}l^-\bar\nu)=
\frac{1}{6}\Gamma(T_{bb\bar{u}\bar{d}}^{-}\to \overline p \Xi_{bc}^{+}l^-\bar\nu).\\
\end{eqnarray*}
    \begin{table}
\caption{Amplitudes for doubly bottom tetraquark $T_{bb\bar{q}\bar{q}}$ decays into a light anti-baryon octet and a charmed and bottomed baryon.}\label{tab:bb_Fbar_Fbc}\begin{tabular}{|cc|cc|}\hline\hline
channel & amplitude &channel & amplitude \tabularnewline\hline
$T_{bb\bar{u}\bar{s}}^{-}\to \overline \Lambda^0  \Xi_{bc}^{0} \ell^-\bar\nu_{\ell}$ & $ -\frac{2 b_9+b_{10}}{\sqrt{6}}V_{cb}$&
$T_{bb\bar{u}\bar{s}}^{-}\to \overline \Sigma^- \Xi_{bc}^{+} \ell^-\bar\nu_{\ell}$ & $ (-2 b_9-b_{10})V_{cb}$\\\hline
$T_{bb\bar{u}\bar{s}}^{-}\to \overline \Sigma^0  \Xi_{bc}^{0} \ell^-\bar\nu_{\ell}$ & $ \frac{2 b_9+b_{10}}{\sqrt{2}}V_{cb}$&
$T_{bb\bar{u}\bar{s}}^{-}\to \overline \Xi^0  \Omega_{bc}^{0} \ell^-\bar\nu_{\ell}$ & $ (-2 b_9-b_{10})V_{cb}$\\\hline
$T_{bb\bar{d}\bar{s}}^{0}\to \overline \Lambda^0  \Xi_{bc}^{+} \ell^-\bar\nu_{\ell}$ & $ \frac{2 b_9+b_{10}}{\sqrt{6}}V_{cb}$&
$T_{bb\bar{d}\bar{s}}^{0}\to \overline \Sigma^0  \Xi_{bc}^{+} \ell^-\bar\nu_{\ell}$ & $ \frac{2 b_9+b_{10}}{\sqrt{2}}V_{cb}$\\\hline
$T_{bb\bar{d}\bar{s}}^{0}\to \overline \Sigma^+  \Xi_{bc}^{0} \ell^-\bar\nu_{\ell}$ & $ (2 b_9+b_{10})V_{cb}$&
$T_{bb\bar{d}\bar{s}}^{0}\to \overline \Xi^+  \Omega_{bc}^{0} \ell^-\bar\nu_{\ell}$ & $ (2 b_9+b_{10})V_{cb}$\\\hline
$T_{bb\bar{u}\bar{d}}^{-}\to \overline \Lambda^0  \Omega_{bc}^{0} \ell^-\bar\nu_{\ell}$ & $ -\sqrt{\frac{2}{3}} \left(2 b_9+b_{10}\right)V_{cb}$&
$T_{bb\bar{u}\bar{d}}^{-}\to \overline p  \Xi_{bc}^{+} \ell^-\bar\nu_{\ell}$ & $ (2 b_9+b_{10})V_{cb}$\\\hline
$T_{bb\bar{u}\bar{d}}^{-}\to \overline n  \Xi_{bc}^{0} \ell^-\bar\nu_{\ell}$ & $( 2 b_9+b_{10})V_{cb}$&&\\
\hline\hline
\end{tabular}
\end{table}

\subsection{$T_{cc\bar{q}\bar{q}}$ decays}

\begin{figure}
\begin{center}
\includegraphics[scale=0.6]{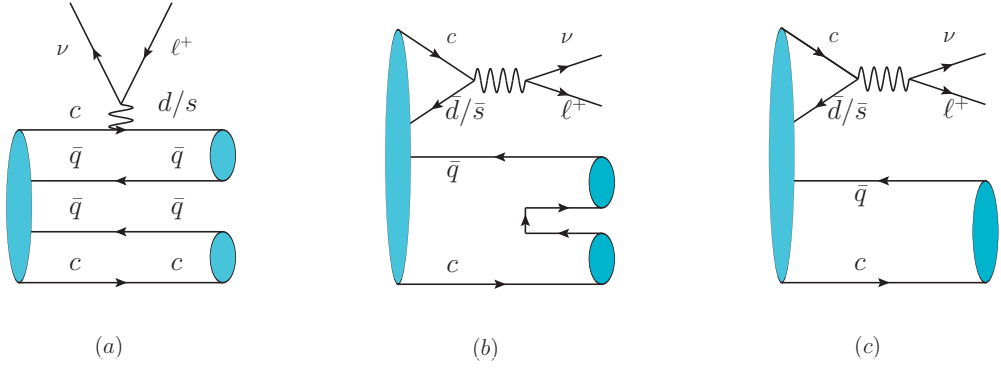}
\end{center}
\caption{Feynman diagrams for semileptonic decays of  doubly charmed tetraquark. Panel (a,b) correspond to the decays into a pair of mesons, and in panel (c), there is only one meson in the final state. In panels (b,c), the two $c\bar d/\bar s$ quarks in the initial state can annihilate, and usually such contributions are power suppressed. }\label{fig:Feynman_semileptonic_tetraquark_2}
\end{figure}

The effective Hamiltonian from the charm semileptonic decays into a light quark is:
\begin{eqnarray}
{\cal H}&=&\frac{G_F}{\sqrt2} \left[V_{cq}^* \bar q  \gamma^\mu(1-\gamma_5)c \bar \nu_{\ell}\gamma_\mu(1-\gamma_5) \ell\right] +h.c.,
\end{eqnarray}
where $q=d,s$. A SU(3) triplet denoted as $H_{  3}$ with the elements $(H_{  3})^1=0,~(H_{  3})^2=V_{cd}^*$ and $(H_{  3})^3=V_{cs}^*$ is introduced for the heavy-to-light quark operators. We plotted the corresponding Feynman diagrams  in Fig.~(\ref{fig:Feynman_semileptonic_tetraquark_2}).
The triplet state $T_{cc3}$ can decay to a charmed meson plus $\ell^+ \nu$ :
\begin{equation}
T_{cc\bar{u}\bar{s}}^{+}/ T_{cc\bar{u}\bar{d}}^{+}\to D^0 l^+\nu  ,\;\;\;
T_{cc\bar{d}\bar{s}}^{++}\to  D^+ l^+\nu ,\;\;\;
T_{cc\bar{d}\bar{s}}^{++}\to  D^+_s l^+\nu, \end{equation}
and their Feynman diagram are given in Fig.~(\ref{fig:Feynman_semileptonic_tetraquark_2}c). Thus we obtained:
\begin{eqnarray*}
    \Gamma(T_{cc\bar{u}\bar{s}}^{+}\to  D^0 l^+\nu)= { }\Gamma(T_{cc\bar{d}\bar{s}}^{++}\to  D^+ l^+\nu), \Gamma(T_{cc\bar{d}\bar{s}}^{++}\to  D^+_s l^+\nu)= { }\Gamma(T_{cc\bar{u}\bar{d}}^{+}\to  D^0 l^+\nu).
\end{eqnarray*}

The effective Hamiltonian  for   decays into a  charmed meson and a light meson is written as:
\begin{eqnarray}
  {\cal H}&=& a_1 (T_{cc3})_{[ij]} (H_{  3})^i  (\overline D)^{k}M^j_k~\bar\nu_\ell \ell+a_2 (T_{cc3})_{[ij]} (H_{  3})^k  (\overline D)^{i}M^j_k~\bar\nu_\ell \ell.\label{eq:ccq_semi}
\end{eqnarray}
 The related Feynman diagrams are plotted in Fig.~\ref{fig:Feynman_semileptonic_tetraquark_2}(a,b), and the related results for the decay width relations are given  in Tab.~\ref{tab:cc_Dand8mson}.
\begin{table}
\caption{Amplitudes for doubly charmed tetraquark $T_{cc\bar{q}\bar{q}}$  decays into a charmed meson and a light meson.}\label{tab:cc_Dand8mson}\begin{tabular}{|cc|cc|}\hline
channel  & amplitude  & channel  & amplitude\tabularnewline
\hline\hline
$T_{cc\bar{u}\bar{s}}^{+}\to  D^0  \pi^0  l^+\nu $ & $ -\frac{a_1V_{cs}^*}{\sqrt{2}}$&
$T_{cc\bar{u}\bar{s}}^{+}\to  D^0  K^0  l^+\nu $ & $ a_2 V_{cd}^*$\\\hline
$T_{cc\bar{u}\bar{s}}^{+}\to  D^0  \eta  l^+\nu $ & $ -\frac{\left(a_1+2 a_2\right)V_{cs}^*}{\sqrt{6}}$&
$T_{cc\bar{u}\bar{s}}^{+}\to  D^+  \pi^-  l^+\nu $ & $ -a_1V_{cs}^*$\\\hline
$T_{cc\bar{u}\bar{s}}^{+}\to  D^+_s  \pi^-  l^+\nu $ & $ -a_2 V_{cd}^*$&
$T_{cc\bar{u}\bar{s}}^{+}\to  D^+_s  K^-  l^+\nu $ & $ -\left(a_1+a_2\right)V_{cs}^*$\\\hline
$T_{cc\bar{d}\bar{s}}^{++}\to  D^0  \pi^+  l^+\nu $ & $ -a_1V_{cs}^*$&
$T_{cc\bar{d}\bar{s}}^{++}\to  D^0  K^+  l^+\nu $ & $ a_1 V_{cd}^*$\\\hline
$T_{cc\bar{d}\bar{s}}^{++}\to  D^+  \pi^0  l^+\nu $ & $ \frac{a_1V_{cs}^*}{\sqrt{2}}$&
$T_{cc\bar{d}\bar{s}}^{++}\to  D^+  K^0  l^+\nu $ & $ \left(a_1+a_2\right) V_{cd}^*$\\\hline
$T_{cc\bar{d}\bar{s}}^{++}\to  D^+  \eta  l^+\nu $ & $ -\frac{\left(a_1+2 a_2\right)V_{cs}^*}{\sqrt{6}}$&
$T_{cc\bar{d}\bar{s}}^{++}\to  D^+_s  \pi^0  l^+\nu $ & $ \frac{a_2 V_{cd}^*}{\sqrt{2}}$\\\hline
$T_{cc\bar{d}\bar{s}}^{++}\to  D^+_s  \overline K^0  l^+\nu $ & $ -\left(a_1+a_2\right)V_{cs}^*$&
$T_{cc\bar{d}\bar{s}}^{++}\to  D^+_s  \eta  l^+\nu $ & $ -\frac{\left(2 a_1+a_2\right) V_{cd}^*}{\sqrt{6}}$\\\hline
$T_{cc\bar{u}\bar{d}}^{+}\to  D^0  \pi^0  l^+\nu $ & $ -\frac{\left(a_1+a_2\right) V_{cd}^*}{\sqrt{2}}$&
$T_{cc\bar{u}\bar{d}}^{+}\to  D^0  \overline K^0  l^+\nu $ & $ a_2V_{cs}^*$\\\hline
$T_{cc\bar{u}\bar{d}}^{+}\to  D^0  \eta  l^+\nu $ & $ \frac{\left(a_2-a_1\right) V_{cd}^*}{\sqrt{6}}$&
$T_{cc\bar{u}\bar{d}}^{+}\to  D^+  \pi^-  l^+\nu $ & $ -\left(a_1+a_2\right) V_{cd}^*$\\\hline
$T_{cc\bar{u}\bar{d}}^{+}\to  D^+  K^-  l^+\nu $ & $ -a_2V_{cs}^*$&
$T_{cc\bar{u}\bar{d}}^{+}\to  D^+_s  K^-  l^+\nu $ & $ -a_1 V_{cd}^*$\\\hline
\hline
\end{tabular}
\end{table}
Thus we have
\begin{eqnarray*}
    \Gamma(T_{cc\bar{u}\bar{s}}^{+}\to  D^0 \pi^0 l^+\nu)=\Gamma(T_{cc\bar{d}\bar{s}}^{++}\to  D^+ \pi^0 l^+\nu)=\frac{1}{2}\Gamma(T_{cc\bar{u}\bar{s}}^{+}\to  D^+ \pi^- l^+\nu)=\frac{1}{2}\Gamma(T_{cc\bar{d}\bar{s}}^{++}\to  D^0 \pi^+ l^+\nu),\\
      \Gamma(T_{cc\bar{d}\bar{s}}^{++}\to  D^0 K^+ l^+\nu)= { }\Gamma(T_{cc\bar{u}\bar{d}}^{+}\to  D^+_s K^- l^+\nu),
      \Gamma(T_{cc\bar{d}\bar{s}}^{++}\to  D^+_s \overline K^0 l^+\nu)= { }\Gamma(T_{cc\bar{u}\bar{s}}^{+}\to  D^+_s K^- l^+\nu),\\
      \Gamma(T_{cc\bar{u}\bar{d}}^{+}\to  D^0 \overline K^0 l^+\nu)= { }\Gamma(T_{cc\bar{u}\bar{d}}^{+}\to  D^+ K^- l^+\nu),
     \Gamma(T_{cc\bar{u}\bar{s}}^{+}\to  D^0 \eta l^+\nu)= { }\Gamma(T_{cc\bar{d}\bar{s}}^{++}\to  D^+ \eta l^+\nu),\\
     \Gamma(T_{cc\bar{u}\bar{s}}^{+}\to  D^+_s \pi^- l^+\nu)=\Gamma(T_{cc\bar{u}\bar{s}}^{+}\to  D^0 K^0 l^+\nu)
     = 2\Gamma(T_{cc\bar{d}\bar{s}}^{++}\to  D^+_s \pi^0 l^+\nu),\\
     \Gamma(T_{cc\bar{u}\bar{d}}^{+}\to  D^0 \pi^0 l^+\nu)= \frac{1}{2}\Gamma(T_{cc\bar{d}\bar{s}}^{++}\to  D^+ K^0 l^+\nu)=\frac{1}{2}\Gamma(T_{cc\bar{u}\bar{d}}^{+}\to  D^+ \pi^- l^+\nu).
\end{eqnarray*}

\subsection{Semileptonic $T_{bc\bar{q}\bar{q}}$  decays }

Both the bottom and charm quarks can decay in the semileptonic $T_{bc\bar{q}\bar{q}}$  decays. For the bottom decay  in $T_{bc\bar{q}\bar{q}}$, one can easily get the decay amplitude  from those for $T_{bb\bar{q}\bar{q}}$ decays with $T_{bb\bar{q}\bar{q}} \to T_{bc\bar{q}\bar{q}}$, $B\to D$. 
 For the charm  decays in $T_{bc\bar{q}\bar{q}}$, one can easily get them from those for $T_{cc\bar{q}\bar{q}}$ decays with the replacement of $T_{cc\bar{q}\bar{q}} \to T_{bc\bar{q}\bar{q}}$, $D\to B$.
Thus we do not need to give the tedious results here.

\section{Non-leptonic $T_{bb\bar{q}\bar{q}}$ decays}
\label{sec:bbq_nonleptonic}

Next, we will study the non-leptonic decay amplitudes. For the  bottom quark decay, there are four types:
\begin{eqnarray}
b\to c\bar c d/s, \;
b\to c \bar u d/s, \;
b\to u \bar c d/s, \;
b\to q_1 \bar q_2 q_3,
\end{eqnarray}
where $q_{i}$ with $i=1,2,3$ denote the light flavors.
We will discuss these decay modes one by one in the following.

\subsection{$b\to c\bar c d/s$ transition}

\subsubsection{ W-exchange Topology}

\begin{figure}
\begin{center}
\includegraphics[scale=0.45]{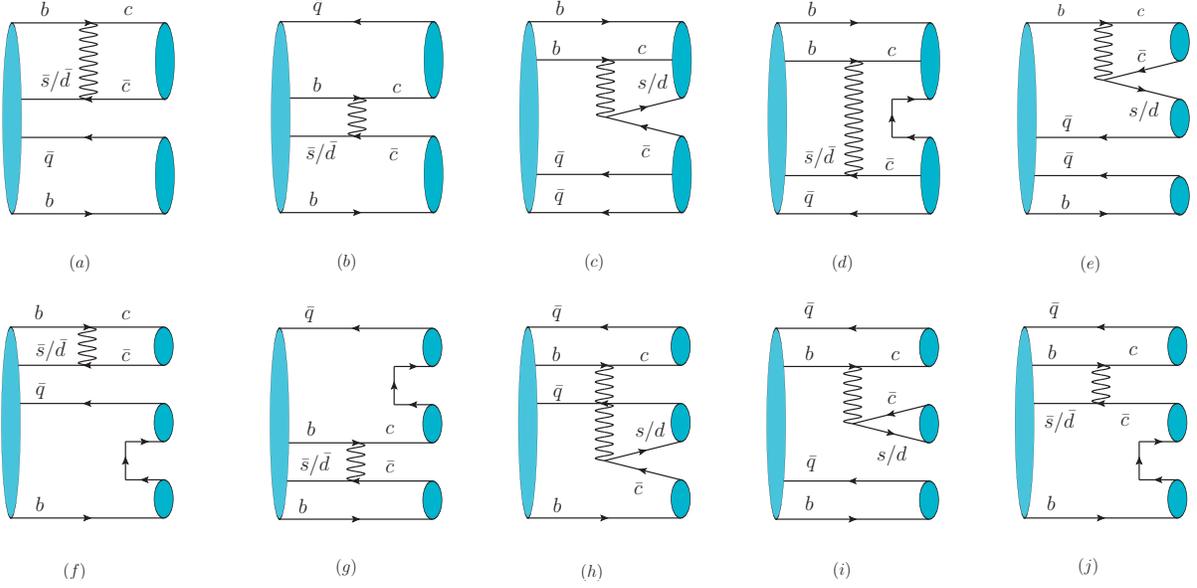}
\end{center}
\caption{Feynman diagrams for nonleptonic decays of  doubly heavy tetraquark. (a,b) are corresponding to the two mesons W-exchange process; (c,d) are corresponding to the baryon and anti-baryon process; (e,f,g,h,i,j) are corresponding to three mesons process ((e,f) match with $J/\psi$ plus B meson and light meson, (g,h) match with  $B_{\bar c}$ plus D and light meson, (i,j) match with B plus $\overline D$ and D ). }\label{fig:Feynman_nonleptonic_tetraquark_1}
\end{figure}

  \begin{table}
\caption{Amplitudes for the W-exchange  $T_{bb\bar{q}\bar{q}}$ decays induced by the $b\to c\bar cd/s$ transition. Note that these amplitudes have
an additional identical CKM factor $V_{cb}$.}\label{tab:W-exchange-b-ccq}
\begin{tabular}{|cc|cc|}\hline
channel & amplitude(/$V_{cb}$) &channel & amplitude(/$V_{cb}$) \tabularnewline\hline\hline
$T_{bb\bar{u}\bar{s}}^{-}\to   B^-  J/\psi $ & $ -f_1V_{cs}^*$&
$T_{bb\bar{d}\bar{s}}^{0}\to   \overline B^0  J/\psi $ & $ -f_1V_{cs}^*$
\\\hline
$T_{bb\bar{d}\bar{s}}^{0}\to   \overline B^0_s  J/\psi $ & $ f_1 V_{cd}^*$&
$T_{bb\bar{u}\bar{d}}^{-}\to   B^-  J/\psi $ & $ -f_1 V_{cd}^*$\\\hline
\hline
$T_{bb\bar{u}\bar{s}}^{-}\to    D^0  B_c^- $ & $ -f_2V_{cs}^*$&
$T_{bb\bar{d}\bar{s}}^{0}\to    D^+  B_c^- $ & $ -f_2V_{cs}^*$\\\hline
$T_{bb\bar{d}\bar{s}}^{0}\to    D^+_s  B_c^- $ & $ f_2 V_{cd}^*$&
$T_{bb\bar{u}\bar{d}}^{-}\to    D^0  B_c^- $ & $ -f_2 V_{cd}^*$\\\hline
\hline
\end{tabular}
\end{table}

The transition $b \to c$ or  $ \bar{d}/\bar{s} \to \bar{c}$ can be signed as W-exchange topology, and we plotted the corresponding Feynman diagrams in Fig.~\ref{fig:Feynman_nonleptonic_tetraquark_1}.  The effective Hamiltonian by this kind of transition is:
  \begin{eqnarray}
  \mathcal{H}&=&f_1 (T_{bb3})_{[ij]}(\overline B)^j (H_3)^i J/\psi+  f_2 (T_{bb3})_{[ij]}(\overline D)^j (H_3)^i \overline B_c,
  \end{eqnarray}
where$(H_{  3})_{2}=V_{cd}^*$ and $(H_{  3})_{3}=V_{cs}^*$.
We gave the decay amplitudes in Tab.~\ref{tab:W-exchange-b-ccq}, from which the relations of decay widths are:
\begin{eqnarray*}
    \Gamma(T_{bb\bar{u}\bar{s}}^{-}\to B^-J/\psi)= { }\Gamma(T_{bb\bar{d}\bar{s}}^{0}\to \overline B^0J/\psi),
     \Gamma(T_{bb\bar{u}\bar{d}}^{-}\to B^-J/\psi)= { }\Gamma(T_{bb\bar{d}\bar{s}}^{0}\to \overline B^0_sJ/\psi),\\
    \Gamma(T_{bb\bar{u}\bar{s}}^{-}\to  D^0B_c^-)= { }\Gamma(T_{bb\bar{d}\bar{s}}^{0}\to  D^+B_c^-),\;\;\;\Gamma(T_{bb\bar{u}\bar{d}}^{-}\to  D^0B_c^-)= { }\Gamma(T_{bb\bar{d}\bar{s}}^{0}\to  D^+_sB_c^-).
\end{eqnarray*}

\subsubsection{Decays into an anti-charmed anti-baryon and a charmed bottom baryon}
The transition $b\to c\bar c d/s$ can lead to the process of an anti-baryon plus a baryon, where the anti-charmed anti-baryons form a triplet or anti-sextet and the charmed bottom baryon form a SU(3) triplet. The effective Hamiltonian is described as:
  \begin{eqnarray}
  \mathcal{H}&=&a_1 (T_{bb3})_{[ij]}  (F_{\bar{c}3})^{[jk]} (H_3)^i (\overline F_{bc})_{k} +a_2 (T_{bb3})_{[ij]}  (F_{\bar{c}3})^{[ij]} (H_3)^k (\overline F_{bc})_{k}\nonumber\\
  &&+a_3 (T_{bb3})_{[ij]}  (F_{\bar{c}\bar{6}})^{\{jk\}} (H_3)^i (\overline F_{bc})_{k}.
  \end{eqnarray}
The related decay amplitudes are given in Tab.\ref{tab:bb_Fbc_Fbarc}, in  which class \uppercase\expandafter{\romannumeral1} represents triplet anti-baryon plus the charmed bottom baryon in the final states, and class \uppercase\expandafter{\romannumeral2} denotes the anti-sextet anti-baryon plus the charmed bottom baryon.

  \begin{table}
\caption{Doubly bottom tetraquark $T_{bb\bar{q}\bar{q}}$ decays into  a anti-charmed anti-baryon triplet(class \uppercase\expandafter{\romannumeral1}) or anti-sextet(class  \uppercase\expandafter{\romannumeral2}) and  a charmed bottom baryon. }\label{tab:bb_Fbc_Fbarc}\begin{tabular}{|cc|cc|}\hline
class  \uppercase\expandafter{\romannumeral1} & amplitude(/$V_{cb}$) &class \uppercase\expandafter{\romannumeral2} & amplitude(/$V_{cb}$) \tabularnewline
\hline\hline
$T_{bb\bar{u}\bar{s}}^{-}\to   \Xi_{bc}^{0}  \Lambda_{\bar{c}}^- $ & $ -a_1V_{cs}^*$&$T_{bb\bar{u}\bar{s}}^{-}\to   \Xi_{bc}^{+}  \Sigma_{\bar{c}}^{--} $ & $ -a_3V_{cs}^*$\\\hline
$T_{bb\bar{u}\bar{s}}^{-}\to   \Xi_{bc}^{0}  \Xi_{\bar{c}}^- $ & $ 2 a_2 V_{cd}^*$ & $T_{bb\bar{u}\bar{s}}^{-}\to   \Xi_{bc}^{0}  \Sigma_{\bar{c}}^{-} $ & $ -\frac{a_3V_{cs}^*}{\sqrt{2}}$\\\hline
$T_{bb\bar{u}\bar{s}}^{-}\to   \Omega_{bc}^{0}  \Xi_{\bar{c}}^- $ & $ -\left(a_1-2 a_2\right)V_{cs}^*$&$T_{bb\bar{u}\bar{s}}^{-}\to   \Omega_{bc}^{0}  \Xi_{\bar{c}}^{\prime-} $ & $ -\frac{a_3V_{cs}^*}{\sqrt{2}}$\\\hline
$T_{bb\bar{d}\bar{s}}^{0}\to   \Xi_{bc}^{+}  \Lambda_{\bar{c}}^- $ & $ a_1V_{cs}^*$&$T_{bb\bar{d}\bar{s}}^{0}\to   \Xi_{bc}^{+}  \Sigma_{\bar{c}}^{-} $ & $ -\frac{a_3V_{cs}^*}{\sqrt{2}}$\\\hline
$T_{bb\bar{d}\bar{s}}^{0}\to   \Xi_{bc}^{+}  \Xi_{\bar{c}}^- $ & $ -a_1 V_{cd}^*$&$T_{bb\bar{d}\bar{s}}^{0}\to   \Xi_{bc}^{+}  \Xi_{\bar{c}}^{\prime-} $ & $ \frac{a_3 V_{cd}^*}{\sqrt{2}}$\\\hline
$T_{bb\bar{d}\bar{s}}^{0}\to   \Xi_{bc}^{0}  \overline \Xi_{\bar{c}}^0 $ & $ -\left(a_1-2 a_2\right) V_{cd}^*$&$T_{bb\bar{d}\bar{s}}^{0}\to   \Xi_{bc}^{0}  \overline \Sigma_{\bar{c}}^{0} $ & $ -a_3V_{cs}^*$\\\hline
$T_{bb\bar{d}\bar{s}}^{0}\to   \Omega_{bc}^{0}  \overline \Xi_{\bar{c}}^0 $ & $ -\left(a_1-2 a_2\right)V_{cs}^*$&$T_{bb\bar{d}\bar{s}}^{0}\to   \Xi_{bc}^{0}  \overline \Xi_{\bar{c}}^{\prime0} $ & $ \frac{a_3 V_{cd}^*}{\sqrt{2}}$\\\hline
$T_{bb\bar{u}\bar{d}}^{-}\to   \Xi_{bc}^{0}  \Lambda_{\bar{c}}^- $ & $ -\left(a_1-2 a_2\right) V_{cd}^*$&$T_{bb\bar{d}\bar{s}}^{0}\to   \Omega_{bc}^{0}  \overline \Xi_{\bar{c}}^{\prime0} $ & $ -\frac{a_3V_{cs}^*}{\sqrt{2}}$\\\hline
$T_{bb\bar{u}\bar{d}}^{-}\to   \Omega_{bc}^{0}  \Lambda_{\bar{c}}^- $ & $ 2 a_2V_{cs}^*$&$T_{bb\bar{d}\bar{s}}^{0}\to   \Omega_{bc}^{0}  \overline \Omega_{\bar{c}}^{0} $ & $ a_3 V_{cd}^*$\\\hline
$T_{bb\bar{u}\bar{d}}^{-}\to   \Omega_{bc}^{0}  \Xi_{\bar{c}}^- $ & $ -a_1 V_{cd}^*$&$T_{bb\bar{u}\bar{d}}^{-}\to   \Xi_{bc}^{+}  \Sigma_{\bar{c}}^{--} $ & $ -a_3 V_{cd}^*$\\\hline
& & $T_{bb\bar{u}\bar{d}}^{-}\to   \Xi_{bc}^{0}  \Sigma_{\bar{c}}^{-} $ & $ -\frac{a_3 V_{cd}^*}{\sqrt{2}}$\\\hline
& &$T_{bb\bar{u}\bar{d}}^{-}\to   \Omega_{bc}^{0}  \Xi_{\bar{c}}^{\prime-} $ & $ -\frac{a_3 V_{cd}^*}{\sqrt{2}}$\\\hline
\hline
\end{tabular}
\end{table}
For class \uppercase\expandafter{\romannumeral1}, we have the relations:
\begin{eqnarray*}
    \Gamma(T_{bb\bar{u}\bar{s}}^{-}\to \Omega_{bc}^{0}\Xi_{\bar{c}}^-)= \Gamma(T_{bb\bar{d}\bar{s}}^{0}\to \Omega_{bc}^{0}\overline \Xi_{\bar{c}}^0),
    \Gamma(T_{bb\bar{u}\bar{d}}^{-}\to \Xi_{bc}^{0}\Lambda_{\bar{c}}^-)= { }\Gamma(T_{bb\bar{d}\bar{s}}^{0}\to \Xi_{bc}^{0}\overline \Xi_{\bar{c}}^0),\\
    \Gamma(T_{bb\bar{d}\bar{s}}^{0}\to \Xi_{bc}^{+}\Lambda_{\bar{c}}^-)= { }\Gamma(T_{bb\bar{u}\bar{s}}^{-}\to \Xi_{bc}^{0}\Lambda_{\bar{c}}^-),
    \Gamma(T_{bb\bar{d}\bar{s}}^{0}\to \Xi_{bc}^{+}\Xi_{\bar{c}}^-)= { }\Gamma(T_{bb\bar{u}\bar{d}}^{-}\to \Omega_{bc}^{0}\Xi_{\bar{c}}^-).
\end{eqnarray*}
For class \uppercase\expandafter{\romannumeral2}, we have:
\begin{eqnarray*}
    \Gamma(T_{bb\bar{u}\bar{s}}^{-}\to \Xi_{bc}^{+}\Sigma_{\bar{c}}^{--})= 2\Gamma(T_{bb\bar{u}\bar{s}}^{-}\to \Xi_{bc}^{0}\Sigma_{\bar{c}}^{-})=2\Gamma(T_{bb\bar{u}\bar{s}}^{-}\to \Omega_{bc}^{0}\Xi_{\bar{c}}^{\prime-})=
2\Gamma(T_{bb\bar{d}\bar{s}}^{0}\to \Xi_{bc}^{+}\Sigma_{\bar{c}}^{-})\\=
\Gamma(T_{bb\bar{d}\bar{s}}^{0}\to \Xi_{bc}^{0}\overline \Sigma_{\bar{c}}^{0})=
2\Gamma(T_{bb\bar{d}\bar{s}}^{0}\to \Omega_{bc}^{0}\overline \Xi_{\bar{c}}^{\prime0}),\\
    \Gamma(T_{bb\bar{d}\bar{s}}^{0}\to \Xi_{bc}^{+}\Xi_{\bar{c}}^{\prime-})=\Gamma(T_{bb\bar{d}\bar{s}}^{0}\to \Xi_{bc}^{0}\overline \Xi_{\bar{c}}^{\prime0})=
\frac{1}{2}\Gamma(T_{bb\bar{d}\bar{s}}^{0}\to \Omega_{bc}^{0}\overline \Omega_{\bar{c}}^{0})=
\frac{1}{2}\Gamma(T_{bb\bar{u}\bar{d}}^{-}\to \Xi_{bc}^{+}\Sigma_{\bar{c}}^{--})\\=
\Gamma(T_{bb\bar{u}\bar{d}}^{-}\to \Xi_{bc}^{0}\Sigma_{\bar{c}}^{-})=
\Gamma(T_{bb\bar{u}\bar{d}}^{-}\to \Omega_{bc}^{0}\Xi_{\bar{c}}^{\prime-}).
\end{eqnarray*}

\subsubsection{Decays into three mesons}


\begin{table}
\caption{Doubly bottom tetraquark $T_{bb\bar{q}\bar{q}}$ decays into a $J/\psi$, a bottom meson and a light meson.}\label{tab:bb_Jpsi_Band8meson}\begin{tabular}{|cc|cc|}
\hline
channel  & amplitude(/$V_{cb}$)  & channel  & amplitude(/$V_{cb}$)\tabularnewline
\hline
\hline
$T_{bb\bar{u}\bar{s}}^{-}\to   B^-  \pi^0   J/\psi $ & $ -\frac{a_2V_{cs}^*}{\sqrt{2}}$&
$T_{bb\bar{u}\bar{s}}^{-}\to   B^-  K^0   J/\psi $ & $ a_1 V_{cd}^*$\\\hline
$T_{bb\bar{u}\bar{s}}^{-}\to   B^-  \eta   J/\psi $ & $ -\frac{\left(2 a_1+a_2\right)V_{cs}^*}{\sqrt{6}}$&
$T_{bb\bar{u}\bar{s}}^{-}\to   \overline B^0  \pi^-   J/\psi $ & $ -a_2V_{cs}^*$\\\hline
$T_{bb\bar{u}\bar{s}}^{-}\to   \overline B^0_s  \pi^-   J/\psi $ & $ -a_1 V_{cd}^*$&
$T_{bb\bar{u}\bar{s}}^{-}\to   \overline B^0_s  K^-   J/\psi $ & $ -\left(a_1+a_2\right)V_{cs}^*$\\\hline
$T_{bb\bar{d}\bar{s}}^{0}\to   B^-  \pi^+   J/\psi $ & $ -a_2V_{cs}^*$&
$T_{bb\bar{d}\bar{s}}^{0}\to   B^-  K^+   J/\psi $ & $ a_2 V_{cd}^*$\\\hline
$T_{bb\bar{d}\bar{s}}^{0}\to   \overline B^0  \pi^0   J/\psi $ & $ \frac{a_2V_{cs}^*}{\sqrt{2}}$&
$T_{bb\bar{d}\bar{s}}^{0}\to   \overline B^0  K^0   J/\psi $ & $ \left(a_1+a_2\right) V_{cd}^*$\\\hline
$T_{bb\bar{d}\bar{s}}^{0}\to   \overline B^0  \eta   J/\psi $ & $ -\frac{\left(2 a_1+a_2\right)V_{cs}^*}{\sqrt{6}}$&
$T_{bb\bar{d}\bar{s}}^{0}\to   \overline B^0_s  \pi^0   J/\psi $ & $ \frac{a_1 V_{cd}^*}{\sqrt{2}}$\\\hline
$T_{bb\bar{d}\bar{s}}^{0}\to   \overline B^0_s  \overline K^0   J/\psi $ & $ -\left(a_1+a_2\right)V_{cs}^*$&
$T_{bb\bar{d}\bar{s}}^{0}\to   \overline B^0_s  \eta   J/\psi $ & $ -\frac{\left(a_1+2 a_2\right) V_{cd}^*}{\sqrt{6}}$\\\hline
$T_{bb\bar{u}\bar{d}}^{-}\to   B^-  \pi^0   J/\psi $ & $ -\frac{\left(a_1+a_2\right) V_{cd}^*}{\sqrt{2}}$&
$T_{bb\bar{u}\bar{d}}^{-}\to   B^-  \overline K^0   J/\psi $ & $ a_1V_{cs}^*$\\\hline
$T_{bb\bar{u}\bar{d}}^{-}\to   B^-  \eta   J/\psi $ & $ \frac{\left(a_1-a_2\right) V_{cd}^*}{\sqrt{6}}$&
$T_{bb\bar{u}\bar{d}}^{-}\to   \overline B^0  \pi^-   J/\psi $ & $ -\left(a_1+a_2\right) V_{cd}^*$\\\hline
$T_{bb\bar{u}\bar{d}}^{-}\to   \overline B^0  K^-   J/\psi $ & $ -a_1V_{cs}^*$&
$T_{bb\bar{u}\bar{d}}^{-}\to   \overline B^0_s  K^-   J/\psi $ & $ -a_2 V_{cd}^*$\\\hline
\hline
\end{tabular}
\end{table}
\begin{table}
\caption{Doubly bottom tetraquark $T_{bb\bar{q}\bar{q}}$ decays into a charmed B meson, a charmed meson and a light meson.}\label{tab:bb_Bc_Dand8meson}\begin{tabular}{|cc|cc|}
\hline
channel  & amplitude(/$V_{cb}$)   & channel  & amplitude(/$V_{cb}$)\tabularnewline
\hline
\hline
$T_{bb\bar{u}\bar{s}}^{-}\to    D^0  \pi^0   B_c^- $ & $ -\frac{a_4V_{cs}^*}{\sqrt{2}}$&
$T_{bb\bar{u}\bar{s}}^{-}\to    D^0  K^0   B_c^- $ & $ a_3 V_{cd}^*$\\\hline
$T_{bb\bar{u}\bar{s}}^{-}\to    D^0  \eta   B_c^- $ & $ -\frac{\left(2 a_3+a_4\right)V_{cs}^*}{\sqrt{6}}$&
$T_{bb\bar{u}\bar{s}}^{-}\to    D^+  \pi^-   B_c^- $ & $ -a_4V_{cs}^*$\\\hline
$T_{bb\bar{u}\bar{s}}^{-}\to    D^+_s  \pi^-   B_c^- $ & $ -a_3 V_{cd}^*$&
$T_{bb\bar{u}\bar{s}}^{-}\to    D^+_s  K^-   B_c^- $ & $ -\left(a_3+a_4\right)V_{cs}^*$\\\hline
$T_{bb\bar{d}\bar{s}}^{0}\to    D^0  \pi^+   B_c^- $ & $ -a_4V_{cs}^*$&
$T_{bb\bar{d}\bar{s}}^{0}\to    D^0  K^+   B_c^- $ & $ a_4 V_{cd}^*$\\\hline
$T_{bb\bar{d}\bar{s}}^{0}\to    D^+  \pi^0   B_c^- $ & $ \frac{a_4V_{cs}^*}{\sqrt{2}}$&
$T_{bb\bar{d}\bar{s}}^{0}\to    D^+  K^0   B_c^- $ & $ \left(a_3+a_4\right) V_{cd}^*$\\\hline
$T_{bb\bar{d}\bar{s}}^{0}\to    D^+  \eta   B_c^- $ & $ -\frac{\left(2 a_3+a_4\right)V_{cs}^*}{\sqrt{6}}$&
$T_{bb\bar{d}\bar{s}}^{0}\to    D^+_s  \pi^0   B_c^- $ & $ \frac{a_3 V_{cd}^*}{\sqrt{2}}$\\\hline
$T_{bb\bar{d}\bar{s}}^{0}\to    D^+_s  \overline K^0   B_c^- $ & $ -\left(a_3+a_4\right)V_{cs}^*$&
$T_{bb\bar{d}\bar{s}}^{0}\to    D^+_s  \eta   B_c^- $ & $ -\frac{\left(a_3+2 a_4\right) V_{cd}^*}{\sqrt{6}}$\\\hline
$T_{bb\bar{u}\bar{d}}^{-}\to    D^0  \pi^0   B_c^- $ & $ -\frac{\left(a_3+a_4\right) V_{cd}^*}{\sqrt{2}}$&
$T_{bb\bar{u}\bar{d}}^{-}\to    D^0  \overline K^0   B_c^- $ & $ a_3V_{cs}^*$\\\hline
$T_{bb\bar{u}\bar{d}}^{-}\to    D^0  \eta   B_c^- $ & $ \frac{\left(a_3-a_4\right) V_{cd}^*}{\sqrt{6}}$&
$T_{bb\bar{u}\bar{d}}^{-}\to    D^+  \pi^-   B_c^- $ & $ -\left(a_3+a_4\right) V_{cd}^*$\\\hline
$T_{bb\bar{u}\bar{d}}^{-}\to    D^+  K^-   B_c^- $ & $ -a_3V_{cs}^*$&
$T_{bb\bar{u}\bar{d}}^{-}\to    D^+_s  K^-   B_c^- $ & $ -a_4 V_{cd}^*$\\\hline
\hline
\end{tabular}
\end{table}
\begin{table}
\caption{Doubly bottom tetraquark $T_{bb\bar{q}\bar{q}}$ decays into a B meson, a charmed meson and an anti-charmed meson.}\label{tab:bb_B_DandantiD}\begin{tabular}{|cc|cc|}\hline
channel & amplitude(/$V_{cb}$) &channel & amplitude(/$V_{cb}$) \\\hline\hline
$T_{bb\bar{u}\bar{s}}^{-}\to    D^0  \overline D^0  B^- $ & $ \left(a_5-a_7\right) V_{cs}^*$&
$T_{bb\bar{u}\bar{s}}^{-}\to    D^0  D^-  \overline B^0 $ & $ a_5 V_{cs}^*$\\\hline
$T_{bb\bar{u}\bar{s}}^{-}\to    D^0  D^-  \overline B^0_s $ & $ a_6 V_{cd}^*$&
$T_{bb\bar{u}\bar{s}}^{-}\to    D^0   D^-_s  \overline B^0_s $ & $ \left(a_5+a_6\right) V_{cs}^*$\\\hline
$T_{bb\bar{u}\bar{s}}^{-}\to    D^+  D^-  B^- $ & $ -a_7 V_{cs}^*$&
$T_{bb\bar{u}\bar{s}}^{-}\to    D^+_s  D^-  B^- $ & $ -a_6 V_{cd}^*$\\\hline
$T_{bb\bar{u}\bar{s}}^{-}\to    D^+_s   D^-_s  B^- $ & $ -\left(a_6+a_7\right) V_{cs}^*$&
$T_{bb\bar{d}\bar{s}}^{0}\to    D^0  \overline D^0  \overline B^0 $ & $ -a_7 V_{cs}^*$\\\hline
$T_{bb\bar{d}\bar{s}}^{0}\to    D^0  \overline D^0  \overline B^0_s $ & $ a_7 V_{cd}^*$&
$T_{bb\bar{d}\bar{s}}^{0}\to    D^+  \overline D^0  B^- $ & $ a_5 V_{cs}^*$\\\hline
$T_{bb\bar{d}\bar{s}}^{0}\to    D^+  D^-  \overline B^0 $ & $ \left(a_5-a_7\right) V_{cs}^*$&
$T_{bb\bar{d}\bar{s}}^{0}\to    D^+  D^-  \overline B^0_s $ & $ \left(a_6+a_7\right) V_{cd}^*$\\\hline
$T_{bb\bar{d}\bar{s}}^{0}\to    D^+   D^-_s  \overline B^0_s $ & $ \left(a_5+a_6\right) V_{cs}^*$&
$T_{bb\bar{d}\bar{s}}^{0}\to    D^+_s  \overline D^0  B^- $ & $ -a_5 V_{cd}^*$\\\hline
$T_{bb\bar{d}\bar{s}}^{0}\to    D^+_s  D^-  \overline B^0 $ & $ -\left(a_5+a_6\right) V_{cd}^*$&
$T_{bb\bar{d}\bar{s}}^{0}\to    D^+_s   D^-_s  \overline B^0 $ & $ -\left(a_6+a_7\right) V_{cs}^*$\\\hline
$T_{bb\bar{d}\bar{s}}^{0}\to    D^+_s   D^-_s  \overline B^0_s $ & $ \left(a_7-a_5\right) V_{cd}^*$&
$T_{bb\bar{u}\bar{d}}^{-}\to    D^0  \overline D^0  B^- $ & $ \left(a_5-a_7\right) V_{cd}^*$\\\hline
$T_{bb\bar{u}\bar{d}}^{-}\to    D^0  D^-  \overline B^0 $ & $ \left(a_5+a_6\right) V_{cd}^*$&
$T_{bb\bar{u}\bar{d}}^{-}\to    D^0   D^-_s  \overline B^0 $ & $ a_6 V_{cs}^*$\\\hline
$T_{bb\bar{u}\bar{d}}^{-}\to    D^0   D^-_s  \overline B^0_s $ & $ a_5 V_{cd}^*$&
$T_{bb\bar{u}\bar{d}}^{-}\to    D^+  D^-  B^- $ & $ -\left(a_6+a_7\right) V_{cd}^*$\\\hline
$T_{bb\bar{u}\bar{d}}^{-}\to    D^+   D^-_s  B^- $ & $ -a_6 V_{cs}^*$&
$T_{bb\bar{u}\bar{d}}^{-}\to    D^+_s   D^-_s  B^- $ & $ -a_7 V_{cd}^*$\\\hline
\hline
\end{tabular}
\end{table}
The transition $b\to c\bar c d/s$ leads to three body decays where the effective Hamiltonian becomes:
\begin{eqnarray}
  {\cal H}&= &  a_1 (T_{bb3})_{[ij]} (H_{  3})^k M^{j}_{k} (\overline B)^{i}~ J/\psi+ a_2 (T_{bb3})_{[ij]} (H_{  3})^i M^{j}_{k} (\overline B)^{k}~ J/\psi\nonumber\\
  &&+  a_3 (T_{bb3})_{[ij]} (H_{  3})^k M^{j}_{k} (\overline D)^{i}~ \overline B_c+ a_4 (T_{bb3})_{[ij]} (H_{  3})^i M^{j}_{k} (\overline D)^{k}~ \overline B_c \nonumber\\
  &&+  a_5 (T_{bb3})_{[ij]} (H_{  3})^j D_{k} (\overline D)^{i}~ \overline B^{k}+a_6 (T_{bb3})_{[ij]} (H_{  3})^k D_{k} (\overline D)^{i}~ \overline B^{j}\nonumber\\
  &&+a_7 (T_{bb3})_{[ij]} (H_{  3})^i D_{k} (\overline D)^{k}~ \overline B^{j}.
\end{eqnarray}
The $T_{bb3}$ decays amplitudes into $J/\psi$ plus a bottom meson and a light meson are given in Tab.~\ref{tab:bb_Jpsi_Band8meson}, from which  we have:
\begin{eqnarray*}
    \Gamma(T_{bb\bar{u}\bar{s}}^{-}\to B^- \pi^0 J/\psi)= \frac{1}{2}\Gamma(T_{bb\bar{u}\bar{s}}^{-}\to \overline B^0 \pi^- J/\psi)=\Gamma(T_{bb\bar{d}\bar{s}}^{0}\to \overline B^0 \pi^0 J/\psi)=\frac{1}{8}\Gamma(T_{bb\bar{d}\bar{s}}^{0}\to B^- \pi^+ J/\psi) ,\\
     \Gamma(T_{bb\bar{d}\bar{s}}^{0}\to B^- K^+ J/\psi)= { }\Gamma(T_{bb\bar{u}\bar{d}}^{-}\to \overline B^0_s K^- J/\psi),
     \Gamma(T_{bb\bar{d}\bar{s}}^{0}\to \overline B^0_s \overline K^0 J/\psi)= { }\Gamma(T_{bb\bar{u}\bar{s}}^{-}\to \overline B^0_s K^- J/\psi),\\
     \Gamma(T_{bb\bar{u}\bar{d}}^{-}\to B^- \overline K^0 J/\psi)= { }\Gamma(T_{bb\bar{u}\bar{d}}^{-}\to \overline B^0 K^- J/\psi),
    \Gamma(T_{bb\bar{u}\bar{s}}^{-}\to B^- \eta J/\psi)= { }\Gamma(T_{bb\bar{d}\bar{s}}^{0}\to \overline B^0 \eta J/\psi),\\
    \Gamma(T_{bb\bar{u}\bar{s}}^{-}\to \overline B^0_s \pi^- J/\psi)= { }\Gamma(T_{bb\bar{u}\bar{s}}^{-}\to B^- K^0 J/\psi)=2\Gamma(T_{bb\bar{d}\bar{s}}^{0}\to \overline B^0_s \pi^0 J/\psi),\\
    \Gamma(T_{bb\bar{u}\bar{d}}^{-}\to B^- \pi^0 J/\psi)= \frac{1}{2}\Gamma(T_{bb\bar{d}\bar{s}}^{0}\to \overline B^0 K^0 J/\psi)=     \frac{1}{2}\Gamma(T_{bb\bar{u}\bar{d}}^{-}\to \overline B^0 \pi^- J/\psi).
     \end{eqnarray*}
Decay amplitudes for $T_{bb3}$ decays into $B_c$ meson plus a  charmed  meson and a light meson are given in Tab.~\ref{tab:bb_Bc_Dand8meson}. The decay width relations are:
\begin{eqnarray*}
    \Gamma(T_{bb\bar{u}\bar{s}}^{-}\to  D^0 \pi^0 B_c^-)= \frac{1}{2}\Gamma(T_{bb\bar{u}\bar{s}}^{-}\to  D^+ \pi^- B_c^-)=\Gamma(T_{bb\bar{d}\bar{s}}^{0}\to  D^+ \pi^0 B_c^-)=\frac{1}{8}\Gamma(T_{bb\bar{d}\bar{s}}^{0}\to  D^0 \pi^+ B_c^-),\\
    \Gamma(T_{bb\bar{d}\bar{s}}^{0}\to  D^0 K^+ B_c^-)= { }\Gamma(T_{bb\bar{u}\bar{d}}^{-}\to  D^+_s K^- B_c^-),
     \Gamma(T_{bb\bar{d}\bar{s}}^{0}\to  D^+_s \overline K^0 B_c^-)= { }\Gamma(T_{bb\bar{u}\bar{s}}^{-}\to  D^+_s K^- B_c^-),\\
     \Gamma(T_{bb\bar{u}\bar{d}}^{-}\to  D^0 \overline K^0 B_c^-)= { }\Gamma(T_{bb\bar{u}\bar{d}}^{-}\to  D^+ K^- B_c^-),
    \Gamma(T_{bb\bar{u}\bar{s}}^{-}\to  D^0 \eta B_c^-)= { }\Gamma(T_{bb\bar{d}\bar{s}}^{0}\to  D^+ \eta B_c^-),\\
    \Gamma(T_{bb\bar{u}\bar{s}}^{-}\to  D^+_s \pi^- B_c^-)= { }\Gamma(T_{bb\bar{u}\bar{s}}^{-}\to  D^0 K^0 B_c^-)=2\Gamma(T_{bb\bar{d}\bar{s}}^{0}\to  D^+_s \pi^0 B_c^-),\\
     \Gamma(T_{bb\bar{u}\bar{d}}^{-}\to  D^0 \pi^0 B_c^-)= \frac{1}{2}\Gamma(T_{bb\bar{d}\bar{s}}^{0}\to  D^+ K^0 B_c^-)=
\frac{1}{2}\Gamma(T_{bb\bar{u}\bar{d}}^{-}\to  D^+ \pi^- B_c^-).
    \end{eqnarray*}

The $T_{bb3}$ decays amplitudes into a bottom meson plus a charmed meson and an anti-charmed meson are given in Tab.~\ref{tab:bb_B_DandantiD}. And the relations of decay widths become:
\begin{eqnarray*}
    \Gamma(T_{bb\bar{u}\bar{s}}^{-}\to  D^0 \overline D^0B^-)= { }\Gamma(T_{bb\bar{d}\bar{s}}^{0}\to  D^+ D^-\overline B^0),
    \Gamma(T_{bb\bar{u}\bar{s}}^{-}\to  D^+ D^-B^-)= { }\Gamma(T_{bb\bar{d}\bar{s}}^{0}\to  D^0 \overline D^0\overline B^0),\\
    \Gamma(T_{bb\bar{u}\bar{s}}^{-}\to  D^+_s D^-B^-)= { }\Gamma(T_{bb\bar{u}\bar{s}}^{-}\to  D^0 D^-\overline B^0_s),
    \Gamma(T_{bb\bar{u}\bar{s}}^{-}\to  D^+_s  D^-_sB^-)= { }\Gamma(T_{bb\bar{d}\bar{s}}^{0}\to  D^+_s  D^-_s\overline B^0),\\
    \Gamma(T_{bb\bar{d}\bar{s}}^{0}\to  D^+ \overline D^0B^-)= { }\Gamma(T_{bb\bar{u}\bar{s}}^{-}\to  D^0 D^-\overline B^0),
    \Gamma(T_{bb\bar{d}\bar{s}}^{0}\to  D^+_s \overline D^0B^-)= { }\Gamma(T_{bb\bar{u}\bar{d}}^{-}\to  D^0  D^-_s\overline B^0_s),\\
    \Gamma(T_{bb\bar{u}\bar{d}}^{-}\to  D^0 \overline D^0B^-)= { }\Gamma(T_{bb\bar{d}\bar{s}}^{0}\to  D^+_s  D^-_s\overline B^0_s),
    \Gamma(T_{bb\bar{u}\bar{d}}^{-}\to  D^+ D^-B^-)= { }\Gamma(T_{bb\bar{d}\bar{s}}^{0}\to  D^+ D^-\overline B^0_s),\\
    \Gamma(T_{bb\bar{u}\bar{d}}^{-}\to  D^+  D^-_sB^-)= { }\Gamma(T_{bb\bar{u}\bar{d}}^{-}\to  D^0  D^-_s\overline B^0),
    \Gamma(T_{bb\bar{u}\bar{d}}^{-}\to  D^+_s  D^-_sB^-)= { }\Gamma(T_{bb\bar{d}\bar{s}}^{0}\to  D^0 \overline D^0\overline B^0_s),\\
    \Gamma(T_{bb\bar{u}\bar{s}}^{-}\to  D^0  D^-_s\overline B^0_s)= { }\Gamma(T_{bb\bar{d}\bar{s}}^{0}\to  D^+  D^-_s\overline B^0_s),
    \Gamma(T_{bb\bar{u}\bar{d}}^{-}\to  D^0 D^-\overline B^0)= { }\Gamma(T_{bb\bar{d}\bar{s}}^{0}\to  D^+_s D^-\overline B^0).
    \end{eqnarray*}
\subsection{$b\to c \bar u d/s$ transition}

\begin{figure}
\begin{center}
\includegraphics[scale=0.5]{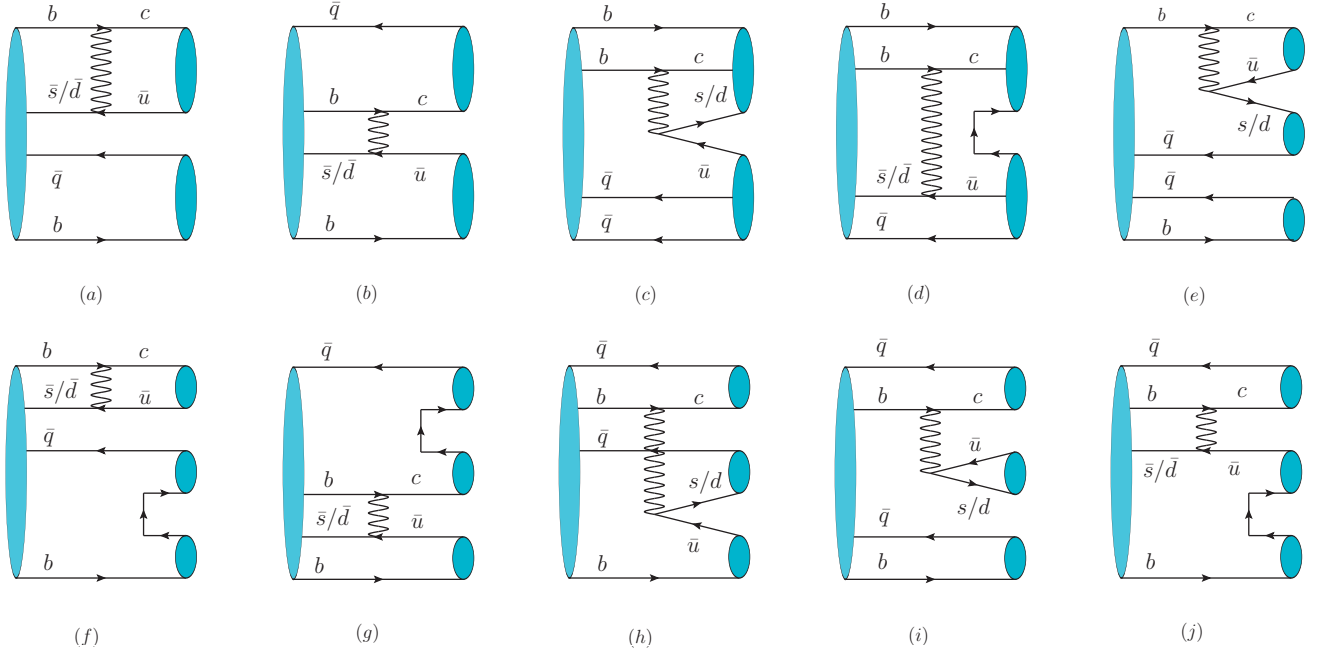}
\end{center}
\caption{Feynman diagrams for nonleptonic decays of  doubly heavy tetraquark. (a,b) are corresponding with two mesons W-exchange process; (c,d) are corresponding with the baryon and anti-baryon process; (e,f,g,h,i,j) are corresponding with the B plus D and light meson process.  }\label{fig:Feynman_nonleptonic_tetraquark_2}
\end{figure}

\subsubsection{Decays into a bottom meson and a charmed meson by W-exchange process}
\begin{table}
\caption{Doubly bottom tetraquark $T_{bb\bar{q}\bar{q}}$ decays into a bottom meson and a charmed meson.}\label{tab:bb_B_D}\begin{tabular}{|cc|cc|}\hline
channel & amplitude(/$V_{cb}$)&channel & amplitude(/$V_{cb}$)\tabularnewline \hline\hline
$T_{bb\bar{u}\bar{s}}^{-}\to   B^-   D^0 $ & $ -\left(f_3+f_4\right) V_{us}^*$&
$T_{bb\bar{d}\bar{s}}^{0}\to   B^-   D^+ $ & $ -f_4 V_{us}^*$\\\hline
$T_{bb\bar{d}\bar{s}}^{0}\to   B^-   D^+_s $ & $ f_4 V_{ud}^*$&
$T_{bb\bar{d}\bar{s}}^{0}\to   \overline B^0   D^0 $ & $ -f_3 V_{us}^*$\\\hline
$T_{bb\bar{d}\bar{s}}^{0}\to   \overline B^0_s   D^0 $ & $ f_3 V_{ud}^*$&
$T_{bb\bar{u}\bar{d}}^{-}\to   B^-   D^0 $ & $ -\left(f_3+f_4\right) V_{ud}^*$\\\hline
\hline
\end{tabular}
\end{table}
For the bottom quark decays to a charm quark, the effective Hamiltonian is given by:
\begin{eqnarray}
{\cal H}_{eff} &=& \frac{G_{F}}{\sqrt{2}}
     V_{cb} V_{uq}^{*} \big[
     C_{1}  O^{\bar cu}_{1}
  +  C_{2}  O^{\bar cu}_{2}\Big] +{\rm h.c.} .
\end{eqnarray}
From the above Hamiltonian, the light quarks reduce an octet where the nonzero component is
$
(H_{{\bf8}})^2_1 =V_{ud}^*
$
for   the $b\to c\bar ud$ or $b\bar d\to c\bar u$   transition, and   $(H_{{\bf8}})^3_1 =V_{us}^*$ for  the $b\to c\bar
us$ or $b\bar s \to c\bar u$ transition. We then obtain the hadron-level effective Hamiltonian
  \begin{eqnarray}
  \mathcal{H}&=&f_3 (T_{bb3})_{[ij]}(\overline B)^j (H_8)^i_k (\overline D)^k+f_4 (T_{bb3})_{[ij]}(\overline B)^k (H_8)^i_k (\overline D)^j.
  \end{eqnarray}
Decays amplitudes are collected in Tab.~\ref{tab:bb_B_D}, from which  the relations of decay widths become:
    \begin{eqnarray*}
    \frac{\Gamma(T_{bb\bar{d}\bar{s}}^{0}\to   B^-   D^+_s )}{\Gamma(T_{bb\bar{d}\bar{s}}^{0}\to   B^-   D^+ )}= \frac{\Gamma(T_{bb\bar{d}\bar{s}}^{0}\to   \overline B^0_s   D^0)}{\Gamma(T_{bb\bar{d}\bar{s}}^{0}\to   \overline B^0   D^0  )}= \frac{\Gamma(T_{bb\bar{u}\bar{d}}^{-}\to   B^-   D^0 )}{\Gamma(T_{bb\bar{u}\bar{s}}^{-}\to   B^-   D^0 )}=\frac{|V_{ud}^*|^2}{|V_{us}^*|^2}.
    \end{eqnarray*}

\subsubsection{Decays into a light anti-baryon and a charmed bottom baryon}
\begin{table}
\caption{Doubly bottom tetraquark $T_{bb\bar{q}\bar{q}}$ decays into a light anti-baryon octet(class \uppercase\expandafter{\romannumeral1}) or anti-decuplet(class \uppercase\expandafter{\romannumeral2})and a charmed bottom baryon.}\label{tab:bb_Fbc_Fbar}\begin{tabular}{|cc|cc|}\hline
class \uppercase\expandafter{\romannumeral1} & amplitude(/$V_{cb}$) &class \uppercase\expandafter{\romannumeral2} & amplitude(/$V_{cb}$) \tabularnewline
\hline\hline
$T_{bb\bar{u}\bar{s}}^{-}\to   \overline \Sigma^- \Xi_{bc}^{0} $ & $ -2 a_7 V_{ud}^*$&$T_{bb\bar{u}\bar{s}}^{-}\to   \overline \Delta^{--}  \Xi_{bc}^{+} $ & $ -a_9 V_{us}^*$\\\hline
$T_{bb\bar{u}\bar{s}}^{-}\to   \overline \Sigma^- \Omega_{bc}^{0} $ & $ \left(a_5+a_6-2 a_7\right) V_{us}^*$&$T_{bb\bar{u}\bar{s}}^{-}\to   \overline \Delta^{-}  \Xi_{bc}^{0} $ & $ -\frac{a_9 V_{us}^*}{\sqrt{3}}$\\\hline
$T_{bb\bar{u}\bar{s}}^{-}\to   \overline p  \Xi_{bc}^{0} $ & $ -\left(a_5+a_6\right) V_{us}^*$&$T_{bb\bar{u}\bar{s}}^{-}\to   \overline \Sigma^{\prime-}  \Omega_{bc}^{0} $ & $ -\frac{a_9 V_{us}^*}{\sqrt{3}}$\\\hline
$T_{bb\bar{d}\bar{s}}^{0}\to   \overline \Lambda^0  \Xi_{bc}^{0} $ & $ \frac{\left(a_4-a_5-2 a_6+2 a_7\right) V_{ud}^*}{\sqrt{6}}$&$T_{bb\bar{d}\bar{s}}^{0}\to   \overline \Delta^{-}  \Xi_{bc}^{+} $ & $ -\frac{a_9 V_{us}^*}{\sqrt{3}}$\\\hline
$T_{bb\bar{d}\bar{s}}^{0}\to   \overline \Lambda^0  \Omega_{bc}^{0} $ & $ \frac{\left(-2 a_4-a_5+a_6+2 a_7\right) V_{us}^*}{\sqrt{6}}$&$T_{bb\bar{d}\bar{s}}^{0}\to   \overline \Delta^{0}  \Xi_{bc}^{0} $ & $ -\frac{a_9 V_{us}^*}{\sqrt{3}}$\\\hline
$T_{bb\bar{d}\bar{s}}^{0}\to   \overline \Sigma^- \Xi_{bc}^{+} $ & $ \left(a_4+a_5\right) V_{ud}^*$&$T_{bb\bar{d}\bar{s}}^{0}\to   \overline \Sigma^{\prime-}  \Xi_{bc}^{+} $ & $ \frac{a_9 V_{ud}^*}{\sqrt{3}}$\\\hline
$T_{bb\bar{d}\bar{s}}^{0}\to   \overline \Sigma^0  \Xi_{bc}^{0} $ & $ -\frac{\left(a_4+a_5-2 a_7\right) V_{ud}^*}{\sqrt{2}}$&$T_{bb\bar{d}\bar{s}}^{0}\to   \overline \Sigma^{\prime0}  \Xi_{bc}^{0} $ & $ \frac{a_9 V_{ud}^*}{\sqrt{6}}$\\\hline
$T_{bb\bar{d}\bar{s}}^{0}\to   \overline \Sigma^0  \Omega_{bc}^{0} $ & $ -\frac{\left(a_5+a_6-2 a_7\right) V_{us}^*}{\sqrt{2}}$&$T_{bb\bar{d}\bar{s}}^{0}\to   \overline \Sigma^{\prime0}  \Omega_{bc}^{0} $ & $ -\frac{a_9 V_{us}^*}{\sqrt{6}}$\\\hline
$T_{bb\bar{d}\bar{s}}^{0}\to   \overline p  \Xi_{bc}^{+} $ & $ \left(a_4+a_5\right) V_{us}^*$&$T_{bb\bar{d}\bar{s}}^{0}\to   \overline \Xi^{\prime0}  \Omega_{bc}^{0} $ & $ \frac{a_9 V_{ud}^*}{\sqrt{3}}$\\\hline
$T_{bb\bar{d}\bar{s}}^{0}\to   \overline n  \Xi_{bc}^{0} $ & $ \left(a_4-a_6\right) V_{us}^*$&$T_{bb\bar{u}\bar{d}}^{-}\to   \overline \Delta^{--}  \Xi_{bc}^{+} $ & $ -a_9 V_{ud}^*$\\\hline
$T_{bb\bar{d}\bar{s}}^{0}\to   \overline \Xi^0  \Omega_{bc}^{0} $ & $ \left(a_4-a_6\right) V_{ud}^*$&$T_{bb\bar{u}\bar{d}}^{-}\to   \overline \Delta^{-}  \Xi_{bc}^{0} $ & $ -\frac{a_9 V_{ud}^*}{\sqrt{3}}$\\\hline
$T_{bb\bar{u}\bar{d}}^{-}\to   \overline \Sigma^- \Omega_{bc}^{0} $ & $ \left(a_5+a_6\right) V_{ud}^*$&$T_{bb\bar{u}\bar{d}}^{-}\to   \overline \Sigma^{\prime-}  \Omega_{bc}^{0} $ & $ -\frac{a_9 V_{ud}^*}{\sqrt{3}}$\\\hline
$T_{bb\bar{u}\bar{d}}^{-}\to   \overline p  \Xi_{bc}^{0} $ & $ -\left(a_5+a_6-2 a_7\right) V_{ud}^*$& &\\\hline
$T_{bb\bar{u}\bar{d}}^{-}\to   \overline p  \Omega_{bc}^{0} $ & $ 2 a_7V_{us}^*$& &\\\hline
\hline
\end{tabular}
\end{table}
There are two kinds of multiplet for the final states, which lead to the Hamiltonian
  \begin{eqnarray}
  \mathcal{H}&=&a_4 (T_{bb3})_{[ij]} \epsilon^{xjk} (F_{8})^{l}_x (H_8)^i_k (\overline F_{bc})_{l}+a_5 (T_{bb3})_{[ij]} \epsilon^{xjl} (F_{8})^{k}_x (H_8)^i_k (\overline F_{bc})_{l}\nonumber\\
  &&+a_6 (T_{bb3})_{[ij]} \epsilon^{xkl} (F_{8})^{j}_x (H_8)^i_k (\overline F_{bc})_{l}+a_7 (T_{bb3})_{[ij]} \epsilon^{xij} (F_{8})^{k}_x (H_8)^l_k (\overline F_{bc})_{l}\nonumber\\
  &&+a_8 (T_{bb3})_{[ij]} (F_{\overline {10}})^{\{jkl\}} (H_8)^i_k (\overline F_{bc})_{l}+\overline a_7 (T_{bb3})_{[ij]} \epsilon^{xik} (F_{8})^{j}_x (H_8)^l_k (\overline F_{bc})_{l}.
  \end{eqnarray}
 Decay amplitudes are presented in Tab.~\ref{tab:bb_Fbc_Fbar}, where different final states are labeled with  class \uppercase\expandafter{\romannumeral1} or \uppercase\expandafter{\romannumeral2}. Note that the factor $2a_7+\overline a_7$ always appear in the results, thus we remove the $\overline a_7$ in the final results.
For  class \uppercase\expandafter{\romannumeral1}, we have the  relations:
    \begin{eqnarray*}
    \Gamma(T_{bb\bar{u}\bar{s}}^{-}\to \Sigma^-\Omega_{bc}^{0})= 2\Gamma(T_{bb\bar{d}\bar{s}}^{0}\to \overline \Sigma^0\Omega_{bc}^{0}).
    \end{eqnarray*}
For class \uppercase\expandafter{\romannumeral2}, we have the  relations:
\begin{eqnarray*}
    \Gamma(T_{bb\bar{u}\bar{s}}^{-}\to \overline \Delta^{--}\Xi_{bc}^{+})= 3\Gamma(T_{bb\bar{u}\bar{s}}^{-}\to \overline \Delta^{-}\Xi_{bc}^{0})=
3\Gamma(T_{bb\bar{u}\bar{s}}^{-}\to \overline \Sigma^{\prime-}\Omega_{bc}^{0})=
3\Gamma(T_{bb\bar{d}\bar{s}}^{0}\to \overline \Delta^{0}\Xi_{bc}^{0})\\=
6\Gamma(T_{bb\bar{d}\bar{s}}^{0}\to \overline \Sigma^{\prime0}\Omega_{bc}^{0})=
3\Gamma(T_{bb\bar{d}\bar{s}}^{0}\to \overline \Delta^{-}\Xi_{bc}^{+}).\\
     \Gamma(T_{bb\bar{d}\bar{s}}^{0}\to \overline \Sigma^{\prime-}\Xi_{bc}^{+})=2\Gamma(T_{bb\bar{d}\bar{s}}^{0}\to \overline \Sigma^{\prime0}\Xi_{bc}^{0})=
\Gamma(T_{bb\bar{d}\bar{s}}^{0}\to \overline \Xi^{\prime0}\Omega_{bc}^{0})=
\Gamma(T_{bb\bar{u}\bar{d}}^{-}\to \overline \Delta^{-}\Xi_{bc}^{0})\\=
\Gamma(T_{bb\bar{u}\bar{d}}^{-}\to \overline \Sigma^{\prime-}\Omega_{bc}^{0})=
\frac{1}{3}\Gamma(T_{bb\bar{u}\bar{d}}^{-}\to \overline \Delta^{--}\Xi_{bc}^{+}).
\end{eqnarray*}
\subsubsection{ Decays into a bottom meson, a charmed meson and a light meson. }
\begin{table}
\caption{Doubly bottom tetraquark $T_{bb\bar{q}\bar{q}}$ decays into a bottom meson, a charmed meson and a light meson.}\label{tab:bb_B_Dand8meson}\begin{tabular}{|cc|cc|}
\hline
channel  & amplitude(/$V_{cb}$)  & channel  & amplitude(/$V_{cb}$)\tabularnewline
\hline
\hline
$T_{bb\bar{u}\bar{s}}^{-}\to   B^-   D^0  \pi^0  $ & $ \frac{\left(a_5-a_8-a_9-a_{10}\right) V_{us}^*}{\sqrt{2}}$&
$T_{bb\bar{u}\bar{s}}^{-}\to   B^-   D^0  K^0  $ & $ \left(a_7+a_{11}\right) V_{ud}^*$\\\hline
$T_{bb\bar{u}\bar{s}}^{-}\to   B^-   D^0  \eta  $ & $ \frac{\left(a_5-2 a_7-a_8-a_9-a_{10}-2 a_{11}\right) V_{us}^*}{\sqrt{6}}$&
$T_{bb\bar{u}\bar{s}}^{-}\to   B^-   D^+  \pi^-  $ & $ \left(a_5-a_9\right) V_{us}^*$\\\hline
$T_{bb\bar{u}\bar{s}}^{-}\to   B^-   D^+_s  \pi^-  $ & $ \left(a_6-a_{11}\right) V_{ud}^*$&
$T_{bb\bar{u}\bar{s}}^{-}\to   B^-   D^+_s  K^-  $ & $ \left(a_5+a_6-a_9-a_{11}\right) V_{us}^*$\\\hline
$T_{bb\bar{u}\bar{s}}^{-}\to   \overline B^0   D^0  \pi^-  $ & $ -\left(a_8+a_{10}\right) V_{us}^*$&
$T_{bb\bar{u}\bar{s}}^{-}\to   \overline B^0_s   D^0
\pi^-  $ & $ -\left(a_6+a_7\right) V_{ud}^*$\\\hline
$T_{bb\bar{u}\bar{s}}^{-}\to   \overline B^0_s   D^0  K^-  $ & $ -\left(a_6+a_7+a_8+a_{10}\right) V_{us}^*$&
$T_{bb\bar{d}\bar{s}}^{0}\to   B^-   D^0  \pi^+  $ & $ -\left(a_9+a_{10}\right) V_{us}^*$\\\hline
$T_{bb\bar{d}\bar{s}}^{0}\to   B^-   D^0  K^+  $ & $ \left(a_9+a_{10}\right) V_{ud}^*$&
$T_{bb\bar{d}\bar{s}}^{0}\to   B^-   D^+  \pi^0  $ & $ \frac{\left(a_9-a_8\right) V_{us}^*}{\sqrt{2}}$\\\hline
$T_{bb\bar{d}\bar{s}}^{0}\to   B^-   D^+  K^0  $ & $ \left(a_9+a_{11}\right) V_{ud}^*$&
$T_{bb\bar{d}\bar{s}}^{0}\to   B^-   D^+  \eta  $ & $ -\frac{\left(a_8+a_9+2 a_{11}\right) V_{us}^*}{\sqrt{6}}$\\\hline
$T_{bb\bar{d}\bar{s}}^{0}\to   B^-   D^+_s  \pi^0  $ & $ \frac{\left(a_8+a_{11}\right) V_{ud}^*}{\sqrt{2}}$&
$T_{bb\bar{d}\bar{s}}^{0}\to   B^-   D^+_s  \overline K^0  $ & $ -\left(a_9+a_{11}\right) V_{us}^*$\\\hline
$T_{bb\bar{d}\bar{s}}^{0}\to   B^-   D^+_s  \eta  $ & $ \frac{\left(a_8-2 a_9-a_{11}\right) V_{ud}^*}{\sqrt{6}}$&
$T_{bb\bar{d}\bar{s}}^{0}\to   \overline B^0   D^0  \pi^0  $ & $ \frac{\left(a_5+a_{10}\right) V_{us}^*}{\sqrt{2}}$\\\hline
$T_{bb\bar{d}\bar{s}}^{0}\to   \overline B^0   D^0  K^0  $ & $ \left(a_7+a_{10}\right) V_{ud}^*$&
$T_{bb\bar{d}\bar{s}}^{0}\to   \overline B^0   D^0  \eta  $ & $ \frac{\left(a_5-2 a_7-a_{10}\right) V_{us}^*}{\sqrt{6}}$\\\hline
$T_{bb\bar{d}\bar{s}}^{0}\to   \overline B^0   D^+  \pi^-  $ & $ \left(a_5-a_8\right) V_{us}^*$&
$T_{bb\bar{d}\bar{s}}^{0}\to   \overline B^0   D^+_s  \pi^-  $ & $ \left(a_6+a_8\right) V_{ud}^*$\\\hline
$T_{bb\bar{d}\bar{s}}^{0}\to   \overline B^0   D^+_s  K^-  $ & $ \left(a_5+a_6\right) V_{us}^*$&
$T_{bb\bar{d}\bar{s}}^{0}\to   \overline B^0_s   D^0  \pi^0  $ & $ \frac{\left(a_7-a_5\right) V_{ud}^*}{\sqrt{2}}$\\\hline
$T_{bb\bar{d}\bar{s}}^{0}\to   \overline B^0_s   D^0  \overline K^0  $ & $ -\left(a_7+a_{10}\right) V_{us}^*$&
$T_{bb\bar{d}\bar{s}}^{0}\to   \overline B^0_s   D^0  \eta  $ & $ -\frac{\left(a_5+a_7+2 a_{10}\right) V_{ud}^*}{\sqrt{6}}$\\\hline
$T_{bb\bar{d}\bar{s}}^{0}\to   \overline B^0_s   D^+  \pi^-  $ & $ -\left(a_5+a_6\right) V_{ud}^*$&
$T_{bb\bar{d}\bar{s}}^{0}\to   \overline B^0_s   D^+  K^-  $ & $ -\left(a_6+a_8\right) V_{us}^*$\\\hline
$T_{bb\bar{d}\bar{s}}^{0}\to   \overline B^0_s   D^+_s  K^-  $ & $ \left(a_8-a_5\right) V_{ud}^*$&
$T_{bb\bar{u}\bar{d}}^{-}\to   B^-   D^0  \pi^0  $ & $ \frac{\left(a_5-a_7-a_8-a_9-a_{10}-a_{11}\right) V_{ud}^*}{\sqrt{2}}$\\\hline
$T_{bb\bar{u}\bar{d}}^{-}\to   B^-   D^0  \overline K^0  $ & $ \left(a_7+a_{11}\right) V_{us}^*$&
$T_{bb\bar{u}\bar{d}}^{-}\to   B^-   D^0  \eta  $ & $ \frac{\left(a_5+a_7-a_8-a_9-a_{10}+a_{11}\right) V_{ud}^*}{\sqrt{6}}$\\\hline
$T_{bb\bar{u}\bar{d}}^{-}\to   B^-   D^+  \pi^-  $ & $ \left(a_5+a_6-a_9-a_{11}\right) V_{ud}^*$&
$T_{bb\bar{u}\bar{d}}^{-}\to   B^-   D^+  K^-  $ & $ \left(a_6-a_{11}\right) V_{us}^*$\\\hline
$T_{bb\bar{u}\bar{d}}^{-}\to   B^-   D^+_s  K^-  $ & $ \left(a_5-a_9\right) V_{ud}^*$&
$T_{bb\bar{u}\bar{d}}^{-}\to   \overline B^0   D^0  \pi^-  $ & $ -\left(a_6+a_7+a_8+a_{10}\right) V_{ud}^*$\\\hline
$T_{bb\bar{u}\bar{d}}^{-}\to   \overline B^0   D^0  K^-  $ & $ -\left(a_6+a_7\right) V_{us}^*$&
$T_{bb\bar{u}\bar{d}}^{-}\to   \overline B^0_s   D^0  K^-  $ & $ -\left(a_8+a_{10}\right) V_{ud}^*$\\\hline
\hline
\end{tabular}
\end{table}
The effective Hamiltonian for decays into a bottom meson, a charmed meson and a light meson is
\begin{eqnarray}
  {\cal H}&=& a_5 (T_{bb3})_{[ij]} (\overline B)^i (\overline D)^l M^k_l (H_{{\bf8}})^j_k+ a_6 (T_{bb3})_{[ij]} (\overline B)^i (\overline D)^j M^k_l (H_{{\bf8}})^l_k\nonumber \\
&&+a_7 (T_{bb3})_{[ij]} (\overline B)^i (\overline D)^k M^j_l (H_{{\bf8}})^l_k+a_8 (T_{bb3})_{[ij]} (\overline B)^l (\overline D)^j M^k_l (H_{{\bf8}})^i_k \nonumber \\
&&+a_9 (T_{bb3})_{[ij]} (\overline B)^l (\overline D)^k M^j_k (H_{{\bf8}})^i_l+a_{10} (T_{bb3})_{[ij]} (\overline B)^k (\overline D)^l M^j_k (H_{{\bf8}})^i_l\nonumber\\
&&+a_{11} (T_{bb3})_{[ij]} (\overline B)^l (\overline D)^i M^j_k (H_{{\bf8}})^k_l.
\end{eqnarray}
Decay amplitudes are collected  in Tab.~\ref{tab:bb_B_Dand8meson}, where no relation for decay widths is found.
\subsection{$b\to u \bar c d/s$ transition}

\begin{figure}
\begin{center}
\includegraphics[scale=0.45]{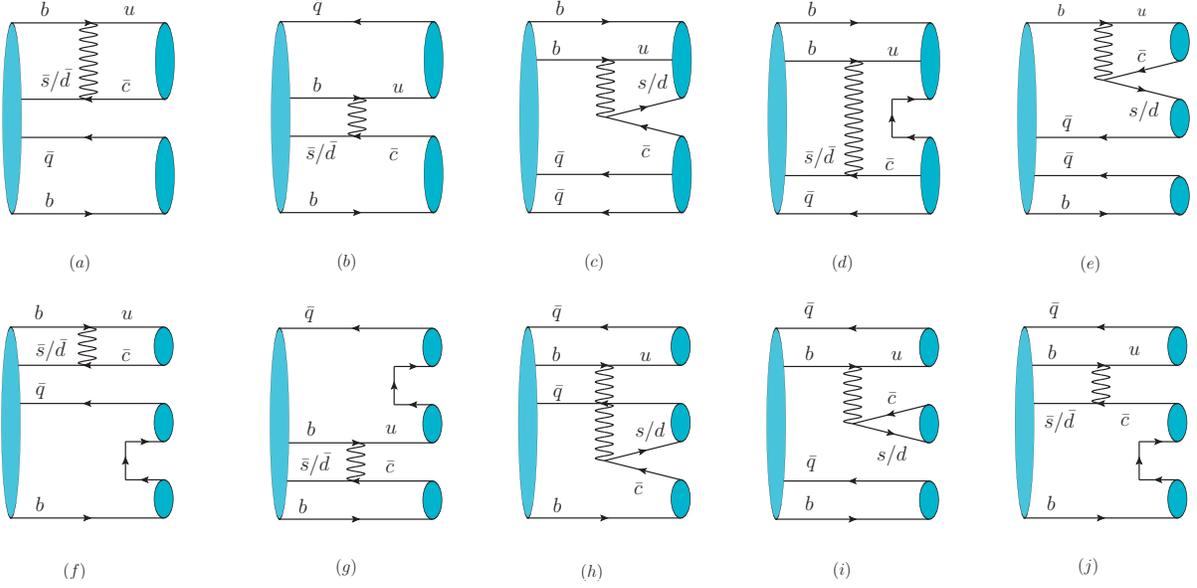}
\end{center}
\caption{Feynman diagrams for nonleptonic decays of  doubly bottom tetraquark. (a,b) are corresponding with B plus $\overline D$ or $B_c$ plus light meson W-exchange process respectively. (c,d) are corresponding with the baryon and anti-baryon process, (e,f,i,j) math with B plus $\overline D$ and light meson. (g,h) match with $B_c$ plus two light mesons. }\label{fig:Feynman_nonleptonic_tetraquark_3}
\end{figure}

\subsubsection{Decays into two mesons by W-exchange process}
\begin{table}
\caption{Doubly bottom tetraquark $T_{bb\bar{q}\bar{q}}$ decays into a bottom meson and an anti-charmed meson. Note that these amplitudes have
an additional identical CKM factor $V_{ub}$.}\label{tab:bb_B_antiD}\begin{tabular}{|cc|cc|}\hline
channel & amplitude(/$V_{ub}$)&channel & amplitude(/$V_{ub}$)\tabularnewline \hline\hline
$T_{bb\bar{u}\bar{s}}^{-}\to   B^-  \overline D^0 $ & $ \left(2 f_5-f_6-f_7\right)V_{cs}^*$&
$T_{bb\bar{u}\bar{s}}^{-}\to   \overline B^0  D^- $ & $ 2 f_5V_{cs}^*$\\\hline
$T_{bb\bar{u}\bar{s}}^{-}\to   \overline B^0_s  D^- $ & $ \left(f_7-f_6\right) V_{cd}^*$&
$T_{bb\bar{u}\bar{s}}^{-}\to   \overline B^0_s   D^-_s $ & $ \left(2 f_5-f_6+f_7\right)V_{cs}^*$\\\hline
$T_{bb\bar{d}\bar{s}}^{0}\to   \overline B^0  \overline D^0 $ & $ -\left(f_6+f_7\right)V_{cs}^*$&
$T_{bb\bar{d}\bar{s}}^{0}\to   \overline B^0_s  \overline D^0 $ & $ \left(f_6+f_7\right) V_{cd}^*$\\\hline
$T_{bb\bar{u}\bar{d}}^{-}\to   B^-  \overline D^0 $ & $ \left(2 f_5-f_6-f_7\right) V_{cd}^*$&
$T_{bb\bar{u}\bar{d}}^{-}\to   \overline B^0  D^- $ & $ \left(2 f_5-f_6+f_7\right) V_{cd}^*$\\\hline
$T_{bb\bar{u}\bar{d}}^{-}\to   \overline B^0   D^-_s $ & $ \left(f_7-f_6\right)V_{cs}^*$&
$T_{bb\bar{u}\bar{d}}^{-}\to   \overline B^0_s   D^-_s $ & $ 2 f_5 V_{cd}^*$\\\hline
\hline
\end{tabular}
\end{table}
    \begin{table}
\caption{Doubly bottom tetraquark $T_{bb\bar{q}\bar{q}}$ decays into an anti-charmed B meson and a light mesons.}\label{tab:bb_Bcandmeson8}\begin{tabular}{|cc|cc|}\hline
channel & amplitude(/$V_{ub}$) &channel & amplitude(/$V_{ub}$) \tabularnewline\hline\hline
$T_{bb\bar{u}\bar{s}}^{-}\to   B_c^-  \pi^0  $ & $ \frac{\left(f_8-f_9\right)V_{cs}^*}{\sqrt{2}}$&
$T_{bb\bar{u}\bar{s}}^{-}\to   B_c^-  K^0  $ & $ \left(f_8+f_9\right) V_{cd}^*$\\\hline
$T_{bb\bar{u}\bar{s}}^{-}\to   B_c^-  \eta  $ & $ -\frac{\left(f_8+3 f_9\right)V_{cs}^*}{\sqrt{6}}$&
$T_{bb\bar{d}\bar{s}}^{0}\to   B_c^-  \pi^+  $ & $ \left(f_8-f_9\right)V_{cs}^*$\\\hline
$T_{bb\bar{d}\bar{s}}^{0}\to   B_c^-  K^+  $ & $ \left(f_9-f_8\right) V_{cd}^*$&
$T_{bb\bar{u}\bar{d}}^{-}\to   B_c^-  \pi^0  $ & $ -\sqrt{2} f_9 V_{cd}^*$\\\hline
$T_{bb\bar{u}\bar{d}}^{-}\to   B_c^-  \overline K^0  $ & $ \left(f_8+f_9\right)V_{cs}^*$&
$T_{bb\bar{u}\bar{d}}^{-}\to   B_c^-  \eta  $ & $ \sqrt{\frac{2}{3}} f_8 V_{cd}^*$\\\hline
\hline
\end{tabular}
\end{table}
We write the effective Hamiltonian for the anti-charm quark production as
\begin{eqnarray}
{\cal H}_{eff} &=& \frac{G_{F}}{\sqrt{2}}
     V_{ub} V_{cq}^{*} \big[
     C_{1}  O^{\bar uc}_{1}
  +  C_{2}  O^{\bar uc}_{2}\Big]+ {\rm h.c.}.
\end{eqnarray}
According to the flavor SU(3) group,
the $H_{\bar 3}''$  is anti-symmetric while the $H_{ 6}$ is symmetric.
The nonzero components are
$  (H_{\bar 3}'')^{13} =- (H_{\bar 3}'')^{31} =V_{cs}^*$ and $ (H_{\bar 6})^{13}=(H_{\bar 6})^{31} =V_{cs}^*,
$
for the $b\to u\bar cs$  transition. When interchange of $2\leftrightarrow 3$  and $s\leftrightarrow d$, we get the  nonzero components
   for the transition $b\to
u\bar cd$.

We get the effective  Hamiltonian
  \begin{eqnarray}
  \mathcal{H}&=&f_5 (T_{bb3})_{[ij]}(\overline B)^k (H_3^{\prime\prime})^{[ij]} (D)_k+f_6 (T_{bb3})_{[ij]}(\overline B)^j (H_3^{\prime\prime})^{[ik]} (D)_k\nonumber\\
  &&+f_7 (T_{bb3})_{[ij]}(\overline B)^j (H_6^{\prime\prime})^{\{ik\}} (D)_k +  f_8 (T_{bb3})_{[ij]}(H_3^{\prime\prime})^{[ik]} M^j_k \overline B_c \nonumber\\
  &&+f_9 (T_{bb3})_{[ij]}(H_6^{\prime\prime})^{\{ik\}} M^j_k \overline B_c.
  \end{eqnarray}
 For a bottom meson and an  anti-charmed meson produced, the amplitudes are given in Tab.~\ref{tab:bb_B_antiD}. And we have:
 \begin{eqnarray*}
    \frac{\Gamma(T_{bb\bar{u}\bar{s}}^{-}\to   \overline B^0  D^- )}{\Gamma(T_{bb\bar{u}\bar{d}}^{-}\to   \overline B^0_s   D^-_s  )}&=& \frac{\Gamma(T_{bb\bar{d}\bar{s}}^{0}\to   \overline B^0  \overline D^0)}{\Gamma(T_{bb\bar{d}\bar{s}}^{0}\to   \overline B^0_s  \overline D^0 )}= \frac{\Gamma(T_{bb\bar{u}\bar{d}}^{-}\to   \overline B^0   D^-_s  )}{\Gamma( T_{bb\bar{u}\bar{s}}^{-}\to   \overline B^0_s  D^- )}\\&=& \frac{\Gamma(T_{bb\bar{u}\bar{s}}^{-}\to   \overline B^0_s   D^-_s )}{\Gamma(T_{bb\bar{u}\bar{d}}^{-}\to   \overline B^0  D^- )}= \frac{\Gamma(T_{bb\bar{u}\bar{s}}^{-}\to   B^-  \overline D^0 )}{\Gamma(T_{bb\bar{u}\bar{d}}^{-}\to   B^-  \overline D^0 )}=\frac{|V_{cs}^*|^2}{|V_{cd}^*|^2}.
    \end{eqnarray*}


$T_{bb3}$ decays amplitudes  into a charmed bottom meson and a light meson for different channels are given in Tab.~\ref{tab:bb_Bcandmeson8}.
And we have
\begin{eqnarray*}
    \Gamma(T_{bb\bar{d}\bar{s}}^{0}\to B_c^-\pi^+ )= 2\Gamma(T_{bb\bar{u}\bar{s}}^{-}\to B_c^-\pi^0 ).
\end{eqnarray*}

\subsubsection{ Decays into an anti-charmed anti-baryon  and a bottom baryon }
    \begin{table}
\caption{Doubly bottom tetraquark $T_{bb\bar{q}\bar{q}}$ decays into an anti-charmed anti-baryon triplet and a bottom baryon anti-triplet(class \uppercase\expandafter{\romannumeral1}) or sextet(class \uppercase\expandafter{\romannumeral2}),an anti-charmed anti-baryon anti-sextet and a bottom baryon anti-triplet(class \uppercase\expandafter{\romannumeral3}) or sextet(class \uppercase\expandafter{\romannumeral4}).}\label{tab:bb_Fbarc_Fb}\begin{tabular}{|cc|cc|}\hline

class \uppercase\expandafter{\romannumeral1}& amplitude(/$V_{ub}$) &class \uppercase\expandafter{\romannumeral4} & amplitude(/$V_{ub}$)\tabularnewline\hline\hline
$T_{bb\bar{u}\bar{s}}^{-}\to   \Lambda_{\bar{c}}^-  \Lambda_b^0 $ & $ \left(4 b_1+b_2-b_4\right)V_{cs}^*$ &$T_{bb\bar{d}\bar{s}}^{0}\to   \overline \Omega_{\bar{c}}^{0}  \Xi_{b}^{\prime0} $ & $ -\frac{b_{11} V_{cd}^*}{\sqrt{2}}$\\\hline
$T_{bb\bar{u}\bar{s}}^{-}\to   \Xi_{\bar{c}}^-  \Lambda_b^0 $ & $ \left(b_2+4 b_3+b_4\right) V_{cd}^*$&$T_{bb\bar{u}\bar{d}}^{-}\to   \Sigma_{\bar{c}}^{--}  \Sigma_{b}^{+} $ & $ \left(2 b_{10}+b_{11}\right) V_{cd}^*$\\\hline
$T_{bb\bar{u}\bar{s}}^{-}\to   \Xi_{\bar{c}}^-  \Xi_b^0 $ & $ 2 \left(2 b_1+b_2+2 b_3\right)V_{cs}^*$&$T_{bb\bar{u}\bar{d}}^{-}\to   \Sigma_{\bar{c}}^{-}  \Sigma_{b}^{0} $ & $ \left(2 b_{10}+b_{11}\right) V_{cd}^*$\\\hline
$T_{bb\bar{u}\bar{s}}^{-}\to   \overline \Xi_{\bar{c}}^0  \Xi_b^- $ & $ \left(4 b_1+b_2+b_4\right)V_{cs}^*$&$T_{bb\bar{u}\bar{d}}^{-}\to   \Sigma_{\bar{c}}^{-}  \Xi_{b}^{\prime0} $ & $ \frac{1}{2} b_{11}V_{cs}^*$\\\hline
$T_{bb\bar{d}\bar{s}}^{0}\to   \overline \Xi_{\bar{c}}^0  \Lambda_b^0 $ & $ \left(b_2+4 b_3-b_4\right) V_{cd}^*$&$T_{bb\bar{u}\bar{d}}^{-}\to   \overline \Sigma_{\bar{c}}^{0}  \Sigma_{b}^{-} $ & $ \left(2 b_{10}+b_{11}\right) V_{cd}^*$\\\hline
$T_{bb\bar{d}\bar{s}}^{0}\to   \overline \Xi_{\bar{c}}^0  \Xi_b^0 $ & $ \left(b_2+4 b_3-b_4\right)V_{cs}^*$&$T_{bb\bar{u}\bar{d}}^{-}\to   \overline \Sigma_{\bar{c}}^{0}  \Xi_{b}^{\prime-} $ & $ \frac{b_{11}V_{cs}^*}{\sqrt{2}}$\\\hline
$T_{bb\bar{u}\bar{d}}^{-}\to   \Lambda_{\bar{c}}^-  \Lambda_b^0 $ & $ 2 \left(2 b_1+b_2+2 b_3\right) V_{cd}^*$&$T_{bb\bar{u}\bar{d}}^{-}\to   \Xi_{\bar{c}}^{\prime-}  \Xi_{b}^{\prime0} $ & $ \frac{1}{2} \left(4 b_{10}+b_{11}\right) V_{cd}^*$\\\hline
$T_{bb\bar{u}\bar{d}}^{-}\to   \Lambda_{\bar{c}}^-  \Xi_b^0 $ & $ \left(b_2+4 b_3+b_4\right)V_{cs}^*$&$T_{bb\bar{u}\bar{d}}^{-}\to   \overline \Xi_{\bar{c}}^{\prime0}  \Xi_{b}^{\prime-} $ & $ \frac{1}{2} \left(4 b_{10}+b_{11}\right) V_{cd}^*$\\\hline
$T_{bb\bar{u}\bar{d}}^{-}\to   \Xi_{\bar{c}}^-  \Xi_b^0 $ & $ \left(4 b_1+b_2-b_4\right) V_{cd}^*$&$T_{bb\bar{u}\bar{d}}^{-}\to   \overline \Xi_{\bar{c}}^{\prime0}  \Omega_{b}^{-} $ & $ \frac{b_{11}V_{cs}^*}{\sqrt{2}}$\\\hline
$T_{bb\bar{u}\bar{d}}^{-}\to   \overline \Xi_{\bar{c}}^0  \Xi_b^- $ & $ \left(4 b_1+b_2+b_4\right) V_{cd}^*$&$T_{bb\bar{u}\bar{d}}^{-}\to   \overline \Omega_{\bar{c}}^{0}  \Omega_{b}^{-} $ & $ 2 b_{10} V_{cd}^*$\\\hline
class \uppercase\expandafter{\romannumeral3}& amplitude(/$V_{ub}$) &$T_{bb\bar{d}\bar{s}}^{0}\to   \overline \Xi_{\bar{c}}^{\prime0}  \Xi_{b}^{\prime0} $ & $ \frac{1}{2} b_{11}V_{cs}^*$\\\hline
$T_{bb\bar{u}\bar{s}}^{-}\to   \Sigma_{\bar{c}}^{-}  \Lambda_b^0 $ & $ \frac{\left(b_8-b_9\right)V_{cs}^*}{\sqrt{2}}$&$T_{bb\bar{u}\bar{s}}^{-}\to   \Sigma_{\bar{c}}^{--}  \Sigma_{b}^{+} $ & $ \left(2 b_{10}+b_{11}\right)V_{cs}^*$\\\hline
$T_{bb\bar{u}\bar{s}}^{-}\to   \Xi_{\bar{c}}^{\prime-}  \Lambda_b^0 $ & $ -\frac{\left(b_8+b_9\right) V_{cd}^*}{\sqrt{2}}$&$T_{bb\bar{u}\bar{s}}^{-}\to   \Sigma_{\bar{c}}^{-}  \Sigma_{b}^{0} $ & $ \frac{1}{2} \left(4 b_{10}+b_{11}\right)V_{cs}^*$\\\hline
$T_{bb\bar{u}\bar{s}}^{-}\to   \Xi_{\bar{c}}^{\prime-}  \Xi_b^0 $ & $ -\sqrt{2} b_9V_{cs}^*$&$T_{bb\bar{u}\bar{s}}^{-}\to   \overline \Sigma_{\bar{c}}^{0}  \Sigma_{b}^{-} $ & $ 2 b_{10}V_{cs}^*$\\\hline
$T_{bb\bar{u}\bar{s}}^{-}\to   \overline \Xi_{\bar{c}}^{\prime0}  \Xi_b^- $ & $ -\frac{\left(b_8+b_9\right)V_{cs}^*}{\sqrt{2}}$&$T_{bb\bar{u}\bar{s}}^{-}\to   \Xi_{\bar{c}}^{\prime-}  \Sigma_{b}^{0} $ & $ \frac{1}{2} b_{11} V_{cd}^*$\\\hline
$T_{bb\bar{u}\bar{s}}^{-}\to   \overline \Omega_{\bar{c}}^{0}  \Xi_b^- $ & $ \left(b_8+b_9\right) V_{cd}^*$&$T_{bb\bar{u}\bar{s}}^{-}\to   \Xi_{\bar{c}}^{\prime-}  \Xi_{b}^{\prime0} $ & $ \left(2 b_{10}+b_{11}\right)V_{cs}^*$\\\hline
$T_{bb\bar{d}\bar{s}}^{0}\to   \overline \Sigma_{\bar{c}}^{0}  \Lambda_b^0 $ & $ \left(b_8-b_9\right)V_{cs}^*$&$T_{bb\bar{u}\bar{s}}^{-}\to   \overline \Xi_{\bar{c}}^{\prime0}  \Sigma_{b}^{-} $ & $ \frac{b_{11} V_{cd}^*}{\sqrt{2}}$\\\hline
$T_{bb\bar{d}\bar{s}}^{0}\to   \overline \Xi_{\bar{c}}^{\prime0}  \Lambda_b^0 $ & $ \frac{\left(b_9-b_8\right) V_{cd}^*}{\sqrt{2}}$&$T_{bb\bar{u}\bar{s}}^{-}\to   \overline \Xi_{\bar{c}}^{\prime0}  \Xi_{b}^{\prime-} $ & $ \frac{1}{2} \left(4 b_{10}+b_{11}\right)V_{cs}^*$\\\hline
$T_{bb\bar{d}\bar{s}}^{0}\to   \overline \Xi_{\bar{c}}^{\prime0}  \Xi_b^0 $ & $ \frac{\left(b_8-b_9\right)V_{cs}^*}{\sqrt{2}}$&$T_{bb\bar{u}\bar{s}}^{-}\to   \overline \Omega_{\bar{c}}^{0}  \Xi_{b}^{\prime-} $ & $ \frac{b_{11} V_{cd}^*}{\sqrt{2}}$\\\hline
$T_{bb\bar{d}\bar{s}}^{0}\to   \overline \Omega_{\bar{c}}^{0}  \Xi_b^0 $ & $ \left(b_9-b_8\right) V_{cd}^*$&$T_{bb\bar{u}\bar{s}}^{-}\to   \overline \Omega_{\bar{c}}^{0}  \Omega_{b}^{-} $ & $ \left(2 b_{10}+b_{11}\right)V_{cs}^*$\\\hline
$T_{bb\bar{u}\bar{d}}^{-}\to   \Sigma_{\bar{c}}^{-}  \Lambda_b^0 $ & $ -\sqrt{2} b_9 V_{cd}^*$&$T_{bb\bar{d}\bar{s}}^{0}\to   \Sigma_{\bar{c}}^{-}  \Sigma_{b}^{+} $ & $ \frac{b_{11}V_{cs}^*}{\sqrt{2}}$\\\hline
$T_{bb\bar{u}\bar{d}}^{-}\to   \Sigma_{\bar{c}}^{-}  \Xi_b^0 $ & $ -\frac{\left(b_8+b_9\right)V_{cs}^*}{\sqrt{2}}$&$T_{bb\bar{d}\bar{s}}^{0}\to   \overline \Sigma_{\bar{c}}^{0}  \Sigma_{b}^{0} $ & $ \frac{b_{11}V_{cs}^*}{\sqrt{2}}$\\\hline
$T_{bb\bar{u}\bar{d}}^{-}\to   \overline \Sigma_{\bar{c}}^{0}  \Xi_b^- $ & $ -\left(b_8+b_9\right)V_{cs}^*$&$T_{bb\bar{d}\bar{s}}^{0}\to   \Xi_{\bar{c}}^{\prime-}  \Sigma_{b}^{+} $ & $ -\frac{b_{11} V_{cd}^*}{\sqrt{2}}$\\\hline
$T_{bb\bar{u}\bar{d}}^{-}\to   \Xi_{\bar{c}}^{\prime-}  \Xi_b^0 $ & $ \frac{\left(b_8-b_9\right) V_{cd}^*}{\sqrt{2}}$&$T_{bb\bar{d}\bar{s}}^{0}\to   \overline \Xi_{\bar{c}}^{\prime0}  \Sigma_{b}^{0} $ & $ -\frac{1}{2} b_{11} V_{cd}^*$\\\hline
$T_{bb\bar{u}\bar{d}}^{-}\to   \overline \Xi_{\bar{c}}^{\prime0}  \Xi_b^- $ & $ \frac{\left(b_8+b_9\right) V_{cd}^*}{\sqrt{2}}$&&\\
\hline
class \uppercase\expandafter{\romannumeral2}& amplitude(/$V_{ub}$) & class \uppercase\expandafter{\romannumeral2} & amplitude(/$V_{ub}$)\tabularnewline
\hline\hline
$T_{bb\bar{u}\bar{s}}^{-}\to   \Lambda_{\bar{c}}^-  \Sigma_{b}^{0} $ & $ \frac{\left(b_5-b_6\right)V_{cs}^*}{\sqrt{2}}$&
$T_{bb\bar{u}\bar{s}}^{-}\to   \Xi_{\bar{c}}^-  \Sigma_{b}^{0} $ & $ -\frac{\left(b_5+b_6-4 b_7\right) V_{cd}^*}{\sqrt{2}}$\\\hline
$T_{bb\bar{u}\bar{s}}^{-}\to   \Xi_{\bar{c}}^-  \Xi_{b}^{\prime0} $ & $ -\sqrt{2} \left(b_6-2 b_7\right)V_{cs}^*$&
$T_{bb\bar{u}\bar{s}}^{-}\to   \overline \Xi_{\bar{c}}^0  \Sigma_{b}^{-} $ & $ -\left(b_5+b_6\right) V_{cd}^*$\\\hline
$T_{bb\bar{u}\bar{s}}^{-}\to   \overline \Xi_{\bar{c}}^0  \Xi_{b}^{\prime-} $ & $ -\frac{\left(b_5+b_6\right)V_{cs}^*}{\sqrt{2}}$&
$T_{bb\bar{d}\bar{s}}^{0}\to   \Lambda_{\bar{c}}^-  \Sigma_{b}^{+} $ & $ \left(b_6-b_5\right)V_{cs}^*$\\\hline
$T_{bb\bar{d}\bar{s}}^{0}\to   \Xi_{\bar{c}}^-  \Sigma_{b}^{+} $ & $ \left(b_5-b_6\right) V_{cd}^*$&
$T_{bb\bar{d}\bar{s}}^{0}\to   \overline \Xi_{\bar{c}}^0  \Sigma_{b}^{0} $ & $ \frac{\left(b_5-b_6+4 b_7\right) V_{cd}^*}{\sqrt{2}}$\\\hline
$T_{bb\bar{d}\bar{s}}^{0}\to   \overline \Xi_{\bar{c}}^0  \Xi_{b}^{\prime0} $ & $ \frac{\left(b_5-b_6+4 b_7\right)V_{cs}^*}{\sqrt{2}}$&
$T_{bb\bar{u}\bar{d}}^{-}\to   \Lambda_{\bar{c}}^-  \Sigma_{b}^{0} $ & $ -\sqrt{2} \left(b_6-2 b_7\right) V_{cd}^*$\\\hline
$T_{bb\bar{u}\bar{d}}^{-}\to   \Lambda_{\bar{c}}^-  \Xi_{b}^{\prime0} $ & $ -\frac{\left(b_5+b_6-4 b_7\right)V_{cs}^*}{\sqrt{2}}$&
$T_{bb\bar{u}\bar{d}}^{-}\to   \Xi_{\bar{c}}^-  \Xi_{b}^{\prime0} $ & $ \frac{\left(b_5-b_6\right) V_{cd}^*}{\sqrt{2}}$\\\hline
$T_{bb\bar{u}\bar{d}}^{-}\to   \overline \Xi_{\bar{c}}^0  \Xi_{b}^{\prime-} $ & $ \frac{\left(b_5+b_6\right) V_{cd}^*}{\sqrt{2}}$&
$T_{bb\bar{u}\bar{d}}^{-}\to   \overline \Xi_{\bar{c}}^0  \Omega_{b}^{-} $ & $ \left(b_5+b_6\right)V_{cs}^*$\\\hline
\hline
\end{tabular}
\end{table}
The effective Hamiltonian for decays into an anti-charmed anti-baryon  and a bottom baryon is
  \begin{eqnarray}
  \mathcal{H}_{eff}&=&b_1 (T_{bb3})_{[ij]} (F_{\bar{c}3})^{[kl]} (H_3^{\prime\prime})^{[ij]} (\overline F_{b\bar{3}})_{[kl]}+b_2 (T_{bb3})_{[ij]} (F_{\bar{c}3})^{[jl]} (H_3^{\prime\prime})^{[ik]} (\overline F_{b\bar{3}})_{[kl]}\nonumber\\
  &&+b_3 (T_{bb3})_{[ij]} (F_{\bar{c}3})^{[ij]} (H_3^{\prime\prime})^{[kl]} (\overline F_{b\bar{3}})_{[kl]}+b_4 (T_{bb3})_{[ij]} (F_{\bar{c}3})^{[jl]} (H_6^{\prime\prime})^{\{ik\}} (\overline F_{b\bar{3}})_{[kl]}\nonumber\\
  &&+b_5 (T_{bb3})_{[ij]} (F_{\bar{c}3})^{[jl]} (H_3^{\prime\prime})^{[ik]} (\overline F_{b6})_{\{kl\}}+b_6 (T_{bb3})_{[ij]} (F_{\bar{c}3})^{[jl]} (H_6^{\prime\prime})^{\{ik\}} (\overline F_{b6})_{\{kl\}}\nonumber\\
  &&+b_7 (T_{bb3})_{[ij]} (F_{\bar{c}3})^{[ij]} (H_6^{\prime\prime})^{\{kl\}} (\overline F_{b6})_{\{kl\}}+b_8 (T_{bb3})_{[ij]} (F_{\bar{c}\bar{6}})^{\{jl\}} (H_3^{\prime\prime})^{[ik]} (\overline F_{b\bar{3}})_{[kl]}\nonumber\\
  &&+b_9 (T_{bb3})_{[ij]} (F_{\bar{c}\bar{6}})^{\{jl\}} (H_6^{\prime\prime})^{\{ik\}} (\overline F_{b\bar{3}})_{[kl]}+b_{10} (T_{bb3})_{[ij]} (F_{\bar{c}\bar{6}})^{\{kl\}} (H_3^{\prime\prime})^{[ij]} (\overline F_{b6})_{\{kl\}}\nonumber\\
  &&+b_{11} (T_{bb3})_{[ij]} (F_{\bar{c}\bar{6}})^{\{jl\}} (H_3^{\prime\prime})^{[ik]} (\overline F_{b6})_{\{kl\}}.
  \end{eqnarray}
Different  decay channel  amplitudes are given in Tab.~\ref{tab:bb_Fbarc_Fb}, where class \uppercase\expandafter{\romannumeral1} corresponds with triplet anti-baryon plus anti-triplet baryon;  class \uppercase\expandafter{\romannumeral2} corresponds with triplet anti-baryon plus sextet baryon; class \uppercase\expandafter{\romannumeral3} corresponds with anti-sextet anti-baryon plus anti-triplet baryon; class \uppercase\expandafter{\romannumeral4} corresponds with anti-sextet anti-baryon plus sextet baryon.

For class \uppercase\expandafter{\romannumeral1}, we obtain the relations of decay widths:
 \begin{eqnarray*}
    \frac{\Gamma(T_{bb\bar{u}\bar{s}}^{-}\to   \Lambda_{\bar{c}}^-  \Lambda_b^0 )}{\Gamma(T_{bb\bar{u}\bar{d}}^{-}\to   \Xi_{\bar{c}}^-  \Xi_b^0 )}&=& \frac{\Gamma(T_{bb\bar{u}\bar{s}}^{-}\to   \Xi_{\bar{c}}^-  \Xi_b^0 )}{\Gamma(T_{bb\bar{u}\bar{d}}^{-}\to   \Lambda_{\bar{c}}^-  \Lambda_b^0  )}= \frac{\Gamma(T_{bb\bar{u}\bar{d}}^{-}\to   \Lambda_{\bar{c}}^-  \Xi_b^0)}{\Gamma( T_{bb\bar{u}\bar{s}}^{-}\to   \Xi_{\bar{c}}^-  \Lambda_b^0)}\\&=& \frac{\Gamma(T_{bb\bar{u}\bar{s}}^{-}\to   \overline \Xi_{\bar{c}}^0  \Xi_b^-  )}{\Gamma(T_{bb\bar{u}\bar{d}}^{-}\to   \overline \Xi_{\bar{c}}^0  \Xi_b^- )}= \frac{\Gamma(T_{bb\bar{d}\bar{s}}^{0}\to   \overline \Xi_{\bar{c}}^0  \Xi_b^0 )}{\Gamma(T_{bb\bar{d}\bar{s}}^{0}\to   \overline \Xi_{\bar{c}}^0  \Lambda_b^0 )}=\frac{|V_{cs}^*|^2}{|V_{cd}^*|^2}.
    \end{eqnarray*}

For class \uppercase\expandafter{\romannumeral2}, we obtain the relations of decay widths:
\begin{eqnarray*}
    \Gamma(T_{bb\bar{u}\bar{s}}^{-}\to \overline \Xi_{\bar{c}}^0\Sigma_{b}^{-})= 2\Gamma(T_{bb\bar{u}\bar{d}}^{-}\to \overline \Xi_{\bar{c}}^0\Xi_{b}^{\prime-}),
    \Gamma(T_{bb\bar{u}\bar{s}}^{-}\to \overline \Xi_{\bar{c}}^0\Xi_{b}^{\prime-})= \frac{1}{2}\Gamma(T_{bb\bar{u}\bar{d}}^{-}\to \overline \Xi_{\bar{c}}^0\Omega_{b}^{-}),\\
    \Gamma(T_{bb\bar{d}\bar{s}}^{0}\to \Lambda_{\bar{c}}^-\Sigma_{b}^{+})= 2\Gamma(T_{bb\bar{u}\bar{s}}^{-}\to \Lambda_{\bar{c}}^-\Sigma_{b}^{0}),
    \Gamma(T_{bb\bar{d}\bar{s}}^{0}\to \Xi_{\bar{c}}^-\Sigma_{b}^{+})= 2\Gamma(T_{bb\bar{u}\bar{d}}^{-}\to \Xi_{\bar{c}}^-\Xi_{b}^{\prime0}).
\end{eqnarray*}
For class \uppercase\expandafter{\romannumeral3}, we obtain the relations of decay widths:
\begin{eqnarray*}
    \Gamma(T_{bb\bar{u}\bar{s}}^{-}\to \Sigma_{\bar{c}}^{-}\Lambda_b^0)= { }\Gamma(T_{bb\bar{d}\bar{s}}^{0}\to \overline \Xi_{\bar{c}}^{\prime0}\Xi_b^0)=\frac{1}{2}\Gamma(T_{bb\bar{d}\bar{s}}^{0}\to \overline \Sigma_{\bar{c}}^{0}\Lambda_b^0).\\
    \Gamma(T_{bb\bar{u}\bar{s}}^{-}\to \Xi_{\bar{c}}^{\prime-}\Lambda_b^0)=\Gamma(T_{bb\bar{u}\bar{d}}^{-}\to \overline \Xi_{\bar{c}}^{\prime0}\Xi_b^-)= \frac{1}{2}\Gamma(T_{bb\bar{u}\bar{s}}^{-}\to \overline \Omega_{\bar{c}}^{0}\Xi_b^-),\\
    \Gamma(T_{bb\bar{d}\bar{s}}^{0}\to \overline \Xi_{\bar{c}}^{\prime0}\Lambda_b^0)=\Gamma(T_{bb\bar{u}\bar{d}}^{-}\to \Xi_{\bar{c}}^{\prime-}\Xi_b^0)= \frac{1}{2}\Gamma(T_{bb\bar{d}\bar{s}}^{0}\to \overline \Omega_{\bar{c}}^{0}\Xi_b^0),\\
    \Gamma(T_{bb\bar{u}\bar{d}}^{-}\to \Sigma_{\bar{c}}^{-}\Xi_b^0)= { }\Gamma(T_{bb\bar{u}\bar{s}}^{-}\to \overline \Xi_{\bar{c}}^{\prime0}\Xi_b^-)=\frac{1}{2}\Gamma(T_{bb\bar{u}\bar{d}}^{-}\to \overline \Sigma_{\bar{c}}^{0}\Xi_b^-).
\end{eqnarray*}
For class \uppercase\expandafter{\romannumeral4}, the results are:
\begin{eqnarray*}
    \Gamma(T_{bb\bar{u}\bar{s}}^{-}\to \Xi_{\bar{c}}^{\prime-}\Sigma_{b}^{0})=\Gamma(T_{bb\bar{d}\bar{s}}^{0}\to \overline \Xi_{\bar{c}}^{\prime0}\Sigma_{b}^{0})= \frac{1}{2}\Gamma(T_{bb\bar{u}\bar{s}}^{-}\to \overline \Xi_{\bar{c}}^{\prime0}\Sigma_{b}^{-})=\frac{1}{2}\Gamma(T_{bb\bar{u}\bar{s}}^{-}\to \overline \Omega_{\bar{c}}^{0}\Xi_{b}^{\prime-})\\=\frac{1}{2}\Gamma(T_{bb\bar{d}\bar{s}}^{0}\to \overline \Omega_{\bar{c}}^{0}\Xi_{b}^{\prime0})=\frac{1}{2}\Gamma(T_{bb\bar{d}\bar{s}}^{0}\to \Xi_{\bar{c}}^{\prime-}\Sigma_{b}^{+}),\\
    \Gamma(T_{bb\bar{d}\bar{s}}^{0}\to \Sigma_{\bar{c}}^{-}\Sigma_{b}^{+})= { }\Gamma(T_{bb\bar{d}\bar{s}}^{0}\to \overline \Sigma_{\bar{c}}^{0}\Sigma_{b}^{0})=2\Gamma(T_{bb\bar{d}\bar{s}}^{0}\to \overline \Xi_{\bar{c}}^{\prime0}\Xi_{b}^{\prime0})=2\Gamma(T_{bb\bar{u}\bar{d}}^{-}\to \Sigma_{\bar{c}}^{-}\Xi_{b}^{\prime0})\\
    =\Gamma(T_{bb\bar{u}\bar{d}}^{-}\to \overline \Sigma_{\bar{c}}^{0}\Xi_{b}^{\prime-})=\Gamma(T_{bb\bar{u}\bar{d}}^{-}\to \overline \Xi_{\bar{c}}^{\prime0}\Omega_{b}^{-}),\\
    \Gamma(T_{bb\bar{u}\bar{d}}^{-}\to \Sigma_{\bar{c}}^{--}\Sigma_{b}^{+})= \Gamma(T_{bb\bar{u}\bar{d}}^{-}\to \Sigma_{\bar{c}}^{-}\Sigma_{b}^{0})=\Gamma(T_{bb\bar{u}\bar{d}}^{-}\to \overline \Sigma_{\bar{c}}^{0}\Sigma_{b}^{-}),\\
    \Gamma(T_{bb\bar{u}\bar{s}}^{-}\to \Sigma_{\bar{c}}^{--}\Sigma_{b}^{+})= { }\Gamma(T_{bb\bar{u}\bar{s}}^{-}\to \Xi_{\bar{c}}^{\prime-}\Xi_{b}^{\prime0})=\Gamma(T_{bb\bar{u}\bar{s}}^{-}\to \overline \Omega_{\bar{c}}^{0}\Omega_{b}^{-}),\\
    \Gamma(T_{bb\bar{u}\bar{d}}^{-}\to \Xi_{\bar{c}}^{\prime-}\Xi_{b}^{\prime0})= { }\Gamma(T_{bb\bar{u}\bar{d}}^{-}\to \overline \Xi_{\bar{c}}^{\prime0}\Xi_{b}^{\prime-}),
        \Gamma(T_{bb\bar{u}\bar{s}}^{-}\to \Sigma_{\bar{c}}^{-}\Sigma_{b}^{0})= { }\Gamma(T_{bb\bar{u}\bar{s}}^{-}\to \overline \Xi_{\bar{c}}^{\prime0}\Xi_{b}^{\prime-}).
\end{eqnarray*}

\subsubsection{ Decays into three mesons  }
\begin{table}
\footnotesize
\caption{Doubly bottom tetraquark $T_{bb\bar{q}\bar{q}}$ decays into a bottom meson, an anti-charmed meson and a light meson.}\label{tab:bb_B_antiDand8meson}\begin{tabular}{|cc|cc|}
\hline
channel  & amplitude(/$V_{ub}$)  & channel  & amplitude(/$V_{ub}$)\tabularnewline
\hline
\hline
$T_{bb\bar{u}\bar{s}}^{-}\to   B^-  \overline D^0  \pi^0  $ & $ \frac{\left(-b_1+2 b_3+b_4+b_5+b_6-b_8-b_9\right)V_{cs}^*}{\sqrt{2}}$&
$T_{bb\bar{u}\bar{s}}^{-}\to   B^-  \overline D^0  K^0  $ & $ \left(-b_2+b_5+b_7+b_9\right) V_{cd}^*$\\\hline
$T_{bb\bar{u}\bar{s}}^{-}\to   B^-  \overline D^0  \eta  $ & $ -\frac{\left(b_1-2 b_2-2 b_3-b_4+b_5-b_6+2 b_7+b_8+3 b_9\right)V_{cs}^*}{\sqrt{6}}$&
$T_{bb\bar{u}\bar{s}}^{-}\to   B^-  D^-  \pi^+  $ & $ \left(-b_1+2 b_3+b_6\right)V_{cs}^*$\\\hline
$T_{bb\bar{u}\bar{s}}^{-}\to   B^-  D^-  K^+  $ & $ \left(b_2+b_4+b_7+b_8\right) V_{cd}^*$&
$T_{bb\bar{u}\bar{s}}^{-}\to   B^-   D^-_s  K^+  $ & $ \left(-b_1+b_2+2 b_3+b_4+b_6+b_7+b_8\right)V_{cs}^*$\\\hline
$T_{bb\bar{u}\bar{s}}^{-}\to   \overline B^0  \overline D^0  \pi^-  $ & $ \left(2 b_3+b_4-b_8\right)V_{cs}^*$&
$T_{bb\bar{u}\bar{s}}^{-}\to   \overline B^0  D^-  \pi^0  $ & $ \frac{\left(-2 b_3+b_5-b_9\right)V_{cs}^*}{\sqrt{2}}$\\\hline
$T_{bb\bar{u}\bar{s}}^{-}\to   \overline B^0  D^-  K^0  $ & $ \left(b_4+b_5+b_8+b_9\right) V_{cd}^*$&
$T_{bb\bar{u}\bar{s}}^{-}\to   \overline B^0  D^-  \eta  $ & $ \frac{\left(2 b_3-b_5-3 b_9\right)V_{cs}^*}{\sqrt{6}}$\\\hline
$T_{bb\bar{u}\bar{s}}^{-}\to   \overline B^0   D^-_s  K^0  $ & $ \left(2 b_3+b_4+b_8\right)V_{cs}^*$&
$T_{bb\bar{u}\bar{s}}^{-}\to   \overline B^0_s  \overline D^0  \pi^-  $ & $ -\left(b_1-b_2+b_6+b_7\right) V_{cd}^*$\\\hline
$T_{bb\bar{u}\bar{s}}^{-}\to   \overline B^0_s  \overline D^0  K^-  $ & $ -\left(b_1-b_2-2 b_3-b_4+b_6+b_7+b_8\right)V_{cs}^*$&
$T_{bb\bar{u}\bar{s}}^{-}\to   \overline B^0_s  D^-  \pi^0  $ & $ \frac{\left(b_1-b_2+b_6-b_7\right) V_{cd}^*}{\sqrt{2}}$\\\hline
$T_{bb\bar{u}\bar{s}}^{-}\to   \overline B^0_s  D^-  \overline K^0  $ & $ -\left(b_1-2 b_3+b_6\right)V_{cs}^*$&
$T_{bb\bar{u}\bar{s}}^{-}\to   \overline B^0_s  D^-  \eta  $ & $ -\frac{\left(b_1+b_2+2 b_4+b_6+b_7+2 b_8\right) V_{cd}^*}{\sqrt{6}}$\\\hline
$T_{bb\bar{u}\bar{s}}^{-}\to   \overline B^0_s   D^-_s  \pi^0  $ & $ -\frac{\left(b_2-b_5+b_7+b_9\right)V_{cs}^*}{\sqrt{2}}$&
$T_{bb\bar{u}\bar{s}}^{-}\to   \overline B^0_s   D^-_s  K^0  $ & $ \left(-b_1+b_5-b_6+b_9\right) V_{cd}^*$\\\hline
$T_{bb\bar{u}\bar{s}}^{-}\to   \overline B^0_s   D^-_s  \eta  $ & $ \frac{\left(2 b_1-b_2-4 b_3-2 b_4-b_5+2 b_6-b_7-2 b_8-3 b_9\right)V_{cs}^*}{\sqrt{6}}$&
$T_{bb\bar{d}\bar{s}}^{0}\to   B^-  \overline D^0  \pi^+  $ & $ \left(b_4+b_5-b_8-b_9\right)V_{cs}^*$\\\hline
$T_{bb\bar{d}\bar{s}}^{0}\to   B^-  \overline D^0  K^+  $ & $ \left(-b_4-b_5+b_8+b_9\right) V_{cd}^*$&
$T_{bb\bar{d}\bar{s}}^{0}\to   \overline B^0  \overline D^0  \pi^0  $ & $ \frac{\left(-b_1-b_4+b_6+b_8\right)V_{cs}^*}{\sqrt{2}}$\\\hline
$T_{bb\bar{d}\bar{s}}^{0}\to   \overline B^0  \overline D^0  K^0  $ & $ \left(-b_2-b_4+b_7+b_8\right) V_{cd}^*$&
$T_{bb\bar{d}\bar{s}}^{0}\to   \overline B^0  \overline D^0  \eta  $ & $ \frac{\left(-b_1+2 b_2+b_4+b_6-2 b_7-b_8\right)V_{cs}^*}{\sqrt{6}}$\\\hline
$T_{bb\bar{d}\bar{s}}^{0}\to   \overline B^0  D^-  \pi^+  $ & $ \left(-b_1+b_5+b_6-b_9\right)V_{cs}^*$&
$T_{bb\bar{d}\bar{s}}^{0}\to   \overline B^0  D^-  K^+  $ & $ \left(b_2-b_5+b_7+b_9\right) V_{cd}^*$\\\hline
$T_{bb\bar{d}\bar{s}}^{0}\to   \overline B^0   D^-_s  K^+  $ & $ \left(-b_1+b_2+b_6+b_7\right)V_{cs}^*$&
$T_{bb\bar{d}\bar{s}}^{0}\to   \overline B^0_s  \overline D^0  \pi^0  $ & $ \frac{\left(b_1-b_2-b_6+b_7\right) V_{cd}^*}{\sqrt{2}}$\\\hline
$T_{bb\bar{d}\bar{s}}^{0}\to   \overline B^0_s  \overline D^0  \overline K^0  $ & $ \left(b_2+b_4-b_7-b_8\right)V_{cs}^*$&
$T_{bb\bar{d}\bar{s}}^{0}\to   \overline B^0_s  \overline D^0  \eta  $ & $ \frac{\left(b_1+b_2+2 b_4-b_6-b_7-2 b_8\right) V_{cd}^*}{\sqrt{6}}$\\\hline
$T_{bb\bar{d}\bar{s}}^{0}\to   \overline B^0_s  D^-  \pi^+  $ & $ \left(b_1-b_2-b_6-b_7\right) V_{cd}^*$&
$T_{bb\bar{d}\bar{s}}^{0}\to   \overline B^0_s   D^-_s  \pi^+  $ & $ -\left(b_2-b_5+b_7+b_9\right)V_{cs}^*$\\\hline
$T_{bb\bar{d}\bar{s}}^{0}\to   \overline B^0_s   D^-_s  K^+  $ & $ \left(b_1-b_5-b_6+b_9\right) V_{cd}^*$&
$T_{bb\bar{u}\bar{d}}^{-}\to   B^-  \overline D^0  \pi^0  $ & $ \frac{\left(-b_1+b_2+2 b_3+b_4+b_6-b_7-b_8-2 b_9\right) V_{cd}^*}{\sqrt{2}}$\\\hline
$T_{bb\bar{u}\bar{d}}^{-}\to   B^-  \overline D^0  \overline K^0  $ & $ \left(-b_2+b_5+b_7+b_9\right)V_{cs}^*$&
$T_{bb\bar{u}\bar{d}}^{-}\to   B^-  \overline D^0  \eta  $ & $ \frac{\left(-b_1-b_2+2 b_3+b_4+2 b_5+b_6+b_7-b_8\right) V_{cd}^*}{\sqrt{6}}$\\\hline
$T_{bb\bar{u}\bar{d}}^{-}\to   B^-  D^-  \pi^+  $ & $ \left(-b_1+b_2+2 b_3+b_4+b_6+b_7+b_8\right) V_{cd}^*$&
$T_{bb\bar{u}\bar{d}}^{-}\to   B^-   D^-_s  \pi^+  $ & $ \left(b_2+b_4+b_7+b_8\right)V_{cs}^*$\\\hline
$T_{bb\bar{u}\bar{d}}^{-}\to   B^-   D^-_s  K^+  $ & $ \left(-b_1+2 b_3+b_6\right) V_{cd}^*$&
$T_{bb\bar{u}\bar{d}}^{-}\to   \overline B^0  \overline D^0  \pi^-  $ & $ -\left(b_1-b_2-2 b_3-b_4+b_6+b_7+b_8\right) V_{cd}^*$\\\hline
$T_{bb\bar{u}\bar{d}}^{-}\to   \overline B^0  \overline D^0  K^-  $ & $ -\left(b_1-b_2+b_6+b_7\right)V_{cs}^*$&
$T_{bb\bar{u}\bar{d}}^{-}\to   \overline B^0  D^-  \pi^0  $ & $ \frac{\left(b_1-b_2-2 b_3-b_4+b_6-b_7-b_8-2 b_9\right) V_{cd}^*}{\sqrt{2}}$\\\hline
$T_{bb\bar{u}\bar{d}}^{-}\to   \overline B^0  D^-  \overline K^0  $ & $ \left(-b_1+b_5-b_6+b_9\right)V_{cs}^*$&
$T_{bb\bar{u}\bar{d}}^{-}\to   \overline B^0  D^-  \eta  $ & $ -\frac{\left(b_1+b_2-2 b_3-b_4-2 b_5+b_6+b_7-b_8\right) V_{cd}^*}{\sqrt{6}}$\\\hline
$T_{bb\bar{u}\bar{d}}^{-}\to   \overline B^0   D^-_s  \pi^0  $ & $ -\frac{\left(b_2+b_4+b_7+b_8\right)V_{cs}^*}{\sqrt{2}}$&
$T_{bb\bar{u}\bar{d}}^{-}\to   \overline B^0   D^-_s  K^0  $ & $ -\left(b_1-2 b_3+b_6\right) V_{cd}^*$\\\hline
$T_{bb\bar{u}\bar{d}}^{-}\to   \overline B^0   D^-_s  \eta  $ & $ \frac{\left(2 b_1-b_2+b_4+2 b_6-b_7+b_8\right)V_{cs}^*}{\sqrt{6}}$&
$T_{bb\bar{u}\bar{d}}^{-}\to   \overline B^0_s  \overline D^0  K^-  $ & $ \left(2 b_3+b_4-b_8\right) V_{cd}^*$\\\hline
$T_{bb\bar{u}\bar{d}}^{-}\to   \overline B^0_s  D^-  \overline K^0  $ & $ \left(2 b_3+b_4+b_8\right) V_{cd}^*$&
$T_{bb\bar{u}\bar{d}}^{-}\to   \overline B^0_s   D^-_s  \pi^0  $ & $ -\sqrt{2} b_9 V_{cd}^*$\\\hline
$T_{bb\bar{u}\bar{d}}^{-}\to   \overline B^0_s   D^-_s  \overline K^0  $ & $ \left(b_4+b_5+b_8+b_9\right)V_{cs}^*$&
$T_{bb\bar{u}\bar{d}}^{-}\to   \overline B^0_s   D^-_s  \eta  $ & $ \sqrt{\frac{2}{3}} \left(b_5-2 b_3\right) V_{cd}^*$\\\hline
\hline
\end{tabular}
\end{table}
\begin{table}
\caption{Doubly bottom tetraquark $T_{bb\bar{q}\bar{q}}$ decays into an anti-charmed B meson and two light mesons.}\label{tab:bb_Bcand2meson8}\begin{tabular}{|cc|cc|}
\hline
channel  & amplitude(/$V_{ub}$)  & channel  & amplitude(/$V_{ub}$)\tabularnewline
\hline
\hline
$T_{bb\bar{u}\bar{s}}^{-}\to   B_c^-  \pi^+   \pi^-  $ & $ \left(4 b_{10}+b_{11}-b_{13}\right)V_{cs}^*$&
$T_{bb\bar{u}\bar{s}}^{-}\to   B_c^-  \pi^0   \pi^0  $ & $ \left(4 b_{10}+b_{11}-b_{13}\right)V_{cs}^*$\\\hline
$T_{bb\bar{u}\bar{s}}^{-}\to   B_c^-  \pi^0   K^0  $ & $ -\frac{\left(b_{11}-2 b_{12}+b_{13}\right) V_{cd}^*}{\sqrt{2}}$&
$T_{bb\bar{u}\bar{s}}^{-}\to   B_c^-  \pi^0   \eta  $ & $ \frac{\left(b_{11}-2 b_{12}-b_{13}\right)V_{cs}^*}{\sqrt{3}}$\\\hline
$T_{bb\bar{u}\bar{s}}^{-}\to   B_c^-  \pi^-   K^+  $ & $ \left(b_{11}-2 b_{12}+b_{13}\right) V_{cd}^*$&
$T_{bb\bar{u}\bar{s}}^{-}\to   B_c^-  K^+   K^-  $ & $ 2 \left(2 b_{10}+b_{11}-b_{12}\right)V_{cs}^*$\\\hline
$T_{bb\bar{u}\bar{s}}^{-}\to   B_c^-  K^0   \overline K^0  $ & $ \left(4 b_{10}+b_{11}+b_{13}\right)V_{cs}^*$&
$T_{bb\bar{u}\bar{s}}^{-}\to   B_c^-  K^0   \eta  $ & $ -\frac{\left(b_{11}-2 b_{12}+b_{13}\right) V_{cd}^*}{\sqrt{6}}$\\\hline
$T_{bb\bar{u}\bar{s}}^{-}\to   B_c^-  \eta   \eta  $ & $ \frac{1}{3} \left(12 b_{10}+5 b_{11}-4 b_{12}+3 b_{13}\right)V_{cs}^*$&
$T_{bb\bar{d}\bar{s}}^{0}\to   B_c^-  \pi^+   K^0  $ & $ \left(-b_{11}+2 b_{12}+b_{13}\right) V_{cd}^*$\\\hline
$T_{bb\bar{d}\bar{s}}^{0}\to   B_c^-  \pi^+   \eta  $ & $ \sqrt{\frac{2}{3}} \left(b_{11}-2 b_{12}-b_{13}\right)V_{cs}^*$&
$T_{bb\bar{d}\bar{s}}^{0}\to   B_c^-  \pi^0   K^+  $ & $ \frac{\left(-b_{11}+2 b_{12}+b_{13}\right) V_{cd}^*}{\sqrt{2}}$\\\hline
$T_{bb\bar{d}\bar{s}}^{0}\to   B_c^-  K^+   \overline K^0  $ & $ \left(b_{11}-2 b_{12}-b_{13}\right)V_{cs}^*$&
$T_{bb\bar{d}\bar{s}}^{0}\to   B_c^-  K^+   \eta  $ & $ \frac{\left(b_{11}-2 b_{12}-b_{13}\right) V_{cd}^*}{\sqrt{6}}$\\\hline
$T_{bb\bar{u}\bar{d}}^{-}\to   B_c^-  \pi^+   \pi^-  $ & $ 2 \left(2 b_{10}+b_{11}-b_{12}\right) V_{cd}^*$&
$T_{bb\bar{u}\bar{d}}^{-}\to   B_c^-  \pi^+   K^-  $ & $ \left(b_{11}-2 b_{12}+b_{13}\right)V_{cs}^*$\\\hline
$T_{bb\bar{u}\bar{d}}^{-}\to   B_c^-  \pi^0   \pi^0  $ & $ 2 \left(2 b_{10}+b_{11}-b_{12}\right) V_{cd}^*$&
$T_{bb\bar{u}\bar{d}}^{-}\to   B_c^-  \pi^0   \overline K^0  $ & $ -\frac{\left(b_{11}-2 b_{12}+b_{13}\right)V_{cs}^*}{\sqrt{2}}$\\\hline
$T_{bb\bar{u}\bar{d}}^{-}\to   B_c^-  \pi^0   \eta  $ & $ -\frac{2 b_{13} V_{cd}^*}{\sqrt{3}}$&
$T_{bb\bar{u}\bar{d}}^{-}\to   B_c^-  K^+   K^-  $ & $ \left(4 b_{10}+b_{11}-b_{13}\right) V_{cd}^*$\\\hline
$T_{bb\bar{u}\bar{d}}^{-}\to   B_c^-  K^0   \overline K^0  $ & $ \left(4 b_{10}+b_{11}+b_{13}\right) V_{cd}^*$&
$T_{bb\bar{u}\bar{d}}^{-}\to   B_c^-  \overline K^0   \eta  $ & $ -\frac{\left(b_{11}-2 b_{12}+b_{13}\right)V_{cs}^*}{\sqrt{6}}$\\\hline
$T_{bb\bar{u}\bar{d}}^{-}\to   B_c^-  \eta   \eta  $ & $ \frac{2}{3} \left(6 b_{10}+b_{11}+b_{12}\right) V_{cd}^*$& & \\\hline
\hline
\end{tabular}
\end{table}
The effective Hamiltonian is written as
\begin{eqnarray}
  {\cal H}&=& b_{1} (T_{bb3})_{[ij]} (\overline B)^{i}   D_l M^l_k (H_{\bar 3}'')^{[jk]}+b_{2} (T_{bb3})_{[ij]} (\overline B)^{i}   D_l M^j_k (H_{\bar 3}'')^{[kl]} \nonumber\\
 &&+b_{3} (T_{bb3})_{[ij]} (\overline B)^{k}   D_l M^l_k (H_{\bar 3}'')^{[ij]} + b_{4} (T_{bb3})_{[ij]} (\overline B)^{k}   D_l M^j_k (H_{\bar 3}'')^{[il]} \nonumber\\
 &&+b_{5} (T_{bb3})_{[ij]} (\overline B)^{l}   D_l M^j_k (H_{\bar 3}'')^{[ik]}+b_{6} (T_{bb3})_{[ij]} (\overline B)^{i}   D_l M^l_k (H_6'')^{\{jk\}}\nonumber\\
 &&+b_{7} (T_{bb3})_{[ij]} (\overline B)^{i}   D_l M^j_k (H_6'')^{\{kl\}} + b_{8} (T_{bb3})_{[ij]} (\overline B)^{k}   D_l M^j_k (H_{6}'')^{\{il\}} \nonumber\\
 &&+b_{9} (T_{bb3})_{[ij]} (\overline B)^{l}   D_l M^j_k (H_{6}'')^{\{ik\}} +  b_{10} (T_{bb3})_{[ij]}   M^l_k M^k_l (H_{\bar 3}'')^{[ij]}~ \overline B_c\nonumber\\
 &&+b_{11} (T_{bb3})_{[ij]}   M^j_k M^k_l (H_{\bar 3}'')^{[il]}~ \overline B_c +b_{12} (T_{bb3})_{[ij]}   M^i_k M^j_l (H_{\bar 3}'')^{[kl]}~ \overline B_c \nonumber\\
 &&+b_{13} (T_{bb3})_{[ij]} M^j_k M^k_l (H_6'')^{\{il\}}~\overline B_c .
\end{eqnarray}
The decay amplitudes into a bottom meson plus an anti-charmed meson and a light meson  are given in Tab.~\ref{tab:bb_B_antiDand8meson}.  Deriving the formulae, we get:
\begin{eqnarray*}
    \Gamma(T_{bb\bar{d}\bar{s}}^{0}\to \overline B^0_s  D^-_s\pi^+ )= 2\Gamma(T_{bb\bar{u}\bar{s}}^{-}\to \overline B^0_s  D^-_s\pi^0 ), \Gamma(T_{bb\bar{u}\bar{d}}^{-}\to B^-  D^-_s\pi^+ )= 2\Gamma(T_{bb\bar{u}\bar{d}}^{-}\to \overline B^0  D^-_s\pi^0 ).
\end{eqnarray*}
The related decay amplitudes of an anti-charmed bottom meson plus two light mesons are given in Tab.~\ref{tab:bb_Bcand2meson8}. Thus we obtain the relations  as follows:
\begin{eqnarray*}
    \Gamma(T_{bb\bar{u}\bar{s}}^{-}\to B_c^- \pi^0 \pi^0 )= \frac{1}{2}\Gamma(T_{bb\bar{u}\bar{s}}^{-}\to B_c^- \pi^+ \pi^- ),
     \Gamma(T_{bb\bar{u}\bar{d}}^{-}\to B_c^- \pi^0 \pi^0 )= \frac{1}{2}\Gamma(T_{bb\bar{u}\bar{d}}^{-}\to B_c^- \pi^+ \pi^- ),\\
    \Gamma(T_{bb\bar{u}\bar{s}}^{-}\to B_c^- \pi^0 K^0 )= 3\Gamma(T_{bb\bar{u}\bar{s}}^{-}\to B_c^- K^0 \eta)=\frac{1}{2}\Gamma(T_{bb\bar{u}\bar{s}}^{-}\to B_c^- \pi^- K^+ ),\\
     \Gamma(T_{bb\bar{d}\bar{s}}^{0}\to B_c^- \pi^+ K^0 )= 6\Gamma(T_{bb\bar{d}\bar{s}}^{0}\to B_c^- K^+ \eta )=2\Gamma(T_{bb\bar{d}\bar{s}}^{0}\to B_c^- \pi^0 K^+ ),\\
     \Gamma(T_{bb\bar{d}\bar{s}}^{0}\to B_c^- K^+ \overline K^0 )= 3\Gamma(T_{bb\bar{u}\bar{s}}^{-}\to B_c^- \pi^0 \eta )=
 \frac{3}{2}\Gamma(T_{bb\bar{d}\bar{s}}^{0}\to B_c^- \pi^+ \eta ),\\
     \Gamma(T_{bb\bar{u}\bar{d}}^{-}\to B_c^- \pi^+ K^- )= 6\Gamma(T_{bb\bar{u}\bar{d}}^{-}\to B_c^- \overline K^0 \eta )= 2\Gamma(T_{bb\bar{u}\bar{d}}^{-}\to B_c^- \pi^0 \overline K^0 ),\\
     \Gamma(T_{bb\bar{u}\bar{d}}^{-}\to B_c^- \pi^0 \overline K^0 )= 3\Gamma(T_{bb\bar{u}\bar{d}}^{-}\to B_c^- \eta \overline K^0 ).
 \end{eqnarray*}
\subsection{Charmless $b\to q_1 \bar q_2 q_3$ transition}

\subsubsection{Decays into a bottom meson and a light meson by W-exchange process}
    \begin{table}
\caption{Doubly bottom tetraquark $T_{bb\bar{q}\bar{q}}$ decays into a bottom meson and a light meson induced by the charmless $b\to d$ transition.}\label{tab:bb_Bandmeson8d}\begin{tabular}{|c|c|c|c|c|c|c|c}\hline\hline
channel & amplitude \\\hline
$T_{bb\bar{u}\bar{s}}^{-}\to   B^-  K^0  $ & $ f_{11}+f_{13}+f_{14}+f_{15}+3 f_{16}$\\\hline
$T_{bb\bar{u}\bar{s}}^{-}\to   \overline B^0_s  \pi^-  $ & $ -f_{11}+f_{13}+f_{14}+3 f_{15}+f_{16}$\\\hline
$T_{bb\bar{d}\bar{s}}^{0}\to   B^-  K^+  $ & $ f_{10}+2 f_{12}-f_{14}+3 f_{16}$\\\hline
$T_{bb\bar{d}\bar{s}}^{0}\to   \overline B^0  K^0  $ & $ f_{10}+f_{11}+2 f_{12}+f_{13}+f_{15}-2 f_{16}$\\\hline
$T_{bb\bar{d}\bar{s}}^{0}\to   \overline B^0_s  \pi^0  $ & $ \frac{f_{11}-f_{13}-f_{14}+5 f_{15}-f_{16}}{\sqrt{2}}$\\\hline
$T_{bb\bar{d}\bar{s}}^{0}\to   \overline B^0_s  \eta  $ & $ -\frac{2 f_{10}+f_{11}+4 f_{12}+3 f_{13}+f_{14}-3 f_{15}-3 f_{16}}{\sqrt{6}}$\\\hline
$T_{bb\bar{u}\bar{d}}^{-}\to   B^-  \pi^0  $ & $ -\frac{f_{10}+f_{11}-2 f_{12}-f_{13}+5 f_{15}+6 f_{16}}{\sqrt{2}}$\\\hline
$T_{bb\bar{u}\bar{d}}^{-}\to   B^-  \eta  $ & $ \frac{-f_{10}+f_{11}+2 f_{12}+3 f_{13}+2 f_{14}-3 f_{15}}{\sqrt{6}}$\\\hline
$T_{bb\bar{u}\bar{d}}^{-}\to   \overline B^0  \pi^-  $ & $ -f_{10}-f_{11}+2 f_{12}+f_{13}+3 f_{15}+2 f_{16}$\\\hline
$T_{bb\bar{u}\bar{d}}^{-}\to   \overline B^0_s  K^-  $ & $ -f_{10}+2 f_{12}-f_{14}+f_{16}$\\\hline
\hline
\end{tabular}
\end{table}
    \begin{table}
\caption{Doubly bottom tetraquark $T_{bb\bar{q}\bar{q}}$ decays into a bottom meson and a light meson induced by the charmless $b\to s$ transition .}\label{tab:bb_Bandmeson8s}\begin{tabular}{|c|c|c|c|c|c|c|c}\hline\hline
channel & amplitude \\\hline
$T_{bb\bar{u}\bar{s}}^{-}\to   B^-  \pi^0  $ & $ \frac{-f_{10}+2 f_{12}+2 f_{13}+f_{14}-4 f_{15}-3 f_{16}}{\sqrt{2}}$\\\hline
$T_{bb\bar{u}\bar{s}}^{-}\to   B^-  \eta  $ & $ -\frac{f_{10}+2 f_{11}-2 f_{12}+f_{14}+6 f_{15}+9 f_{16}}{\sqrt{6}}$\\\hline
$T_{bb\bar{u}\bar{s}}^{-}\to   \overline B^0  \pi^-  $ & $ -f_{10}+2 f_{12}-f_{14}+f_{16}$\\\hline
$T_{bb\bar{u}\bar{s}}^{-}\to   \overline B^0_s  K^-  $ & $ -f_{10}-f_{11}+2 f_{12}+f_{13}+3 f_{15}+2 f_{16}$\\\hline
$T_{bb\bar{d}\bar{s}}^{0}\to   B^-  \pi^+  $ & $ -f_{10}-2 f_{12}+f_{14}-3 f_{16}$\\\hline
$T_{bb\bar{d}\bar{s}}^{0}\to   \overline B^0  \pi^0  $ & $ \frac{f_{10}+2 f_{12}+2 f_{13}+f_{14}-4 f_{15}-f_{16}}{\sqrt{2}}$\\\hline
$T_{bb\bar{d}\bar{s}}^{0}\to   \overline B^0  \eta  $ & $ -\frac{f_{10}+2 f_{11}+2 f_{12}-f_{14}+6 f_{15}-3 f_{16}}{\sqrt{6}}$\\\hline
$T_{bb\bar{d}\bar{s}}^{0}\to   \overline B^0_s  \overline K^0  $ & $ -f_{10}-f_{11}-2 f_{12}-f_{13}-f_{15}+2 f_{16}$\\\hline
$T_{bb\bar{u}\bar{d}}^{-}\to   B^-  \overline K^0  $ & $ f_{11}+f_{13}+f_{14}+f_{15}+3 f_{16}$\\\hline
$T_{bb\bar{u}\bar{d}}^{-}\to   \overline B^0  K^-  $ & $ -f_{11}+f_{13}+f_{14}+3 f_{15}+f_{16}$\\\hline
\hline
\end{tabular}
\end{table}
The bottom to light quark transition leads to the effective Hamiltonian:
 \begin{eqnarray}
 {\cal H}_{eff} &=& \frac{G_{F}}{\sqrt{2}}
     \bigg\{ V_{ub} V_{uq}^{*} \big[
     C_{1}  O^{\bar uu}_{1}
  +  C_{2}  O^{\bar uu}_{2}\Big]- V_{tb} V_{tq}^{*} \big[{\sum\limits_{i=3}^{10}} C_{i}  O_{i} \Big]\bigg\}+ \mbox{h.c.} ,
 \label{eq:hamiltonian}
\end{eqnarray}
where  $O_{i}$ is the weak four-fermion effective operators. The tree  operators are described as a vector $H_{\bf 3}$, a
tensor  $H_{\bf\overline6}$, and  a tensor $H_{\bf{15}}$. The penguin operators are described as another vector $H_{\bf 3}$.
The nonzero components of these operators are
\begin{eqnarray}
 (H_3)^2=1,\;\;\;(H_{\overline6})^{12}_1=-(H_{\overline6})^{21}_1=(H_{\overline6})^{23}_3=-(H_{\overline6})^{32}_3=1,\nonumber\\
 2(H_{15})^{12}_1= 2(H_{15})^{21}_1=-3(H_{15})^{22}_2=
 -6(H_{15})^{23}_3=-6(H_{15})^{32}_3=6,\label{eq:H3615_bb}
\end{eqnarray}
for  the  non-strange decays. After doing the exchange of $2\leftrightarrow 3$, we will  get the formulae for the $\Delta S=1(b\to s)$
decays.
We get the effective hadron-level Hamiltonian for decays into the bottom anti-triplet
  \begin{eqnarray}
  \mathcal{H}_{eff}&=&f_{10} (T_{bb3})_{[ij]}(\overline B)^k (H_3)^i M^j_k  +f_{11} (T_{bb3})_{[ij]}(\overline B)^i (H_3)^k M^j_k\nonumber\\
  &&+f_{12}(T_{bb3})_{[ij]}(\overline B)^l (H_{\overline 6})^{[ij]}_k M^k_l+f_{13}(T_{bb3})_{[ij]}(\overline B)^j (H_{\overline 6})^{[il]}_k M^k_l\nonumber\\
  &&+f_{14}(T_{bb3})_{[ij]}(\overline B)^k (H_{\overline 6})^{[il]}_k M^j_l+f_{15}(T_{bb3})_{[ij]}(\overline B)^j (H_{15})^{\{il\}}_k M^k_l\nonumber\\
  &&+f_{16}(T_{bb3})_{[ij]}(\overline B)^k (H_{15})^{\{il\}}_k M^j_l.
  \end{eqnarray}
The amplitudes are given in Tab.~\ref{tab:bb_Bandmeson8d} for the $\Delta S=0(b\to d)$ decays and Tab.~\ref{tab:bb_Bandmeson8s} for the $\Delta S=1(b\to s)$ decays.
\subsubsection{ Decays into a light anti-baryon and a bottom baryon}
    \begin{table}
    \footnotesize
\caption{Doubly bottom tetraquark $T_{bb\bar{q}\bar{q}}$ decays into a light anti-baryon octet and a bottom baryon anti-triplet(class \uppercase\expandafter{\romannumeral1}) or sextet(class \uppercase\expandafter{\romannumeral2}), a light anti-baryon anti-decuplet and a bottom baryon anti-triplet(class \uppercase\expandafter{\romannumeral3}) or sextet(class \uppercase\expandafter{\romannumeral4}) induced by the charmless $b\to d$ transition .}\label{tab:bb_Fbar_Fb_bd}\begin{tabular}{|cc |cc|}\hline
\multicolumn{1}{|c}{ class \uppercase\expandafter{\romannumeral1}}&\multicolumn{1}{c}{amplitude}&&\\
\hline\hline
$T_{bb\bar{u}\bar{s}}^{-}\to   \overline \Sigma^- \Lambda_b^0 $ & $ 2 c_2+c_4-c_5+2 c_{6}+3 c_{7}-c_{8}$&
$T_{bb\bar{u}\bar{s}}^{-}\to   \overline \Xi^0  \Xi_b^- $ & $ -2 c_2+c_4-c_5-2 c_{6}+c_{7}-3 c_{8}$\\\hline
$T_{bb\bar{d}\bar{s}}^{0}\to   \overline \Lambda^0  \Lambda_b^0 $ & $ \frac{2 c_1-2 c_2+4 c_3+c_4+c_5+3 c_{7}-3 c_{8}}{\sqrt{6}}$&
$T_{bb\bar{d}\bar{s}}^{0}\to   \overline \Sigma^0  \Lambda_b^0 $ & $ \frac{-2 c_2-c_4+c_5+6 c_{6}+5 c_{7}+c_{8}}{\sqrt{2}}$\\\hline
$T_{bb\bar{d}\bar{s}}^{0}\to   \overline \Xi^+  \Xi_b^- $ & $ -c_1+2 c_2-2 c_3-2 c_4+c_5-3 c_{6}-3 c_{8}$&
$T_{bb\bar{d}\bar{s}}^{0}\to   \overline \Xi^0  \Xi_b^0 $ & $ c_1+2 c_3+c_4-3 c_{6}-c_{7}-2 c_{8}$\\\hline
$T_{bb\bar{u}\bar{d}}^{-}\to   \overline \Lambda^0  \Xi_b^- $ & $ \frac{c_1-4 c_2-2 c_3+c_4-2 c_5-3 c_{6}-3 c_{7}}{\sqrt{6}}$&
$T_{bb\bar{u}\bar{d}}^{-}\to   \overline \Sigma^- \Xi_b^0 $ & $ -c_1+2 c_3+c_4-c_{6}-3 c_{7}+2 c_{8}$\\\hline
$T_{bb\bar{u}\bar{d}}^{-}\to   \overline \Sigma^0  \Xi_b^- $ & $ \frac{c_1-2 c_3-c_4+c_{6}-5 c_{7}+6 c_{8}}{\sqrt{2}}$&
$T_{bb\bar{u}\bar{d}}^{-}\to   \overline p  \Lambda_b^0 $ & $ c_1-2 c_2-2 c_3-2 c_4+c_5-c_{6}-c_{8}$\\\hline\hline
\multicolumn{1}{|c}{ class \uppercase\expandafter{\romannumeral2}}&\multicolumn{1}{c}{amplitude}&&\\\hline
$T_{bb\bar{u}\bar{s}}^{-}\to   \overline \Lambda^0  \Sigma_{b}^{-} $ & $ -\frac{2 d_2+3 d_3+3 d_4-2 d_5+5 d_6+7 d_{7}-4 d_{8}}{\sqrt{6}}$&
$T_{bb\bar{u}\bar{s}}^{-}\to   \overline \Sigma^- \Sigma_{b}^{0} $ & $ \frac{-2 d_2+d_3+d_4+2 d_5+3 d_6+d_{7}-12 d_{8}}{\sqrt{2}}$\\\hline
$T_{bb\bar{u}\bar{s}}^{-}\to   \overline \Sigma^0  \Sigma_{b}^{-} $ & $ \frac{2 d_2-d_3-d_4-2 d_5-3 d_6-d_{7}-4 d_{8}}{\sqrt{2}}$&
$T_{bb\bar{u}\bar{s}}^{-}\to   \overline \Xi^0  \Xi_{b}^{\prime-} $ & $ -\frac{2 d_2+d_3+d_4-2 d_5+d_6+3 d_{7}-4 d_{8}}{\sqrt{2}}$\\\hline
$T_{bb\bar{d}\bar{s}}^{0}\to   \overline \Lambda^0  \Sigma_{b}^{0} $ & $ \frac{2 d_2+3 \left(d_3+d_4+2 d_5-d_6-3 d_{7}+4 d_{8}\right)}{2 \sqrt{3}}$&
$T_{bb\bar{d}\bar{s}}^{0}\to   \overline \Sigma^- \Sigma_{b}^{+} $ & $ d_1-d_3+3 d_5+3 d_6$\\\hline
$T_{bb\bar{d}\bar{s}}^{0}\to   \overline \Sigma^0  \Sigma_{b}^{0} $ & $ \frac{1}{2} \left(-2 d_1+2 d_2+d_3-d_4-d_6-d_{7}+12 d_{8}\right)$&
$T_{bb\bar{d}\bar{s}}^{0}\to   \overline \Sigma^+  \Sigma_{b}^{-} $ & $ -d_1+2 d_2-d_4+3 d_5+2 d_6-d_{7}-4 d_{8}$\\\hline
$T_{bb\bar{d}\bar{s}}^{0}\to   \overline \Xi^+  \Xi_{b}^{\prime-} $ & $ -\frac{d_1-2 d_2+d_4-3 d_5-2 d_6+d_{7}+4 d_{8}}{\sqrt{2}}$&
$T_{bb\bar{d}\bar{s}}^{0}\to   \overline \Xi^0  \Xi_{b}^{\prime0} $ & $ \frac{d_1+d_3+2 d_4+3 d_5-d_6-4 d_{7}}{\sqrt{2}}$\\\hline
 $T_{bb\bar{u}\bar{d}}^{-}\to   \overline \Lambda^0  \Xi_{b}^{\prime-} $ & $ \frac{3 d_1-4 d_2+3 d_3+d_5-d_6-2 d_{7}+8 d_{8}}{2 \sqrt{3}}$&
$T_{bb\bar{u}\bar{d}}^{-}\to   \overline \Sigma^- \Xi_{b}^{\prime0} $ & $ \frac{d_1-d_3-2 d_4-d_5+3 d_6+4 d_{7}}{\sqrt{2}}$\\\hline
$T_{bb\bar{u}\bar{d}}^{-}\to   \overline \Sigma^0  \Xi_{b}^{\prime-} $ & $ \frac{1}{2} \left(-d_1+d_3+2 d_4+d_5+5 d_6+4 d_{7}\right)$&
$T_{bb\bar{u}\bar{d}}^{-}\to   \overline p  \Sigma_{b}^{0} $ & $ \frac{-d_1+2 d_2+d_4-d_5-6 d_6-5 d_{7}+12 d_{8}}{\sqrt{2}}$\\\hline
$T_{bb\bar{u}\bar{d}}^{-}\to   \overline n  \Sigma_{b}^{-} $ & $ -d_1+2 d_2+d_4-d_5+2 d_6+3 d_{7}-4 d_{8}$&
$T_{bb\bar{u}\bar{d}}^{-}\to   \overline \Xi^0  \Omega_{b}^{-} $ & $ d_1+d_3-d_5-d_6$\\\hline\hline
\multicolumn{1}{|c}{ class \uppercase\expandafter{\romannumeral3}}&\multicolumn{1}{c}{amplitude}&&\\\hline
$T_{bb\bar{u}\bar{s}}^{-}\to   \overline \Sigma^{\prime-}  \Lambda_b^0 $ & $ \frac{2 \left(a_1+2 a_2\right)}{\sqrt{3}}$&
$T_{bb\bar{u}\bar{s}}^{-}\to   \overline \Xi^{\prime0}  \Xi_b^- $ & $ -\frac{2 \left(a_1+2 a_2\right)}{\sqrt{3}}$\\\hline
$T_{bb\bar{d}\bar{s}}^{0}\to   \overline \Sigma^{\prime0}  \Lambda_b^0 $ & $ \sqrt{\frac{2}{3}} \left(a_1-2 a_2\right)$&
$T_{bb\bar{d}\bar{s}}^{0}\to   \overline \Xi^{\prime0}  \Xi_b^0 $ & $ \frac{2 \left(a_1-2 a_2\right)}{\sqrt{3}}$\\\hline
$T_{bb\bar{u}\bar{d}}^{-}\to   \overline \Delta^{-}  \Lambda_b^0 $ & $ \frac{8 a_2}{\sqrt{3}}$&
$T_{bb\bar{u}\bar{d}}^{-}\to   \overline \Sigma^{\prime-}  \Xi_b^0 $ & $ -\frac{2 \left(a_1-2 a_2\right)}{\sqrt{3}}$\\\hline
$T_{bb\bar{u}\bar{d}}^{-}\to   \overline \Sigma^{\prime0}  \Xi_b^- $ & $ -\sqrt{\frac{2}{3}} \left(a_1+2 a_2\right)$&&\\
\hline\hline
\multicolumn{1}{|c}{ class \uppercase\expandafter{\romannumeral4}}&\multicolumn{1}{c}{amplitude}&&\\\hline
$T_{bb\bar{u}\bar{s}}^{-}\to   \overline \Sigma^{\prime-}  \Sigma_{b}^{0} $ & $ \sqrt{\frac{2}{3}} \left(b_3+2 b_4\right)$&
$T_{bb\bar{u}\bar{s}}^{-}\to   \overline \Sigma^{\prime0}  \Sigma_{b}^{-} $ & $ \sqrt{\frac{2}{3}} \left(b_3+2 b_4\right)$\\\hline
$T_{bb\bar{u}\bar{s}}^{-}\to   \overline \Xi^{\prime0}  \Xi_{b}^{\prime-} $ & $ \sqrt{\frac{2}{3}} \left(b_3+2 b_4\right)$&
$T_{bb\bar{d}\bar{s}}^{0}\to   \overline \Sigma^{\prime-}  \Sigma_{b}^{+} $ & $ \frac{b_1+2 b_2-b_3+3 b_4}{\sqrt{3}}$\\\hline
$T_{bb\bar{d}\bar{s}}^{0}\to   \overline \Sigma^{\prime0}  \Sigma_{b}^{0} $ & $ \frac{b_1+2 b_2+b_4}{\sqrt{3}}$&
$T_{bb\bar{d}\bar{s}}^{0}\to   \overline \Sigma^{\prime+}  \Sigma_{b}^{-} $ & $ \frac{b_1+2 b_2+b_3-b_4}{\sqrt{3}}$\\\hline
$T_{bb\bar{d}\bar{s}}^{0}\to   \overline \Xi^{\prime0}  \Xi_{b}^{\prime0} $ & $ \sqrt{\frac{2}{3}} \left(b_1+2 b_2+b_4\right)$&
$T_{bb\bar{d}\bar{s}}^{0}\to   \overline \Xi^{\prime+}  \Xi_{b}^{\prime-} $ & $ \sqrt{\frac{2}{3}} \left(b_1+2 b_2+b_3-b_4\right)$\\\hline
$T_{bb\bar{d}\bar{s}}^{0}\to   \overline \Omega^+  \Omega_{b}^{-} $ & $ b_1+2 b_2+b_3-b_4$&
$T_{bb\bar{u}\bar{d}}^{-}\to   \overline \Delta^{--}  \Sigma_{b}^{+} $ & $ -b_1+2 b_2+b_3-3 b_4$\\\hline
$T_{bb\bar{u}\bar{d}}^{-}\to   \overline \Delta^{-}  \Sigma_{b}^{0} $ & $ \sqrt{\frac{2}{3}} \left(-b_1+2 b_2+b_3+b_4\right)$&
$T_{bb\bar{u}\bar{d}}^{-}\to   \overline \Delta^{0}  \Sigma_{b}^{-} $ & $ \frac{-b_1+2 b_2+b_3+5 b_4}{\sqrt{3}}$\\\hline
$T_{bb\bar{u}\bar{d}}^{-}\to   \overline \Sigma^{\prime-}  \Xi_{b}^{\prime0} $ & $ -\sqrt{\frac{2}{3}} \left(b_1-2 b_2+b_4\right)$&
$T_{bb\bar{u}\bar{d}}^{-}\to   \overline \Sigma^{\prime0}  \Xi_{b}^{\prime-} $ & $ \frac{-b_1+2 b_2+3 b_4}{\sqrt{3}}$\\\hline
$T_{bb\bar{u}\bar{d}}^{-}\to   \overline \Xi^{\prime0}  \Omega_{b}^{-} $ & $ -\frac{b_1-2 b_2+b_3-b_4}{\sqrt{3}}$&&\\
\hline\hline
\end{tabular}
\end{table}
    \begin{table}
    \footnotesize
\caption{Doubly bottom tetraquark $T_{bb\bar{q}\bar{q}}$ decays into a light anti-baryon octet and a bottom baryon anti-triplet(class \uppercase\expandafter{\romannumeral1}) or sextet(class \uppercase\expandafter{\romannumeral2}), a light anti-baryon anti-decuplet and a bottom baryon anti-triplet(class \uppercase\expandafter{\romannumeral3}) or sextet(class \uppercase\expandafter{\romannumeral4}) induced by the charmless $b\to s$ transition.}\label{tab:bb_Fbar_Fb_bs}\begin{tabular}{|cc|cc|}\hline\hline
\multicolumn{1}{|c}{ class \uppercase\expandafter{\romannumeral1}}&\multicolumn{1}{c}{amplitude}&&\\
\hline
$T_{bb\bar{u}\bar{s}}^{-}\to   \overline \Lambda^0  \Xi_b^- $ & $ \frac{c_1+2 c_2-2 c_3-2 c_4+c_5+3 c_{6}-6 c_{7}+9 c_{8}}{\sqrt{6}}$&
$T_{bb\bar{u}\bar{s}}^{-}\to   \overline \Sigma^- \Xi_b^0 $ & $ -c_1+2 c_2+2 c_3+2 c_4-c_5+c_{6}+c_{8}$\\\hline
$T_{bb\bar{u}\bar{s}}^{-}\to   \overline \Sigma^0  \Xi_b^- $ & $ \frac{c_1-2 c_2-2 c_3-c_5-c_{6}-4 c_{7}+3 c_{8}}{\sqrt{2}}$&
$T_{bb\bar{u}\bar{s}}^{-}\to   \overline p  \Lambda_b^0 $ & $ c_1-2 c_3-c_4+c_{6}+3 c_{7}-2 c_{8}$\\\hline
$T_{bb\bar{d}\bar{s}}^{0}\to   \overline \Lambda^0  \Xi_b^0 $ & $ \frac{-c_1-2 c_2-2 c_3-2 c_4+c_5+9 c_{6}+6 c_{7}+3 c_{8}}{\sqrt{6}}$&
$T_{bb\bar{d}\bar{s}}^{0}\to   \overline \Sigma^0  \Xi_b^0 $ & $ \frac{c_1-2 c_2+2 c_3+c_5+3 c_{6}+4 c_{7}-c_{8}}{\sqrt{2}}$\\\hline
$T_{bb\bar{d}\bar{s}}^{0}\to   \overline \Sigma^+  \Xi_b^- $ & $ c_1-2 c_2+2 c_3+2 c_4-c_5+3 c_{6}+3 c_{8}$&
$T_{bb\bar{d}\bar{s}}^{0}\to   \overline n  \Lambda_b^0 $ & $ c_1+2 c_3+c_4-3 c_{6}-c_{7}-2 c_{8}$\\\hline
$T_{bb\bar{u}\bar{d}}^{-}\to   \overline p  \Xi_b^0 $ & $ -2 c_2-c_4+c_5-2 c_{6}-3 c_{7}+c_{8}$&
$T_{bb\bar{u}\bar{d}}^{-}\to   \overline n  \Xi_b^- $ & $ -2 c_2+c_4-c_5-2 c_{6}+c_{7}-3 c_{8}$\\\hline
\hline
\multicolumn{1}{|c}{ class \uppercase\expandafter{\romannumeral2}}&\multicolumn{1}{c}{amplitude}&&\\\hline
$T_{bb\bar{u}\bar{s}}^{-}\to   \overline \Lambda^0  \Xi_{b}^{\prime-} $ & $ \frac{3 d_1-2 d_2-3 d_4-d_5-8 d_6-7 d_{7}+4 d_{8}}{2 \sqrt{3}}$&
$T_{bb\bar{u}\bar{s}}^{-}\to   \overline \Sigma^- \Xi_{b}^{\prime0} $ & $ \frac{d_1-2 d_2-d_4+d_5+6 d_6+5 d_{7}-12 d_{8}}{\sqrt{2}}$\\\hline
$T_{bb\bar{u}\bar{s}}^{-}\to   \overline \Sigma^0  \Xi_{b}^{\prime-} $ & $ \frac{-d_1+2 d_2-2 d_3-d_4-d_5-2 d_6-d_{7}-4 d_{8}}{2}$&
$T_{bb\bar{u}\bar{s}}^{-}\to   \overline p  \Sigma_{b}^{0} $ & $ \frac{-d_1+d_3+2 d_4+d_5-3 d_6-4 d_{7}}{\sqrt{2}}$\\\hline
$T_{bb\bar{u}\bar{s}}^{-}\to   \overline n  \Sigma_{b}^{-} $ & $ -d_1-d_3+d_5+d_6$&
$T_{bb\bar{u}\bar{s}}^{-}\to   \overline \Xi^0  \Omega_{b}^{-} $ & $ d_1-2 d_2-d_4+d_5-2 d_6-3 d_{7}+4 d_{8}$\\\hline
$T_{bb\bar{d}\bar{s}}^{0}\to   \overline \Lambda^0  \Xi_{b}^{\prime0} $ & $ \frac{-3 d_1+2 d_2-3 d_4-3 d_5+3 d_{7}+12 d_{8}}{2 \sqrt{3}}$&
$T_{bb\bar{d}\bar{s}}^{0}\to   \overline \Sigma^0  \Xi_{b}^{\prime0} $ & $ \frac{-d_1+2 d_2+2 d_3+d_4+3 d_5-2 d_6-5 d_{7}+12 d_{8}}{2} $\\\hline
$T_{bb\bar{d}\bar{s}}^{0}\to   \overline \Sigma^+  \Xi_{b}^{\prime-} $ & $ -\frac{d_1-2 d_2+d_4-3 d_5-2 d_6+d_{7}+4 d_{8}}{\sqrt{2}}$&
$T_{bb\bar{d}\bar{s}}^{0}\to   \overline p  \Sigma_{b}^{+} $ & $ d_1-d_3+3 d_5+3 d_6$\\\hline
$T_{bb\bar{d}\bar{s}}^{0}\to   \overline n  \Sigma_{b}^{0} $ & $ \frac{d_1+d_3+2 d_4+3 d_5-d_6-4 d_{7}}{\sqrt{2}}$&
$T_{bb\bar{d}\bar{s}}^{0}\to   \overline \Xi^+  \Omega_{b}^{-} $ & $ -d_1+2 d_2-d_4+3 d_5+2 d_6-d_{7}-4 d_{8}$\\\hline
$T_{bb\bar{u}\bar{d}}^{-}\to   \overline \Lambda^0  \Omega_{b}^{-} $ & $ \sqrt{\frac{2}{3}} \left(-2 d_2+2 d_5+d_6-d_{7}+4 d_{8}\right)$&
$T_{bb\bar{u}\bar{d}}^{-}\to   \overline \Sigma^0  \Omega_{b}^{-} $ & $ \sqrt{2} \left(d_3+d_4+2 \left(d_6+d_{7}\right)\right)$\\\hline
$T_{bb\bar{u}\bar{d}}^{-}\to   \overline p  \Xi_{b}^{\prime0} $ & $ \frac{2 d_2-d_3-d_4-2 d_5-3 d_6-d_{7}+12 d_{8}}{\sqrt{2}}$&
$T_{bb\bar{u}\bar{d}}^{-}\to   \overline n  \Xi_{b}^{\prime-} $ & $ \frac{2 d_2+d_3+d_4-2 d_5+d_6+3 d_{7}-4 d_{8}}{\sqrt{2}}$\\\hline
\hline
\multicolumn{1}{|c}{ class \uppercase\expandafter{\romannumeral3}}&\multicolumn{1}{c}{amplitude}&&\\\hline
$T_{bb\bar{u}\bar{s}}^{-}\to   \overline \Delta^{-}  \Lambda_b^0 $ & $ -\frac{2 \left(a_1-2 a_2\right)}{\sqrt{3}}$&
$T_{bb\bar{u}\bar{s}}^{-}\to   \overline \Sigma^{\prime-}  \Xi_b^0 $ & $ \frac{8 a_2}{\sqrt{3}}$\\\hline
$T_{bb\bar{u}\bar{s}}^{-}\to   \overline \Sigma^{\prime0}  \Xi_b^- $ & $ \sqrt{\frac{2}{3}} \left(a_1+2 a_2\right)$&
$T_{bb\bar{d}\bar{s}}^{0}\to   \overline \Delta^{0}  \Lambda_b^0 $ & $ -\frac{2 \left(a_1-2 a_2\right)}{\sqrt{3}}$\\\hline
$T_{bb\bar{d}\bar{s}}^{0}\to   \overline \Sigma^{\prime0}  \Xi_b^0 $ & $ -\sqrt{\frac{2}{3}} \left(a_1-2 a_2\right)$&
$T_{bb\bar{u}\bar{d}}^{-}\to   \overline \Delta^{-}  \Xi_b^0 $ & $ \frac{2 \left(a_1+2 a_2\right)}{\sqrt{3}}$\\\hline
$T_{bb\bar{u}\bar{d}}^{-}\to   \overline \Delta^{0}  \Xi_b^- $ & $ \frac{2 \left(a_1+2 a_2\right)}{\sqrt{3}}$&&\\
\hline\hline
\multicolumn{1}{|c}{ class \uppercase\expandafter{\romannumeral4}}&\multicolumn{1}{c}{amplitude}&&\\\hline
$T_{bb\bar{u}\bar{s}}^{-}\to   \overline \Delta^{--}  \Sigma_{b}^{+} $ & $ -b_1+2 b_2+b_3-3 b_4$&
$T_{bb\bar{u}\bar{s}}^{-}\to   \overline \Delta^{-}  \Sigma_{b}^{0} $ & $ -\sqrt{\frac{2}{3}} \left(b_1-2 b_2+b_4\right)$\\\hline
$T_{bb\bar{u}\bar{s}}^{-}\to   \overline \Delta^{0}  \Sigma_{b}^{-} $ & $ -\frac{b_1-2 b_2+b_3-b_4}{\sqrt{3}}$&
$T_{bb\bar{u}\bar{s}}^{-}\to   \overline \Sigma^{\prime-}  \Xi_{b}^{\prime0} $ & $ \sqrt{\frac{2}{3}} \left(-b_1+2 b_2+b_3+b_4\right)$\\\hline
$T_{bb\bar{u}\bar{s}}^{-}\to   \overline \Sigma^{\prime0}  \Xi_{b}^{\prime-} $ & $ \frac{-b_1+2 b_2+3 b_4}{\sqrt{3}}$&
$T_{bb\bar{u}\bar{s}}^{-}\to   \overline \Xi^{\prime0}  \Omega_{b}^{-} $ & $ \frac{-b_1+2 b_2+b_3+5 b_4}{\sqrt{3}}$\\\hline
$T_{bb\bar{d}\bar{s}}^{0}\to   \overline \Delta^{-}  \Sigma_{b}^{+} $ & $ -\frac{b_1+2 b_2-b_3+3 b_4}{\sqrt{3}}$&
$T_{bb\bar{d}\bar{s}}^{0}\to   \overline \Delta^{0}  \Sigma_{b}^{0} $ & $ -\sqrt{\frac{2}{3}} \left(b_1+2 b_2+b_4\right)$\\\hline
$T_{bb\bar{d}\bar{s}}^{0}\to   \overline \Delta^{+}  \Sigma_{b}^{-} $ & $ -b_1-2 b_2-b_3+b_4$&
$T_{bb\bar{d}\bar{s}}^{0}\to   \overline \Sigma^{\prime0}  \Xi_{b}^{\prime0} $ & $ -\frac{b_1+2 b_2+b_4}{\sqrt{3}}$\\\hline
$T_{bb\bar{d}\bar{s}}^{0}\to   \overline \Sigma^{\prime+}  \Xi_{b}^{\prime-} $ & $ -\sqrt{\frac{2}{3}} \left(b_1+2 b_2+b_3-b_4\right)$&
$T_{bb\bar{d}\bar{s}}^{0}\to   \overline \Xi^{\prime+}  \Omega_{b}^{-} $ & $ -\frac{b_1+2 b_2+b_3-b_4}{\sqrt{3}}$\\\hline
$T_{bb\bar{u}\bar{d}}^{-}\to   \overline \Delta^{-}  \Xi_{b}^{\prime0} $ & $ \sqrt{\frac{2}{3}} \left(b_3+2 b_4\right)$&
$T_{bb\bar{u}\bar{d}}^{-}\to   \overline \Delta^{0}  \Xi_{b}^{\prime-} $ & $ \sqrt{\frac{2}{3}} \left(b_3+2 b_4\right)$\\\hline
$T_{bb\bar{u}\bar{d}}^{-}\to   \overline \Sigma^{\prime0}  \Omega_{b}^{-} $ & $ \sqrt{\frac{2}{3}} \left(b_3+2 b_4\right)$&&\\
\hline\hline
\end{tabular}
\end{table}
There are four kinds of different final states which are light octet or anti-decuplet anti-baryon plus anti-triplet or sextet baryon respectively.
Thus the  Hamiltonian become
  \begin{eqnarray}
  \mathcal{H}_{eff}&=&c_1 (T_{bb3})_{[ij]}\epsilon^{xjk} (F_{8})^{l}_x (H_3)^{i} (\overline F_{b\bar{3}})_{[kl]}+\overline c_1 (T_{bb3})_{[ij]}\epsilon^{xkl} (F_{8})^{j}_x (H_3)^{i} (\overline F_{b\bar{3}})_{[kl]}\nonumber\\
  &&+c_2 (T_{bb3})_{[ij]}\epsilon^{xij} (F_{8})^{l}_x (H_3)^{k} (\overline F_{b\bar{3}})_{[kl]}+\overline c_2 (T_{bb3})_{[ij]}\epsilon^{xik} (F_{8})^{j}_x (H_3)^{l} (\overline F_{b\bar{3}})_{[kl]}\nonumber\\
  &&+c_3 (T_{bb3})_{[ij]}\epsilon^{xkl} (F_{8})^{m}_x (H_{\bar 6})^{[ij]}_k (\overline F_{b\bar{3}})_{[lm]}+\overline c_3 (T_{bb3})_{[ij]}\epsilon^{xlm} (F_{8})^{k}_x (H_{\bar 6})^{[ij]}_k (\overline F_{b\bar{3}})_{[lm]}\nonumber\\
  &&+\overline c_{3^{\prime}} (T_{bb3})_{[ij]}\epsilon^{xjk} (F_{8})^{m}_x (H_{\bar 6})^{[il]}_k (\overline F_{b\bar{3}})_{[lm]}+c_4 (T_{bb3})_{[ij]}\epsilon^{xjk} (F_{8})^{m}_x (H_{\bar 6})^{[il]}_m (\overline F_{b\bar{3}})_{[kl]}\nonumber\\
  &&+c_5 (T_{bb3})_{[ij]}\epsilon^{xlm} (F_{8})^{j}_x (H_{\bar 6})^{[ik]}_l (\overline F_{b\bar{3}})_{[km]}+\overline c_{5} (T_{bb3})_{[ij]}\epsilon^{xij} (F_{8})^{m}_x (H_{\bar 6})^{[kl]}_m (\overline F_{b\bar{3}})_{[kl]}\nonumber\\
  &&+\overline c_{5^{\prime}} (T_{bb3})_{[ij]}\epsilon^{xim} (F_{8})^{j}_x (H_{\bar 6})^{[kl]}_m (\overline F_{b\bar{3}})_{[kl]}+c_{6} (T_{bb3})_{[ij]}\epsilon^{xjk} (F_{8})^{m}_x (H_{15})^{\{il\}}_k (\overline F_{b\bar{3}})_{[lm]}\nonumber\\
  &&+c_{7} (T_{bb3})_{[ij]}\epsilon^{xjk} (F_{8})^{m}_x (H_{15})^{\{il\}}_m (\overline F_{b\bar{3}})_{[kl]}+c_{8} (T_{bb3})_{[ij]}\epsilon^{xlm} (F_{8})^{j}_x (H_{15})^{\{ik\}}_l (\overline F_{b\bar{3}})_{[km]}\nonumber\\
  &&+d_1 (T_{bb3})_{[ij]}\epsilon^{xjk} (F_{8})^{l}_x (H_{3})^{i} (\overline F_{b6})_{\{kl\}}+\overline d_1 (T_{bb3})_{[ij]}\epsilon^{xkl} (F_{8})^{m}_x (H_{\bar{6}})^{[ij]}_k (\overline F_{b6})_{\{lm\}}\nonumber\\
  &&+\overline d_{1^{\prime}} (T_{bb3})_{[ij]}\epsilon^{xjk} (F_{8})^{m}_x (H_{\bar{6}})^{[il]}_k (\overline F_{b6})_{\{lm\}}+d_2 (T_{bb3})_{[ij]}\epsilon^{xij} (F_{8})^{l}_x (H_{3})^{k} (\overline F_{b6})_{\{kl\}}\nonumber\\
  &&+\overline d_2 (T_{bb3})_{[ij]}\epsilon^{xik} (F_{8})^{j}_x (H_{3})^{l} (\overline F_{b6})_{\{kl\}}+
  d_3 (T_{bb3})_{[ij]}\epsilon^{xjk} (F_{8})^{m}_x (H_{\bar{6}})^{[il]}_m (\overline F_{b6})_{\{kl\}}\nonumber\\
  &&+d_4 (T_{bb3})_{[ij]}\epsilon^{xlm} (F_{8})^{j}_x (H_{\bar{6}})^{[ik]}_l (\overline F_{b6})_{\{km\}}+d_5 (T_{bb3})_{[ij]}\epsilon^{xjk} (F_{8})^{m}_x (H_{15})^{\{il\}}_k (\overline F_{b6})_{\{lm\}}\nonumber\\
  &&+d_6 (T_{bb3})_{[ij]}\epsilon^{xjk} (F_{8})^{m}_x (H_{15})^{\{il\}}_m (\overline F_{b6})_{\{kl\}}+d_{7} (T_{bb3})_{[ij]}\epsilon^{xlm} (F_{8})^{j}_x (H_{15})^{\{ik\}}_l (\overline F_{b6})_{\{km\}}\nonumber\\
  &&+d_{8} (T_{bb3})_{[ij]}\epsilon^{xij} (F_{8})^{m}_x (H_{15})^{\{kl\}}_m (\overline F_{b6})_{\{kl\}}+\overline d_{8} (T_{bb3})_{[ij]}\epsilon^{xim} (F_{8})^{j}_x (H_{15})^{\{kl\}}_m (\overline F_{b6})_{\{kl\}}\nonumber\\
  &&+a_1 (T_{bb3})_{[ij]}(F_{\overline {10}})^{\{jkl\}} (H_{\bar{6}})^{[im]}_k (\overline F_{b\bar{3}})_{[lm]}+a_2 (T_{bb3})_{[ij]}(F_{\overline {10}})^{\{jkl\}} (H_{15})^{\{im\}}_k (\overline F_{b\bar{3}})_{[lm]}\nonumber\\
  &&+b_1 (T_{bb3})_{[ij]}(F_{\overline {10}})^{\{jkl\}} (H_{3})^{i}_k (\overline F_{b6})_{\{kl\}}+b_2 (T_{bb3})_{[ij]}(F_{\overline {10}})^{\{klm\}} (H_{\bar{6}})^{[ij]}_k (\overline F_{b6})_{\{lm\}}\nonumber\\
  &&+b_3 (T_{bb3})_{[ij]}(F_{\overline {10}})^{\{jkl\}} (H_{\bar{6}})^{[im]}_k (\overline F_{b6})_{\{lm\}}+b_4 (T_{bb3})_{[ij]}(F_{\overline {10}})^{\{jkl\}} (H_{15})^{\{im\}}_k (\overline F_{b6})_{\{lm\}}.
  \end{eqnarray}
Decay amplitudes are given in Tab.~\ref{tab:bb_Fbar_Fb_bd} for the transition $b\to d$ , Tab.~\ref{tab:bb_Fbar_Fb_bs} for the transition $b\to s$. We remove the similar contributions in the amplitudes, such as $c_1-2\overline c_1$, $2c_2-\overline c_2$, $c_3-2\overline c_3+\overline c_{3^{\prime}}$, $c_5+2\overline c_{5}+\overline c_{5^{\prime}}$, $2d_2+\overline d_2$, $d_1-2\overline d_1+\overline d_{1^{\prime}}$, $2d_{8}+\overline d_{8}$.
There is no relation of decay widths for class \uppercase\expandafter{\romannumeral1}.
The relations of decay widths for class \uppercase\expandafter{\romannumeral2} become:
\begin{eqnarray*}
    \Gamma(T_{bb\bar{d}\bar{s}}^{0}\to \overline \Sigma^+\Sigma_{b}^{-})= 2\Gamma(T_{bb\bar{d}\bar{s}}^{0}\to \overline \Xi^+\Xi_{b}^{\prime-}).
\end{eqnarray*}
The relations of decay widths for class \uppercase\expandafter{\romannumeral3} become:
\begin{eqnarray*}
    \Gamma(T_{bb\bar{u}\bar{s}}^{-}\to \overline \Sigma^{\prime-}\Lambda_b^0)= { }\Gamma(T_{bb\bar{u}\bar{s}}^{-}\to \overline \Xi^{\prime0}\Xi_b^-), \Gamma(T_{bb\bar{u}\bar{s}}^{-}\to \overline \Sigma^{\prime-}\Lambda_b^0)= 2\Gamma(T_{bb\bar{u}\bar{d}}^{-}\to \overline \Sigma^{\prime0}\Xi_b^-),\\ \Gamma(T_{bb\bar{d}\bar{s}}^{0}\to \overline \Sigma^{\prime0}\Lambda_b^0)= \frac{1}{2}\Gamma(T_{bb\bar{d}\bar{s}}^{0}\to \overline \Xi^{\prime0}\Xi_b^0),
    \Gamma(T_{bb\bar{d}\bar{s}}^{0}\to \overline \Sigma^{\prime0}\Lambda_b^0)= \frac{1}{2}\Gamma(T_{bb\bar{u}\bar{d}}^{-}\to \overline \Sigma^{\prime-}\Xi_b^0),\\
    \Gamma(T_{bb\bar{u}\bar{d}}^{-}\to \overline \Sigma^{\prime-}\Xi_b^0)= { }\Gamma(T_{bb\bar{d}\bar{s}}^{0}\to \overline \Xi^{\prime0}\Xi_b^0), \Gamma(T_{bb\bar{u}\bar{d}}^{-}\to \overline \Sigma^{\prime0}\Xi_b^-)= \frac{1}{2}\Gamma(T_{bb\bar{u}\bar{s}}^{-}\to \overline \Xi^{\prime0}\Xi_b^-).
\end{eqnarray*}
The relations of decay widths for class \uppercase\expandafter{\romannumeral4} become:
\begin{eqnarray*}
    \Gamma(T_{bb\bar{u}\bar{s}}^{-}\to \overline \Sigma^{\prime-}\Sigma_{b}^{0})= { }\Gamma(T_{bb\bar{u}\bar{s}}^{-}\to \overline \Sigma^{\prime0}\Sigma_{b}^{-}), \Gamma(T_{bb\bar{u}\bar{s}}^{-}\to \overline \Sigma^{\prime-}\Sigma_{b}^{0})= { }\Gamma(T_{bb\bar{u}\bar{s}}^{-}\to \overline \Xi^{\prime0}\Xi_{b}^{\prime-}),\\ \Gamma(T_{bb\bar{u}\bar{s}}^{-}\to \overline \Sigma^{\prime0}\Sigma_{b}^{-})= { }\Gamma(T_{bb\bar{u}\bar{s}}^{-}\to \overline \Xi^{\prime0}\Xi_{b}^{\prime-}),
    \Gamma(T_{bb\bar{d}\bar{s}}^{0}\to \overline \Sigma^{\prime0}\Sigma_{b}^{0})= \frac{1}{2}\Gamma(T_{bb\bar{d}\bar{s}}^{0}\to \overline \Xi^{\prime0}\Xi_{b}^{\prime0}),\\
    \Gamma(T_{bb\bar{d}\bar{s}}^{0}\to \overline \Sigma^{\prime+}\Sigma_{b}^{-})= \frac{1}{2}\Gamma(T_{bb\bar{d}\bar{s}}^{0}\to \overline \Xi^{\prime+}\Xi_{b}^{\prime-}), \Gamma(T_{bb\bar{d}\bar{s}}^{0}\to \overline \Sigma^{\prime+}\Sigma_{b}^{-})= \frac{1}{3}\Gamma(T_{bb\bar{d}\bar{s}}^{0}\to \overline \Omega^+\Omega_{b}^{-}),\\ \Gamma(T_{bb\bar{d}\bar{s}}^{0}\to \overline \Xi^{\prime+}\Xi_{b}^{\prime-})= \frac{2}{3}\Gamma(T_{bb\bar{d}\bar{s}}^{0}\to \overline \Omega^+\Omega_{b}^{-}).
\end{eqnarray*}
There are no relation of decay widths for class \uppercase\expandafter{\romannumeral1}.

The relations of decay widths for class \uppercase\expandafter{\romannumeral2} become:
\begin{eqnarray*}
    \Gamma(T_{bb\bar{d}\bar{s}}^{0}\to \overline \Sigma^+\Xi_{b}^{\prime-})= \frac{1}{2}\Gamma(T_{bb\bar{d}\bar{s}}^{0}\to \overline \Xi^+\Omega_{b}^{-}).
\end{eqnarray*}
The relations of decay widths for class \uppercase\expandafter{\romannumeral3} are:
\begin{eqnarray*}
    \Gamma(T_{bb\bar{u}\bar{s}}^{-}\to \overline \Delta^{-}\Lambda_b^0)= 2\Gamma(T_{bb\bar{d}\bar{s}}^{0}\to \overline \Sigma^{\prime0}\Xi_b^0), \Gamma(T_{bb\bar{d}\bar{s}}^{0}\to \overline \Delta^{0}\Lambda_b^0)= 2\Gamma(T_{bb\bar{d}\bar{s}}^{0}\to \overline \Sigma^{\prime0}\Xi_b^0),\\ \Gamma(T_{bb\bar{u}\bar{d}}^{-}\to \overline \Delta^{-}\Xi_b^0)= 2\Gamma(T_{bb\bar{u}\bar{s}}^{-}\to \overline \Sigma^{\prime0}\Xi_b^-), \Gamma(T_{bb\bar{u}\bar{d}}^{-}\to \overline \Delta^{-}\Xi_b^0)= { }\Gamma(T_{bb\bar{u}\bar{d}}^{-}\to \overline \Delta^{0}\Xi_b^-),\\
    \Gamma(T_{bb\bar{u}\bar{s}}^{-}\to \overline \Delta^{-}\Lambda_b^0)= { }\Gamma(T_{bb\bar{d}\bar{s}}^{0}\to \overline \Delta^{0}\Lambda_b^0), \Gamma(T_{bb\bar{u}\bar{d}}^{-}\to \overline \Delta^{0}\Xi_b^-)= 2\Gamma(T_{bb\bar{u}\bar{s}}^{-}\to \overline \Sigma^{\prime0}\Xi_b^-).
\end{eqnarray*}
The relations of decay widths for class \uppercase\expandafter{\romannumeral4} become:
\begin{eqnarray*}
    \Gamma(T_{bb\bar{d}\bar{s}}^{0}\to \overline \Delta^{0}\Sigma_{b}^{0})= 2\Gamma(T_{bb\bar{d}\bar{s}}^{0}\to \overline \Sigma^{\prime0}\Xi_{b}^{\prime0}),
    \Gamma(T_{bb\bar{d}\bar{s}}^{0}\to \overline \Delta^{+}\Sigma_{b}^{-})= \frac{3}{2}\Gamma(T_{bb\bar{d}\bar{s}}^{0}\to \overline \Sigma^{\prime+}\Xi_{b}^{\prime-}),\\ \Gamma(T_{bb\bar{d}\bar{s}}^{0}\to \overline \Delta^{+}\Sigma_{b}^{-})= 3\Gamma(T_{bb\bar{d}\bar{s}}^{0}\to \overline \Xi^{\prime+}\Omega_{b}^{-}), \Gamma(T_{bb\bar{d}\bar{s}}^{0}\to \overline \Sigma^{\prime+}\Xi_{b}^{\prime-})= 2\Gamma(T_{bb\bar{d}\bar{s}}^{0}\to \overline \Xi^{\prime+}\Omega_{b}^{-}),\\ \Gamma(T_{bb\bar{u}\bar{d}}^{-}\to \overline \Delta^{-}\Xi_{b}^{\prime0})= { }\Gamma(T_{bb\bar{u}\bar{d}}^{-}\to \overline \Delta^{0}\Xi_{b}^{\prime-}), \Gamma(T_{bb\bar{u}\bar{d}}^{-}\to \overline \Delta^{-}\Xi_{b}^{\prime0})= { }\Gamma(T_{bb\bar{u}\bar{d}}^{-}\to \overline \Sigma^{\prime0}\Omega_{b}^{-}),\\ \Gamma(T_{bb\bar{u}\bar{d}}^{-}\to \overline \Delta^{0}\Xi_{b}^{\prime-})= { }\Gamma(T_{bb\bar{u}\bar{d}}^{-}\to \overline \Sigma^{\prime0}\Omega_{b}^{-}).
    \end{eqnarray*}

\subsubsection{ Decays into a bottom meson and two light mesons}
\begin{table}
\caption{Doubly bottom tetraquark $T_{bb\bar{q}\bar{q}}$ decays into a bottom meson and two light mesons induced by the charmless $b\to d$ transition.}\label{tab:bb_Band2meson8_1}\begin{tabular}{|c|c|c|c|c|c|c|c}\hline\hline
channel & amplitude \\\hline
$T_{bb\bar{u}\bar{s}}^{-}\to   B^-  \pi^0   K^0  $ & $ \frac{-c_2+c_4+2 c_5+c_6-c_7-c_8+c_{11}+5 c_{12}+c_{13}+3 c_{14}-3 c_{15}-c_{16}}{\sqrt{2}}$\\\hline
$T_{bb\bar{u}\bar{s}}^{-}\to   B^-  \pi^-   K^+  $ & $ c_2-c_4-2 c_5-c_6-c_7-c_8-c_{11}+3 c_{12}-c_{13}-3 c_{14}-c_{15}-3 c_{16}$\\\hline
$T_{bb\bar{u}\bar{s}}^{-}\to   B^-  K^0   \eta  $ & $ \frac{-c_2+c_4+2 c_5+c_6-c_7-c_8+c_{11}+5 c_{12}+c_{13}+3 c_{14}-3 c_{15}-c_{16}}{\sqrt{6}}$\\\hline
$T_{bb\bar{u}\bar{s}}^{-}\to   \overline B^0  \pi^-   K^0  $ & $ -2 \left(c_7+c_8+2 \left(c_{15}+c_{16}\right)\right)$\\\hline
$T_{bb\bar{u}\bar{s}}^{-}\to   \overline B^0_s  \pi^0   \pi^-  $ & $ -4 \sqrt{2} c_{12}$\\\hline
$T_{bb\bar{u}\bar{s}}^{-}\to   \overline B^0_s  \pi^-   \eta  $ & $ -\sqrt{\frac{2}{3}} \left(c_2-c_4+2 c_5+c_6-c_7-c_8+c_{11}+3 c_{12}+3 c_{13}+c_{14}-c_{15}-3 c_{16}\right)$\\\hline
$T_{bb\bar{u}\bar{s}}^{-}\to   \overline B^0_s  K^0   K^-  $ & $ -c_2+c_4-2 c_5-c_6-c_7-c_8-c_{11}+c_{12}-3 c_{13}-c_{14}-3 c_{15}-c_{16}$\\\hline
$T_{bb\bar{d}\bar{s}}^{0}\to   B^-  \pi^+   K^0  $ & $ c_3+c_4+2 c_5+c_6-c_8+2 c_{10}-3 c_{14}+2 c_{15}-c_{16}$\\\hline
$T_{bb\bar{d}\bar{s}}^{0}\to   B^-  \pi^0   K^+  $ & $ \frac{c_3+c_4+2 c_5+c_6+2 c_7+c_8+2 c_{10}-3 c_{14}-2 c_{15}-5 c_{16}}{\sqrt{2}}$\\\hline
$T_{bb\bar{d}\bar{s}}^{0}\to   B^-  K^+   \eta  $ & $ -\frac{c_3+c_4+2 c_5+c_6-2 c_7-3 c_8+2 c_{10}-3 c_{14}+6 c_{15}+3 c_{16}}{\sqrt{6}}$\\\hline
$T_{bb\bar{d}\bar{s}}^{0}\to   \overline B^0  \pi^0   K^0  $ & $ \frac{-c_2-c_3+c_7+2 c_8-2 c_{10}+c_{11}+5 c_{12}+c_{13}-2 c_{14}-c_{15}-4 c_{16}}{\sqrt{2}}$\\\hline
$T_{bb\bar{d}\bar{s}}^{0}\to   \overline B^0  \pi^-   K^+  $ & $ c_2+c_3+c_7+2 c_{10}-c_{11}+3 c_{12}-c_{13}+2 c_{14}-3 c_{15}$\\\hline
$T_{bb\bar{d}\bar{s}}^{0}\to   \overline B^0  K^0   \eta  $ & $ \frac{-c_2-c_3+c_7+2 c_8-2 c_{10}+c_{11}+5 c_{12}+c_{13}-2 c_{14}-c_{15}-4 c_{16}}{\sqrt{6}}$\\\hline
$T_{bb\bar{d}\bar{s}}^{0}\to   \overline B^0_s  \pi^+   \pi^-  $ & $ -2 c_1-c_2-c_6+4 c_9+c_{11}-3 c_{12}-c_{13}-c_{14}$\\\hline
$T_{bb\bar{d}\bar{s}}^{0}\to   \overline B^0_s  \pi^0   \pi^0  $ & $ -2 c_1-c_2-c_6+4 c_9+c_{11}+5 c_{12}-c_{13}-c_{14}$\\\hline
$T_{bb\bar{d}\bar{s}}^{0}\to   \overline B^0_s  \pi^0   \eta  $ & $ \frac{c_2-c_4+2 c_5+c_6-c_7-c_8+c_{11}-c_{12}-5 c_{13}+c_{14}-c_{15}+5 c_{16}}{\sqrt{3}}$\\\hline
$T_{bb\bar{d}\bar{s}}^{0}\to   \overline B^0_s  K^+   K^-  $ & $ -2 c_1+c_3-c_6+c_7+4 c_9+2 c_{10}-2 c_{13}+c_{14}-3 c_{15}$\\\hline
$T_{bb\bar{d}\bar{s}}^{0}\to   \overline B^0_s  K^0   \overline K^0  $ & $ -2 c_1-c_2+c_3+c_4-2 c_5-2 c_6-c_8+4 c_9+2 c_{10}-c_{11}+c_{12}+3 c_{13}+2 c_{15}-c_{16}$\\\hline
$T_{bb\bar{d}\bar{s}}^{0}\to   \overline B^0_s  \eta   \eta  $ & $ \frac{-6 c_1-c_2+4 c_3+2 c_4-4 c_5-5 c_6-2 c_7-6 c_8+12 c_9+8 c_{10}-3 c_{11}-3 c_{12}+3 c_{13}+3 c_{14}+6 c_{15}+6 c_{16}}{3}$\\\hline
$T_{bb\bar{u}\bar{d}}^{-}\to   B^-  \pi^+   \pi^-  $ & $ 2 c_1+c_2-c_3-c_4-2 c_5-2 c_6-c_8+4 c_9+2 c_{10}-c_{11}+3 c_{12}+c_{13}-2 c_{15}-3 c_{16}$\\\hline
$T_{bb\bar{u}\bar{d}}^{-}\to   B^-  \pi^0   \pi^0  $ & $ 2 c_1+c_2-c_3-c_4-2 c_5-2 c_6-c_8+4 c_9+2 c_{10}-c_{11}-5 c_{12}+c_{13}+6 c_{15}+5 c_{16}$\\\hline
$T_{bb\bar{u}\bar{d}}^{-}\to   B^-  \pi^0   \eta  $ & $ -\frac{c_2+c_3+c_7+2 c_8-2 c_{10}+c_{11}-c_{12}-5 c_{13}-6 c_{14}-3 c_{15}-4 c_{16}}{\sqrt{3}}$\\\hline
$T_{bb\bar{u}\bar{d}}^{-}\to   B^-  K^+   K^-  $ & $ 2 c_1-c_3-c_6+c_7+4 c_9+2 c_{10}+2 c_{13}+3 c_{14}-c_{15}$\\\hline
$T_{bb\bar{u}\bar{d}}^{-}\to   B^-  K^0   \overline K^0  $ & $ 2 c_1+c_2-c_6+4 c_9+c_{11}-c_{12}-3 c_{13}-3 c_{14}$\\\hline
$T_{bb\bar{u}\bar{d}}^{-}\to   B^-  \eta   \eta  $ & $ \frac{6 c_1+c_2-c_3+c_4+2 c_5-2 c_6-2 c_7-3 c_8+12 c_9+2 c_{10}+3 c_{11}+3 c_{12}-3 c_{13}+3 c_{16}}{3} $\\\hline
$T_{bb\bar{u}\bar{d}}^{-}\to   \overline B^0  \pi^0   \pi^-  $ & $ 4 \sqrt{2} \left(-c_{12}+c_{15}+c_{16}\right)$\\\hline
$T_{bb\bar{u}\bar{d}}^{-}\to   \overline B^0  \pi^-   \eta  $ & $ -\sqrt{\frac{2}{3}} \left(c_2+c_3+c_7+2 c_8-2 c_{10}+c_{11}+3 c_{12}+3 c_{13}+2 c_{14}+c_{15}\right)$\\\hline
$T_{bb\bar{u}\bar{d}}^{-}\to   \overline B^0  K^0   K^-  $ & $ -c_2-c_3+c_7+2 c_{10}-c_{11}+c_{12}-3 c_{13}-2 c_{14}-c_{15}$\\\hline
$T_{bb\bar{u}\bar{d}}^{-}\to   \overline B^0_s  \pi^0   K^-  $ & $ -\frac{c_3+c_4-2 c_5-c_6+c_8-2 c_{10}+c_{14}-6 c_{15}-5 c_{16}}{\sqrt{2}}$\\\hline
$T_{bb\bar{u}\bar{d}}^{-}\to   \overline B^0_s  \pi^-   \overline K^0  $ & $ -c_3-c_4+2 c_5+c_6-c_8+2 c_{10}-c_{14}-2 c_{15}-3 c_{16}$\\\hline
$T_{bb\bar{u}\bar{d}}^{-}\to   \overline B^0_s  K^-   \eta  $ & $ \frac{c_3+c_4-2 c_5-c_6-4 c_7-3 c_8-2 c_{10}+c_{14}+2 c_{15}+3 c_{16}}{\sqrt{6}}$\\\hline
\hline
\end{tabular}
\end{table}

\begin{table}
\caption{Doubly bottom tetraquark $T_{bb\bar{q}\bar{q}}$ decays into a bottom meson and two light mesons induced by the charmless $b\to s$ transition.}\label{tab:bb_Band2meson8_2}\begin{tabular}{|c|c|c|c|c|c|c|c}\hline\hline
channel & amplitude \\\hline
$T_{bb\bar{u}\bar{s}}^{-}\to   B^-  \pi^+   \pi^-  $ & $ 2 c_1-c_3-c_6+c_7+4 c_9+2 c_{10}+2 c_{13}+3 c_{14}-c_{15}$\\\hline
$T_{bb\bar{u}\bar{s}}^{-}\to   B^-  \pi^0   \pi^0  $ & $ 2 c_1-c_3-c_6-c_7-2 c_8+4 c_9+2 c_{10}+2 c_{13}+3 c_{14}+3 c_{15}+4 c_{16}$\\\hline
$T_{bb\bar{u}\bar{s}}^{-}\to   B^-  \pi^0   \eta  $ & $ -\frac{c_3+c_4+2 c_5+c_6+c_8-2 c_{10}+2 c_{11}+4 c_{12}-4 c_{13}-3 c_{14}-6 c_{15}-5 c_{16}}{\sqrt{3}}$\\\hline
$T_{bb\bar{u}\bar{s}}^{-}\to   B^-  K^+   K^-  $ & $ 2 c_1+c_2-c_3-c_4-2 c_5-2 c_6-c_8+4 c_9+2 c_{10}-c_{11}+3 c_{12}+c_{13}-2 c_{15}-3 c_{16}$\\\hline
$T_{bb\bar{u}\bar{s}}^{-}\to   B^-  K^0   \overline K^0  $ & $ 2 c_1+c_2-c_6+4 c_9+c_{11}-c_{12}-3 c_{13}-3 c_{14}$\\\hline
$T_{bb\bar{u}\bar{s}}^{-}\to   B^-  \eta   \eta  $ & $ \frac{6 c_1+4 c_2-c_3-2 c_4-4 c_5-5 c_6+c_7+12 c_9+2 c_{10}-12 c_{12}-6 c_{13}-9 c_{14}+9 c_{15}+6 c_{16}}{3} $\\\hline
$T_{bb\bar{u}\bar{s}}^{-}\to   \overline B^0  \pi^0   \pi^-  $ & $ -\sqrt{2} \left(c_7+c_8-2 \left(c_{15}+c_{16}\right)\right)$\\\hline
$T_{bb\bar{u}\bar{s}}^{-}\to   \overline B^0  \pi^-   \eta  $ & $ \sqrt{\frac{2}{3}} \left(-c_3-c_4+2 c_5+c_6+c_7+2 c_{10}-c_{14}+4 c_{15}+3 c_{16}\right)$\\\hline
$T_{bb\bar{u}\bar{s}}^{-}\to   \overline B^0  K^0   K^-  $ & $ -c_3-c_4+2 c_5+c_6-c_8+2 c_{10}-c_{14}-2 c_{15}-3 c_{16}$\\\hline
$T_{bb\bar{u}\bar{s}}^{-}\to   \overline B^0_s  \pi^0   K^-  $ & $ -\frac{c_2+c_3+c_7+2 c_8-2 c_{10}+c_{11}+7 c_{12}+3 c_{13}+2 c_{14}-3 c_{15}-4 c_{16}}{\sqrt{2}}$\\\hline
$T_{bb\bar{u}\bar{s}}^{-}\to   \overline B^0_s  \pi^-   \overline K^0  $ & $ -c_2-c_3+c_7+2 c_{10}-c_{11}+c_{12}-3 c_{13}-2 c_{14}-c_{15}$\\\hline
$T_{bb\bar{u}\bar{s}}^{-}\to   \overline B^0_s  K^-   \eta  $ & $ \frac{c_2+c_3+c_7+2 c_8-2 c_{10}+c_{11}-9 c_{12}+3 c_{13}+2 c_{14}+13 c_{15}+12 c_{16}}{\sqrt{6}}$\\\hline
$T_{bb\bar{d}\bar{s}}^{0}\to   B^-  \pi^+   \pi^0  $ & $ -\sqrt{2} \left(c_7+c_8-2 \left(c_{15}+c_{16}\right)\right)$\\\hline
$T_{bb\bar{d}\bar{s}}^{0}\to   B^-  \pi^+   \eta  $ & $ -\sqrt{\frac{2}{3}} \left(c_3+c_4+2 c_5+c_6+c_7+2 c_{10}-3 c_{14}-3 c_{16}\right)$\\\hline
$T_{bb\bar{d}\bar{s}}^{0}\to   B^-  K^+   \overline K^0  $ & $ -c_3-c_4-2 c_5-c_6+c_8-2 c_{10}+3 c_{14}-2 c_{15}+c_{16}$\\\hline
$T_{bb\bar{d}\bar{s}}^{0}\to   \overline B^0  \pi^+   \pi^-  $ & $ 2 c_1-c_3+c_6-c_7-4 c_9-2 c_{10}+2 c_{13}-c_{14}+3 c_{15}$\\\hline
$T_{bb\bar{d}\bar{s}}^{0}\to   \overline B^0  \pi^0   \pi^0  $ & $ 2 c_1-c_3+c_6+c_7+2 c_8-4 c_9-2 c_{10}+2 c_{13}-c_{14}-c_{15}-4 c_{16}$\\\hline
$T_{bb\bar{d}\bar{s}}^{0}\to   \overline B^0  \pi^0   \eta  $ & $ \frac{c_3+c_4-2 c_5-c_6-c_8+2 c_{10}-2 c_{11}-4 c_{12}+4 c_{13}+c_{14}+2 c_{15}-c_{16}}{\sqrt{3}}$\\\hline
$T_{bb\bar{d}\bar{s}}^{0}\to   \overline B^0  K^+   K^-  $ & $ 2 c_1+c_2+c_6-4 c_9-c_{11}+3 c_{12}+c_{13}+c_{14}$\\\hline
$T_{bb\bar{d}\bar{s}}^{0}\to   \overline B^0  K^0   \overline K^0  $ & $ 2 c_1+c_2-c_3-c_4+2 c_5+2 c_6+c_8-4 c_9-2 c_{10}+c_{11}-c_{12}-3 c_{13}-2 c_{15}+c_{16}$\\\hline
$T_{bb\bar{d}\bar{s}}^{0}\to   \overline B^0  \eta   \eta  $ & $ \frac{ 6 c_1+4 c_2-c_3-2 c_4+4 c_5+5 c_6-c_7-12 c_9-2 c_{10}-12 c_{12}-6 c_{13}+3 c_{14}-3 c_{15}+6 c_{16}}{3}$\\\hline
$T_{bb\bar{d}\bar{s}}^{0}\to   \overline B^0_s  \pi^+   K^-  $ & $ -c_2-c_3-c_7-2 c_{10}+c_{11}-3 c_{12}+c_{13}-2 c_{14}+3 c_{15}$\\\hline
$T_{bb\bar{d}\bar{s}}^{0}\to   \overline B^0_s  \pi^0   \overline K^0  $ & $ \frac{c_2+c_3-c_7-2 c_8+2 c_{10}-c_{11}-5 c_{12}-c_{13}+2 c_{14}+c_{15}+4 c_{16}}{\sqrt{2}}$\\\hline
$T_{bb\bar{d}\bar{s}}^{0}\to   \overline B^0_s  \overline K^0   \eta  $ & $ \frac{c_2+c_3-c_7-2 c_8+2 c_{10}-c_{11}-5 c_{12}-c_{13}+2 c_{14}+c_{15}+4 c_{16}}{\sqrt{6}}$\\\hline
$T_{bb\bar{u}\bar{d}}^{-}\to   B^-  \pi^+   K^-  $ & $ c_2-c_4-2 c_5-c_6-c_7-c_8-c_{11}+3 c_{12}-c_{13}-3 c_{14}-c_{15}-3 c_{16}$\\\hline
$T_{bb\bar{u}\bar{d}}^{-}\to   B^-  \pi^0   \overline K^0  $ & $ \frac{-c_2+c_4+2 c_5+c_6-c_7-c_8+c_{11}+5 c_{12}+c_{13}+3 c_{14}-3 c_{15}-c_{16}}{\sqrt{2}}$\\\hline
$T_{bb\bar{u}\bar{d}}^{-}\to   B^-  \overline K^0   \eta  $ & $ \frac{-c_2+c_4+2 c_5+c_6-c_7-c_8+c_{11}+5 c_{12}+c_{13}+3 c_{14}-3 c_{15}-c_{16}}{\sqrt{6}}$\\\hline
$T_{bb\bar{u}\bar{d}}^{-}\to   \overline B^0  \pi^0   K^-  $ & $ -\frac{c_2-c_4+2 c_5+c_6-c_7-c_8+c_{11}+7 c_{12}+3 c_{13}+c_{14}-c_{15}-3 c_{16}}{\sqrt{2}}$\\\hline
$T_{bb\bar{u}\bar{d}}^{-}\to   \overline B^0  \pi^-   \overline K^0  $ & $ -c_2+c_4-2 c_5-c_6-c_7-c_8-c_{11}+c_{12}-3 c_{13}-c_{14}-3 c_{15}-c_{16}$\\\hline
$T_{bb\bar{u}\bar{d}}^{-}\to   \overline B^0  K^-   \eta  $ & $ \frac{c_2-c_4+2 c_5+c_6-c_7-c_8+c_{11}-9 c_{12}+3 c_{13}+c_{14}-c_{15}-3 c_{16}}{\sqrt{6}}$\\\hline
$T_{bb\bar{u}\bar{d}}^{-}\to   \overline B^0_s  \overline K^0   K^-  $ & $ -2 \left(c_7+c_8+2 \left(c_{15}+c_{16}\right)\right)$\\\hline
\hline
\end{tabular}
\end{table}
The  Hamiltonian for decays into a bottom meson and two light mesons  is
\begin{eqnarray}
 {\cal H}_{eff}&=&c_1(T_{bb3})_{[ij]}  (\overline B)^{i} M^{k}_{l} M^{l}_{k} (H_{3})^j +c_2(T_{bb3})_{[ij]}  (\overline B)^{i} M^{j}_{k} M^{k}_{l} (H_{3})^l  \nonumber\\
 && + c_3(T_{bb3})_{[ij]}  (\overline B)^{l} M^{j}_{k} M^{k}_{l} (H_{3})^i+c_4(T_{bb3})_{[ij]}  (\overline B)^{k} M^{i}_{k} M^{j}_{l} (H_{3})^l\nonumber\\
 &&+  c_5(T_{bb3})_{[ij]}  (\overline B)^{m}M^{i}_{k}M^{j}_{l}  (H_{\overline6})^{[kl]}_{m} + c_6(T_{bb3})_{[ij]}  (\overline B)^{l}M^{i}_{m}M^{m}_{k}  (H_{\overline6})^{[jk]}_{l}\nonumber\\
 &&+ c_7(T_{bb3})_{[ij]}  (\overline B)^{m}M^{i}_{k}M^{l}_{m}  (H_{\overline6})^{[jk]}_{l} +c_8(T_{bb3})_{[ij]}  (\overline B)^{m}M^{i}_{m}M^{l}_{k}  (H_{\overline6})^{[jk]}_{l}\nonumber\\
 &&+ c_9(T_{bb3})_{[ij]}  (\overline B)^{k}M^{l}_{m}M^{m}_{l}  (H_{\overline6})^{[ij]}_{k} + c_{10}(T_{bb3})_{[ij]}  (\overline B)^{m}M^{k}_{l}M^{l}_{m}  (H_{\overline6})^{[ij]}_{k}\nonumber\\
 &&+c_{11}(T_{bb3})_{[ij]}  (\overline B)^{i}M^{l}_{m}M^{m}_{k}  (H_{\overline6})^{[jk]}_{l}+c_{12}(T_{bb3})_{[ij]}  (\overline B)^{i}M^{j}_{k}M^{m}_{l}  (H_{15})^{\{kl\}}_{m}\nonumber\\
 &&+ c_{13}(T_{bb3})_{[ij]}  (\overline B)^{i}M^{l}_{m}M^{m}_{k}  (H_{15})^{\{jk\}}_{l}+ c_{14}(T_{bb3})_{[ij]}  (\overline B)^{l}M^{i}_{m}M^{m}_{k}  (H_{15})^{\{jk\}}_{l}\nonumber\\
 &&+ c_{15}(T_{bb3})_{[ij]}  (\overline B)^{m}M^{i}_{k}M^{l}_{m}  (H_{15})^{\{jk\}}_{l}+ c_{16}(T_{bb3})_{[ij]}  (\overline B)^{m}M^{i}_{m}M^{l}_{k}  (H_{15})^{\{jk\}}_{l}\nonumber\\.
 &&+\overline{c_{11}}(T_{bb3})_{[ij]}  (\overline B)^{i}M^{j}_{k}M^{m}_{l}  (H_{\overline6})^{[kl]}_{m}.
\end{eqnarray}
The $\overline {c_{11}}$ and $c_{11}$ terms give the same contribution which always contain the factor $c_{11}-\overline {c_{11}}$. We remove the $\overline {c_{11}}$ term in the expanded amplitudes. The amplitudes are given in Tab.~\ref{tab:bb_Band2meson8_1} for the transition $b\to d$ and Tab.~\ref{tab:bb_Band2meson8_2} for the transition $b\to s$. The relations are:
\begin{eqnarray*}
    \Gamma(T_{bb\bar{u}\bar{s}}^{-}\to B^- \pi^0 K^0 )= 3\Gamma(T_{bb\bar{u}\bar{s}}^{-}\to B^- K^0 \eta ),
    \Gamma(T_{bb\bar{d}\bar{s}}^{0}\to \overline B^0 \pi^0 K^0 )= 3\Gamma(T_{bb\bar{d}\bar{s}}^{0}\to \overline B^0 K^0 \eta ).
 \end{eqnarray*}
\begin{eqnarray*}
    \Gamma(T_{bb\bar{d}\bar{s}}^{0}\to B^- \pi^+ \pi^0 )= { }\Gamma(T_{bb\bar{u}\bar{s}}^{-}\to \overline B^0 \pi^0 \pi^- ), \Gamma(T_{bb\bar{d}\bar{s}}^{0}\to \overline B^0_s \pi^0 \overline K^0 )= 3\Gamma(T_{bb\bar{d}\bar{s}}^{0}\to \overline B^0_s \overline K^0 \eta ),\\ \Gamma(T_{bb\bar{u}\bar{d}}^{-}\to B^- \pi^0 \overline K^0 )= 3\Gamma(T_{bb\bar{u}\bar{d}}^{-}\to B^- \overline K^0 \eta ).
\end{eqnarray*}
\section{Non-Leptonic $T_{bc\bar{q}\bar{q}}$ decays}
\label{sec:bcq_nonleptonic}
The charm decays or (and) bottom decays can be present in the decays of $T_{bc\bar{q}\bar{q}}$.
For the bottom decay, there is a new decay channel in which two heavy quarks of tetraquark interact by a virtual W-boson. The others can be obtained  from those for $T_{bb\bar{q}\bar{q}}$  decays with $B\to D$ and $B_c \to J/\psi$.
For the charm  decays, the decay amplitudes can be obtained from those for $T_{cc\bar{q}\bar{q}}$ decays with the replacement of $D\to B$ and $J/\psi \to B_c$.
 Thus we do not present those  results again.

\subsection{$bc\to ud/s$ or $bc \to cd/s$ transition}
\begin{figure}
\begin{center}
\includegraphics[scale=0.5]{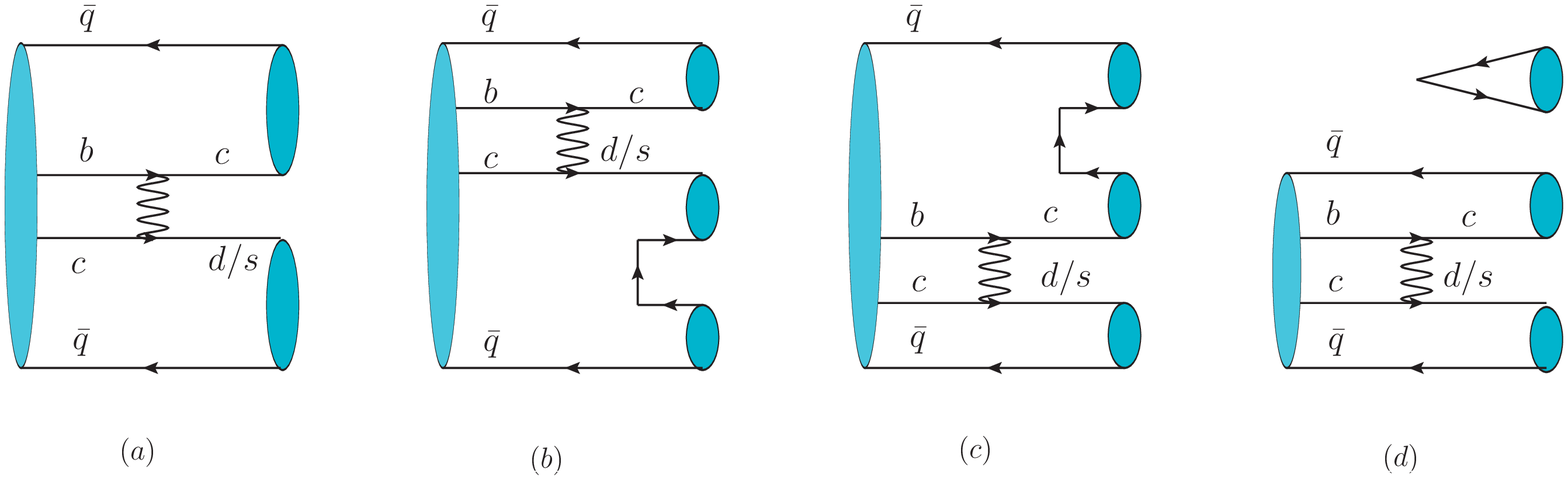}
\end{center}
\caption{Feynman diagrams for nonleptonic decays of  doubly heavy tetraquark $T_{bc\bar{q}\bar{q}}$. (a) is corresponding with two mesons by the new W-exchange process. (b,c,d) are corresponding with three mesons by the new W-exchange process. }\label{fig:Feynman_nonleptonic_tetraquark_4}
\end{figure}
\subsubsection{decays into two mesons}
\begin{table}
\caption{Doubly heavy tetraquark $T_{bc\bar{q}\bar{q}}$ decays into a charmed meson and a light meson .}\label{tab:bc_D_meson8}\begin{tabular}{|cc|cc|}\hline
channel & amplitude(/$V_{cb}$)&channel & amplitude(/$V_{cb}$)\tabularnewline\hline\hline
$T_{bc\bar{u}\bar{s}}^{0}\to    D^0  \pi^0  $ & $ -\frac{f_2V_{cs}^*}{\sqrt{2}}$&
$T_{bc\bar{u}\bar{s}}^{0}\to    D^0  K^0  $ & $ f_1 V_{cd}^*$\\\hline
$T_{bc\bar{u}\bar{s}}^{0}\to    D^0  \eta  $ & $ -\frac{\left(2 f_1+f_2\right)V_{cs}^*}{\sqrt{6}}$&
$T_{bc\bar{u}\bar{s}}^{0}\to    D^+  \pi^-  $ & $ -f_2V_{cs}^*$\\\hline
$T_{bc\bar{u}\bar{s}}^{0}\to    D^+_s  \pi^-  $ & $ -f_1 V_{cd}^*$&
$T_{bc\bar{u}\bar{s}}^{0}\to    D^+_s  K^-  $ & $ -\left(f_1+f_2\right)V_{cs}^*$\\\hline
$T_{bc\bar{d}\bar{s}}^{+}\to    D^0  \pi^+  $ & $ -f_2V_{cs}^*$&
$T_{bc\bar{d}\bar{s}}^{+}\to    D^0  K^+  $ & $ f_2 V_{cd}^*$\\\hline
$T_{bc\bar{d}\bar{s}}^{+}\to    D^+  \pi^0  $ & $ \frac{f_2V_{cs}^*}{\sqrt{2}}$&
$T_{bc\bar{d}\bar{s}}^{+}\to    D^+  K^0  $ & $ \left(f_1+f_2\right) V_{cd}^*$\\\hline
$T_{bc\bar{d}\bar{s}}^{+}\to    D^+  \eta  $ & $ -\frac{\left(2 f_1+f_2\right)V_{cs}^*}{\sqrt{6}}$&
$T_{bc\bar{d}\bar{s}}^{+}\to    D^+_s  \pi^0  $ & $ \frac{f_1 V_{cd}^*}{\sqrt{2}}$\\\hline
$T_{bc\bar{d}\bar{s}}^{+}\to    D^+_s  \overline K^0  $ & $ -\left(f_1+f_2\right)V_{cs}^*$&
$T_{bc\bar{d}\bar{s}}^{+}\to    D^+_s  \eta  $ & $ -\frac{\left(f_1+2 f_2\right) V_{cd}^*}{\sqrt{6}}$\\\hline
$T_{bc\bar{u}\bar{d}}^{0}\to    D^0  \pi^0  $ & $ -\frac{\left(f_1+f_2\right) V_{cd}^*}{\sqrt{2}}$&
$T_{bc\bar{u}\bar{d}}^{0}\to    D^0  \overline K^0  $ & $ f_1V_{cs}^*$\\\hline
$T_{bc\bar{u}\bar{d}}^{0}\to    D^0  \eta  $ & $ \frac{\left(f_1-f_2\right) V_{cd}^*}{\sqrt{6}}$&
$T_{bc\bar{u}\bar{d}}^{0}\to    D^+  \pi^-  $ & $ -\left(f_1+f_2\right) V_{cd}^*$\\\hline
$T_{bc\bar{u}\bar{d}}^{0}\to    D^+  K^-  $ & $ -f_1V_{cs}^*$&
$T_{bc\bar{u}\bar{d}}^{0}\to    D^+_s  K^-  $ & $ -f_2 V_{cd}^*$\\\hline
\hline
\end{tabular}
\end{table}
\begin{table}
\caption{Doubly heavy tetraquark $T_{bc\bar{q}\bar{q}}$ decays into two light mesons.}\label{tab:bc_2meson8}\begin{tabular}{|cc|cc|}\hline
channel & amplitude(/$V_{cb}$) &channel & amplitude(/$V_{cb}$)\tabularnewline\hline\hline
$T_{bc\bar{u}\bar{s}}^{0}\to   \pi^+   \pi^-  $ & $ \left(-f_3+4 f_5+f_6\right)V_{cs}^*$&
$T_{bc\bar{u}\bar{s}}^{0}\to   \pi^0   \pi^0  $ & $ \left(-f_3+4 f_5+f_6\right)V_{cs}^*$\\\hline
$T_{bc\bar{u}\bar{s}}^{0}\to   \pi^0   K^0  $ & $ \frac{\left(f_3+2 f_4+f_6\right) V_{cd}^*}{\sqrt{2}}$&
$T_{bc\bar{u}\bar{s}}^{0}\to   \pi^0   \eta  $ & $ \frac{\left(-f_3-2 f_4+f_6\right)V_{cs}^*}{\sqrt{3}}$\\\hline
$T_{bc\bar{u}\bar{s}}^{0}\to   \pi^-   K^+  $ & $ -\left(f_3+2 f_4+f_6\right) V_{cd}^*$&
$T_{bc\bar{u}\bar{s}}^{0}\to   K^+   K^-  $ & $ -2 \left(f_3+f_4-2 f_5\right)V_{cs}^*$\\\hline
$T_{bc\bar{u}\bar{s}}^{0}\to   K^0   \overline K^0  $ & $ -\left(f_3-4 f_5+f_6\right)V_{cs}^*$&
$T_{bc\bar{u}\bar{s}}^{0}\to   K^0   \eta  $ & $ \frac{\left(f_3+2 f_4+f_6\right) V_{cd}^*}{\sqrt{6}}$\\\hline
$T_{bc\bar{u}\bar{s}}^{0}\to   \eta   \eta  $ & $ -\frac{1}{3} \left(5 f_3+4 f_4+3 \left(f_6-4 f_5\right)\right)V_{cs}^*$&
$T_{bc\bar{d}\bar{s}}^{+}\to   \pi^+   K^0  $ & $ \left(f_3+2 f_4-f_6\right) V_{cd}^*$\\\hline
$T_{bc\bar{d}\bar{s}}^{+}\to   \pi^+   \eta  $ & $ \sqrt{\frac{2}{3}} \left(-f_3-2 f_4+f_6\right)V_{cs}^*$&
$T_{bc\bar{d}\bar{s}}^{+}\to   \pi^0   K^+  $ & $ \frac{\left(f_3+2 f_4-f_6\right) V_{cd}^*}{\sqrt{2}}$\\\hline
$T_{bc\bar{d}\bar{s}}^{+}\to   K^+   \overline K^0  $ & $ \left(-f_3-2 f_4+f_6\right)V_{cs}^*$&
$T_{bc\bar{d}\bar{s}}^{+}\to   K^+   \eta  $ & $ \frac{\left(-f_3-2 f_4+f_6\right) V_{cd}^*}{\sqrt{6}}$\\\hline
$T_{bc\bar{u}\bar{d}}^{0}\to   \pi^+   \pi^-  $ & $ -2 \left(f_3+f_4-2 f_5\right) V_{cd}^*$&
$T_{bc\bar{u}\bar{d}}^{0}\to   \pi^+   K^-  $ & $ -\left(f_3+2 f_4+f_6\right)V_{cs}^*$\\\hline
$T_{bc\bar{u}\bar{d}}^{0}\to   \pi^0   \pi^0  $ & $ -2 \left(f_3+f_4-2 f_5\right) V_{cd}^*$&
$T_{bc\bar{u}\bar{d}}^{0}\to   \pi^0   \overline K^0  $ & $ \frac{\left(f_3+2 f_4+f_6\right)V_{cs}^*}{\sqrt{2}}$\\\hline
$T_{bc\bar{u}\bar{d}}^{0}\to   \pi^0   \eta  $ & $ \frac{2 f_6 V_{cd}^*}{\sqrt{3}}$&
$T_{bc\bar{u}\bar{d}}^{0}\to   K^+   K^-  $ & $ \left(-f_3+4 f_5+f_6\right) V_{cd}^*$\\\hline
$T_{bc\bar{u}\bar{d}}^{0}\to   K^0   \overline K^0  $ & $ -\left(f_3-4 f_5+f_6\right) V_{cd}^*$&
$T_{bc\bar{u}\bar{d}}^{0}\to   \overline K^0   \eta  $ & $ \frac{\left(f_3+2 f_4+f_6\right)V_{cs}^*}{\sqrt{6}}$\\\hline
$T_{bc\bar{u}\bar{d}}^{0}\to   \eta   \eta  $ & $ \frac{2}{3} \left(-f_3+f_4+6 f_5\right) V_{cd}^*$&&\\
\hline\hline
\end{tabular}
\end{table}
The transitions $bc\to cd/s$ and $bc\to ud/s$ can lead to the tetraquark to decays to two mesons, whose corresponding Feynman diagrams are given in Fig.~(\ref{fig:Feynman_nonleptonic_tetraquark_4}a). The Hamiltonian becomes
\begin{eqnarray}
 {\cal H}_{eff}&=&f_1(T_{bc3})_{[ij]}  (\overline D)^{i} M^{j}_{k} (H_{3})^k +f_2(T_{bc3})_{[ij]}  (\overline D)^{k} M^{j}_{k} (H_{3})^i\nonumber\\
 &&+f_3(T_{bc3})_{[ij]}  M^i_k M^{k}_{l} (H_{\bar3})^{[jl]} +f_4(T_{bc3})_{[ij]}  M^i_k M^{j}_{l} (H_{\bar3})^{[kl]}\nonumber\\
 &&+f_5(T_{bc3})_{[ij]}  M^l_k M^{k}_{l} (H_{\bar3})^{[ij]}+f_6(T_{bc3})_{[ij]}  M^i_k M^{k}_{l} (H_{6})^{\{jl\}}.
\end{eqnarray}
The decay amplitudes for $T_{bc3}$ decays into a charmed meson and a light meson are given in Tab.~\ref{tab:bc_D_meson8}. We have the relations:
\begin{eqnarray*}
    \Gamma(T_{bc\bar{u}\bar{s}}^{0}\to  D^0\pi^0 )= \frac{1}{2}\Gamma(T_{bc\bar{u}\bar{s}}^{0}\to  D^+\pi^- )=\frac{1}{2}\Gamma(T_{bc\bar{d}\bar{s}}^{+}\to  D^0\pi^+ )=\Gamma(T_{bc\bar{d}\bar{s}}^{+}\to  D^+\pi^0 ),\\
     \Gamma(T_{bc\bar{d}\bar{s}}^{+}\to  D^0K^+ )= { }\Gamma(T_{bc\bar{u}\bar{d}}^{0}\to  D^+_sK^- ),
       \Gamma(T_{bc\bar{d}\bar{s}}^{+}\to  D^+_s\overline K^0 )= { }\Gamma(T_{bc\bar{u}\bar{s}}^{0}\to  D^+_sK^- ),\\
       \Gamma(T_{bc\bar{u}\bar{d}}^{0}\to  D^0\overline K^0 )= { }\Gamma(T_{bc\bar{u}\bar{d}}^{0}\to  D^+K^- ),
     \Gamma(T_{bc\bar{u}\bar{s}}^{0}\to  D^0\eta )= { }\Gamma(T_{bc\bar{d}\bar{s}}^{+}\to  D^+\eta ),\\
          \Gamma(T_{bc\bar{u}\bar{s}}^{0}\to  D^+_s\pi^- )= { }\Gamma(T_{bc\bar{u}\bar{s}}^{0}\to  D^0K^0 )=2\Gamma(T_{bc\bar{d}\bar{s}}^{+}\to  D^+_s\pi^0 ),\\
     \Gamma(T_{bc\bar{u}\bar{d}}^{0}\to  D^0\pi^0 )= \frac{1}{2}\Gamma(T_{bc\bar{d}\bar{s}}^{+}\to  D^+K^0 )=\frac{1}{2}\Gamma(T_{bc\bar{u}\bar{d}}^{0}\to  D^+\pi^- ).
\end{eqnarray*}
The decay amplitudes for $T_{bc3}$ decays into two light mesons are given in Tab.~\ref{tab:bc_2meson8}. The relations of decay widths are:
\begin{eqnarray*}
    \Gamma(T_{bc\bar{u}\bar{d}}^{0}\to \pi^0 \overline K^0 )= \frac{1}{2}\Gamma(T_{bc\bar{u}\bar{d}}^{0}\to \pi^+ K^- )= 3\Gamma(T_{bc\bar{u}\bar{d}}^{0}\to \eta \overline K^0 ),\\
    \Gamma(T_{bc\bar{d}\bar{s}}^{+}\to \pi^0 K^+ )= \frac{1}{2}\Gamma(T_{bc\bar{d}\bar{s}}^{+}\to \pi^+ K^0 )=3\Gamma(T_{bc\bar{d}\bar{s}}^{+}\to \eta K^+ ),\\
    \Gamma(T_{bc\bar{d}\bar{s}}^{+}\to K^+ \overline K^0 )= \frac{3}{2}\Gamma(T_{bc\bar{d}\bar{s}}^{+}\to \pi^+ \eta )= 3\Gamma(T_{bc\bar{u}\bar{s}}^{0}\to \pi^0 \eta ),\\
    \Gamma(T_{bc\bar{u}\bar{d}}^{0}\to \pi^0 \pi^0 )= \frac{1}{2}\Gamma(T_{bc\bar{u}\bar{d}}^{0}\to \pi^+ \pi^- ),
    \Gamma(T_{bc\bar{u}\bar{s}}^{0}\to \pi^0 K^0 )= 3\Gamma(T_{bc\bar{u}\bar{s}}^{0}\to \eta K^0 ),\\
        \Gamma(T_{bc\bar{u}\bar{s}}^{0}\to \pi^0 \pi^0 )= \frac{1}{2}\Gamma(T_{bc\bar{u}\bar{s}}^{0}\to \pi^+ \pi^- ).
\end{eqnarray*}
\subsubsection{decays into three mesons}
\begin{table}
\caption{Doubly heavy tetraquark $T_{bc\bar{q}\bar{q}}$ decays into a charmed meson and two light mesons.}\label{tab:bc_D_2meson8}\begin{tabular}{|cc|cc|}\hline\hline
channel & amplitude(/$V_{cb}$) &channel &amplitude(/$V_{cb}$)\\\hline
$T_{bc\bar{u}\bar{s}}^{0}\to    D^0  \pi^+   \pi^-  $ & $ \left(2 d_2+d_4\right) V_{\text{cs}}^*$&
$T_{bc\bar{u}\bar{s}}^{0}\to    D^0  \pi^0   \pi^0  $ & $ \left(2 d_2+d_4\right) V_{\text{cs}}^*$\\\hline
$T_{bc\bar{u}\bar{s}}^{0}\to    D^0  \pi^0   K^0  $ & $ \frac{\left(d_3-d_1\right) V_{\text{cd}}^*}{\sqrt{2}}$&
$T_{bc\bar{u}\bar{s}}^{0}\to    D^0  \pi^0   \eta  $ & $ \frac{\left(d_4-d_3\right) V_{\text{cs}}^*}{\sqrt{3}}$\\\hline
$T_{bc\bar{u}\bar{s}}^{0}\to    D^0  \pi^-   K^+  $ & $ \left(d_1-d_3\right) V_{\text{cd}}^*$&
$T_{bc\bar{u}\bar{s}}^{0}\to    D^0  K^+   K^-  $ & $ \left(d_1+2 d_2-d_3+d_4\right) V_{\text{cs}}^*$\\\hline
$T_{bc\bar{u}\bar{s}}^{0}\to    D^0  K^0   \overline K^0  $ & $ \left(d_1+2 d_2\right) V_{\text{cs}}^*$&
$T_{bc\bar{u}\bar{s}}^{0}\to    D^0  K^0   \eta  $ & $ \frac{\left(d_3-d_1\right) V_{\text{cd}}^*}{\sqrt{6}}$\\\hline
$T_{bc\bar{u}\bar{s}}^{0}\to    D^0  \eta   \eta  $ & $ \frac{1}{3} \left(4 d_1+6 d_2-2 d_3+d_4\right) V_{\text{cs}}^*$&
$T_{bc\bar{u}\bar{s}}^{0}\to    D^+  \pi^-   \eta  $ & $ \sqrt{\frac{2}{3}} \left(d_4-d_3\right) V_{\text{cs}}^*$\\\hline
$T_{bc\bar{u}\bar{s}}^{0}\to    D^+  K^0   K^-  $ & $ \left(d_4-d_3\right) V_{\text{cs}}^*$&
$T_{bc\bar{u}\bar{s}}^{0}\to    D^+_s  \pi^0   K^-  $ & $ \frac{\left(d_4-d_1\right) V_{\text{cs}}^*}{\sqrt{2}}$\\\hline
$T_{bc\bar{u}\bar{s}}^{0}\to    D^+_s  \pi^-   \overline K^0  $ & $ \left(d_4-d_1\right) V_{\text{cs}}^*$&
$T_{bc\bar{u}\bar{s}}^{0}\to    D^+_s  \pi^-   \eta  $ & $ \sqrt{\frac{2}{3}} \left(d_3-d_1\right) V_{\text{cd}}^*$\\\hline
$T_{bc\bar{u}\bar{s}}^{0}\to    D^+_s  K^0   K^-  $ & $ \left(d_3-d_1\right) V_{\text{cd}}^*$&
$T_{bc\bar{u}\bar{s}}^{0}\to    D^+_s  K^-   \eta  $ & $ \frac{\left(d_1-d_4\right) V_{\text{cs}}^*}{\sqrt{6}}$\\\hline
$T_{bc\bar{d}\bar{s}}^{+}\to    D^0  \pi^+   K^0  $ & $ \left(d_3-d_4\right) V_{\text{cd}}^*$&
$T_{bc\bar{d}\bar{s}}^{+}\to    D^0  \pi^+   \eta  $ & $ \sqrt{\frac{2}{3}} \left(d_4-d_3\right) V_{\text{cs}}^*$\\\hline
$T_{bc\bar{d}\bar{s}}^{+}\to    D^0  \pi^0   K^+  $ & $ \frac{\left(d_3-d_4\right) V_{\text{cd}}^*}{\sqrt{2}}$&
$T_{bc\bar{d}\bar{s}}^{+}\to    D^0  K^+   \overline K^0  $ & $ \left(d_4-d_3\right) V_{\text{cs}}^*$\\\hline
$T_{bc\bar{d}\bar{s}}^{+}\to    D^0  K^+   \eta  $ & $ \frac{\left(d_4-d_3\right) V_{\text{cd}}^*}{\sqrt{6}}$&
$T_{bc\bar{d}\bar{s}}^{+}\to    D^+  \pi^+   \pi^-  $ & $ \left(2 d_2+d_4\right) V_{\text{cs}}^*$\\\hline
$T_{bc\bar{d}\bar{s}}^{+}\to    D^+  \pi^0   \pi^0  $ & $ \left(2 d_2+d_4\right) V_{\text{cs}}^*$&
$T_{bc\bar{d}\bar{s}}^{+}\to    D^+  \pi^0   K^0  $ & $ \frac{\left(d_4-d_1\right) V_{\text{cd}}^*}{\sqrt{2}}$\\\hline
$T_{bc\bar{d}\bar{s}}^{+}\to    D^+  \pi^0   \eta  $ & $ \frac{\left(d_3-d_4\right) V_{\text{cs}}^*}{\sqrt{3}}$&
$T_{bc\bar{d}\bar{s}}^{+}\to    D^+  \pi^-   K^+  $ & $ \left(d_1-d_4\right) V_{\text{cd}}^*$\\\hline
$T_{bc\bar{d}\bar{s}}^{+}\to    D^+  K^+   K^-  $ & $ \left(d_1+2 d_2\right) V_{\text{cs}}^*$&
$T_{bc\bar{d}\bar{s}}^{+}\to    D^+  K^0   \overline K^0  $ & $ \left(d_1+2 d_2-d_3+d_4\right) V_{\text{cs}}^*$\\\hline
$T_{bc\bar{d}\bar{s}}^{+}\to    D^+  K^0   \eta  $ & $ \frac{\left(d_4-d_1\right) V_{\text{cd}}^*}{\sqrt{6}}$&
$T_{bc\bar{d}\bar{s}}^{+}\to    D^+  \eta   \eta  $ & $ \frac{1}{3} \left(4 d_1+6 d_2-2 d_3+d_4\right) V_{\text{cs}}^*$\\\hline
$T_{bc\bar{d}\bar{s}}^{+}\to    D^+_s  \pi^+   \pi^-  $ & $ -\left(d_1+2 d_2\right) V_{\text{cd}}^*$&
$T_{bc\bar{d}\bar{s}}^{+}\to    D^+_s  \pi^+   K^-  $ & $ \left(d_4-d_1\right) V_{\text{cs}}^*$\\\hline
$T_{bc\bar{d}\bar{s}}^{+}\to    D^+_s  \pi^0   \pi^0  $ & $ -\left(d_1+2 d_2\right) V_{\text{cd}}^*$&
$T_{bc\bar{d}\bar{s}}^{+}\to    D^+_s  \pi^0   \overline K^0  $ & $ \frac{\left(d_1-d_4\right) V_{\text{cs}}^*}{\sqrt{2}}$\\\hline
$T_{bc\bar{d}\bar{s}}^{+}\to    D^+_s  \pi^0   \eta  $ & $ \frac{\left(d_1-d_3\right) V_{\text{cd}}^*}{\sqrt{3}}$&
$T_{bc\bar{d}\bar{s}}^{+}\to    D^+_s  K^+   K^-  $ & $ -\left(2 d_2+d_4\right) V_{\text{cd}}^*$\\\hline
$T_{bc\bar{d}\bar{s}}^{+}\to    D^+_s  K^0   \overline K^0  $ & $ -\left(d_1+2 d_2-d_3+d_4\right) V_{\text{cd}}^*$&
$T_{bc\bar{d}\bar{s}}^{+}\to    D^+_s  \overline K^0   \eta  $ & $ \frac{\left(d_1-d_4\right) V_{\text{cs}}^*}{\sqrt{6}}$\\\hline
$T_{bc\bar{d}\bar{s}}^{+}\to    D^+_s  \eta   \eta  $ & $ -\frac{1}{3} \left(d_1+6 d_2-2 d_3+4 d_4\right) V_{\text{cd}}^*$&
$T_{bc\bar{u}\bar{d}}^{0}\to    D^0  \pi^+   \pi^-  $ & $ \left(d_1+2 d_2-d_3+d_4\right) V_{\text{cd}}^*$\\\hline
$T_{bc\bar{u}\bar{d}}^{0}\to    D^0  \pi^+   K^-  $ & $ \left(d_1-d_3\right) V_{\text{cs}}^*$&
$T_{bc\bar{u}\bar{d}}^{0}\to    D^0  \pi^0   \pi^0  $ & $ \left(d_1+2 d_2-d_3+d_4\right) V_{\text{cd}}^*$\\\hline
$T_{bc\bar{u}\bar{d}}^{0}\to    D^0  \pi^0   \overline K^0  $ & $ \frac{\left(d_3-d_1\right) V_{\text{cs}}^*}{\sqrt{2}}$&
$T_{bc\bar{u}\bar{d}}^{0}\to    D^0  \pi^0   \eta  $ & $ \frac{\left(d_4-d_1\right) V_{\text{cd}}^*}{\sqrt{3}}$\\\hline
$T_{bc\bar{u}\bar{d}}^{0}\to    D^0  K^+   K^-  $ & $ \left(2 d_2+d_4\right) V_{\text{cd}}^*$&
$T_{bc\bar{u}\bar{d}}^{0}\to    D^0  K^0   \overline K^0  $ & $ \left(d_1+2 d_2\right) V_{\text{cd}}^*$\\\hline
$T_{bc\bar{u}\bar{d}}^{0}\to    D^0  \overline K^0   \eta  $ & $ \frac{\left(d_3-d_1\right) V_{\text{cs}}^*}{\sqrt{6}}$&
$T_{bc\bar{u}\bar{d}}^{0}\to    D^0  \eta   \eta  $ & $ \frac{1}{3} \left(d_1+6 d_2+d_3+d_4\right) V_{\text{cd}}^*$\\\hline
$T_{bc\bar{u}\bar{d}}^{0}\to    D^+  \pi^0   K^-  $ & $ \frac{\left(d_3-d_1\right) V_{\text{cs}}^*}{\sqrt{2}}$&
$T_{bc\bar{u}\bar{d}}^{0}\to    D^+  \pi^-   \overline K^0  $ & $ \left(d_3-d_1\right) V_{\text{cs}}^*$\\\hline
$T_{bc\bar{u}\bar{d}}^{0}\to    D^+  \pi^-   \eta  $ & $ \sqrt{\frac{2}{3}} \left(d_4-d_1\right) V_{\text{cd}}^*$&
$T_{bc\bar{u}\bar{d}}^{0}\to    D^+  K^0   K^-  $ & $ \left(d_4-d_1\right) V_{\text{cd}}^*$\\\hline
$T_{bc\bar{u}\bar{d}}^{0}\to    D^+  K^-   \eta  $ & $ \frac{\left(d_1-d_3\right) V_{\text{cs}}^*}{\sqrt{6}}$&
$T_{bc\bar{u}\bar{d}}^{0}\to    D^+_s  \pi^0   K^-  $ & $ \frac{\left(d_4-d_3\right) V_{\text{cd}}^*}{\sqrt{2}}$\\\hline
$T_{bc\bar{u}\bar{d}}^{0}\to    D^+_s  \pi^-   \overline K^0  $ & $ \left(d_4-d_3\right) V_{\text{cd}}^*$&
$T_{bc\bar{u}\bar{d}}^{0}\to    D^+_s  K^-   \eta  $ & $ \frac{\left(d_3-d_4\right) V_{\text{cd}}^*}{\sqrt{6}}$\\\hline
\hline
\end{tabular}
\end{table}
\begin{table}
\footnotesize
\caption{Doubly heavy tetraquark $T_{bc\bar{q}\bar{q}}$ decays into three light mesons.}\label{tab:bc_3meson8}\begin{tabular}{|cc|cc|}\hline\hline
channel & amplitude(/$V_{cb}$) &channel & amplitude(/$V_{cb}$)\\\hline
$T_{bc\bar{u}\bar{s}}^{0}\to   \pi^+   \pi^0   \pi^-  $ & $ \frac{\left(-d_6-2 d_7+d_{10}+2 d_{11}\right) V_{\text{cs}}^*}{\sqrt{2}}$&
$T_{bc\bar{u}\bar{s}}^{0}\to   \pi^+   \pi^-   K^0  $ & $ -\left(d_6+2 d_7+d_{10}+2 d_{11}\right) V_{\text{cd}}^*$\\\hline
$T_{bc\bar{u}\bar{s}}^{0}\to   \pi^+   \pi^-   \eta  $ & $ \frac{\left(-2 d_5-3 d_6+2 d_7+12 d_8-2 d_9+3 d_{10}+6 d_{11}\right) V_{\text{cs}}^*}{\sqrt{6}}$&
$T_{bc\bar{u}\bar{s}}^{0}\to   \pi^+   K^0   K^-  $ & $ -\left(d_5+2 d_6-6 d_8+d_9\right) V_{\text{cs}}^*$\\\hline
$T_{bc\bar{u}\bar{s}}^{0}\to   \pi^0   \pi^0   \pi^0  $ & $ \frac{3 \left(-d_6-2 d_7+d_{10}+2 d_{11}\right) V_{\text{cs}}^*}{\sqrt{2}}$&
$T_{bc\bar{u}\bar{s}}^{0}\to   \pi^0   \pi^0   K^0  $ & $ -\left(d_6+2 d_7-2 d_9+d_{10}+2 d_{11}\right) V_{\text{cd}}^*$\\\hline
$T_{bc\bar{u}\bar{s}}^{0}\to   \pi^0   \pi^0   \eta  $ & $ \frac{\left(-2 d_5-3 d_6+2 d_7+12 d_8-2 d_9+3 d_{10}+6 d_{11}\right) V_{\text{cs}}^*}{\sqrt{6}}$&
$T_{bc\bar{u}\bar{s}}^{0}\to   \pi^0   \pi^-   K^+  $ & $ -\sqrt{2} d_9 V_{\text{cd}}^*$\\\hline
$T_{bc\bar{u}\bar{s}}^{0}\to   \pi^0   K^+   K^-  $ & $ -\frac{\left(d_5+3 d_6+2 d_7-6 d_8+d_9-d_{10}-2 d_{11}\right) V_{\text{cs}}^*}{\sqrt{2}}$&
$T_{bc\bar{u}\bar{s}}^{0}\to   \pi^0   K^0   \overline K^0  $ & $ \frac{\left(d_5+d_6-2 d_7-6 d_8-d_9+d_{10}+2 d_{11}\right) V_{\text{cs}}^*}{\sqrt{2}}$\\\hline
$T_{bc\bar{u}\bar{s}}^{0}\to   \pi^0   K^0   \eta  $ & $ \frac{2 d_9 V_{\text{cd}}^*}{\sqrt{3}}$&
$T_{bc\bar{u}\bar{s}}^{0}\to   \pi^0   \eta   \eta  $ & $ -\frac{\left(3 d_6+6 d_7+8 d_9-3 d_{10}-6 d_{11}\right) V_{\text{cs}}^*}{3 \sqrt{2}}$\\\hline
$T_{bc\bar{u}\bar{s}}^{0}\to   \pi^-   K^+   \overline K^0  $ & $ \left(-d_5-2 d_6+6 d_8+d_9\right) V_{\text{cs}}^*$&
$T_{bc\bar{u}\bar{s}}^{0}\to   \pi^-   K^+   \eta  $ & $ \sqrt{\frac{2}{3}} d_9 V_{\text{cd}}^*$\\\hline
$T_{bc\bar{u}\bar{s}}^{0}\to   K^+   K^0   K^-  $ & $ -\left(d_6+2 d_7-2 d_9+d_{10}+2 d_{11}\right) V_{\text{cd}}^*$&
$T_{bc\bar{u}\bar{s}}^{0}\to   K^+   K^-   \eta  $ & $ \frac{\left(d_5+3 d_6+2 d_7-6 d_8-3 d_9+3 d_{10}+6 d_{11}\right) V_{\text{cs}}^*}{\sqrt{6}}$\\\hline
$T_{bc\bar{u}\bar{s}}^{0}\to   K^0   K^0   \overline K^0  $ & $ -2 \left(d_6+2 d_7+d_{10}+2 d_{11}\right) V_{\text{cd}}^*$&
$T_{bc\bar{u}\bar{s}}^{0}\to   K^0   \overline K^0   \eta  $ & $ \frac{\left(d_5+3 d_6+2 d_7-6 d_8-d_9+3 d_{10}+6 d_{11}\right) V_{\text{cs}}^*}{\sqrt{6}}$\\\hline
$T_{bc\bar{u}\bar{s}}^{0}\to   K^0   \eta   \eta  $ & $ -\frac{ \left(3 d_6+6 d_7-2 d_9+3 d_{10}+6 d_{11}\right) V_{\text{cd}}^*}{3}$&
$T_{bc\bar{u}\bar{s}}^{0}\to   \eta   \eta   \eta  $ & $ \frac{\left(2 d_5+7 d_6+6 d_7-12 d_8-6 d_9+9 d_{10}+18 d_{11}\right) V_{\text{cs}}^*}{\sqrt{6}}$\\\hline
$T_{bc\bar{d}\bar{s}}^{+}\to   \pi^+   \pi^+   \pi^-  $ & $ 2 \left(-d_6-2 d_7+d_{10}+2 d_{11}\right) V_{\text{cs}}^*$&
$T_{bc\bar{d}\bar{s}}^{+}\to   \pi^+   \pi^0   \pi^0  $ & $ \left(-d_6-2 d_7+d_{10}+2 d_{11}\right) V_{\text{cs}}^*$\\\hline
$T_{bc\bar{d}\bar{s}}^{+}\to   \pi^+   \pi^0   K^0  $ & $ \sqrt{2} d_9 V_{\text{cd}}^*$&
$T_{bc\bar{d}\bar{s}}^{+}\to   \pi^+   \pi^-   K^+  $ & $ \left(d_6+2 d_7-d_{10}-2 d_{11}\right) V_{\text{cd}}^*$\\\hline
$T_{bc\bar{d}\bar{s}}^{+}\to   \pi^+   K^+   K^-  $ & $ \left(-d_6-2 d_7+d_{10}+2 d_{11}\right) V_{\text{cs}}^*$&
$T_{bc\bar{d}\bar{s}}^{+}\to   \pi^+   K^0   \overline K^0  $ & $ \left(-d_6-2 d_7-2 d_9+d_{10}+2 d_{11}\right) V_{\text{cs}}^*$\\\hline
$T_{bc\bar{d}\bar{s}}^{+}\to   \pi^+   K^0   \eta  $ & $ \sqrt{\frac{2}{3}} d_9 V_{\text{cd}}^*$&
$T_{bc\bar{d}\bar{s}}^{+}\to   \pi^+   \eta   \eta  $ & $ -\frac{\left(3 d_6+6 d_7+8 d_9-3 d_{10}-6 d_{11}\right) V_{\text{cs}}^*}{3} $\\\hline
$T_{bc\bar{d}\bar{s}}^{+}\to   \pi^0   \pi^0   K^+  $ & $ \left(d_6+2 d_7+2 d_9-d_{10}-2 d_{11}\right) V_{\text{cd}}^*$&
$T_{bc\bar{d}\bar{s}}^{+}\to   \pi^0   K^+   \overline K^0  $ & $ -\sqrt{2} d_9 V_{\text{cs}}^*$\\\hline
$T_{bc\bar{d}\bar{s}}^{+}\to   \pi^0   K^+   \eta  $ & $ -\frac{2 d_9 V_{\text{cd}}^*}{\sqrt{3}}$&
$T_{bc\bar{d}\bar{s}}^{+}\to   K^+   K^+   K^-  $ & $ 2 \left(d_6+2 d_7-d_{10}-2 d_{11}\right) V_{\text{cd}}^*$\\\hline
$T_{bc\bar{d}\bar{s}}^{+}\to   K^+   K^0   \overline K^0  $ & $ \left(d_6+2 d_7+2 d_9-d_{10}-2 d_{11}\right) V_{\text{cd}}^*$&
$T_{bc\bar{d}\bar{s}}^{+}\to   K^+   \overline K^0   \eta  $ & $ -\sqrt{\frac{2}{3}} d_9 V_{\text{cs}}^*$\\\hline
$T_{bc\bar{d}\bar{s}}^{+}\to   K^+   \eta   \eta  $ & $ \frac{ \left(3 d_6+6 d_7+2 d_9-3 d_{10}-6 d_{11}\right) V_{\text{cd}}^*}{3}$&
$T_{bc\bar{u}\bar{d}}^{0}\to   \pi^+   \pi^0   \pi^-  $ & $ \sqrt{2} \left(-d_9+d_{10}+2 d_{11}\right) V_{\text{cd}}^*$\\\hline
$T_{bc\bar{u}\bar{d}}^{0}\to   \pi^+   \pi^-   \overline K^0  $ & $ -\left(d_6+2 d_7-2 d_9+d_{10}+2 d_{11}\right) V_{\text{cs}}^*$&
$T_{bc\bar{u}\bar{d}}^{0}\to   \pi^+   \pi^-   \eta  $ & $ -\sqrt{\frac{2}{3}} \left(d_5+3 d_6+2 d_7-6 d_8\right) V_{\text{cd}}^*$\\\hline
$T_{bc\bar{u}\bar{d}}^{0}\to   \pi^+   K^0   K^-  $ & $ \left(-d_5-2 d_6+6 d_8+d_9\right) V_{\text{cd}}^*$&
$T_{bc\bar{u}\bar{d}}^{0}\to   \pi^+   K^-   \eta  $ & $ -2 \sqrt{\frac{2}{3}} d_9 V_{\text{cs}}^*$\\\hline
$T_{bc\bar{u}\bar{d}}^{0}\to   \pi^0   \pi^0   \pi^0  $ & $ 3 \sqrt{2} \left(-d_9+d_{10}+2 d_{11}\right) V_{\text{cd}}^*$&
$T_{bc\bar{u}\bar{d}}^{0}\to   \pi^0   \pi^0   \overline K^0  $ & $ -\left(d_6+2 d_7-2 d_9+d_{10}+2 d_{11}\right) V_{\text{cs}}^*$\\\hline
$T_{bc\bar{u}\bar{d}}^{0}\to   \pi^0   \pi^0   \eta  $ & $ -\sqrt{\frac{2}{3}} \left(d_5+3 d_6+2 d_7-6 d_8\right) V_{\text{cd}}^*$&
$T_{bc\bar{u}\bar{d}}^{0}\to   \pi^0   K^+   K^-  $ & $ -\frac{\left(d_5+2 d_6-6 d_8+d_9-2 d_{10}-4 d_{11}\right) V_{\text{cd}}^*}{\sqrt{2}}$\\\hline
$T_{bc\bar{u}\bar{d}}^{0}\to   \pi^0   K^0   \overline K^0  $ & $ \frac{\left(d_5+2 d_6-6 d_8-d_9+2 d_{10}+4 d_{11}\right) V_{\text{cd}}^*}{\sqrt{2}}$&
$T_{bc\bar{u}\bar{d}}^{0}\to   \pi^0   \overline K^0   \eta  $ & $ \frac{2 d_9 V_{\text{cs}}^*}{\sqrt{3}}$\\\hline
$T_{bc\bar{u}\bar{d}}^{0}\to   \pi^0   \eta   \eta  $ & $ \frac{ \sqrt{2} \left(d_9+3 d_{10}+6 d_{11}\right) V_{\text{cd}}^*}{3}$&
$T_{bc\bar{u}\bar{d}}^{0}\to   \pi^-   K^+   \overline K^0  $ & $ -\left(d_5+2 d_6-6 d_8+d_9\right) V_{\text{cd}}^*$\\\hline
$T_{bc\bar{u}\bar{d}}^{0}\to   K^+   \overline K^0   K^-  $ & $ -\left(d_6+2 d_7+d_{10}+2 d_{11}\right) V_{\text{cs}}^*$&
$T_{bc\bar{u}\bar{d}}^{0}\to   K^+   K^-   \eta  $ & $ \frac{\left(d_5-4 d_7-6 d_8+d_9\right) V_{\text{cd}}^*}{\sqrt{6}}$\\\hline
$T_{bc\bar{u}\bar{d}}^{0}\to   K^0   \overline K^0   \overline K^0  $ & $ -2 \left(d_6+2 d_7+d_{10}+2 d_{11}\right) V_{\text{cs}}^*$&
$T_{bc\bar{u}\bar{d}}^{0}\to   K^0   \overline K^0   \eta  $ & $ \frac{\left(d_5-4 d_7-6 d_8-d_9\right) V_{\text{cd}}^*}{\sqrt{6}}$\\\hline
$T_{bc\bar{u}\bar{d}}^{0}\to   \overline K^0   \eta   \eta  $ & $ -\frac{\left(3 d_6+6 d_7-2 d_9+3 d_{10}+6 d_{11}\right) V_{\text{cs}}^*}{3} $&
$T_{bc\bar{u}\bar{d}}^{0}\to   \eta   \eta   \eta  $ & $ \sqrt{\frac{2}{3}} \left(d_5-d_6-6 \left(d_7+d_8\right)\right) V_{\text{cd}}^*$\\\hline
\hline
\end{tabular}
\end{table}
The effective Hamiltonian which leads to the tetraquark to decays to three mesons is
\begin{eqnarray}
 {\cal H}_{eff}&=&d_1(T_{bc3})_{[ij]}  (\overline D)^{i} M^{j}_{k}M^{k}_{l} (H_{3})^l +d_2(T_{bc3})_{[ij]}  (\overline D)^{i} M^{k}_{l}M^{l}_{k} (H_{3})^j\nonumber\\
 &&+d_3(T_{bc3})_{[ij]}  (\overline D)^{k} M^{i}_{k}M^{j}_{l} (H_{3})^l+d_4(T_{bc3})_{[ij]}  (\overline D)^{k} M^{i}_{l}M^{l}_{k} (H_{3})^j\nonumber\\
 &&+d_5(T_{bc3})_{[ij]}  M^i_k M^{j}_{m} M^{k}_{l} (H_{\bar3})^{[lm]} +d_6(T_{bc3})_{[ij]}  M^i_k M^{k}_{l} M^{l}_{m} (H_{\bar3})^{[jm]}\nonumber\\
 &&+d_7(T_{bc3})_{[ij]}  M^i_m M^{k}_{l} M^{l}_{k} (H_{\bar3})^{[jm]}+d_8(T_{bc3})_{[ij]}  M^k_m M^{l}_{k} M^{m}_{l} (H_{\bar3})^{[ij]}\nonumber\\
 &&+d_9(T_{bc3})_{[ij]}  M^i_k M^{j}_{m} M^{k}_{l} (H_{6})^{\{lm\}} +d_{10}(T_{bc3})_{[ij]}  M^i_k M^{k}_{l} M^{l}_{m} (H_{6})^{\{jm\}}\nonumber\\
 &&+d_{11}(T_{bc3})_{[ij]}  M^i_m M^{k}_{l} M^{l}_{k} (H_{6})^{\{jm\}}.
\end{eqnarray}
The corresponding Feynman diagrams are given in Fig.~\ref{fig:Feynman_nonleptonic_tetraquark_4}(b,c,d). We give the amplitudes in Tab.~\ref{tab:bc_D_2meson8} for a D meson plus two light mesons, and Tab.~\ref{tab:bc_3meson8} for three light mesons. Then the relations of decay widths for a D meson plus two light mesons are as follows:
\begin{eqnarray*}
    \Gamma(T_{bc\bar{u}\bar{s}}^{0}\to  D^0 \pi^0 \pi^0 )= \frac{1}{2}\Gamma(T_{bc\bar{u}\bar{s}}^{0}\to  D^0 \pi^+ \pi^- )=
 \frac{1}{2}\Gamma(T_{bc\bar{d}\bar{s}}^{+}\to  D^+ \pi^+ \pi^- )=\Gamma(T_{bc\bar{d}\bar{s}}^{+}\to  D^+ \pi^0 \pi^0 ) ,\\
     \Gamma(T_{bc\bar{u}\bar{s}}^{0}\to  D^0 \pi^0 K^0 )= 3\Gamma(T_{bc\bar{u}\bar{s}}^{0}\to  D^0 K^0 \eta )=\frac{3}{4}\Gamma(T_{bc\bar{u}\bar{s}}^{0}\to  D^+_s \pi^- \eta )=\frac{1}{2}\Gamma(T_{bc\bar{u}\bar{s}}^{0}\to  D^+_s K^0 K^- )\\=\frac{3}{2}\Gamma(T_{bc\bar{d}\bar{s}}^{+}\to  D^+_s \pi^0 \eta )=\frac{1}{2}\Gamma(T_{bc\bar{u}\bar{s}}^{0}\to  D^0 \pi^- K^+ ),\\
     \Gamma(T_{bc\bar{u}\bar{s}}^{0}\to  D^+ K^0 K^- )= 3\Gamma(T_{bc\bar{u}\bar{s}}^{0}\to  D^0 \pi^0 \eta )=\frac{3}{2}\Gamma(T_{bc\bar{u}\bar{s}}^{0}\to  D^+ \pi^- \eta )=\frac{3}{2}\Gamma(T_{bc\bar{d}\bar{s}}^{+}\to  D^0 \pi^+ \eta )\\=3\Gamma(T_{bc\bar{d}\bar{s}}^{+}\to  D^+ \pi^0 \eta )=\Gamma(T_{bc\bar{d}\bar{s}}^{+}\to  D^0 K^+ \overline K^0 ),\\
     \Gamma(T_{bc\bar{u}\bar{s}}^{0}\to  D^+_s \pi^0 K^- )=
3\Gamma(T_{bc\bar{d}\bar{s}}^{+}\to  D^+_s \overline K^0 \eta )=
\frac{1}{2}\Gamma(T_{bc\bar{u}\bar{s}}^{0}\to  D^+_s \pi^- \overline K^0 )=
3\Gamma(T_{bc\bar{u}\bar{s}}^{0}\to  D^+_s \eta K^- )\\=
\frac{1}{2}\Gamma(T_{bc\bar{d}\bar{s}}^{+}\to  D^+_s \pi^+ K^- )=
\Gamma(T_{bc\bar{d}\bar{s}}^{+}\to  D^+_s \pi^0 \overline K^0 ),\\
     \Gamma(T_{bc\bar{d}\bar{s}}^{+}\to  D^0 \pi^+ K^0 )= 6\Gamma(T_{bc\bar{d}\bar{s}}^{+}\to  D^0 K^+ \eta )=
 2\Gamma(T_{bc\bar{u}\bar{d}}^{0}\to  D^+_s \pi^0 K^- )=
\Gamma(T_{bc\bar{u}\bar{d}}^{0}\to  D^+_s \pi^- \overline K^0 )\\=
6\Gamma(T_{bc\bar{u}\bar{d}}^{0}\to  D^+_s K^- \eta )=
2\Gamma(T_{bc\bar{d}\bar{s}}^{+}\to  D^0 \pi^0 K^+ ),\\
      \Gamma(T_{bc\bar{d}\bar{s}}^{+}\to  D^+ \pi^0 K^0 )= 3\Gamma(T_{bc\bar{d}\bar{s}}^{+}\to  D^+ K^0 \eta )=
\frac{3}{2}\Gamma(T_{bc\bar{u}\bar{d}}^{0}\to  D^0 \pi^0 \eta )=
\frac{3}{4}\Gamma(T_{bc\bar{u}\bar{d}}^{0}\to  D^+ \pi^- \eta )\\=
 \frac{1}{2}\Gamma(T_{bc\bar{u}\bar{d}}^{0}\to  D^+ K^0 K^- )=
 \frac{1}{2}\Gamma(T_{bc\bar{d}\bar{s}}^{+}\to  D^+ \pi^- K^+ ),\\
     \Gamma(T_{bc\bar{u}\bar{d}}^{0}\to  D^0 \pi^+ K^- )= 6\Gamma(T_{bc\bar{u}\bar{d}}^{0}\to  D^0 \overline K^0 \eta )=
6\Gamma(T_{bc\bar{u}\bar{d}}^{0}\to  D^+ K^- \eta )=
2\Gamma(T_{bc\bar{u}\bar{d}}^{0}\to  D^0 \pi^0 \overline K^0 )\\=
2\Gamma(T_{bc\bar{u}\bar{d}}^{0}\to  D^+ \pi^0 K^- )=
\Gamma(T_{bc\bar{u}\bar{d}}^{0}\to  D^+ \pi^- \overline K^0 ),\\
   \Gamma(T_{bc\bar{d}\bar{s}}^{+}\to  D^+_s \pi^+ \pi^- )=\Gamma(T_{bc\bar{u}\bar{d}}^{0}\to  D^0 \overline K^0 K^0 )=2\Gamma(T_{bc\bar{d}\bar{s}}^{+}\to  D^+_s \pi^0 \pi^0 ),\\
    \Gamma(T_{bc\bar{u}\bar{d}}^{0}\to  D^0 \pi^+ \pi^- )=
\Gamma(T_{bc\bar{d}\bar{s}}^{+}\to  D^+_s \overline K^0 K^0 )=2\Gamma(T_{bc\bar{u}\bar{d}}^{0}\to  D^0 \pi^0 \pi^0 ),\\
      \Gamma(T_{bc\bar{u}\bar{s}}^{0}\to  D^0 \eta \eta )= { }\Gamma(T_{bc\bar{d}\bar{s}}^{+}\to  D^+ \eta \eta ),
      \Gamma(T_{bc\bar{u}\bar{d}}^{0}\to  D^0 K^+ K^- )= { }\Gamma(T_{bc\bar{d}\bar{s}}^{+}\to  D^+_s K^+ K^- ),\\
      \Gamma(T_{bc\bar{d}\bar{s}}^{+}\to  D^+ K^0 \overline K^0 )= { }\Gamma(T_{bc\bar{u}\bar{s}}^{0}\to  D^0 K^+ K^- ),
      \Gamma(T_{bc\bar{u}\bar{s}}^{0}\to  D^0 K^0 \overline K^0 )= { }\Gamma(T_{bc\bar{d}\bar{s}}^{+}\to  D^+ K^+ K^- ),\\
\end{eqnarray*}
The relations of decay widths for three light mesons can be written as:
\begin{eqnarray*}
  2\Gamma(T_{bc\bar{u}\bar{s}}^{0}\to   \pi^+   \pi^0   \pi^-  )=\frac{4}{3}\Gamma(T_{bc\bar{u}\bar{s}}^{0}\to   \pi^0   \pi^0   \pi^0  )=\frac{1}{2}\Gamma(T_{bc\bar{d}\bar{s}}^{+}\to   \pi^+   \pi^+   \pi^-  )=2\Gamma(T_{bc\bar{d}\bar{s}}^{+}\to   \pi^+   \pi^0   \pi^0  )\\
  =\Gamma(T_{bc\bar{d}\bar{s}}^{+}\to   \pi^+   K^+   K^-  ),\\
  \Gamma(T_{bc\bar{u}\bar{s}}^{0}\to   \pi^0   \pi^-   K^+  )=\frac{3}{2}\Gamma(T_{bc\bar{u}\bar{s}}^{0}\to   \pi^0   K^0   \eta  )=3\Gamma(T_{bc\bar{u}\bar{s}}^{0}\to   \pi^-   K^+   \eta  )=\Gamma(T_{bc\bar{d}\bar{s}}^{+}\to   \pi^+   \pi^0   K^0  )\\
  =3\Gamma(T_{bc\bar{d}\bar{s}}^{+}\to   \pi^+   K^0   \eta  )=\frac{3}{2}\Gamma(T_{bc\bar{d}\bar{s}}^{+}\to   \pi^0   K^+   \eta  ),\\
  \Gamma(T_{bc\bar{d}\bar{s}}^{+}\to   \pi^0   K^+   \overline K^0  )=3\Gamma(T_{bc\bar{d}\bar{s}}^{+}\to   K^+   \overline K^0   \eta  )=\frac{3}{4}\Gamma(T_{bc\bar{u}\bar{d}}^{0}\to   \pi^+   K^-   \eta  )=\frac{3}{2}\Gamma(T_{bc\bar{u}\bar{d}}^{0}\to   \pi^0   \overline K^0   \eta  ),\\
  \Gamma(T_{bc\bar{d}\bar{s}}^{+}\to   \pi^+   \pi^-   K^+  )=\frac{1}{2}\Gamma(T_{bc\bar{d}\bar{s}}^{+}\to   K^+   K^+   K^-  ),
  \Gamma(T_{bc\bar{u}\bar{s}}^{0}\to   \pi^+   \pi^-   K^0  )=\frac{1}{2}\Gamma(T_{bc\bar{u}\bar{s}}^{0}\to   K^0   K^0   \overline K^0  ),\\
  \Gamma(T_{bc\bar{u}\bar{d}}^{0}\to   K^+   \overline K^0   K^-  )=\frac{1}{2}\Gamma(T_{bc\bar{u}\bar{d}}^{0}\to   K^0   \overline K^0   \overline K^0  ),
  \Gamma(T_{bc\bar{u}\bar{s}}^{0}\to   \pi^+   \pi^-   \eta  )=2\Gamma(T_{bc\bar{u}\bar{s}}^{0}\to   \pi^0   \pi^0   \eta  ),\\
  2\Gamma(T_{bc\bar{u}\bar{s}}^{0}\to   \pi^0   \pi^0   K^0  )=\Gamma(T_{bc\bar{u}\bar{s}}^{0}\to   K^+   K^0   K^-  ),
  \Gamma(T_{bc\bar{u}\bar{d}}^{0}\to   \pi^+   \pi^-   \overline K^0  )=2\Gamma(T_{bc\bar{u}\bar{d}}^{0}\to   \pi^0   \pi^0   \overline K^0  ),\\
  \Gamma(T_{bc\bar{u}\bar{s}}^{0}\to   \pi^0   \eta   \eta  )=\frac{1}{2}\Gamma(T_{bc\bar{d}\bar{s}}^{+}\to   \pi^+   \eta   \eta  ),
  \Gamma(T_{bc\bar{d}\bar{s}}^{+}\to   \pi^0   \pi^0   K^+  )=\frac{1}{2}\Gamma(T_{bc\bar{d}\bar{s}}^{+}\to   K^+   K^0   \overline K^0  ),\\
  \Gamma(T_{bc\bar{u}\bar{d}}^{0}\to   \pi^+   \pi^0   \pi^-  )=\frac{2}{3}\Gamma(T_{bc\bar{u}\bar{d}}^{0}\to   \pi^0   \pi^0   \pi^0  ),
  \Gamma(T_{bc\bar{u}\bar{d}}^{0}\to   \pi^+   \pi^-   \eta  )=2\Gamma(T_{bc\bar{u}\bar{d}}^{0}\to   \pi^0   \pi^0   \eta  ),\\
  \Gamma(T_{bc\bar{u}\bar{d}}^{0}\to   K^+   K^-   \eta  )=\Gamma(T_{bc\bar{u}\bar{d}}^{0}\to   K^0   \overline K^0   \eta  ),\\
  \end{eqnarray*}

\section{Non-Leptonic $T_{cc\bar{q}\bar{q}}$ decays}
\begin{figure}
\begin{center}
\includegraphics[scale=0.5]{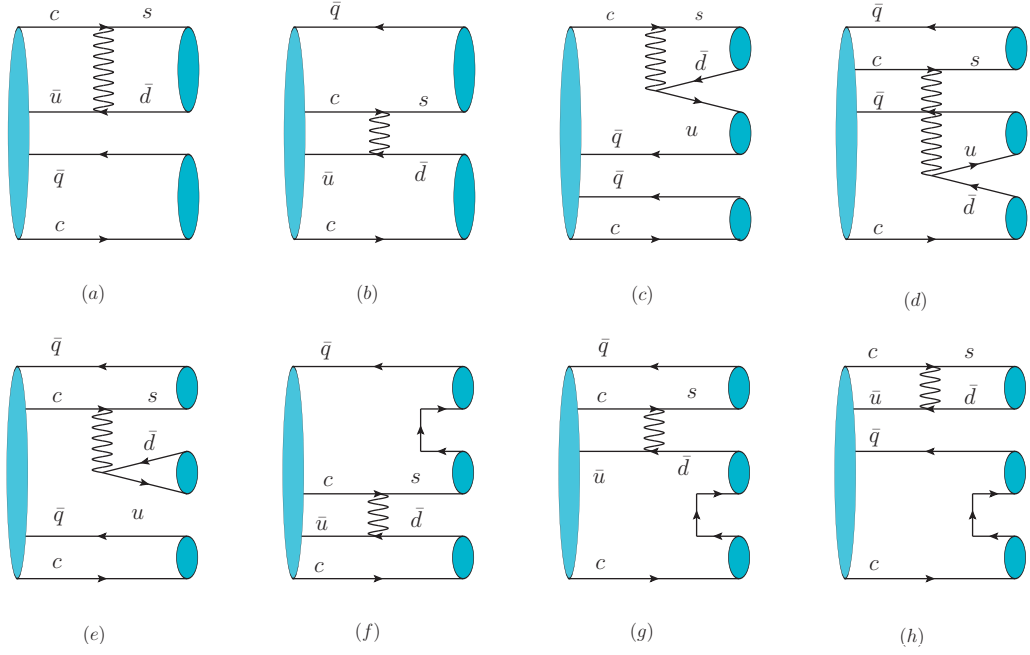}
\end{center}
\caption{Feynman diagrams for nonleptonic decays of  doubly charmed tetraquark $T_{cc\bar{q}\bar{q}}$. (a,b) are corresponding with a charmed meson and a light meson final states by W-exchange process. (c,d,e,f,g,h) are corresponding with a charmed meson and two light mesons final states process. }\label{fig:Feynman_nonleptonic_tetraquark_5}
\end{figure}

For the charm quark decays, we classified them into: Cabibbo allowed, singly Cabibbo suppressed, and doubly Cabibbo suppressed, i.e.
\begin{eqnarray}
 c\to s \bar d u,  \;\;\; c\to u \bar dd/\bar ss, \;\;\; c\to  d \bar s u.
\end{eqnarray}

For the Cabibbo allowed decays, the nonzero components are:
\begin{eqnarray}
(H_{\overline 6})^{31}_2=-(H_{\overline 6})^{13}_2=1,\;\;\;
 (H_{15})^{31}_2= (H_{15})^{13}_2=1.\label{eq:H3615_c_allowed}
\end{eqnarray}
For the singly Cabibbo-suppressed decays,  the nonzero components become:
\begin{eqnarray}
(H_{\overline 6})^{31}_3 =-(H_{\overline 6})^{13}_3 =(H_{\overline 6})^{12}_2 =-(H_{\overline 6})^{21}_2 =\sin(\theta_C),\nonumber\\
 (H_{15})^{31}_3= (H_{15})^{13}_3=-(H_{15})^{12}_2=-(H_{15})^{21}_2= \sin(\theta_C).\label{eq:H3615_cc_singly_suppressed}
\end{eqnarray}
At last, for the doubly Cabibbo suppressed decays, the nonzero formulae become:
\begin{eqnarray}
(H_{\overline 6})^{21}_3=-(H_{\overline 6})^{12}_3=\sin^2\theta_C,\;\;
 (H_{15})^{21}_3= (H_{15})^{12}_3=\sin^2\theta_C. \label{eq:H3615_c_doubly_suprressed}
\end{eqnarray}

\subsubsection{Decays into a charmed meson and a light meson by W-exchange process}
Following the above formulae,  the effective Hamiltonian for decays involving a charmed meson and a light meson is
\begin{eqnarray}
 {\cal H}_{eff}&=& f_1(T_{cc3})_{[ij]}  (\overline D)^{i}M^{k}_{l}  (H_{\overline6})^{[jl]}_{k}+f_2(T_{cc3})_{[ij]}  (\overline D)^{l}M^{k}_{l}  (H_{\overline6})^{[ij]}_{k}\nonumber\\
&&+f_3(T_{cc3})_{[ij]}  (\overline D)^{k}M^{j}_{l}  (H_{\overline6})^{[il]}_{k}+f_4(T_{cc3})_{[ij]}  (\overline D)^{i}M^{k}_{l}  (H_{15})^{\{jl\}}_{k}\nonumber\\
&&+f_5(T_{cc3})_{[ij]}  (\overline D)^{k}M^{j}_{l}  (H_{15})^{\{il\}}_{k}.
\end{eqnarray}
The corresponding Feynman diagrams are shown in Fig.~\ref{fig:Feynman_nonleptonic_tetraquark_5}(a,b). We expand the Hamiltonian to obtain the decay amplitudes which are given in Tab.~\ref{tab:cc_Dandmeson_8}.
\begin{table}
\caption{Doubly charmed tetraquark $T_{cc\bar{q}\bar{q}}$ decays into a charmed meson and a light meson. sC is the abbreviation of $sin (\theta_C)$. }\label{tab:cc_Dandmeson_8}\begin{tabular}{|cc|cc|}\hline
channel & amplitude &channel& amplitude \tabularnewline\hline\hline
$T_{cc\bar{u}\bar{s}}^{+}\to    D^0  \pi^+  $ & $ f_1-2 f_2+f_4$&$T_{cc\bar{u}\bar{s}}^{+}\to    D^0  K^+  $ & $ \left(f_1-2 f_2+f_4\right) \text{sC}$\\\hline
$T_{cc\bar{u}\bar{s}}^{+}\to    D^+  \pi^0  $ & $ \frac{2 f_2-f_3-f_5}{\sqrt{2}}$&$T_{cc\bar{u}\bar{s}}^{+}\to    D^+  K^0  $ & $ \left(-2 f_2+f_3-f_5\right) \text{sC}$\\\hline
$T_{cc\bar{u}\bar{s}}^{+}\to    D^+  \eta  $ & $ \frac{-2 f_2+f_3-3 f_5}{\sqrt{6}}$&$T_{cc\bar{u}\bar{s}}^{+}\to    D^+_s  \pi^0  $ & $ \frac{\left(f_1-f_3-f_4-f_5\right) \text{sC}}{\sqrt{2}}$\\\hline
$T_{cc\bar{u}\bar{s}}^{+}\to    D^+_s  \overline K^0  $ & $ f_1-2 f_2-f_4$&$T_{cc\bar{u}\bar{s}}^{+}\to    D^+_s  \eta  $ & $ \frac{\left(-3 f_1+4 f_2+f_3+3 f_4-3 f_5\right) \text{sC}}{\sqrt{6}}$\\\hline
$T_{cc\bar{d}\bar{s}}^{++}\to    D^+  \pi^+  $ & $ f_1-f_3+f_4-f_5$&$T_{cc\bar{d}\bar{s}}^{++}\to    D^+  K^+  $ & $ \left(f_1-f_3+f_4-f_5\right) \text{sC}$\\\hline
$T_{cc\bar{u}\bar{d}}^{+}\to    D^+  \overline K^0  $ & $ f_1-f_3-f_4+f_5$&$T_{cc\bar{d}\bar{s}}^{++}\to    D^+_s  \pi^+  $ & $ \left(f_1-f_3+f_4-f_5\right) \text{sC}$\\\hline
\hline
$T_{cc\bar{u}\bar{s}}^{+}\to    D^+_s  K^0  $ & $ \left(f_1-f_3-f_4+f_5\right) \text{sC}^2$&$T_{cc\bar{u}\bar{d}}^{+}\to    D^0  \pi^+  $ & $ \left(f_1-2 f_2+f_4\right) (-\text{sC})$\\\hline
$T_{cc\bar{d}\bar{s}}^{++}\to    D^+_s  K^+  $ & $ \left(-f_1+f_3-f_4+f_5\right) \text{sC}^2$&$T_{cc\bar{u}\bar{d}}^{+}\to    D^+  \pi^0  $ & $ \frac{\left(f_1-2 f_2-f_4+2 f_5\right) \text{sC}}{\sqrt{2}}$\\\hline
$T_{cc\bar{u}\bar{d}}^{+}\to    D^0  K^+  $ & $ \left(f_1-2 f_2+f_4\right) \text{sC}^2$&$T_{cc\bar{u}\bar{d}}^{+}\to    D^+  \eta  $ & $ \frac{\left(-3 f_1+2 f_2+2 f_3+3 f_4\right) \text{sC}}{\sqrt{6}}$\\\hline
$T_{cc\bar{u}\bar{d}}^{+}\to    D^+  K^0  $ & $ \left(f_1-2 f_2-f_4\right) \text{sC}^2$&$T_{cc\bar{u}\bar{d}}^{+}\to    D^+_s  \overline K^0  $ & $ \left(2 f_2-f_3+f_5\right) \text{sC}$\\\hline
$T_{cc\bar{u}\bar{d}}^{+}\to    D^+_s  \pi^0  $ & $ -\sqrt{2} f_5 \text{sC}^2$&&\\\hline
$T_{cc\bar{u}\bar{d}}^{+}\to    D^+_s  \eta  $ & $ \sqrt{\frac{2}{3}} \left(2 f_2-f_3\right) \text{sC}^2$&&\\\hline
\hline
\end{tabular}
\end{table}
The relations of decay widths become:
\begin{eqnarray*}
    \Gamma(T_{cc\bar{u}\bar{s}}^{+}\to  D^+K^0 )= { }\Gamma(T_{cc\bar{u}\bar{d}}^{+}\to  D^+_s\overline K^0 ), \Gamma(T_{cc\bar{d}\bar{s}}^{++}\to  D^+_s\pi^+ )= { }\Gamma(T_{cc\bar{d}\bar{s}}^{++}\to  D^+K^+ ),\\ \Gamma(T_{cc\bar{u}\bar{d}}^{+}\to  D^0\pi^+ )= { }\Gamma(T_{cc\bar{u}\bar{s}}^{+}\to  D^0K^+ ).
\end{eqnarray*}
\subsubsection{Decays into a charmed meson and two light mesons}
For decays involving a charmed meson and two light mesons, the effective Hamiltonian is
\begin{eqnarray}
 {\cal H}_{eff}&=& b_1(T_{cc3})_{[ij]}  (\overline D)^{i}M^{l}_{m}M^{m}_{k}  (H_{\overline6})^{[jk]}_{l}+  b_2(T_{cc3})_{[ij]}  (\overline D)^{m}M^{i}_{k}M^{j}_{l}  (H_{\overline6})^{[kl]}_{m} \nonumber\\
 &&+ b_3(T_{cc3})_{[ij]}  (\overline D)^{l}M^{i}_{m}M^{m}_{k}  (H_{\overline6})^{[jk]}_{l}+ b_4(T_{cc3})_{[ij]}  (\overline D)^{m}M^{i}_{k}M^{l}_{m}  (H_{\overline6})^{[jk]}_{l} \nonumber\\
 &&+ b_5(T_{cc3})_{[ij]}  (\overline D)^{m}M^{i}_{m}M^{l}_{k}  (H_{\overline6})^{[jk]}_{l}+ b_6(T_{cc3})_{[ij]}  (\overline D)^{k}M^{l}_{m}M^{m}_{l}  (H_{\overline6})^{[ij]}_{k} \nonumber\\
 &&+ b_7(T_{cc3})_{[ij]}  (\overline D)^{m}M^{k}_{l}M^{l}_{m}  (H_{\overline6})^{[ij]}_{k}+b_8(T_{cc3})_{[ij]}  (\overline D)^{i}M^{j}_{k}M^{m}_{l}  (H_{15})^{\{kl\}}_{m}\nonumber\\
 &&+ b_9(T_{cc3})_{[ij]}  (\overline D)^{i}M^{l}_{m}M^{m}_{k}  (H_{15})^{\{jk\}}_{l}+ b_{10}(T_{cc3})_{[ij]}  (\overline D)^{l}M^{i}_{m}M^{m}_{k}  (H_{15})^{\{jk\}}_{l}\nonumber\\
 &&+ b_{11}(T_{cc3})_{[ij]}  (\overline D)^{m}M^{i}_{k}M^{l}_{m}  (H_{15})^{\{jk\}}_{l} + b_{12}(T_{cc3})_{[ij]}  (\overline D)^{m}M^{i}_{m}M^{l}_{k}  (H_{15})^{\{jk\}}_{l}\nonumber\\
 && +\overline {b_1}(T_{cc3})_{[ij]}  (\overline D)^{i}M^{j}_{k}M^{m}_{l}  (H_{\overline6})^{[kl]}_{m}.
\end{eqnarray}
These related Feynman diagrams  with the three final states are shown in Fig.~\ref{fig:Feynman_nonleptonic_tetraquark_5}. For the production of  two light mesons, some terms contain one QCD coupling while the others contain two QCD couplings. We found that $b_1$ and $\overline{b_1}$ terms give the same contribution which always contain the factor $b_1-\overline{b_1}$. Then $\overline{b_1}$ terms are removed in the final results. The decay amplitudes are given in Tab.~\ref{tab:cc_Dand2meson_8_1} and Tab.~\ref{tab:cc_Dand2meson_8_2}.
\begin{table}
\footnotesize
\caption{Doubly charmed tetraquark $T_{cc\bar{q}\bar{q}}$ decays into a charmed meson and two light mesons. sC is the abbreviation of $sin (\theta_C)$. }\label{tab:cc_Dand2meson_8_1}\begin{tabular}{|cc|cc|}
\hline
channel  & amplitude  & channel  & amplitude\tabularnewline
\hline
\hline
$T_{cc\bar{u}\bar{s}}^{+}\to    D^0  \pi^+   \pi^0  $ & $ \frac{b_4+b_5+b_{11}+b_{12}}{\sqrt{2}}$&$T_{cc\bar{u}\bar{s}}^{+}\to    D^0  K^+   K^0  $ & $ \left(b_4+b_5+2 b_8-b_{11}-b_{12}\right) \text{sC}^2$\\\hline
$T_{cc\bar{u}\bar{s}}^{+}\to    D^0  \pi^+   \eta  $ & $ \frac{2 b_1-b_4+b_5-4 b_7-2 b_8+2 b_9+3 b_{11}+b_{12}}{\sqrt{6}}$&$T_{cc\bar{u}\bar{s}}^{+}\to    D^+  K^0   K^0  $ & $ 2 \left(b_4+b_5-b_{11}-b_{12}\right) \text{sC}^2$\\\hline
$T_{cc\bar{u}\bar{s}}^{+}\to    D^0  K^+   \overline K^0  $ & $ b_1+b_5-2 b_7+b_8+b_9-b_{12}$&$T_{cc\bar{u}\bar{s}}^{+}\to    D^+_s  \pi^0   K^0  $ & $ -\frac{\left(b_1+2 b_2+b_3+b_8-b_9-b_{10}\right) \text{sC}^2}{\sqrt{2}}$\\\hline
$T_{cc\bar{u}\bar{s}}^{+}\to    D^+  \pi^+   \pi^-  $ & $ b_3+b_5-4 b_6-2 b_7+b_{10}+b_{12}$&$T_{cc\bar{u}\bar{s}}^{+}\to    D^+_s  \pi^-   K^+  $ & $ \left(b_1+2 b_2+b_3-b_8-b_9-b_{10}\right) \text{sC}^2$\\\hline
$T_{cc\bar{u}\bar{s}}^{+}\to    D^+  \pi^0   \pi^0  $ & $ b_3-b_4-4 b_6-2 b_7+b_{10}-b_{11}$&$T_{cc\bar{u}\bar{s}}^{+}\to    D^+_s  K^0   \eta  $ & $ -\frac{\left(b_1+2 b_2+b_3+2 b_4+2 b_5+b_8-b_9-b_{10}-2 b_{11}-2 b_{12}\right) \text{sC}^2}{\sqrt{6}}$\\\hline
$T_{cc\bar{u}\bar{s}}^{+}\to    D^+  \pi^0   \eta  $ & $ \frac{2 b_2+b_3+b_4+2 b_7+b_{10}-b_{11}}{\sqrt{3}}$&$T_{cc\bar{d}\bar{s}}^{++}\to    D^0  K^+   K^+  $ & $ -2 \left(b_4+b_5+b_{11}+b_{12}\right) \text{sC}^2$\\\hline
$T_{cc\bar{u}\bar{s}}^{+}\to    D^+  K^+   K^-  $ & $ 2 \left(b_2+b_3-2 b_6\right)$&$T_{cc\bar{d}\bar{s}}^{++}\to    D^+  K^+   K^0  $ & $ \left(b_4+b_5-2 b_8+b_{11}+b_{12}\right) \left(-\text{sC}^2\right)$\\\hline
$T_{cc\bar{u}\bar{s}}^{+}\to    D^+  K^0   \overline K^0  $ & $ b_3+b_5-4 b_6-2 b_7-b_{10}-b_{12}$&$T_{cc\bar{d}\bar{s}}^{++}\to    D^+_s  \pi^+   K^0  $ & $ \left(b_1+2 b_2+b_3+b_8+b_9+b_{10}\right) \left(-\text{sC}^2\right)$\\\hline
$T_{cc\bar{u}\bar{s}}^{+}\to    D^+  \eta   \eta  $ & $ \frac{4 b_2+5 b_3-b_4-12 b_6-2 b_7-3 b_{10}+3 b_{11}}{3}$&$T_{cc\bar{d}\bar{s}}^{++}\to    D^+_s  \pi^0   K^+  $ & $ -\frac{\left(b_1+2 b_2+b_3-b_8+b_9+b_{10}\right) \text{sC}^2}{\sqrt{2}}$\\\hline
$T_{cc\bar{u}\bar{s}}^{+}\to    D^+_s  \pi^+   K^-  $ & $ b_1+b_5-2 b_7-b_8-b_9+b_{12}$&$T_{cc\bar{d}\bar{s}}^{++}\to    D^+_s  K^+   \eta  $ & $ \frac{\left(b_1+2 b_2+b_3+2 b_4+2 b_5-b_8+b_9+b_{10}+2 b_{11}+2 b_{12}\right) \text{sC}^2}{\sqrt{6}}$\\\hline
$T_{cc\bar{u}\bar{s}}^{+}\to    D^+_s  \pi^0   \overline K^0  $ & $ \frac{-b_1+b_4+2 b_7-b_8+b_9+b_{11}}{\sqrt{2}}$&$T_{cc\bar{u}\bar{d}}^{+}\to    D^0  \pi^+   K^0  $ & $ \left(b_1+b_5-2 b_7+b_8+b_9-b_{12}\right) \text{sC}^2$\\\hline
$T_{cc\bar{u}\bar{s}}^{+}\to    D^+_s  \overline K^0   \eta  $ & $ -\frac{b_1+b_4+2 b_5-2 b_7+b_8-b_9-3 b_{11}-2 b_{12}}{\sqrt{6}}$&$T_{cc\bar{u}\bar{d}}^{+}\to    D^0  \pi^0   K^+  $ & $ \frac{\left(b_1+b_5-2 b_7-b_8+b_9+2 b_{11}+b_{12}\right) \text{sC}^2}{\sqrt{2}}$\\\hline
$T_{cc\bar{d}\bar{s}}^{++}\to    D^0  \pi^+   \pi^+  $ & $ 2 \left(b_4+b_5+b_{11}+b_{12}\right)$&$T_{cc\bar{u}\bar{d}}^{+}\to    D^0  K^+   \eta  $ & $ -\frac{\left(b_1-2 b_4-b_5-2 b_7-b_8+b_9-b_{12}\right) \text{sC}^2}{\sqrt{6}}$\\\hline
$T_{cc\bar{d}\bar{s}}^{++}\to    D^+  \pi^+   \pi^0  $ & $ -\frac{b_4+b_5+b_{11}+b_{12}}{\sqrt{2}}$&$T_{cc\bar{u}\bar{d}}^{+}\to    D^+  \pi^0   K^0  $ & $ -\frac{\left(b_1+b_5-2 b_7+b_8-b_9-2 b_{11}-b_{12}\right) \text{sC}^2}{\sqrt{2}}$\\\hline
$T_{cc\bar{d}\bar{s}}^{++}\to    D^+  \pi^+   \eta  $ & $ \frac{2 b_1+4 b_2+2 b_3+b_4+b_5-2 b_8+2 b_9+2 b_{10}+b_{11}+b_{12}}{\sqrt{6}}$&$T_{cc\bar{u}\bar{d}}^{+}\to    D^+  \pi^-   K^+  $ & $ \left(b_1+b_5-2 b_7-b_8-b_9+b_{12}\right) \text{sC}^2$\\\hline
$T_{cc\bar{d}\bar{s}}^{++}\to    D^+  K^+   \overline K^0  $ & $ b_1+2 b_2+b_3+b_8+b_9+b_{10}$&$T_{cc\bar{u}\bar{d}}^{+}\to    D^+  K^0   \eta  $ & $ -\frac{\left(b_1-2 b_4-b_5-2 b_7+b_8-b_9+b_{12}\right) \text{sC}^2}{\sqrt{6}}$\\\hline
$T_{cc\bar{d}\bar{s}}^{++}\to    D^+_s  \pi^+   \overline K^0  $ & $ b_4+b_5-2 b_8+b_{11}+b_{12}$&$T_{cc\bar{u}\bar{d}}^{+}\to    D^+_s  \pi^+   \pi^-  $ & $ 2 \left(b_2+b_3-2 b_6\right) \text{sC}^2$\\\hline
$T_{cc\bar{u}\bar{d}}^{+}\to    D^0  \pi^+   \overline K^0  $ & $ b_4+b_5+2 b_8-b_{11}-b_{12}$&$T_{cc\bar{u}\bar{d}}^{+}\to    D^+_s  \pi^0   \pi^0  $ & $ 2 \left(b_2+b_3-2 b_6\right) \text{sC}^2$\\\hline
$T_{cc\bar{u}\bar{d}}^{+}\to    D^+  \pi^+   K^-  $ & $ b_1+2 b_2+b_3-b_8-b_9-b_{10}$&$T_{cc\bar{u}\bar{d}}^{+}\to    D^+_s  \pi^0   \eta  $ & $ \frac{2 \left(b_{10}-b_{11}\right) \text{sC}^2}{\sqrt{3}}$\\\hline
$T_{cc\bar{u}\bar{d}}^{+}\to    D^+  \pi^0   \overline K^0  $ & $ -\frac{b_1+2 b_2+b_3+b_4+b_5+b_8-b_9-b_{10}-b_{11}-b_{12}}{\sqrt{2}}$&$T_{cc\bar{u}\bar{d}}^{+}\to    D^+_s  K^+   K^-  $ & $ \left(b_3+b_5-4 b_6-2 b_7+b_{10}+b_{12}\right) \text{sC}^2$\\\hline
$T_{cc\bar{u}\bar{d}}^{+}\to    D^+  \overline K^0   \eta  $ & $ -\frac{b_1+2 b_2+b_3-b_4-b_5+b_8-b_9-b_{10}+b_{11}+b_{12}}{\sqrt{6}}$&$T_{cc\bar{u}\bar{d}}^{+}\to    D^+_s  K^0   \overline K^0  $ & $ \left(b_3+b_5-4 b_6-2 b_7-b_{10}-b_{12}\right) \text{sC}^2$\\\hline
$T_{cc\bar{u}\bar{d}}^{+}\to    D^+_s  \overline K^0   \overline K^0  $ & $ 2 \left(b_4+b_5-b_{11}-b_{12}\right)$&$T_{cc\bar{u}\bar{d}}^{+}\to    D^+_s  \eta   \eta  $ & $ -\frac{2}{3} \left(b_2-b_3+2 b_4+6 b_6+4 b_7\right) \text{sC}^2$\\\hline
\hline
\end{tabular}
\end{table}
\begin{table}
\caption{Doubly charmed tetraquark $T_{cc\bar{q}\bar{q}}$ decays into a charmed meson and two light mesons. sC is the abbreviation of $sin (\theta_C)$. }\label{tab:cc_Dand2meson_8_2}\begin{tabular}{|c|c|c|c|c|c|c|c}\hline\hline
channel & amplitude \\\hline
\hline$T_{cc\bar{u}\bar{s}}^{+}\to    D^0  \pi^+   K^0  $ & $ \left(b_1-b_4-2 b_7-b_8+b_9+b_{11}\right) \text{sC}$\\\hline
$T_{cc\bar{u}\bar{s}}^{+}\to    D^0  \pi^0   K^+  $ & $ \frac{\left(b_1+b_4+2 b_5-2 b_7+b_8+b_9+b_{11}\right) \text{sC}}{\sqrt{2}}$\\\hline
$T_{cc\bar{u}\bar{s}}^{+}\to    D^0  K^+   \eta  $ & $ -\frac{\left(b_1+b_4+2 b_5-2 b_7+5 b_8+b_9-3 b_{11}-4 b_{12}\right) \text{sC}}{\sqrt{6}}$\\\hline
$T_{cc\bar{u}\bar{s}}^{+}\to    D^+  \pi^0   K^0  $ & $ \frac{\left(2 b_2+b_3+2 b_4+b_5+2 b_7-b_{10}-b_{12}\right) \text{sC}}{\sqrt{2}}$\\\hline
$T_{cc\bar{u}\bar{s}}^{+}\to    D^+  \pi^-   K^+  $ & $ \left(2 b_2+b_3-b_5+2 b_7-b_{10}-b_{12}\right) (-\text{sC})$\\\hline
$T_{cc\bar{u}\bar{s}}^{+}\to    D^+  K^0   \eta  $ & $ \frac{\left(2 b_2+b_3-2 b_4-3 b_5+2 b_7-b_{10}+4 b_{11}+3 b_{12}\right) \text{sC}}{\sqrt{6}}$\\\hline
$T_{cc\bar{u}\bar{s}}^{+}\to    D^+_s  \pi^+   \pi^-  $ & $ \left(-b_1+b_3-4 b_6+b_8+b_9+b_{10}\right) \text{sC}$\\\hline
$T_{cc\bar{u}\bar{s}}^{+}\to    D^+_s  \pi^0   \pi^0  $ & $ \left(b_1-b_3+4 b_6+b_8-b_9-b_{10}\right) (-\text{sC})$\\\hline
$T_{cc\bar{u}\bar{s}}^{+}\to    D^+_s  \pi^0   \eta  $ & $ \frac{\left(b_1+2 b_2+b_3-b_4-b_5+b_8-b_9+b_{10}-b_{11}+b_{12}\right) \text{sC}}{\sqrt{3}}$\\\hline
$T_{cc\bar{u}\bar{s}}^{+}\to    D^+_s  K^+   K^-  $ & $ \left(b_1+2 b_2+2 b_3+b_5-4 b_6-2 b_7-b_8-b_9+b_{12}\right) \text{sC}$\\\hline
$T_{cc\bar{u}\bar{s}}^{+}\to    D^+_s  K^0   \overline K^0  $ & $ \left(b_3-b_4-4 b_6-2 b_7-b_{10}+b_{11}\right) \text{sC}$\\\hline
$T_{cc\bar{u}\bar{s}}^{+}\to    D^+_s  \eta   \eta  $ & $ \frac{1}{3} \left(3 b_1+4 b_2+5 b_3+2 b_4+6 b_5-12 b_6-8 b_7+3 b_8-3 b_9-3 b_{10}-6 b_{11}-6 b_{12}\right) \text{sC}$\\\hline
$T_{cc\bar{d}\bar{s}}^{++}\to    D^0  \pi^+   K^+  $ & $ 2 \left(b_4+b_5+b_{11}+b_{12}\right) \text{sC}$\\\hline
$T_{cc\bar{d}\bar{s}}^{++}\to    D^+  \pi^+   K^0  $ & $ \left(b_1+2 b_2+b_3+b_4+b_5-b_8+b_9+b_{10}+b_{11}+b_{12}\right) \text{sC}$\\\hline
$T_{cc\bar{d}\bar{s}}^{++}\to    D^+  \pi^0   K^+  $ & $ \frac{\left(b_1+2 b_2+b_3-b_4-b_5+b_8+b_9+b_{10}-b_{11}-b_{12}\right) \text{sC}}{\sqrt{2}}$\\\hline
$T_{cc\bar{d}\bar{s}}^{++}\to    D^+  K^+   \eta  $ & $ -\frac{\left(b_1+2 b_2+b_3-b_4-b_5+5 b_8+b_9+b_{10}-b_{11}-b_{12}\right) \text{sC}}{\sqrt{6}}$\\\hline
$T_{cc\bar{d}\bar{s}}^{++}\to    D^+_s  \pi^+   \pi^0  $ & $ -\sqrt{2} b_8 \text{sC}$\\\hline
$T_{cc\bar{d}\bar{s}}^{++}\to    D^+_s  \pi^+   \eta  $ & $ \sqrt{\frac{2}{3}} \left(b_1+2 b_2+b_3-b_4-b_5+2 b_8+b_9+b_{10}-b_{11}-b_{12}\right) \text{sC}$\\\hline
$T_{cc\bar{d}\bar{s}}^{++}\to    D^+_s  K^+   \overline K^0  $ & $ \left(b_1+2 b_2+b_3+b_4+b_5-b_8+b_9+b_{10}+b_{11}+b_{12}\right) \text{sC}$\\\hline
$T_{cc\bar{u}\bar{d}}^{+}\to    D^0  \pi^+   \pi^0  $ & $ \sqrt{2} \left(b_8-b_{11}-b_{12}\right) \text{sC}$\\\hline
$T_{cc\bar{u}\bar{d}}^{+}\to    D^0  \pi^+   \eta  $ & $ -\sqrt{\frac{2}{3}} \left(b_1+b_4+2 b_5-2 b_7+2 b_8+b_9-b_{12}\right) \text{sC}$\\\hline
$T_{cc\bar{u}\bar{d}}^{+}\to    D^0  K^+   \overline K^0  $ & $ \left(b_1-b_4-2 b_7-b_8+b_9+b_{11}\right) (-\text{sC})$\\\hline
$T_{cc\bar{u}\bar{d}}^{+}\to    D^+  \pi^+   \pi^-  $ & $ \left(b_1+2 b_2+2 b_3+b_5-4 b_6-2 b_7-b_8-b_9+b_{12}\right) (-\text{sC})$\\\hline
$T_{cc\bar{u}\bar{d}}^{+}\to    D^+  \pi^0   \pi^0  $ & $ \left(b_1+2 b_2+2 b_3+b_5-4 b_6-2 b_7+b_8-b_9-2 b_{11}-b_{12}\right) (-\text{sC})$\\\hline
$T_{cc\bar{u}\bar{d}}^{+}\to    D^+  \pi^0   \eta  $ & $ \frac{\left(b_1+b_4+2 b_5-2 b_7+b_8-b_9-2 b_{10}-b_{11}-2 b_{12}\right) \text{sC}}{\sqrt{3}}$\\\hline
$T_{cc\bar{u}\bar{d}}^{+}\to    D^+  K^+   K^-  $ & $ \left(b_1-b_3+4 b_6-b_8-b_9-b_{10}\right) \text{sC}$\\\hline
$T_{cc\bar{u}\bar{d}}^{+}\to    D^+  K^0   \overline K^0  $ & $ \left(b_3-b_4-4 b_6-2 b_7-b_{10}+b_{11}\right) (-\text{sC})$\\\hline
$T_{cc\bar{u}\bar{d}}^{+}\to    D^+  \eta   \eta  $ & $ \frac{1}{3} \left(3 b_1+2 b_2-2 b_3-2 b_4-3 b_5+12 b_6+2 b_7+3 b_8-3 b_9+3 b_{12}\right) \text{sC}$\\\hline
$T_{cc\bar{u}\bar{d}}^{+}\to    D^+_s  \pi^+   K^-  $ & $ \left(2 b_2+b_3-b_5+2 b_7-b_{10}-b_{12}\right) \text{sC}$\\\hline
$T_{cc\bar{u}\bar{d}}^{+}\to    D^+_s  \pi^0   \overline K^0  $ & $ -\frac{\left(2 b_2+b_3-b_5+2 b_7-b_{10}+2 b_{11}+b_{12}\right) \text{sC}}{\sqrt{2}}$\\\hline
$T_{cc\bar{u}\bar{d}}^{+}\to    D^+_s  \overline K^0   \eta  $ & $ -\frac{\left(2 b_2+b_3+4 b_4+3 b_5+2 b_7-b_{10}-2 b_{11}-3 b_{12}\right) \text{sC}}{\sqrt{6}}$\\\hline
\hline
\end{tabular}
\end{table}
Based on the expanded amplitudes, the  relations become:
\begin{eqnarray*}
    \Gamma(T_{cc\bar{u}\bar{s}}^{+}\to  D^0 \pi^+ K^0 )= { }\Gamma(T_{cc\bar{u}\bar{d}}^{+}\to  D^0 K^+ \overline K^0 ),
    \Gamma(T_{cc\bar{u}\bar{s}}^{+}\to  D^+ \pi^- K^+ )= { }\Gamma(T_{cc\bar{u}\bar{d}}^{+}\to  D^+_s \pi^+ K^- ),\\
    \Gamma(T_{cc\bar{u}\bar{s}}^{+}\to  D^+_s \pi^+ \pi^- )= { }\Gamma(T_{cc\bar{u}\bar{d}}^{+}\to  D^+ K^+ K^- ) ,
     \Gamma(T_{cc\bar{d}\bar{s}}^{++}\to  D^+ \pi^+ K^0 )= { }\Gamma(T_{cc\bar{d}\bar{s}}^{++}\to  D^+_s K^+ \overline K^0 ),\\ \Gamma(T_{cc\bar{u}\bar{d}}^{+}\to  D^+ \pi^+ \pi^- )= { }\Gamma(T_{cc\bar{u}\bar{s}}^{+}\to  D^+_s K^+ K^- ) ,
     \Gamma(T_{cc\bar{u}\bar{d}}^{+}\to  D^+_s \pi^0 \pi^0 )= \frac{1}{2}\Gamma(T_{cc\bar{u}\bar{d}}^{+}\to  D^+_s \pi^+ \pi^- ),\\
        \Gamma(T_{cc\bar{d}\bar{s}}^{++}\to  D^0 \pi^+ \pi^+ )= 4\Gamma(T_{cc\bar{u}\bar{s}}^{+}\to  D^0 \pi^+ \pi^0 ) =4\Gamma(T_{cc\bar{d}\bar{s}}^{++}\to  D^+ \pi^+ \pi^0 ),\\
            \Gamma(T_{cc\bar{u}\bar{d}}^{+}\to  D^+ K^0 \overline K^0 )= { }\Gamma(T_{cc\bar{u}\bar{s}}^{+}\to  D^+_s K^0 \overline K^0 ).
\end{eqnarray*}

\section{Golden Decay Channels}
\label{sec:golden_channels}

In order to hunt for the doubly heavy tetraquarks,
 the golden decay channels are very useful. So we  list them and give an estimation of the decay branching fractions in this section.
In the following lists, a hadron  is a general particle and can be replaced by the states with the identical quark contents. The light pseudoscalar   may replace by the light vector meson such as replacing the $\overline K^0$ by $\overline K^{*0}$ which decays into  $K^-\pi^+$.  The modes involving the neutral mesons $\pi^0, \eta, \rho^0, \omega$ are removed because a neutral meson is  difficult to be reconstructed at hadron-hadron colliders.

\subsection{$T_{cc\bar{q}\bar{q}}$ }

For the $T_{cc\bar{q}\bar{q}}$  decays, we collected  Cabibbo allowed decays in Tab.~\ref{tab:Tccqq_golden}. From the
data of the D meson decays, we conclude that these Cabibbo allowed decay channels in Tab.~\ref{tab:Tccqq_golden} may lead to  branching
fractions at a few percent level. Note that to reconstruct the final charm meson, another factor of $10^{-3}$ is required.


\begin{table}
 \caption{Cabibbo allowed $T_{cc\bar{q}\bar{q}}$ decays. $\bar{K}^0$ can be replaced by vector meson $\bar{K}^{*0}$.  }\label{tab:Tccqq_golden}\begin{tabular}{|c   c  c  c c|}\hline\hline

 \multicolumn{5}{|l|}{\qquad \textbf{Two body decays}}\\\hline
$T_{cc\bar{u}\bar{s}}^{+}\to  D^0 l^+\nu $ &
$T_{cc\bar{u}\bar{s}}^{+}\to    D^0  \pi^+  $&
$T_{cc\bar{u}\bar{s}}^{+}\to    D^+_s  \overline K^0  $& &\\\hline
$T_{cc\bar{d}\bar{s}}^{++}\to  D^+ l^+\nu $ &
$T_{cc\bar{d}\bar{s}}^{++}\to    D^+  \pi^+  $ &&&\\\hline
$T_{cc\bar{u}\bar{d}}^{+}\to    D^+  \overline K^0  $ &&&&\\\hline
 \hline
 \multicolumn{5}{|l|}{\qquad \textbf{Three body decays}}\\\hline
$T_{cc\bar{u}\bar{s}}^{+}\to  D^+  \pi^-  l^+\nu $ 
&$T_{cc\bar{u}\bar{s}}^{+}\to  D^+_s  K^-  l^+\nu $ 
&$T_{cc\bar{u}\bar{s}}^{+}\to    D^0  K^+   \overline K^0  $ 
&$T_{cc\bar{u}\bar{s}}^{+}\to    D^+  \pi^+   \pi^-  $ 
&$T_{cc\bar{u}\bar{s}}^{+}\to    D^+  K^+   K^-  $ 
\\
$T_{cc\bar{u}\bar{s}}^{+}\to    D^+  K^0   \overline K^0  $ 
&$T_{cc\bar{u}\bar{s}}^{+}\to    D^+_s  \pi^+   K^-  $ 
&&&\\\hline
$T_{cc\bar{d}\bar{s}}^{++}\to  D^0  \pi^+  l^+\nu $ 
&$T_{cc\bar{d}\bar{s}}^{++}\to  D^+_s  \overline K^0  l^+\nu $ 
&$T_{cc\bar{d}\bar{s}}^{++}\to    D^0  \pi^+   \pi^+  $ 
&$T_{cc\bar{d}\bar{s}}^{++}\to    D^+  K^+   \overline K^0  $ 
&$T_{cc\bar{d}\bar{s}}^{++}\to    D^+_s  \pi^+   \overline K^0  $ 
\\\hline
$T_{cc\bar{u}\bar{d}}^{+}\to  D^0  \overline K^0  l^+\nu $ 
&$T_{cc\bar{u}\bar{d}}^{+}\to  D^+  K^-  l^+\nu $ 
&$T_{cc\bar{u}\bar{d}}^{+}\to    D^0  \pi^+   \overline K^0  $ 
&$T_{cc\bar{u}\bar{d}}^{+}\to    D^+  \pi^+   K^-  $ 
&$T_{cc\bar{u}\bar{d}}^{+}\to    D^+_s  \overline K^0   \overline K^0  $
\\\hline
\end{tabular}
\end{table}
\begin{table}
 \caption{Cabibbo allowed $T_{bc\bar{q}\bar{q}}$ decays. $\bar{K}^0$ can be replaced by vector meson $\bar{K}^{*0}$.  }\label{tab:Tbcqq_golden}\begin{tabular}{|c  c   c  c c|}\hline\hline
& \multicolumn{3}{c}{ \textbf{Two body decays (charm decays)}}  & \\\hline
$T_{bc\bar{u}\bar{s}}^{0}\to B^- l^+\nu $&
$T_{bc\bar{u}\bar{s}}^{0}\to    B^-  \pi^+  $&
$T_{bc\bar{u}\bar{s}}^{0}\to    \overline B^0_s  \overline K^0  $& &\\
\hline
$T_{bc\bar{d}\bar{s}}^{+}\to \overline B^0 l^+\nu $&
$T_{bc\bar{d}\bar{s}}^{+}\to    \overline B^0  \pi^+  $ &&&\\\hline
$T_{bc\bar{u}\bar{d}}^{0}\to    \overline B^0  \overline K^0  $ &&&&\\\hline
\hline
& \multicolumn{3}{c}{ \textbf{Three body decays (charm decays)}}  & \\\hline
$T_{bc\bar{u}\bar{s}}^{0}\to \overline B^0  \pi^-  l^+\nu $ 
&$T_{bc\bar{u}\bar{s}}^{0}\to \overline B^0_s  K^-  l^+\nu $ 
&$T_{bc\bar{u}\bar{s}}^{0}\to    B^-  K^+   \overline K^0  $ 
&$T_{bc\bar{u}\bar{s}}^{0}\to    \overline B^0  \pi^+   \pi^-  $ 
&$T_{bc\bar{u}\bar{s}}^{0}\to    \overline B^0  K^+   K^-  $ 
\\
$T_{bc\bar{u}\bar{s}}^{0}\to    \overline B^0  K^0   \overline K^0  $ 
&$T_{bc\bar{u}\bar{s}}^{0}\to    \overline B^0_s  \pi^+   K^-  $ 
&&&\\\hline

$T_{bc\bar{d}\bar{s}}^{+}\to B^-  \pi^+  l^+\nu $ 
&$T_{bc\bar{d}\bar{s}}^{+}\to \overline B^0_s  \overline K^0  l^+\nu $ 
&$T_{bc\bar{d}\bar{s}}^{+}\to    B^-  \pi^+   \pi^+  $ 
&$T_{bc\bar{d}\bar{s}}^{+}\to    \overline B^0 K^+   \overline K^0  $ 
&$T_{bc\bar{d}\bar{s}}^{+}\to    \overline B^0_s  \pi^+   \overline K^0  $ 
\\\hline
$T_{bc\bar{u}\bar{d}}^{0}\to \overline B^0  K^-  l^+\nu $ 
&$T_{bc\bar{u}\bar{d}}^{0}\to B^-  \overline K^0  l^+\nu $ 
&$T_{bc\bar{u}\bar{d}}^{0}\to    B^-  \pi^+   \overline K^0  $ 
&$T_{bc\bar{u}\bar{d}}^{0}\to    \overline B^0  \pi^+   K^-  $ 
&$T_{bc\bar{u}\bar{d}}^{0}\to    \overline B^0_s  \overline K^0   \overline K^0  $
\\\hline\hline

& \multicolumn{3}{c}{\textbf{Two body decays  (bottom decays)}}  & \\\hline
 $T_{bc\bar{u}\bar{s}}^{0}\to      D^0  J/\psi $ &
 $T_{bc\bar{u}\bar{s}}^{0}\to   \overline D^0   D^0 $ &
 $T_{bc\bar{u}\bar{s}}^{0}\to    D^+  D^- $ &
 $T_{bc\bar{u}\bar{s}}^{0}\to    D^+_s   D^-_s $& 
 $T_{bc\bar{u}\bar{s}}^{0}\to   \Xi_{cc}^{+}  \Lambda_{\bar{c}}^- $ 
  \\
 $T_{bc\bar{u}\bar{s}}^{0}\to   \Lambda_{\bar{c}}^-  \Lambda_c^+ $ &
  $T_{bc\bar{u}\bar{s}}^{0}\to   \overline \Xi_{\bar{c}}^{\prime0}  \Xi_{c}^{\prime0} $& 
 $T_{bc\bar{u}\bar{s}}^{0}\to   \Xi_{cc}^{+}  \Sigma_{\bar{c}}^{-} $& 
 $T_{bc\bar{u}\bar{s}}^{0}\to   \Omega_{cc}^{+}  \Xi_{\bar{c}}^- $ &
 $T_{bc\bar{u}\bar{s}}^{0}\to   \Omega_{cc}^{+}  \Xi_{\bar{c}}^{\prime-} $ 
 \\
 $T_{bc\bar{u}\bar{s}}^{0}\to   \Xi_{cc}^{++}  \Sigma_{\bar{c}}^{--} $ &
 $T_{bc\bar{u}\bar{s}}^{0}\to   \overline \Sigma^- \Xi_{cc}^{+} $ &
 $T_{bc\bar{u}\bar{s}}^{0}\to   \Lambda_{\bar{c}}^-  \Sigma_{c}^{+} $& 
 $T_{bc\bar{u}\bar{s}}^{0}\to   \Xi_{\bar{c}}^-  \Xi_c^+ $ &
 $T_{bc\bar{u}\bar{s}}^{0}\to   \Xi_{\bar{c}}^-  \Xi_{c}^{\prime+} $ 
 \\
 $T_{bc\bar{u}\bar{s}}^{0}\to   \overline \Xi_{\bar{c}}^0  \Xi_c^0 $ &
 $T_{bc\bar{u}\bar{s}}^{0}\to   \overline \Xi_{\bar{c}}^0  \Xi_{c}^{\prime0} $ &
 $T_{bc\bar{u}\bar{s}}^{0}\to   \Sigma_{\bar{c}}^{-}  \Lambda_c^+ $ &
 $T_{bc\bar{u}\bar{s}}^{0}\to   \Sigma_{\bar{c}}^{--}  \Sigma_{c}^{++} $& 
 $T_{bc\bar{u}\bar{s}}^{0}\to   \Sigma_{\bar{c}}^{-}  \Sigma_{c}^{+} $ \\
 $T_{bc\bar{u}\bar{s}}^{0}\to   \Xi_{\bar{c}}^{\prime-}  \Xi_c^+ $ &
 $T_{bc\bar{u}\bar{s}}^{0}\to   \overline \Sigma_{\bar{c}}^{0}  \Sigma_{c}^{-} $ &
 $T_{bc\bar{u}\bar{s}}^{0}\to   \overline \Xi_{\bar{c}}^{\prime0}  \Xi_c^0 $ &
 $T_{bc\bar{u}\bar{s}}^{0}\to   \Xi_{\bar{c}}^{\prime-}  \Xi_{c}^{\prime+} $&
 $T_{bc\bar{u}\bar{s}}^{0}\to   \overline \Omega_{\bar{c}}^{0}  \Omega_{c}^{0} $ 
\\\hline

 $T_{bc\bar{d}\bar{s}}^{+}\to   D^+  J/\psi $ &
 $T_{bc\bar{d}\bar{s}}^{+}\to    D^+_s   D^0 $ &
 $T_{bc\bar{d}\bar{s}}^{+}\to   D^+  \overline D^0 $ &
 $T_{bc\bar{d}\bar{s}}^{+}\to   J/\psi  \pi^+  $ &
 $T_{bc\bar{d}\bar{s}}^{+}\to   \overline \Lambda^0  \Xi_{cc}^{+} $ 
 \\
  $T_{bc\bar{d}\bar{s}}^{+}\to   \Omega_{cc}^{+}  \overline \Xi_{\bar{c}}^{\prime0} $ &
  $T_{bc\bar{d}\bar{s}}^{+}\to   \overline \Xi_{\bar{c}}^{\prime0}  \Xi_c^+ $ &
 $T_{bc\bar{d}\bar{s}}^{+}\to   \overline \Sigma^- \Xi_{cc}^{++} $ &
 $T_{bc\bar{d}\bar{s}}^{+}\to   \overline \Sigma^{\prime-}  \Xi_{cc}^{++} $ &
 $T_{bc\bar{d}\bar{s}}^{+}\to   \Omega_{cc}^{+}  \overline \Xi_{\bar{c}}^0 $ 
 \\
 $T_{bc\bar{d}\bar{s}}^{+}\to   \Xi_{cc}^{+}  \overline \Sigma_{\bar{c}}^{0} $ &
 $T_{bc\bar{d}\bar{s}}^{+}\to   \Xi_{cc}^{++}  \Lambda_{\bar{c}}^- $ &
 $T_{bc\bar{d}\bar{s}}^{+}\to   \Xi_{cc}^{++}  \Sigma_{\bar{c}}^{-} $ &
 $T_{bc\bar{d}\bar{s}}^{+}\to   \overline \Sigma^0  \Xi_{cc}^{+} $ &
 $T_{bc\bar{d}\bar{s}}^{+}\to   \overline \Sigma^{\prime0}  \Xi_{cc}^{+} $
 \\
 $T_{bc\bar{d}\bar{s}}^{+}\to   \overline \Xi^{\prime0}  \Omega_{cc}^{+} $ &
 $T_{bc\bar{d}\bar{s}}^{+}\to   \overline \Xi^0  \Omega_{cc}^{+} $ &
 $T_{bc\bar{d}\bar{s}}^{+}\to   \overline \Xi_{\bar{c}}^0  \Xi_c^+ $& 
 $T_{bc\bar{d}\bar{s}}^{+}\to   \Lambda_{\bar{c}}^-  \Sigma_{c}^{++} $& 
 $T_{bc\bar{d}\bar{s}}^{+}\to   \overline \Xi_{\bar{c}}^0  \Xi_{c}^{\prime+} $ \\
 $T_{bc\bar{d}\bar{s}}^{+}\to   \Sigma_{\bar{c}}^{-}  \Sigma_{c}^{++} $ &
 $T_{bc\bar{d}\bar{s}}^{+}\to   \overline \Sigma_{\bar{c}}^{0}  \Sigma_{c}^{+} $& 
 $T_{bc\bar{d}\bar{s}}^{+}\to   \overline \Xi_{\bar{c}}^{\prime0}  \Xi_{c}^{\prime+} $& 
 $T_{bc\bar{d}\bar{s}}^{+}\to   \overline \Sigma_{\bar{c}}^{0}  \Lambda_c^+ $& \\
\hline

  $T_{bc\bar{u}\bar{d}}^{0}\to   D^+   D^-_s $ &
  $T_{bc\bar{u}\bar{d}}^{0}\to   J/\psi  \overline K^0  $ &
  $T_{bc\bar{u}\bar{d}}^{0}\to    D^0   D^0 $ &
  $T_{bc\bar{u}\bar{d}}^{0}\to   \overline \Delta^{--}  \Xi_{cc}^{++} $& 
  $T_{bc\bar{u}\bar{d}}^{0}\to   \overline \Delta^{-}  \Xi_{cc}^{+} $ 
  \\
  $T_{bc\bar{u}\bar{d}}^{0}\to   \overline \Sigma^- \Omega_{cc}^{+} $ &
  $T_{bc\bar{u}\bar{d}}^{0}\to   \overline \Sigma^{\prime-}  \Omega_{cc}^{+} $ &
  $T_{bc\bar{u}\bar{d}}^{0}\to   \overline p  \Xi_{cc}^{+} $& 
  $T_{bc\bar{u}\bar{d}}^{0}\to   \Omega_{cc}^{+}  \Lambda_{\bar{c}}^- $ &
  $T_{bc\bar{u}\bar{d}}^{0}\to   \Lambda_{\bar{c}}^-  \Xi_{c}^{\prime+} $ 
  \\
  $T_{bc\bar{u}\bar{d}}^{0}\to   \overline \Xi_{\bar{c}}^0  \Omega_{c}^{0} $& 
  $T_{bc\bar{u}\bar{d}}^{0}\to   \Sigma_{\bar{c}}^{-}  \Xi_c^+ $ &
  $T_{bc\bar{u}\bar{d}}^{0}\to   \overline \Sigma_{\bar{c}}^{0}  \Xi_c^0 $& 
  $T_{bc\bar{u}\bar{d}}^{0}\to   \Sigma_{\bar{c}}^{-}  \Xi_{c}^{\prime+} $ &
  $T_{bc\bar{u}\bar{d}}^{0}\to   \overline \Sigma_{\bar{c}}^{0}  \Xi_{c}^{\prime0} $ 
  \\
  $T_{bc\bar{u}\bar{d}}^{0}\to   \overline \Xi_{\bar{c}}^{\prime0}  \Omega_{c}^{0} $ &&&&
\\\hline

\hline\hline
& \multicolumn{3}{c}{\textbf{Three body decays (bottom decays)} }  & \\\hline
$T_{bc\bar{u}\bar{s}}^{0}\to   D^+  \pi^-   J/\psi $ 
&$T_{bc\bar{u}\bar{s}}^{0}\to   D^+_s  K^-   J/\psi $ 
&$T_{bc\bar{u}\bar{s}}^{0}\to    D^0  \overline D^0  D^0 $ 
&$T_{bc\bar{u}\bar{s}}^{0}\to    D^0  D^- D^+ $ 
&$T_{bc\bar{u}\bar{s}}^{0}\to   D^0   D^0  K^0  $ 
\\
$T_{bc\bar{u}\bar{s}}^{0}\to   D^0   D^+_s  \pi^-  $ 
&$T_{bc\bar{u}\bar{s}}^{0}\to   D^0  D^-  \pi^+  $ 
&$T_{bc\bar{u}\bar{s}}^{0}\to    D^0   D^-_s  D^+_s $ 
&$T_{bc\bar{u}\bar{s}}^{0}\to   D^0   D^-_s  K^+  $ 
&$T_{bc\bar{u}\bar{s}}^{0}\to   D^+  \overline D^0  \pi^-  $ 
\\
$T_{bc\bar{u}\bar{s}}^{0}\to   D^+   D^-_s  K^0  $ 
&$T_{bc\bar{u}\bar{s}}^{0}\to   D^+_s  \overline D^0  K^-  $ 
&$T_{bc\bar{u}\bar{s}}^{0}\to   D^+_s  D^-  \overline K^0  $ 
&$T_{bc\bar{u}\bar{s}}^{0}\to   J/\psi  \pi^+   \pi^-  $ 
&$T_{bc\bar{u}\bar{s}}^{0}\to   J/\psi  K^0   \overline K^0  $ 
\\
$T_{bc\bar{u}\bar{s}}^{0}\to  J/\psi  K^+   K^-  $ 
&$T_{bc\bar{u}\bar{s}}^{0}\to    D^0  \pi^+   \pi^-  $ &
$T_{bc\bar{u}\bar{s}}^{0}\to    D^0  K^+   K^-  $ & 
$T_{bc\bar{u}\bar{s}}^{0}\to    D^0  K^0   \overline K^0  $ & 
$T_{bc\bar{u}\bar{s}}^{0}\to    D^+  K^0   K^-  $ 
\\
$T_{bc\bar{u}\bar{s}}^{0}\to    D^+_s  \pi^-   \overline K^0  $ &
$T_{bc\bar{u}\bar{s}}^{0}\to   \pi^+   K^0   K^-  $ &
$T_{bc\bar{u}\bar{s}}^{0}\to   \pi^-   K^+   \overline K^0  $ &&
\\\hline
$T_{bc\bar{d}\bar{s}}^{+}\to   D^0  \pi^+   J/\psi $ 
&$T_{bc\bar{d}\bar{s}}^{+}\to   D^+_s  \overline K^0   J/\psi $
&$T_{bc\bar{d}\bar{s}}^{+}\to    D^0  \overline D^0  D^+ $ 
&$T_{bc\bar{d}\bar{s}}^{+}\to    D^+  D^-  D^+ $ 
&$T_{bc\bar{d}\bar{s}}^{+}\to   D^0   D^0  K^+  $ 
\\
$T_{bc\bar{d}\bar{s}}^{+}\to   D^0   D^+  K^0  $ 
&$T_{bc\bar{d}\bar{s}}^{+}\to   D^+   D^+_s  \pi^-  $ 
&$T_{bc\bar{d}\bar{s}}^{+}\to    D^+   D^-_s  D^+_s $ 
&$T_{bc\bar{d}\bar{s}}^{+}\to   D^+_s   D^+_s  K^-  $ 
&$T_{bc\bar{d}\bar{s}}^{+}\to   D^0  \overline D^0  \pi^+  $ 
\\
$T_{bc\bar{d}\bar{s}}^{+}\to   D^+  D^-  \pi^+  $ 
&$T_{bc\bar{d}\bar{s}}^{+}\to   D^+   D^-_s  K^+  $ 
&$T_{bc\bar{d}\bar{s}}^{+}\to   D^+_s  \overline D^0  \overline K^0  $ 
&$T_{bc\bar{d}\bar{s}}^{+}\to   D^+_s   D^-_s  \pi^+  $ 
&$T_{bc\bar{d}\bar{s}}^{+}\to   J/\psi  K^+   \overline K^0  $ 
\\
$T_{bc\bar{d}\bar{s}}^{+}\to    D^0  K^+   \overline K^0  $ & 
$T_{bc\bar{d}\bar{s}}^{+}\to    D^+  \pi^+   \pi^-  $ & 
$T_{bc\bar{d}\bar{s}}^{+}\to    D^+  K^+   K^-  $ &
$T_{bc\bar{d}\bar{s}}^{+}\to    D^+  K^0   \overline K^0  $ &
$T_{bc\bar{d}\bar{s}}^{+}\to    D^+_s  \pi^+   K^-  $ 
\\
$T_{bc\bar{d}\bar{s}}^{+}\to   \pi^+   \pi^+   \pi^-  $ &
$T_{bc\bar{d}\bar{s}}^{+}\to   \pi^+   K^+   K^-  $ &
$T_{bc\bar{d}\bar{s}}^{+}\to   \pi^+   K^0   \overline K^0  $ & &
\\\hline

$T_{bc\bar{u}\bar{d}}^{0}\to   D^0  \overline K^0   J/\psi $ 
&$T_{bc\bar{u}\bar{d}}^{0}\to   D^+  K^-   J/\psi $ 
&$T_{bc\bar{u}\bar{d}}^{0}\to    D^0   D^-_s  D^+ $ 
&$T_{bc\bar{u}\bar{d}}^{0}\to   D^0   D^+  \pi^-  $ 
&$T_{bc\bar{u}\bar{d}}^{0}\to   D^0   D^+_s  K^-  $ 
\\
$T_{bc\bar{u}\bar{d}}^{0}\to   D^0  \overline D^0  \overline K^0  $ 
&$T_{bc\bar{u}\bar{d}}^{0}\to   D^+  \overline D^0  K^-  $ 
&$T_{bc\bar{u}\bar{d}}^{0}\to   D^+  D^-  \overline K^0  $ 
&$T_{bc\bar{u}\bar{d}}^{0}\to   D^0   D^-_s  \pi^+  $ 
&$T_{bc\bar{u}\bar{d}}^{0}\to   D^+_s   D^-_s  \overline K^0  $ 
\\
$T_{bc\bar{u}\bar{d}}^{0}\to   J/\psi  \pi^+   K^-  $ 
&$T_{bc\bar{u}\bar{d}}^{0}\to    D^0  \pi^+   K^-  $  
&$T_{bc\bar{u}\bar{d}}^{0}\to    D^+  \pi^-   \overline K^0  $ &
$T_{bc\bar{u}\bar{d}}^{0}\to   \pi^+   \pi^-   \overline K^0  $ & 
$T_{bc\bar{u}\bar{d}}^{0}\to   K^+   \overline K^0   K^-  $  
\\
$T_{bc\bar{u}\bar{d}}^{0}\to   K^0   \overline K^0   \overline K^0  $ &&&&
\\\hline
\end{tabular}
\end{table}
\begin{table}
 \caption{Cabibbo allowed $T_{bb\bar{q}\bar{q}}$ decays. $\bar{K}^0$ can be replaced by vector meson $\bar{K}^{*0}$.  }\label{tab:Tbbqq_golden}\begin{tabular}{|c  c   c  c c|}\hline\hline
 \multicolumn{5}{|l|}{\qquad \textbf{Two body decays}}\\\hline
 $T_{bb\bar{u}\bar{s}}^{-}\to   B^-  J/\psi $ &
 $T_{bb\bar{u}\bar{s}}^{-}\to    D^0  B_c^- $ &
 $T_{bb\bar{u}\bar{s}}^{-}\to   B^-  \overline D^0 $ &
 $T_{bb\bar{u}\bar{s}}^{-}\to   \overline B^0  D^- $ &
 $T_{bb\bar{u}\bar{s}}^{-}\to   \overline B^0_s   D^-_s $\\ 
 $T_{bb\bar{u}\bar{s}}^{-}\to   \Xi_{bc}^{0}  \Lambda_{\bar{c}}^- $ &
 $T_{bb\bar{u}\bar{s}}^{-}\to   \Lambda_{\bar{c}}^-  \Lambda_b^0 $ &
 $T_{bb\bar{u}\bar{s}}^{-}\to   \Xi_{bc}^{0}  \Sigma_{\bar{c}}^{-} $& 
 $T_{bb\bar{u}\bar{s}}^{-}\to   \Omega_{bc}^{0}  \Xi_{\bar{c}}^- $ &
 $T_{bb\bar{u}\bar{s}}^{-}\to   \Omega_{bc}^{0}  \Xi_{\bar{c}}^{\prime-} $ \\
 $T_{bb\bar{u}\bar{s}}^{-}\to   \Xi_{bc}^{+}  \Sigma_{\bar{c}}^{--} $ &
 $T_{bb\bar{u}\bar{s}}^{-}\to   \overline \Sigma^- \Xi_{bc}^{0} $ &
 $T_{bb\bar{u}\bar{s}}^{-}\to   \Lambda_{\bar{c}}^-  \Sigma_{b}^{0} $& 
 $T_{bb\bar{u}\bar{s}}^{-}\to   \Xi_{\bar{c}}^-  \Xi_b^0 $ &
 $T_{bb\bar{u}\bar{s}}^{-}\to   \Xi_{\bar{c}}^-  \Xi_{b}^{\prime0} $ \\
 $T_{bb\bar{u}\bar{s}}^{-}\to   \overline \Xi_{\bar{c}}^0  \Xi_b^- $ &
 $T_{bb\bar{u}\bar{s}}^{-}\to   \overline \Xi_{\bar{c}}^0  \Xi_{b}^{\prime-} $ &
 $T_{bb\bar{u}\bar{s}}^{-}\to   \Sigma_{\bar{c}}^{-}  \Lambda_b^0 $ &
 $T_{bb\bar{u}\bar{s}}^{-}\to   \Sigma_{\bar{c}}^{--}  \Sigma_{b}^{+} $& 
 $T_{bb\bar{u}\bar{s}}^{-}\to   \Sigma_{\bar{c}}^{-}  \Sigma_{b}^{0} $ \\
 $T_{bb\bar{u}\bar{s}}^{-}\to   \Xi_{\bar{c}}^{\prime-}  \Xi_b^0 $ &
 $T_{bb\bar{u}\bar{s}}^{-}\to   \overline \Sigma_{\bar{c}}^{0}  \Sigma_{b}^{-} $ &
 $T_{bb\bar{u}\bar{s}}^{-}\to   \overline \Xi_{\bar{c}}^{\prime0}  \Xi_b^- $ &
 $T_{bb\bar{u}\bar{s}}^{-}\to   \Xi_{\bar{c}}^{\prime-}  \Xi_{b}^{\prime0} $&
 $T_{bb\bar{u}\bar{s}}^{-}\to   \overline \Xi_{\bar{c}}^{\prime0}  \Xi_{b}^{\prime-} $\\ 
 $T_{bb\bar{u}\bar{s}}^{-}\to   \overline \Omega_{\bar{c}}^{0}  \Omega_{b}^{-} $ &
& &&\\\hline

 $T_{bb\bar{d}\bar{s}}^{0}\to    D^+  B_c^- $ &
 $T_{bb\bar{d}\bar{s}}^{0}\to   \overline B^0  J/\psi $ &
 $T_{bb\bar{d}\bar{s}}^{0}\to   B^-   D^+_s $ &
 $T_{bb\bar{d}\bar{s}}^{0}\to   \overline B^0_s   D^0 $ &
 $T_{bb\bar{d}\bar{s}}^{0}\to   \overline B^0  \overline D^0 $ \\
 $T_{bb\bar{d}\bar{s}}^{0}\to   B_c^-  \pi^+  $ &
 $T_{bb\bar{d}\bar{s}}^{0}\to   \overline \Lambda^0  \Xi_{bc}^{0} $ &
 $T_{bb\bar{d}\bar{s}}^{0}\to   \overline \Sigma^- \Xi_{bc}^{+} $ &
 $T_{bb\bar{d}\bar{s}}^{0}\to   \overline \Sigma^{\prime-}  \Xi_{bc}^{+} $ &
 $T_{bb\bar{d}\bar{s}}^{0}\to   \Omega_{bc}^{0}  \overline \Xi_{\bar{c}}^0 $ \\
 $T_{bb\bar{d}\bar{s}}^{0}\to   \Xi_{bc}^{0}  \overline \Sigma_{\bar{c}}^{0} $ &
 $T_{bb\bar{d}\bar{s}}^{0}\to   \Xi_{bc}^{+}  \Lambda_{\bar{c}}^- $ &
 $T_{bb\bar{d}\bar{s}}^{0}\to   \Xi_{bc}^{+}  \Sigma_{\bar{c}}^{-} $ &

 $T_{bb\bar{d}\bar{s}}^{0}\to   \overline \Sigma^0  \Xi_{bc}^{0} $ &
 $T_{bb\bar{d}\bar{s}}^{0}\to   \overline \Sigma^{\prime0}  \Xi_{bc}^{0} $\\ 
 $T_{bb\bar{d}\bar{s}}^{0}\to   \overline \Xi^{\prime0}  \Omega_{bc}^{0} $ &
 $T_{bb\bar{d}\bar{s}}^{0}\to   \overline \Xi^0  \Omega_{bc}^{0} $ &
 $T_{bb\bar{d}\bar{s}}^{0}\to   \overline \Xi_{\bar{c}}^0  \Xi_b^0 $& 
 $T_{bb\bar{d}\bar{s}}^{0}\to   \Lambda_{\bar{c}}^-  \Sigma_{b}^{+} $& 
 $T_{bb\bar{d}\bar{s}}^{0}\to   \overline \Xi_{\bar{c}}^0  \Xi_{b}^{\prime0} $ \\
 $T_{bb\bar{d}\bar{s}}^{0}\to   \Sigma_{\bar{c}}^{-}  \Sigma_{b}^{+} $ &
 $T_{bb\bar{d}\bar{s}}^{0}\to   \overline \Sigma_{\bar{c}}^{0}  \Sigma_{b}^{0} $& 
 $T_{bb\bar{d}\bar{s}}^{0}\to   \overline \Xi_{\bar{c}}^{\prime0}  \Xi_{b}^{\prime0} $& 
 $T_{bb\bar{d}\bar{s}}^{0}\to   \overline \Sigma_{\bar{c}}^{0}  \Lambda_b^0 $ &
 $T_{bb\bar{d}\bar{s}}^{0}\to   \overline \Xi_{\bar{c}}^{\prime0}  \Xi_b^0 $ \\
 $T_{bb\bar{d}\bar{s}}^{0}\to   \Omega_{bc}^{0}  \overline \Xi_{\bar{c}}^{\prime0} $ &&&&

\\\hline

  $T_{bb\bar{u}\bar{d}}^{-}\to   \overline B^0   D^-_s $ &
  $T_{bb\bar{u}\bar{d}}^{-}\to   B_c^-  \overline K^0  $ &
  $T_{bb\bar{u}\bar{d}}^{-}\to   B^-   D^0 $ &
  $T_{bb\bar{u}\bar{d}}^{-}\to   \overline \Delta^{--}  \Xi_{bc}^{+} $& 
  $T_{bb\bar{u}\bar{d}}^{-}\to   \overline \Delta^{-}  \Xi_{bc}^{0} $ \\
  $T_{bb\bar{u}\bar{d}}^{-}\to   \overline \Sigma^- \Omega_{bc}^{0} $ &
  $T_{bb\bar{u}\bar{d}}^{-}\to   \overline \Sigma^{\prime-}  \Omega_{bc}^{0} $ &
  $T_{bb\bar{u}\bar{d}}^{-}\to   \overline p  \Xi_{bc}^{0} $& 
  $T_{bb\bar{u}\bar{d}}^{-}\to   \Omega_{bc}^{0}  \Lambda_{\bar{c}}^- $ &
  $T_{bb\bar{u}\bar{d}}^{-}\to   \Lambda_{\bar{c}}^-  \Xi_{b}^{\prime0} $\\ 
  $T_{bb\bar{u}\bar{d}}^{-}\to   \overline \Xi_{\bar{c}}^0  \Omega_{b}^{-} $& 
  $T_{bb\bar{u}\bar{d}}^{-}\to   \Sigma_{\bar{c}}^{-}  \Xi_b^0 $ &
  $T_{bb\bar{u}\bar{d}}^{-}\to   \overline \Sigma_{\bar{c}}^{0}  \Xi_b^- $& 
  $T_{bb\bar{u}\bar{d}}^{-}\to   \Sigma_{\bar{c}}^{-}  \Xi_{b}^{\prime0} $ &
  $T_{bb\bar{u}\bar{d}}^{-}\to   \overline \Sigma_{\bar{c}}^{0}  \Xi_{b}^{\prime-} $\\
  $T_{bb\bar{u}\bar{d}}^{-}\to   \overline \Xi_{\bar{c}}^{\prime0}  \Omega_{b}^{-} $ &&&&
\\\hline
 \hline
 \multicolumn{5}{|l|}{\qquad \textbf{Three body decays}}\\\hline
$T_{bb\bar{u}\bar{s}}^{-}\to   \overline B^0  \pi^-   J/\psi $ 
&$T_{bb\bar{u}\bar{s}}^{-}\to   \overline B^0_s  K^-   J/\psi $ 
&$T_{bb\bar{u}\bar{s}}^{-}\to    D^+  \pi^-   B_c^- $ 
&$T_{bb\bar{u}\bar{s}}^{-}\to    D^+_s  K^-   B_c^- $ 
&$T_{bb\bar{u}\bar{s}}^{-}\to   B^-   D^0  K^0  $ 
\\
$T_{bb\bar{u}\bar{s}}^{-}\to   B^-   D^+_s  \pi^-  $ 
&$T_{bb\bar{u}\bar{s}}^{-}\to   \overline B^0_s   D^0  \pi^-  $ 
&$T_{bb\bar{u}\bar{s}}^{-}\to   B^-  D^-  \pi^+  $ 
&$T_{bb\bar{u}\bar{s}}^{-}\to   B^-   D^-_s  K^+  $ 
&$T_{bb\bar{u}\bar{s}}^{-}\to   \overline B^0  \overline D^0  \pi^-  $ 
\\
$T_{bb\bar{u}\bar{s}}^{-}\to   \overline B^0   D^-_s  K^0  $ 
&$T_{bb\bar{u}\bar{s}}^{-}\to   \overline B^0_s  \overline D^0  K^-  $ 
&$T_{bb\bar{u}\bar{s}}^{-}\to   \overline B^0_s  D^-  \overline K^0  $ 
&$T_{bb\bar{u}\bar{s}}^{-}\to   B_c^-  \pi^+   \pi^-  $ 
&$T_{bb\bar{u}\bar{s}}^{-}\to   B_c^-  K^0   \overline K^0  $ 
\\
$T_{bb\bar{u}\bar{s}}^{-}\to   B_c^-  K^+   K^-  $ 
&$T_{bb\bar{u}\bar{s}}^{-}\to    D^0  \overline D^0  B^- $ 
&$T_{bb\bar{u}\bar{s}}^{-}\to    D^0  D^-  \overline B^0 $ 
&$T_{bb\bar{u}\bar{s}}^{-}\to    D^0   D^-_s  \overline B^0_s $ 
&$T_{bb\bar{u}\bar{s}}^{-}\to    D^+  D^-  B^- $ 
\\
$T_{bb\bar{u}\bar{s}}^{-}\to    D^+_s   D^-_s  B^- $ &&&&
\\\hline
$T_{bb\bar{d}\bar{s}}^{0}\to   B^-  \pi^+   J/\psi $ 
&$T_{bb\bar{d}\bar{s}}^{0}\to   \overline B^0_s  \overline K^0   J/\psi $
&$T_{bb\bar{d}\bar{s}}^{0}\to    D^0  \pi^+   B_c^- $ 
&$T_{bb\bar{d}\bar{s}}^{0}\to    D^+_s  \overline K^0   B_c^- $
&$T_{bb\bar{d}\bar{s}}^{0}\to   B^-   D^0  K^+  $ 
\\
$T_{bb\bar{d}\bar{s}}^{0}\to   B^-   D^+  K^0  $ 
&$T_{bb\bar{d}\bar{s}}^{0}\to   \overline B^0   D^0  K^0  $
&$T_{bb\bar{d}\bar{s}}^{0}\to   \overline B^0   D^+_s  \pi^-  $ 
&$T_{bb\bar{d}\bar{s}}^{0}\to   \overline B^0_s   D^+  \pi^-  $ 
&$T_{bb\bar{d}\bar{s}}^{0}\to   \overline B^0_s   D^+_s  K^-  $ 
\\
$T_{bb\bar{d}\bar{s}}^{0}\to   B^-  \overline D^0  \pi^+  $ 
&$T_{bb\bar{d}\bar{s}}^{0}\to   \overline B^0  D^-  \pi^+  $ 
&$T_{bb\bar{d}\bar{s}}^{0}\to   \overline B^0   D^-_s  K^+  $ 
&$T_{bb\bar{d}\bar{s}}^{0}\to   \overline B^0_s  \overline D^0  \overline K^0  $ 
&$T_{bb\bar{d}\bar{s}}^{0}\to   \overline B^0_s   D^-_s  \pi^+  $ 
\\
$T_{bb\bar{d}\bar{s}}^{0}\to   B_c^-  K^+   \overline K^0  $ 
&$T_{bb\bar{d}\bar{s}}^{0}\to    D^0  \overline D^0  \overline B^0 $ 
&$T_{bb\bar{d}\bar{s}}^{0}\to    D^+  \overline D^0  B^- $ 
&$T_{bb\bar{d}\bar{s}}^{0}\to    D^+  D^-  \overline B^0 $ 
&$T_{bb\bar{d}\bar{s}}^{0}\to    D^+   D^-_s  \overline B^0_s $ 
\\
$T_{bb\bar{d}\bar{s}}^{0}\to    D^+_s   D^-_s  \overline B^0 $ &&&&
\\\hline

$T_{bb\bar{u}\bar{d}}^{-}\to   B^-  \overline K^0   J/\psi $ 
&$T_{bb\bar{u}\bar{d}}^{-}\to   \overline B^0  K^-   J/\psi $ 
&$T_{bb\bar{u}\bar{d}}^{-}\to    D^0  \overline K^0   B_c^- $ 
&$T_{bb\bar{u}\bar{d}}^{-}\to    D^+  K^-   B_c^- $ 
&$T_{bb\bar{u}\bar{d}}^{-}\to   B^-   D^+  \pi^-  $ 
\\
$T_{bb\bar{u}\bar{d}}^{-}\to   B^-   D^+_s  K^-  $ 
&$T_{bb\bar{u}\bar{d}}^{-}\to   \overline B^0   D^0  \pi^-  $ 
&$T_{bb\bar{u}\bar{d}}^{-}\to   \overline B^0_s   D^0  K^-  $ 
&$T_{bb\bar{u}\bar{d}}^{-}\to   B^-  \overline D^0  \overline K^0  $ 
&$T_{bb\bar{u}\bar{d}}^{-}\to   \overline B^0  \overline D^0  K^-  $ 
\\
$T_{bb\bar{u}\bar{d}}^{-}\to   \overline B^0  D^-  \overline K^0  $ 
&$T_{bb\bar{u}\bar{d}}^{-}\to   B^-   D^-_s  \pi^+  $ 
&$T_{bb\bar{u}\bar{d}}^{-}\to   \overline B^0_s   D^-_s  \overline K^0  $ 
&$T_{bb\bar{u}\bar{d}}^{-}\to   B_c^-  \pi^+   K^-  $ 
&$T_{bb\bar{u}\bar{d}}^{-}\to    D^0   D^-_s  \overline B^0 $ 
\\
$T_{bb\bar{u}\bar{d}}^{-}\to    D^+   D^-_s  B^- $ &&&& 
\\\hline

\end{tabular}
\end{table}

\subsection{$T_{bc\bar{q}\bar{q}}$}

To reconstruct the $T_{bc\bar{q}\bar{q}}$  tetraquark,  lists of possible modes are given in Tab.~\ref{tab:Tbcqq_golden}. The decay width is dominant by the charm quark decay from the estimation of the magnitudes of CKM matrix elements. For the charm quark decay, the typical branching fractions in Tab.~\ref{tab:Tbcqq_golden} are estimated to be  a few percents.
If  the bottom quark decays, the branching fraction might be smaller than $10^{-3}$. Note that to reconstruct the final charm or bottom meson, another factor of $10^{-3}$ is also required.

\subsection{$T_{bb\bar{q}\bar{q}}$}

To reconstruct the $T_{bb\bar{q}\bar{q}}$, we listed the gold decay channels in Tab.~\ref{tab:Tbbqq_golden}. The  branching fractions are estimated  at the order $10^{-3}$. In these decay channels, the reconstruction of the final bottom meson or bottom baryon state requires another factor of $10^{-3}$. The reconstruction of $J/\psi$ or $D$ or charmed baryons,  the corresponding factor becomes to be $10^{-2}$. Thus to reconstruct the doubly bottom tetraquark, the channel $T_{bb\bar{u}\bar{d}}^{-}\to   \overline p  \Xi_{bc}^{0}(\to p\Sigma+ B^-)$ is a wonderful tool and whose branching fraction is the order of $10^{-6}$.

\section{Conclusions}
\label{sec:conclusions}
Most theoretical works including the Lattice QCD simulations supported the possibility of the stable doubly heavy tetraquarks.
In this work, we have given the spectra of the doubly heavy tetraquarks by Sakharov-Zeldovich formula. We found that $T^{-}_{bb\bar{u}\bar{d}}({\bf  3})$ is abount 73MeV below the $BB^*$ threshold. In order to hunting for these stable doubly heavy tetraquarks, we investigated systematically the semileptonic and nonleptonic weak decay amplitudes of the stable doubly heavy tetraquarks under the flavor SU(3) symmetry which is a powerful tool to analyze the general decay properties.  The ratios between decay widths of different channels were also given.   We have given the Cabibbo allowed two-body and three-body decay channels of the stable doubly heavy tetraquarks, which have large branching ratios and shall be employed  as the ``discovery channels" in the reconstructions at future LHCb and Bell II experiments.

\section*{Acknowledgments}

We thank Prof. Wei Wang for useful discussions  and the collaboration at the early stage of this work.
 This work was
supported in part by the National Natural Science Foundation of
China under Grant No.~11575110, 11655002, 11705092, by Natural
Science Foundation of Shanghai under Grant No.~15DZ2272100 and
No.~15ZR1423100,  by Natural Science Foundation of Jiangsu under
Grant No.~BK20171471, by the Young Thousand Talents Plan,   by Key
Laboratory for Particle Physics, Astrophysics and Cosmology,
Ministry of Education.

\end{document}